# MODERN
# INFORMATION TECHNOLOGIES
## IN SCIENTIFIC RESEARCH AND EDUCATIONAL ACTIVITIES

Sergii Kotlyk
*Editor-in-chief*

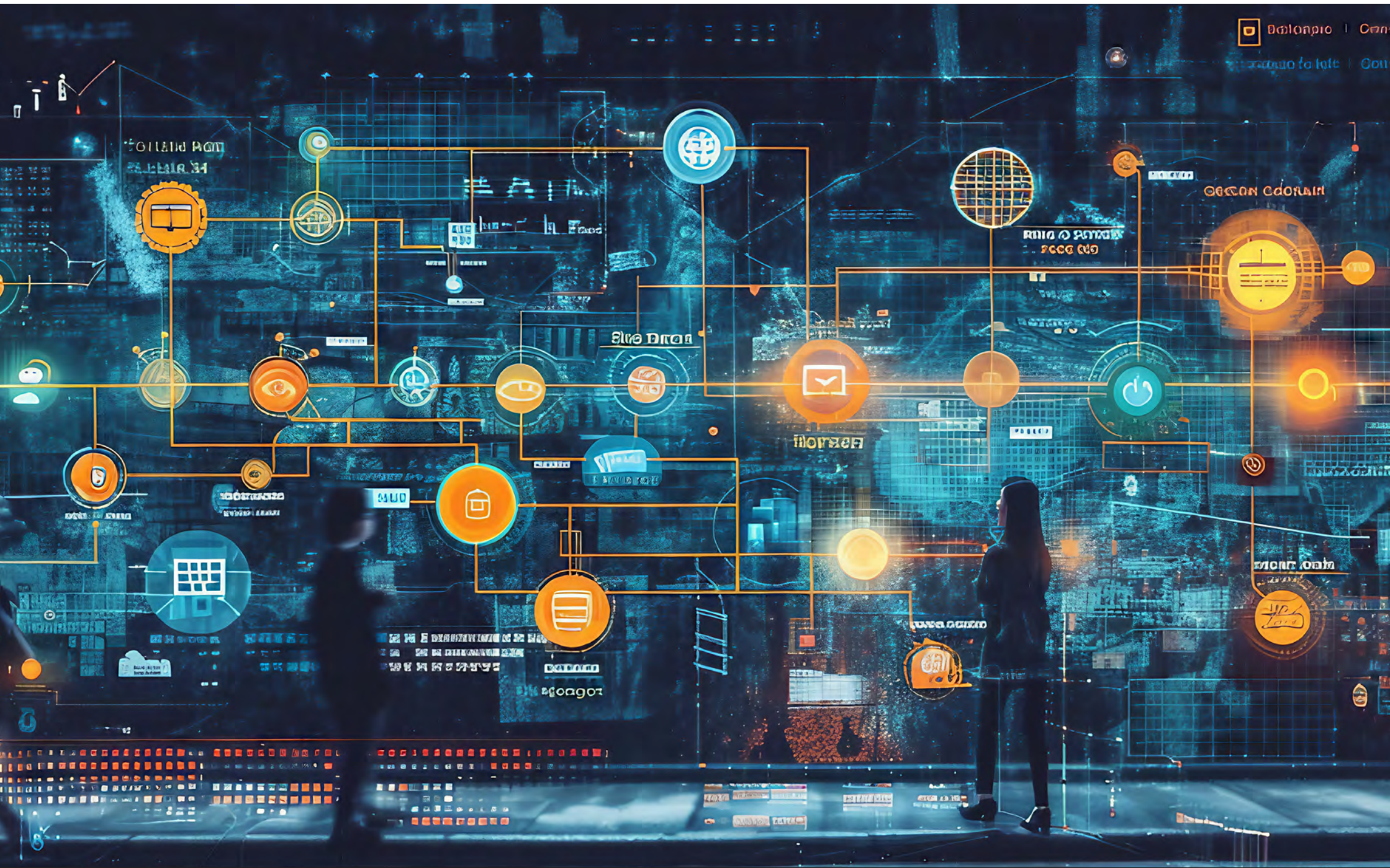

MINISTRY OF EDUCATION AND SCIENCE OF UKRAINE

Odesa National University of Technology

# MODERN INFORMATION TECHNOLOGIES IN SCIENTIFIC RESEARCH AND EDUCATIONAL ACTIVITIES

Monograph
Scientific publication (issue)

Editor-in-Chief
**Sergii Kotlyk**

Team of Authors: ***Liliia Ivanova***, ***Vladislav Kaverinskiy***, ***Sergii Kotlyk***, ***Kyrylo Malakhov***, ***Oleksandr Romanyuk***, ***Oksana Romaniuk***, ***Oksana Sokolova***, ***Svitlana Voinova***

The monograph summarizes and analyzes the current state of scientific research in the field of interactive artificial intelligence systems, text generation systems, diagnostics of the competitiveness of specialists, in the areas of correct color rendering in image formation, informatization of the work of graduate students, accessible technology for creating 3D models.

The monograph will be useful both to specialists and employees of companies working in the IT field, as well as teachers, masters, students and graduate students of higher educational institutions, as well as anyone interested in issues related to information technology.

The monograph was compiled based on the results of the XVI international scientific and practical conference "Information technologies and automation — 2023", which took place in October 2023 at Odessa National University of Technology.

# Content

# Preface

Nowadays, there is a rapid development of information technology, which entails the need to constantly improve and expand the capabilities of interactive artificial intelligence systems. This monograph combines several current topics related to the field of information technology.

One of the key topics is the methodology for enhancing the capabilities of conversational systems, with a focus on ChatGPT, which represents the latest advance in the field of artificial intelligence. The monograph also discusses text generation systems based on ontological representations, which open up wide opportunities for creating high-quality content.

A special place in the work is given to an automated computer system for diagnosing the competitiveness of specialists in the field of information technology. This helps to effectively assess the professionalism of specialists and determine the need for advanced training.

Theoretical aspects of correct color rendering and informatization of educational and research work of graduate students are important in ensuring the quality of education and scientific research.

And finally, the use of technology for creating 3D models has becomean integral part of the modern information environment, whichmakes it possible to bring the most daring ideas and projects to life.

Research and development in these areas contribute to the improvement of information technologies, finding application in various fields of activity. The purpose of our monograph is to conduct analysis and research in these areas in order to promote the development of information technologies and increase their efficiency.

The monograph was compiled based on the results of the XVI international scientific and practical conference “Information technologies and automation — 2023”, which took place in October 2023 at Odessa National University of Technology.

# 1. EXPLORING RESEARCH-RELATED DESIGN: A COMPREHENSIVE INFORMATION SYSTEM FOR KNOWLEDGE PRODUCTION — ONTOCHATGPT

***Kyrylo Malakhov***

*This article presents a comprehensive methodology for enhancing the capabilities of conversational AI systems, focusing on ChatGPT, through the integration of ontology-driven structured prompts and meta-learning techniques. The proposed approach, exemplified by the RDWE OntoChatGPT system, demonstrates significant potential for improving productivity and expanding knowledge bases in specific subject domains and languages. By formalizing methods for creating and configuring structured prompts and context ontologies, the system enables ChatGPT to provide specific information using selected contexts, thereby enhancing its performance in handling data across diverse linguistic and domain contexts. Through empirical testing, the study evaluates the system's effectiveness in providing accurate responses while also highlighting areas for further refinement. The methodology's adaptability to other conversational AI systems based on large language models underscores its potential for broader application in natural language processing and dialogue systems. Future enhancements aim to refine structured prompts and incorporate additional input data to further improve system performance and accuracy. Overall, this research contributes to advancing the capabilities of conversational AI systems, paving the way for more sophisticated and effective human-computer interactions in various domains.*

*To Mark, Liudmyla, and my Mother Tetiana*

## INTRODUCTION

In the modern technological world, the ability to integrate cutting-edge scientific research and technological innovations becomes critical for advancing intellectual achievements in the field of Research and Development (R&D). The sphere of research and development is transforming under the influence of intensive technological progress, requiring deeper integration and convergence across different disciplines.

This section focuses on the development of complex information systems for knowledge production in the Research and Development Workstation Environment (RDWE) class [1], aimed at optimizing scientific research processes. This includes a detailed analysis of existing systems and technologies, the development of new methodologies and techniques, and experimental validation of proposals.

The current context of research development demands a transdisciplinary approach and deep intellectualization, achievable through integrated knowledge systems such as RDWE systems. Supporting methodological processes and knowledge convergence becomes a crucial factor in the development of information systems for scientific research.

Investigating current trends and standards present in the scientific community leads to the need for complex information systems, such as RDWE systems — OntoChatGPT [2], capable of embodying progressive technological solutions and supporting scientific research at all stages of its lifecycle.

Such systems aim to address key challenges associated with creating an efficient, flexible, and intelligent environment for scientific research, fostering the integration, convergence, and unification of transdisciplinary knowledge and operations.

This section presents detailed research, analyses, and conclusions regarding the development, implementation, and optimization of the RDWE system OntoChatGPT, opening new perspectives in the field of scientific research and development, and enhancing the effectiveness of scientific and technical programs through the synergistic effect of industry research.

Automated interactive systems, such as OntoChatGPT, are essential components of RDWE systems, contributing significantly to the optimization of scientific research. The RDWE system OntoChatGPT, developed based on advanced technologies such as GPT-3, GPT-3.5, and GPT-4 from OpenAI, represents a highly efficient tool for processing and understanding natural language.

This system not only opens up new opportunities for intuitive human-machine interaction but also serves as a strategic tool within RDWE, supporting the development of innovative information systems for scientific research.

OntoChatGPT enables users to express their thoughts and needs in natural language, simplifying interaction and reducing the need for deep interface learning. Its unique functional capabilities, such as deductive reasoning and logical inference, make it a powerful tool for expanding and structuring knowledge.

In practice, OntoChatGPT can be integrated into various software systems to address specific tasks, such as in the field of rehabilitation, where it can support efficiency and universality through structured prompts.

OntoChatGPT becomes a key tool for generating new knowledge in the field of scientific research and development. The application of this technology allows for the integration of an ontological approach, facilitating the creation and use of structured prompts in interaction with ChatGPT. This, in turn, may contribute to higher accuracy and efficiency in natural language processing, ensuring more precise understanding and logical inference based on knowledge gathered from various sources. Integrating the ontological approach with ChatGPT not only allows the RDWE system OntoChatGPT to effectively structure and analyze information but also ensures flexibility and adaptability to align with diverse user tasks and needs, thereby enhancing the system's ability for self-learning and self-improvement, leading to continuous knowledge expansion and optimization.

This research aims to thoroughly analyze and validate OntoChatGPT in the context of RDWE systems, confirming its suitability and effectiveness in generating new knowledge and creating innovative solutions for scientific research and development.

## 1.1. COMPLEX INFORMATION SYSTEM FOR KNOWLEDGE PRODUCTION IN THE RDWE CLASS — ONTOCHATGPT

The concept of using natural language for interaction between humans and machines has always been highly relevant. It presupposes convenience and ease compared to the necessity of learning a specialized language and adhering to pre-defined instructions. The introduction of text interfaces, and later window-based and menu-based interfaces, marked a significant breakthrough, making computers accessible and widely used tools by various users.

However, despite their practicality and convenience, these interfaces still lack the necessary perfection and efficiency. They are often characterized by a rigid, pre-defined structure and can be complex for the general user population. As a result, users often have to spend significant time and effort familiarizing themselves with all the features of such interfaces. It would be ideal if users could simply express their desired actions using natural language or text input, thereby eliminating the need for thorough interface learning.

The integration of natural language processing and understanding capabilities into interactive systems would allow users to interact with computers more intuitively and conveniently, enhancing the usability and efficiency of these systems, providing users with a more comfortable working environment, and reducing the learning curve typically associated with interface complexity.

Currently, there is a wide range of virtual assistants utilizing natural language processing. Prominent examples include artificial intelligence (AI) systems such as Apple Siri, Google Assistant, Amazon Alexa, and Microsoft Cortana. However, the development of the ChatGPT system, based on large language models (LLMs) GPT-3, GPT-3.5, and GPT-4 by OpenAI, and further refined for conversational applications using supervised and reinforcement learning methods, marked a significant breakthrough in the field of AI, particularly in natural language processing (NLP) and natural language understanding (NLU). GPT, short for Generative Pre-trained Transformers, is one type of large language model. ChatGPT is a purely text-based system and lacks the ability to recognize and generate speech or interact with physical objects in the material world. Nevertheless, its potential as a powerful natural language system is impressive, offering a wide range of capabilities for information provision, generation, and structuring. ChatGPT can be implemented for various software functions and more.

Thus, ChatGPT not only serves as a useful standalone virtual assistant and "companion" but also has significant potential for integration into other software systems, utilizing its capabilities to achieve specific goals. This perspective opens up new avenues for research, particularly in exploring how ChatGPT can be utilized to achieve specific goals necessary for system functioning.

Researchers now have the opportunity to delve into methods of effectively utilizing and optimizing ChatGPT capabilities in specific domains. This includes adapting and fine-tuning ChatGPT to perform tasks tailored to the unique requirements of different software systems. Effectively "taming and mastering" ChatGPT, researchers can leverage its strengths and align functional capabilities with various application tasks, leading to further progress in natural language processing and expanding the boundaries of what can be achieved through intelligent information systems (IIS).

The OntoChatGPT system enables information provision and logical inference (as defined in [4] — expanding the knowledge base by obtaining new information from existing knowledge units; this process includes various operations, among which logical inference is an important case) based on a specific set of contexts, functioning as a dialog system. Logical inference involves deriving new information, including new RDF triples in Semantic Web technologies, based on established facts, rules, and logical principles, enabling the system to establish connections between different parts of textual information. Using deductive reasoning, the system can expand its understanding (of texts). This process of obtaining new information units from previously known ones plays a crucial role in expanding the system's knowledge and increasing its overall functionality.

The implementation of the proposed technology is demonstrated using a pragmatically oriented corpus of texts — a dataset written in the Ukrainian language, in the field of physical and rehabilitation medicine (PRM), particularly e-rehabilitation [5].

Applying a comprehensive approach in the research ensures an effectively structured foundation for organizing and presenting knowledge in a systematic manner. The models encompass informational and functional aspects, laying the groundwork for the efficient integration of an ontological approach with ChatGPT.

Furthermore, the research focuses on the practical application of the developed technology in the field of PRM. By implementing and testing OntoChatGPT using the Ukrainian language, the research showcases the potential and versatility of the approach. Specifically, it demonstrates how an ontology-driven system of structured prompts, combined with ChatGPT, can enhance information provision and conclusions in the context of rehabilitation-related discussions.

The research results demonstrate the applicability and relevance of the proposed methodology in the specific field, providing a clearer understanding of the potential benefits of integrating ontology-driven systems of structured prompts with modern language models, paving the way for further developments and applications in the field of IISs and knowledge manipulation technologies.

The aim of this research is to create informational and functional models and methodological principles using an ontology-driven system of structured prompts in interaction with ChatGPT. The developed RDWE system is named OntoChatGPT.

### *1.1.1. Current state of development of service- and knowledge-related information systems: the influence of large language models*

At present, ChatGPT does not offer a flexible Application Program Interface (API); its interface operates through a limited set of URLs and corresponding commands, with predefined actions attached to them. However, there is the capability to convey commands in natural language, making it more accessible for developers and users. This approach comes with certain specifications and limitations. Firstly, although ChatGPT can process multiple languages, its effectiveness and understanding of each may vary significantly. The system is primarily optimized for the English language, so all commands and instructions aimed at optimizing interaction should predominantly be formulated in English, regardless of the user's language. This particularity may pose additional challenges for those proficient in other languages and requires extra attention when developing applications and services based on ChatGPT in linguistically diverse environments.

Another key requirement is presenting instructions to ChatGPT in a well-structured, clear, and specific format. Since ChatGPT can process a limited number of tokens at a time, instructions need to be concise yet comprehensive and substantive. Empirical research on ChatGPT [2] indicates that formatting instructions in JSON format is one of the effective strategies for presenting short but meaningful commands and directives. The JSON format ensures structure, accessibility, readability, and optimal information transmission for ChatGPT, enabling users to precisely articulate their requests and expectations, thus optimizing the interaction between the user and ChatGPT [6]. This method of information presentation facilitates the most accurate and efficient interaction, allowing users to express their intentions and expectations from the system in the most optimal and understandable form, thereby ensuring interaction between the user and the system at an appropriate level.

In the open GitHub repository "Mr. Ranedeer: your personalized tutor!" [7], samples of instructions for transforming ChatGPT into a virtual learning system can be found. The instructions are presented as nested dictionaries, where key words represent short representations of basic concepts or tasks and may be associated with values in the form of more detailed dictionaries or phrases in English, providing full and clear explanations.

To activate the functionality of the virtual tutor, users simply need to copy and paste the corresponding instructions into the ChatGPT interface. This allows adapting ChatGPT to perform the duties of a virtual tutor in a variety of disciplines, considering its extensive knowledge base. The strategy of using structured instructions to direct ChatGPT behavior demonstrates significant potential and opens up new possibilities for expanding the system's functionality, making it even more useful and flexible in various usage contexts.

Additionally, ChatGPT has the capability to integrate plugins stored on external resources. Links to these resources can also be included in the instructions, thereby expanding the range of functionality and possibilities. This opens up new research directions, referred to as hint engineering and meta-learning. Recent studies [8] underscore the importance and relevance of exploring these directions.

The primary objective of prompt engineering is to guide ChatGPT towards providing accurate responses, particularly in tasks necessitating logical deductions. Instructions can be formulated to clarify task intricacies and break it down into sequential steps, directing the AI towards the desired outcome. Prompt engineering resembles an art, involving meticulous selection of specific words, phrase constructions, and their arrangement to elicit the desired behavior from ChatGPT. Various strategies have been devised, including the utilization of imperatives to define the AI's role, planned sequences, structured data formats (such as JSON, XML, YAML), chains of self-critique, and others.

Determining the most effective prompt engineering strategy remains an open question, but promising approaches are described in [9]. The importance of phrase structure and the use of specific words and expressions are emphasized in the paper. Combining the findings of this research with other relevant studies can yield significant results and contribute to the development in this field.

Prompt engineering opens up opportunities for targeted and specific training of ChatGPT, allowing it to gain deeper understanding of subjects it may lack sufficient knowledge in. While mechanisms like model training and fine-tuning exist in ChatGPT, they can be costly and require extensive, carefully curated datasets. In some cases, such an approach may be impractical or unwarranted. Instead, crucial information can be conveyed in textual form or through data structures in conjunction with JSON prompts, guiding ChatGPT algorithms on how to process the input data. This expands ChatGPT's knowledge and capabilities, orienting it towards applications for dialogue systems or even management systems.

Although employing a rigid structure similar to that described in [10] may be a functional approach, there is room for further development and exploration. Systems interacting with ChatGPT can utilize various instructions or templates tailored for different purposes. The prompts themselves can be made more flexible by including optional fields and providing different explanations (values) for each field. Such a system should include instructions on when and how to use templates with ChatGPT and what specific values to use in different scenarios. These instructions for creating and using structured prompts in ChatGPT can be organized within an ontology, ultimately creating an ontology-driven system.

The use of ontologies, or meta-ontologies, as repositories of system behavior rules is discussed in [11], although without direct reference to ChatGPT or similar applications and information systems. In this approach, the ontology serves as a decision-making module, orienting the system on how to handle certain types of data and represent them in the user interface.

Fundamental concepts of information systems with ontology-driven architecture are widely discussed in works [12]-[14]. In the "Explanatory Ontographic Dictionary of Knowledge Engineering" [4], ontology-driven architecture is defined as a systematic architecture based on two main components: an "active" computer ontology and a "problem solver." These components work together to manage the information processing process, paying special attention to solving users' practical tasks and supporting goal-oriented activities. Additionally, an ontology-driven information system [4] is characterized as a complex system consisting of several key elements. These include a knowledge base closely linked to ontologies (typically presented as a finite set of systematically integrated knowledge bases within specific subject areas), an inference mechanism, a query processing subsystem, and interfaces (User Interface, API) for user interaction and/or integration with external environments. Together, these components facilitate the effective utilization of ontological knowledge in the information system.

The approach used in this research differs from the methodology employed in our previous research works, particularly in [15] and [16]. In those, the ontology served as the main information repository in the dialogue system rather than a container for rules organized using a different approach. Nonetheless, certain elements from those previous developments are still applicable to our current work. For example, methods such as Named Entity Recognition (NER), analysis of related context, and automatic generation of formal SPARQL queries based on natural language phrases provided by the user are utilized. Addressing these issues is important and necessary in the development of such RDWE systems.

It is important to note that resolving issues related to methods for ensuring correct entity extraction and generating accurate SPARQL queries is complex and requires further research.

### *1.1.2. Models and methods for implementing the composition of problem-related web services*

In the modern world, where information technologies play a crucial role in knowledge development and production, the composition of web services is becoming increasingly critical for creating effective information systems. Complex knowledge production systems require the integration of various services and technologies for knowledge processing, analysis, exchange, and management. All of this is aimed at solving specific problems, improving decision-making, and fostering innovation.

Such systems often require specialized integration approaches that provide flexibility, scalability, and efficiency to meet the high demands of knowledge production ecosystems. Organizations and developers are adopting advanced models and methods for composing problem-oriented web services aimed at supporting the integrity and interoperability of services, as well as providing robust and balanced solutions for users.

This section will discuss key models and methods that facilitate the implementation of web service composition within the context of complex knowledge production systems. The discussion will include Service-Oriented Architecture (SOA), RESTful web services, Semantic web services, cloud technologies, and other modern approaches and technologies that enable the development of flexible, scalable, and integrated systems for knowledge production.

1. **Service-Oriented Architecture (SOA) Model:** The SOA model is critical for designing and implementing web services that are loosely coupled but can communicate seamlessly. It allows for the creation of services that work independently but can be combined to achieve more complex functionalities.

- **Method:** Orchestration and Choreography. These methods facilitate the coordination and management of interactions between various web services in the SOA model. Orchestration involves centralized management of interactions, while choreography focuses on decentralized interactions.

2. **RESTful Web Service Model:** RESTful services, using standard HTTP methods, ensure that web services are scalable and stateless, facilitating easy integration and improved performance.

• **Method:** CRUD Operations. Create, Read, Update, Delete (CRUD) operations are the fundamental methods for managing resources in RESTful services, allowing manipulation and retrieval of resources in the system.

3. **Semantic Web Service Model:** This model includes semantic annotations to describe the functionality and capabilities of web services, enhancing the ability of services to interact and integrate.

• **Method:** Ontology-Based Integration. By utilizing ontology, semantic web services can achieve higher levels of integration, enabling the unification of terms and understanding of relationships between different concepts, ensuring more consistent and intelligent interactions between services.

4. **Composite Web Service Model:** This model allows for the combination of multiple web services to provide comprehensive solutions to more complex and diverse problems.

• **Method:** Composition Based on Workflow Processes. This method allows for the structuring of interactions between different web services, enabling systematic execution of services based on predefined workflow processes, ensuring smooth and logical flow of operations.

5. **Machine Learning Models:** Incorporating machine learning models can help optimize the composition of web services by predicting the best possible combination of services for a given problem based on historical data and patterns.

• **Method:** Reinforcement Learning. By using reinforcement learning algorithms, the system can learn optimal policies for composing web services, improving the efficiency and effectiveness of composition over time through reward-based learning.

6. **Cloud-Oriented Model:** Utilizing cloud computing enables scalable and flexible deployment of web services, allowing for efficient resource utilization based on demand.

• **Method:** Microservices Deployment. Deploying web services as microservices in the cloud facilitates independent scaling, development, and deployment of each service, enhancing flexibility and agility.

**Implementation Strategy:**

1. **Service Identification:** Identify atomic and composite services necessary to address specific problems, ensuring each service has clear and defined functionality.

2. **Service Composition Design:** Design the workflow and interactions between identified services, considering dependencies, sequence, and data flow between them.

3. **Semantic Annotation:** Annotate services with relevant semantic metadata to facilitate intelligent and meaningful interactions between them.

4. **Integration and Testing:** Integrate composed services and conduct thorough testing to ensure reliability, efficiency, and correctness of composition.

5. **Optimization:** Continuously monitor and optimize service composition based on performance data, user feedback, and evolving requirements.

6. **Deployment:** Deploy composed services in a scalable and flexible environment, ensuring high availability and resilience.

The development of models and methods for composing problem-oriented web services is crucial for creating specialized, flexible, and adaptive solutions capable of addressing a wide range of problems. By leveraging advanced models such as SOA, RESTful services, and Semantic Web Services, and implementing methods such as orchestration, CRUD operations, and ontology-based integration, developers can create powerful composite services that provide enhanced value and functionality to end-users. Continuous optimization and adaptation of these compositions are vital to maintaining relevance and effectiveness in a rapidly changing technological landscape.

### *1.1.3. Models, methods, and services of the RDWE system OntoChatGPT*

#### *1.1.3.1. Information model of the RDWE system OntoChatGPT*

The OntoChatGPT information system utilizes an information model based on the Composite Service (CS) with a three-component tuple [1]. It serves as the foundation of OntoChatGPT's functionality, allowing the integration of diverse services within the system.

$$CS_{OntoChatGPT} = \langle D_{evkit}, F_{unc}, E_{nv} \rangle \tag{1.1}$$

where:

$D_{evkit} = \left\{ ws_w, as_d \,\middle|\, w = \overline{1,k}, d = \overline{1,l} \right\}_{k,l \in \mathbb{N}}$ — is a comprehensive set of web services and application software available for developers which and enables the development of various applications and services within the system — OntoChatGPT development kit. $\mathbb{N}$ denotes a set of nonnegative integer numbers.

The formalization of the web service, denoted as $ws$, is an extension of the $Service$ formalism introduced in [17]. This specialized representation incorporates additional properties, namely $m_{read}$, $h_{read}$ and $r_{est}$, which enhance the descriptive power and characteristics of the formal model:

$$ws = \{p_{re}, e_{ff}, i_{nput}, o_{utput}, p_{rovider}, c_{aller}, d_{esc}, r_{est}\} \quad (1.2)$$

where:

$c_{aller}$ — *caller* is the consumer or user of the web service.

$p_{re}$ — in the context of web services $ws$, *preconditions* refer to the conditions that must be satisfied before a web service can be consumed. They define the prerequisites that need to be met by the caller $c_{aller}$ before invoking the service.

$e_{ff}$ — *effects* represent the conditions or changes in the world that can be expected to be true after the web service $ws$ has been executed. They indicate the outcomes or results of performing the service. Within the preconditions $p_{re}$ and effects $e_{ff}$ framework, there are special subclasses known as input $i_{nput}$ and output $o_{utput}$.

$i_{nput}$ — *input* conditions correspond to preconditions $p_{re}$, specifying the necessary input data or parameters required by the web service $ws$.

$o_{utput}$ — *output* conditions, on the other hand, align with effects $e_{ff}$, denoting the expected output or outcomes produced by the web service $ws$.

$p_{rovider}$ — is the *provider* entity responsible for offering the web service $ws$.

$d_{esc} = \{m_{read}, h_{read}\}$ — is a *description* of the particular web service $ws$ is provided in both machine-readable $m_{read}$ and human-readable $h_{read}$ formats. This description, known as $d_{esc}$, serves as a resource accessible to the caller $c_{aller}$, providing information about the web service $ws$ and its functionality.

Additionally, the creation of web services $ws$ adheres to a set of constraints $r_{est}$, influenced by the RESTful architectural style as outlined in [18]. These constraints include:

- *Client/Server*: This constraint emphasizes the separation of concerns by adopting a client-server architecture. It allows for independent evolution of different components, enabling the client's user interface to evolve separately from the server and promoting simplicity in the server's design.
- *Stateless*: The client-server interaction is designed to be stateless, meaning that the server does not store any client-specific context. Instead, the client maintains any necessary session information, ensuring that each request can be treated independently.
- *Cacheable*: Data within a response can be labeled as cacheable or non-cacheable. If a response is cacheable, the client or intermediary can reuse it for similar future requests, reducing the need for redundant interactions with the server.

- *Uniform Interface*: The uniform interface constraint ensures that there is a consistent and standardized interface between components. This uniformity facilitates interoperability and allows clients, servers, and network-based intermediaries to depend on the predictability of the interface's behavior.
- *Layered System*: Components are organized into hierarchical layers, where each component is only aware of the layer with which it directly interacts. This layered approach promotes modularity and scalability, as components can operate within their designated layers without requiring knowledge of other layers.
- *Code on Demand*: This constraint is optional and provides support for extending client functionality through the downloading and execution of scripts. Clients can dynamically enhance their capabilities by acquiring and running code components from the server.

The formalization of application software $as$, encompassing both desktop applications and utilities that feature graphical or command-line user interfaces, can be considered a specialization within the broader $Service$ formalism discussed in [17]. In this specialized context, an additional property $d_{esc}$ is introduced, denoted by the human-readable $h_{read}$ *description* of the particular desktop application:

$$as = \left\{ p_{re}, e_{ff}, i_{nput}, o_{utput}, p_{rovider}, c_{aller}, d_{esc}\left\{ h_{read} \right\} \right\} \tag{1.3}$$

where:

$d_{esc} = \left\{ h_{read} \right\}$ — is a human-readable $h_{read}$ description of a particular desktop application service, accessible for the caller $c_{aller}$. All other elements in the formalization remain the same as described in equation (2). The elements such as preconditions $p_{re}$, effects $e_{ff}$, input $i_{nput}$, output $o_{utput}$, provider $p_{rovider}$, and caller $c_{aller}$ continue to hold their respective meanings and definitions as previously stated.

Additionally, in the context of application software $as$ formalization, there is no specific set of constraints imposed. Unlike web services $ws$, which adhere to the RESTful architectural style with a defined set of constraints [17], application software $as$ does not have a predetermined set of constraints that govern its design and behavior. Instead, the constraints applicable to application software $as$ may vary depending on the specific requirements, platform, and design principles employed during its development. Therefore, the formalization of application software $as$ allows for greater flexibility and adaptability, as it can encompass a wide range of applications with different constraints and design considerations. This flex-ibility enables developers to tailor the software to meet the unique needs of users and provide a seamless user experience, whether through a graphical or command-line interface.

$F_{unc} = D_{evkit} : \left\{ C_j \middle| j = \overline{1,n} \right\}_{n \in \mathbb{N}}$ — is a set of functions that encompass the functional aspects of OntoChatGPT's information technology. Each function corresponds to a specific knowledge management pipeline or process, which arises from the integration and interaction of the elements within the $D_{evkit}$. $\mathbb{N}$ denotes a set of nonnegative integer numbers.

$C_j \subseteq D_{evkit}, C_j = \left\{ ws_o, as_p \middle| o, p \geq 0, o \leq k, p \leq l \right\}_{o,p \in \mathbb{N}}$ — is a subset of web services and application software that are required for the successful implementation of the *j*-th function within the $D_{evkit}$. This subset specifically caters to the requirements of the respective function. $\mathbb{N}$ denotes a set of nonnegative integer numbers.

$E_{nv} = \{ prl, os, floss \}$ — is a set of elements that come together as layers forming the *Knowledge Integrated Development Environment* (K-IDE). Each element within this set contributes to the overall functionality and capabilities of the K-IDE.

The element $prl$, which stands for the physical resource layer, represents the physical hardware and facility resources as defined in [18]. It encompasses the tangible components that form the foundation for the K-IDE infrastructure. The $prl$ layer ensures the availability and proper functioning of the necessary physical resources required to support the K-IDE framework.

The element $os$, which refers to the operating system layer, represents the guest operating system within the K-IDE. The operating system layer is designed to utilize Unix-like operating systems, such as Ubuntu Server for x86 systems and DietPi Debian-based lightweight operating system for ARM-based single board computers. It supports various light-weight desktop environments including LXDE, XFCE, or LXQt. This layer provides the foundation for running the K-IDE framework and ensures compatibility with the selected operating system environments and desktop environments.

The FLOSS layer, denoted as $floss = \{ in, ex \}$, represents the Free/Libre and open-source software (FLOSS) component within the K-IDE. This layer encompasses both internal software components, represented by the set $in = \left\{ ws_w, as_d \middle| w = \overline{1,k}, d = \overline{1,l} \right\}_{k,l \in \mathbb{N}}$, and external software components, represented by the set $ex = \{ ws_i \}, i \in \left\{ \overline{1,n} \right\}, n \in \mathbb{N}$. $\mathbb{N}$ denotes a set of nonnegative integer numbers.

The internal software components $in$ include a comprehensive application suite tailored for the scientific research and development lifecycle, along with additional application software $as$ and web services $ws$. On the other hand, the external software components $ex$ refer to specific web services $ws$.

It is important to note that the $D_{evkit}$ subset is part of the FLOSS layer $D_{evkit} \subset floss$, signifying that the development kit is built upon and aligned with the principles of Free/Libre and open-source software.

### *1.1.3.2. Problem-related set of services and toolkit of the RDWE system OntoChatGPT*

• In the current stage of OntoChatGPT information technology, the development kit $D_{evkit}$ set consists of the following comprehensive collection of problem-oriented web services $ws$ and application software $as$:

• $ws_1$ — WebProtégé [19] — is an external web service $ws_1 \in ex$, $ws_1 \in D_{evkit}$. It serves as a free and open-source ontology development environment designed for the Web. With WebProtégé, users can effortlessly create, upload, modify, and collaborate on ontologies, enabling seamless collaborative viewing and editing experiences.

• $ws_2$ — Apache Jena Fuseki [20] — is a FLOSS that provides an HTTP interface for working with RDF data. Fuseki is a part of the Apache Jena Java framework and offers robust support for SPARQL, enabling seamless querying and updating of RDF data through its SPARQL server engine [21]. Fuseki can be locally deployed within the K-IDE environment $E_{nv}$ as an internal component $ws_2 \in in$, or it can be externally deployed via the Software-as-a-Service application delivery model (SaaS) $ws_2 \in ex$ [22], also $ws_2 \in D_{evkit}$.

$ws_3$ — KEn[1] [23] — is an NLP-powered network toolkit (web service with API) for contextual and semantic analysis with document taxonomy building feature. The KEn web service supports processing of English, Ukrainian and Norwegian (Bokmal). The KEn web service offers comprehensive coverage of essential stages in NLP. These stages include: text data extraction; text preprocessing, spell checking and automatic correction, sentence/word tokenization, part-of-speech tagging, lemmatization, word stemming, shallow parsing, JSON/XML-structure generation. KEn web service can be deployed locally as a part of K-IDE $E_{nv}$, as $ws_3 \in in$, or can be deployed externally as $ws_3 \in ex$ via SaaS [35], $ws_3 \in D_{evkit}$.

[1] https://github.com/malakhovks/ken

$ws_4$ — natural language phrase analysis network service [16] — is a specialized web service that supports natural language text in both Ukrainian and English, enabling the construction of semantic trees for phrases. These semantic trees is a key part in facilitating SPARQL queries to form connections with ontologies. Each semantic tree is defined by marker words and expression types, providing valuable insights into the structure of the sentence. In certain cases, multiple semantic trees can be identified within an initial sentence, allowing for the generation of suitable SPARQL queries for each specific tree. This web service can be deployed locally as a part of K-IDE $E_{nv}$, as $ws_4 \in in$, or can be deployed externally as $ws_4 \in ex$ via SaaS [18], $ws_4 \in D_{evkit}$.

$ws_5$ — OpenAI ChatGPT Playground [24] — is an interactive web-based platform that allows users to experiment with the capabilities of the ChatGPT language model. It provides a user-friendly interface where individuals can input text prompts and receive responses generated by ChatGPT in real-time. The Playground offers a range of features to enhance the user experience, including options to adjust the model's temperature and sampling settings. Playground is an external web service $ws_5 \in ex$, $ws_5 \in D_{evkit}$.

$ws_6$ — UkrVectōrēs[1] [25] — an NLU-powered tool for knowledge discovery, classification, diagnostics and prediction. UkrVectōrēs can be described as a "cognitive-semantic calculator" that serves as a powerful tool for distributional analysis. This web service encompasses several essential elements, including: semantic similarity calculation (UkrVectōrēs allows for the computation of semantic similarity between pairs of entities; this feature provides insights into the relatedness and proximity of words in a semantic space); word nearest neighbors (this functionality aids in exploring words with similar meanings or associations); algebraic operations on word vectors (UkrVectōrēs supports various algebraic operations on word vectors, such as addition and subtraction); semantic mapping (users can generate semantic maps that depict the relations between input words; these maps are valuable for visualizing clusters, oppositions, and exploring hypotheses related to semantic relationships); access to raw vectors and visualizations features; use of third-party prognostic models.

$as_1$ — Apache Jena ARQ [26] is a SPARQL query engine Java-based command-line utility. ARQ is a part of FLOSS Java framework Apache Jena. The main ARQ features are: SPARQL 1.1 support; client-support for remote access to any SPARQL endpoint (including usage of SPARQL 1.1

[1] https://github.com/malakhovks/docsim

SERVICE keyword); support for federated query; access and extension of the SPARQL algebra. $as_1 \in in$ , $as_1 \in D_{evkit}$ .

$as_2$ — nlp_api [27] — is a collection of scripts (NLP API from Language Tool) designed for essential text preprocessing tasks specifically tailored to Ukrainian language. $as_2 \in in$ , $as_2 \in D_{evkit}$ .

$as_3$ — is a desktop application service that enables the semi-automatic and fully automatic generation of an OWL ontology [28] from natural language text. It also supports the semi-automatic import of knowledge from a dataset, capturing it as RDF triples, and storing it in an RDF triplestore (TDB or TDB2 component of Apache Jena for RDF storage and query; Apache Jena Fuseki) or in the graph database (Neo4j).

*1.1.3.3. Functional model of the RDWE system OntoChatGPT.*

The functional enrichment of the OntoChatGPT information system is represented by the following set $F$ of functions synthesized from the $D_{evkit}$ :

$$F = \{C_1, C_2, C_3\} \tag{1.4}$$

where:

$C_1$ — semi-automatic and fully automatic generation of an OWL ontology from natural language text.

$C_2$ — ontology-driven dialogue function that integrates ChatGPT and the structured prompts.

$C_3$ — structured prompts for ChatGPT meta-learning function.

### *1.1.4. Methodology for using the RDWE system OntoChatGPT*

The presented methodology can be divided into two key components, each serving a distinct purpose in the development of the OntoChatGPT system. Firstly, we focus on the technique of prompts-based meta-learning and the creation of structured prompts for ChatGPT. This approach involves leveraging prompts to instruct the meta-learning process of ChatGPT, enabling it to generate more contextually relevant and accurate responses. We delve into the methodology behind designing and implementing these prompts, highlighting their significance in enhancing the conversational capabilities of ChatGPT.

The second part of the methodology centers around the development of an automatic ontology-driven dialogue system that integrates ChatGPT and the structured prompts. The core idea behind this system is to incorporate specific subject areas and their associated contexts, which may contain domain-specific information not fully covered in ChatGPT's knowledge base. These contexts are stored in a database, such as MongoDB or a relational database model, and are linked to sets of named entities with their own ontology-like structure. Additionally, sentiment analysis can be used to categorize the contexts. The binding of named entities to their corresponding contexts includes semantic components that elucidate the entity's role within the context. These additional features aim to improve the relevance and clarity of the selected context for subsequent processing. To automate these processes, we utilize our previously developed tools and incorporate transformer pre-trained BERT-based models like [29].

For semantic analysis and named entity extraction from user-provided phrases, ChatGPT proves to be a valuable resource. Special prompts are created specifically for this purpose. Furthermore, ChatGPT is utilized for intent analysis of user phrases. The defined intents, along with extracted named entities annotated with their semantic roles, and the selected list of contexts are then provided as input to ChatGPT. Accompanying these inputs are the appropriate structured prompts that clarify the information to be extracted and the desired representation format. To provide a visual representation of the overall system scheme, we present a context/container C4 model diagram [30] as depicted in Fig. 1.1. This diagram offers a comprehensive overview of the system architecture, showcasing the interplay between various components and their relationships. It serves as a visual aid in understanding the underlying structure and functionality of the OntoChatGPT system.

One of the most outstanding features of the system is the flexibility of its structured prompts for ChatGPT. Instead of rigid prompts, they are dynamically generated based on the specific situation using instructions provided in the form of a meta-ontology. This meta-ontology outlines the fields to be included in the JSON (or XML) structure and the corresponding prompt phrases to be inserted. Each instruction or structured prompt for ChatGPT has its own set of fields and predefined values that can be incorporated. Additionally, the prompt includes a template structure for the response, which ensures consistency and simplifies subsequent processing. The creation of prompt phrases uses proven techniques from [9] to achieve effective and coherent prompts.

The operation of the OntoChatGPT system can be conceptualized through a Petri Net-like scheme, specifically in the form of a modified System Net Marking Graph [31]. This formalization provides a structured representation of the system's functioning, capturing the flow and interactions between various components.

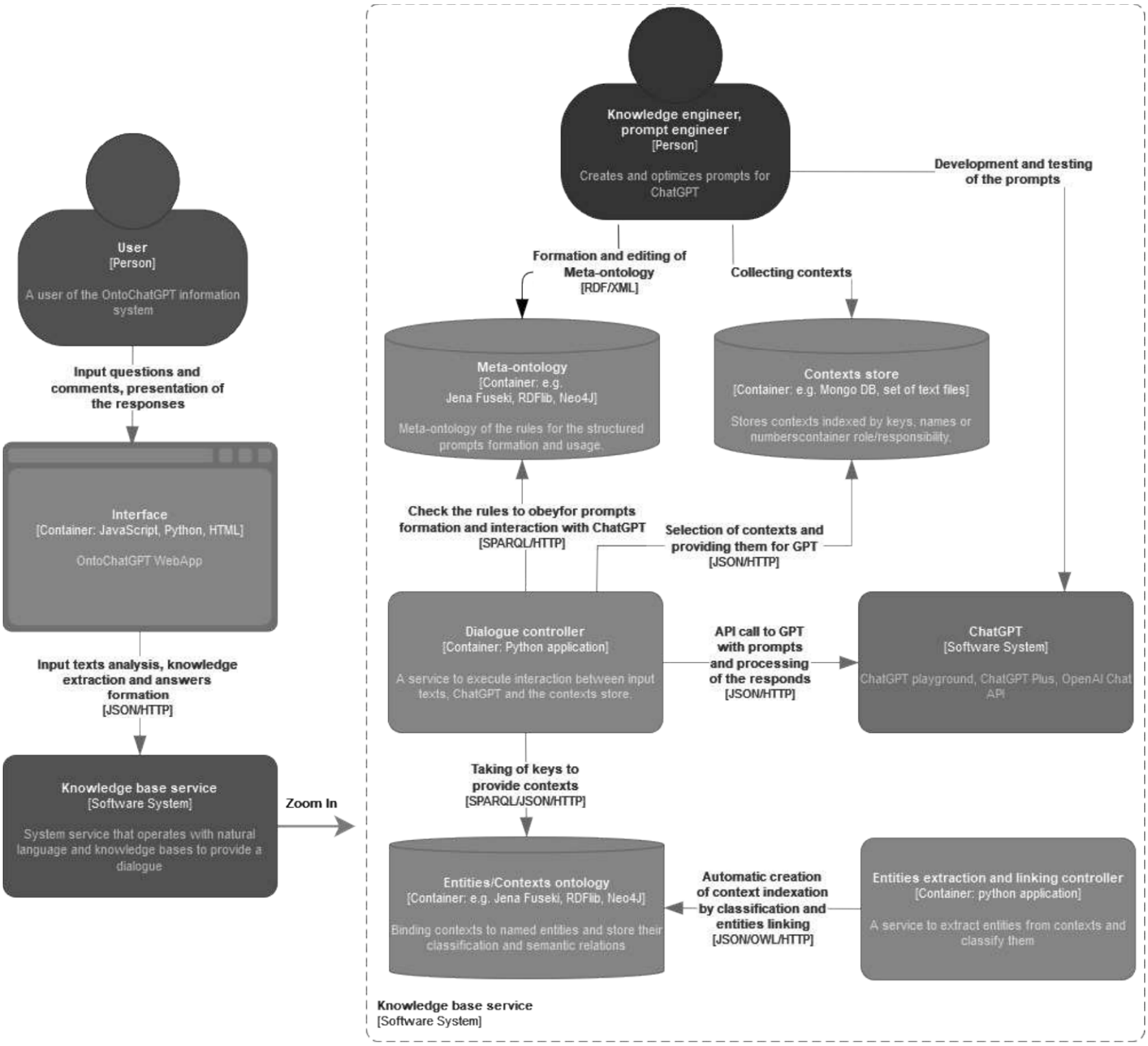

Fig. 1.1. Context/container C4 model diagram of OntoChatGPT information system

A more detailed formulation is given in Eq. (5), (6), (7).

Various methods and approaches can be applied to automate the creation of a context ontology, depending on the nature of the source data and the characteristics of the system as a whole. Below is one possible example of automated ontology creation based on a set of source files in PDF format with a regular, predefined structure. The basic structure of the ontology in this case is also known. The specified component $C_1$ can be described using the following Petri net schema:

States for component $C_1$ (from $S_0$ to $S_{10}$):

$S_0$ — initial state — a set of files in PDF format;

$S_1$ — list of text representations of PDF documents;

$S_2$ — set of documents in structured JSON format;

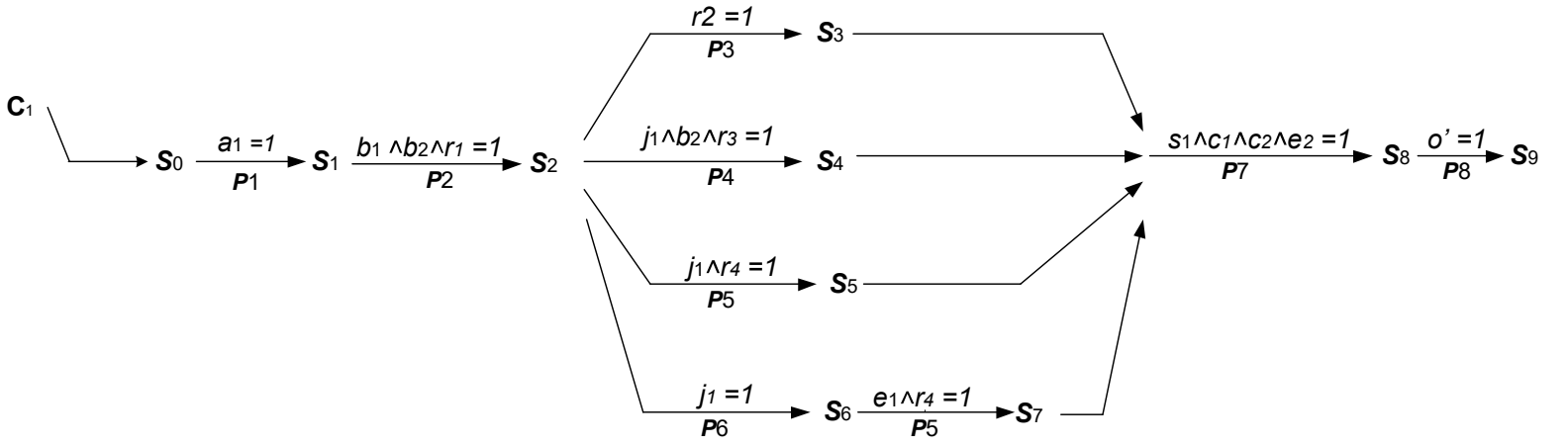

(5)

$S_3$ — programmatic representation of basic ontology components independent of source data. This includes descriptions of declarations and predicates, as well as top-level classes and properties that are basic elements. In this case, for example, such properties include those responsible for associating contexts with specific articles (documents) and named entities with corresponding contexts;

$S_4$ — programmatic representation of ontology classes;

$S_5$ — programmatic representation of instances (Named Individuals) of the ontology;

$S_6$ — named entities extracted from respective contexts;

$S_7$ — programmatic representation of named entities as instances (Named Individuals) of the ontology;

$S_8$ — overall programmatic representation of the created ontology;

$S_9$ — file containing the OWL description of the context ontology.

Procedures:

$P_1$ — extracting text data from source PDF files. For each file, the text is extracted as a list of strings;

$P_2$ — obtaining JSON structures based on the text data with filling corresponding contexts;

$P_3$ — constructing a programmatic representation of the basic ontological structure of the specified form;

$P_4$ — creating a programmatic representation of the set of ontology classes. Classes are mainly created based on the keys of JSON dictionaries representing documents, considering the specified template structure;

$P_5$ — creating a programmatic representation of the set of instances (Named Individuals) of the ontology based on contexts or named entities.

$P_6$ — extracting named entities from contexts;

$P_7$ — compiling a programmatic representation of the ontology structure from the obtained classes, properties, and instances;

$P_8$ — serializing the representation of the created ontology structure into an OWL file.

Variables:

$a_1$ — original set of PDF documents;

$b_1$ — list of text representations of PDF documents;

$b_2$ — specified structure of created JSON representations;

$r_2$ — set of rules and instructions for parsing text representations of documents;

$r_3$ — set of rules and instructions for constructing the overall ontology structure;

$j_1$ — set of JSON representations of source documents;

$r_3$ — set of rules and instructions for constructing programmatic representations of ontology classes;

$r_4$ — set of rules and instructions for constructing programmatic representations of ontology instances;

$e_1$ — set of named entities extracted from contexts;

$s_1$ — programmatic representation of the overall ontology structure;

$c_1$ — programmatic representation of ontology classes;

$c_2$ — programmatic representation of ontology context instances;

$e_2$ — programmatic representation of named entity instances of the ontology;

o' — programmatic representation of the context ontology.

Component C2 (as given in Eq. 6) is an ontology-driven dialogue function that integrates ChatGPT and structured prompts.

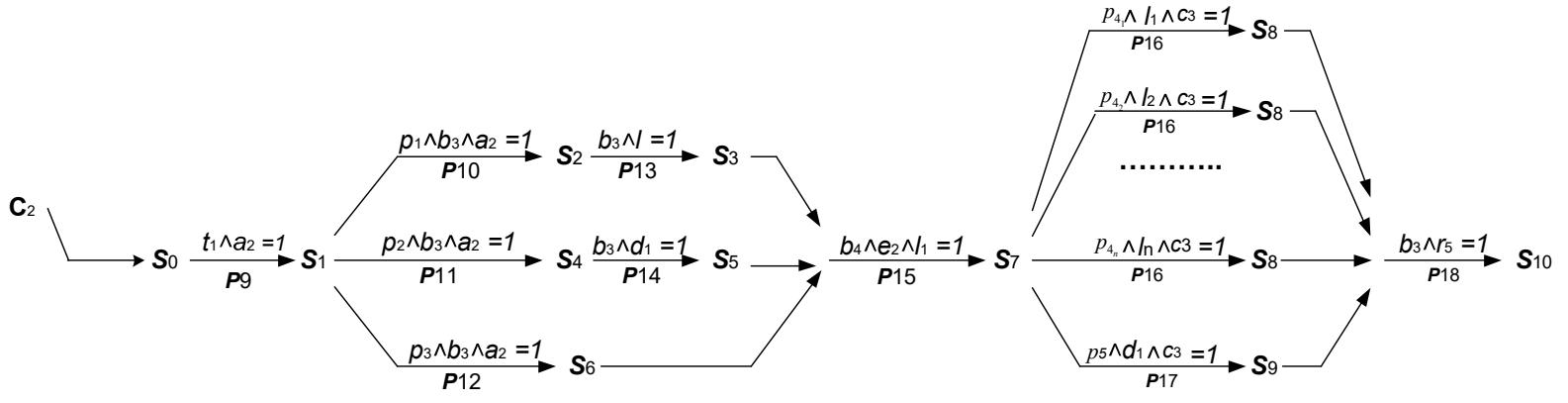

(6)

States for component $C_2$ (from $S_0$ to $S_{10}$):

$S_0$ — initial state of receiving a text message from the user.

$S_1$ — state of preprocessed text prepared for further operations.

$S_2$ — state of the list of identified intentions.

$S_3$ — state of the formed query template for extracting information.

$S_4$ — output state indicating the action to be performed.

$S_5$ — state of the formed template for prompting output conclusions.

$S_6$ — state of the extracted list of named entities.

$S_7$ — state of the selected contexts.

$S_8$ — state of the extracted information from the contexts.
$S_9$ — state of the conclusions obtained from the contexts.
$S_{10}$ — state of the obtained results presented in a user-friendly form.
Procedures:
$P_9$ — initial text processing.
$P_{10}$ — determination of intentions expressed in the input text.
$P_{11}$ — determination of whether the intention involves obtaining conclusions, prompting ChatGPT to provide relevant information from the provided contexts or related knowledge.
$P_{12}$ — extraction of named entities from the input text.
$P_{13}$ — formation of prompts for intention determination.
$P_{14}$ — formation of a prompt for conclusion generation.
$P_{15}$ — selection of relevant contexts based on identified named entities and intentions.
$P_{16}$ — extraction of information from the selected contexts according to the intention.
$P_{17}$ — output of conclusions from the selected contexts.
$P_{18}$ — formatting and presentation of results to the user.
Variables:
$t_1$ — set of rules and operations to be applied during initial processing of the input text.
$a_2$ — initial text provided by the user.
$a_3$ — preprocessed and prepared text for further operations.
$b_3$ — meta-ontology comprising rules of operation and instructions for preparing operational messages.
$b_4$ — ontology of contexts associated with the system.
$p_1$ — prompt for intention determination.
$p_2$ — prompt for analysis to determine if action is required.
$p_3$ — prompt for named entity extraction.
$p_{4_{n,n=\overline{1,N}}}$ — prompts for information extraction based on specific intention.
$p_5$ — prompt for ChatGPT to generate conclusions from the provided contexts.
$I$ — set of intentions identified in the input text triggered with certain entities.
$I^n$ — specific single intention, $n=\overline{1,N}$.
$d_1$ — indicates whether conclusions will be made and if so, what is expected to be concluded.
$e_2$ — set of named entities found in the input text.
$c_3$ — selected contexts.

$r_5$ — data structures representing the obtained results from ChatGPT.

The procedure of the dialog act can be described as follows: after receiving textual information (request) from the user, it undergoes intention analysis using ChatGPT with the help of a specific prompt. The result is a list of dictionaries containing the following keys:

- "name": the name of the intention from the provided list in the request;
- "type": a more general classification of the intention, such as statement, query, or imperative;
- "probability": a floating-point value ranging from 0 to 1 indicating the probability of the intention's presence in the user's text;
- "subject": the subject associated with the intention, if possible;
- "object": the object associated with the intentions, if possible.

The request may contain various possible intentions, such as "quantity," "method of execution," "object," "subject," "action," "location," "direction," "place of action," "conditions," "instrument," "participant," "relationship," "cause," "sequence," "origin," and others. These intentions represent semantic categories and provide the basis for understanding the user's request.

Additionally, the structured request contains fields for providing information, language, and other technical details related to data input and output.

Simultaneously, with the help of ChatGPT and another prompt, named entities are extracted. The result should contain lemmatized words grouped by entities, the type of each group (nominal or verbal), and the headword in each group.

Based on the extracted named entities and their semantic roles (from intentions), the system selects corresponding contexts from the ontology of contexts.

Next, a query is made to ChatGPT, which includes pre-selected contexts and lists of intentions, as well as relevant entities (subjects and objects) updated in the prompts. The form of this query can vary significantly and depends on the specific intention or list of intentions. The rules for its construction, including mandatory fields and prompt phrases, are defined in the metao-ntology.

The meta-ontology also covers rules for the final presentation of results, depending on the types of fields involved. This can include plain text, numbers, dates, lists, tables, or links to external resources, among other possibilities.

Creating the meta-ontology involves meta-learning, which encompasses the development and refinement of relevant prompts. In this iterative process, knowledge engineers (as separate actors) and the ChatGPT playground are involved. Initially, the knowledge engineer formulates a structured prompt in JSON format (XML, YAML, etc.), consisting of keys and prompt phrases designed for a specific purpose, based on previous experience in prompt development and general knowledge. Then this prompt is provided to ChatGPT, and the response is analyzed to determine if it meets the intended goal.

If the ChatGPT response is completely incorrect or has deficiencies, changes are made to the initial prompt. These changes may include adding additional fields, removing unnecessary or ineffective aspects of the prompt, as well as editing prompt phrases to improve task understanding. The enhanced prompt is then passed back to ChatGPT for further iterations. This iterative process continues until the obtained response fully meets the desired and expected result.

After developing the appropriate set of prompts and studying the behavior of ChatGPT on these prompts, they are consolidated into the format of the meta-ontology. Creating the meta-ontology is a manual process where the structure of prompts and the properties of their fields are described based on various anticipated situations and task-solving goals.

The formalized graphical representation of this component $C_3$ of the meta-learning process is shown in Eq. 7.

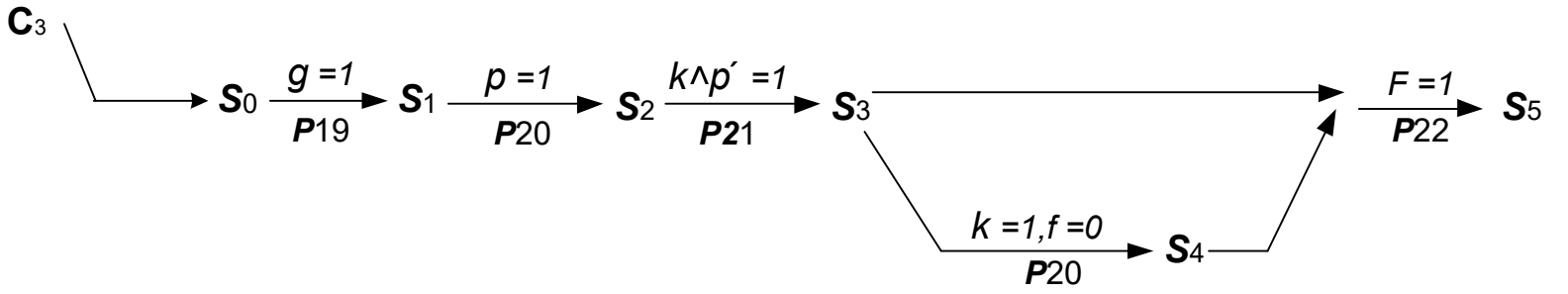

(7)

States for component $C_3$ (from $S_0$ to $S_5$):

$S_0$ — goal of creating a prompt;

$S_1$ — initial prompt created based on guesses and common sense;

$S_2$ — ChatGPT's response to the initial prompt;

$S_3$ — corrected and edited prompt;

$S_4$ — ChatGPT's response to the corrected prompt;

$S_5$ — meta-ontology with instructions and rules for the new prompt.

Procedures:

$P_{19}$ — expressing reasoned guesses and considerations for the new prompt;

$P_{20}$ — initialization of ChatGPT;

$P_{21}$ — correction and editing of the prompt;

$P_{22}$ — integration of the new prompt into the meta-ontology.

Variables:

g — initial idea for creating a new prompt;

p — initial prompt;
p' — revised version of the prompt;
f — final version of the prompt;
k — ChatGPT's response to the initial query;
k' — ChatGPT's response to the modified query.

## 1.2. EXPERIMENTAL VALIDATION OF THE RDWE SYSTEM ONTOCHATGPT AND ANALYSIS OF RESULTS

A prototype of the proposed system has been developed, which includes all the main components. Although the structure of the meta-ontology may be complex for intuitive understanding, including it in this document is not feasible. However, the meta-ontology can be found in the public GitHub repository of the RDWE OntoChatGPT system [28], along with other related materials. Prompt formulations in the system are based on provided rules and presented in the form of JSON structures. The JSON schema provided below represents an example prompt focused on extracting intents and their connection to relevant entities, if possible:

The above example provides a concise and basic prompt covering several fields, allowing for a comprehensive explanation of the main features. An important aspect is the indication of "information to provide," which declares the main task goals. The input message to be analyzed is specified in the "text" field. To enhance productivity, it is useful to specify the language of the input text, such as Ukrainian. However, any other relevant natural language can be used depending on the content processed by ChatGPT. To instruct ChatGPT and limit its ability to detect intents, desired intents should be explicitly indicated in the "possible intents" field. Several boolean fields are provided to provide technical information about the output. In this case, they include "multiple intents," "intent probability," and "show intent subject." The first field allows for identifying multiple intents in the given text, the second field allows ChatGPT to assess the probability of each intent, and the last field instructs ChatGPT on the subjects that activate the intents. To prevent the detection of an excessive number of intents, their quantity can be limited (in this work — up to 4 intents) and sorted by probability. It is always useful to define a template for the structure of the output data, specifying the format in which the information should be returned as a result. Therefore, the "output data representation template" field has been included.

```
{
    “information to provide”: [
        “define intents”,
        “find subjects”,
        “find objects”
    ],
    “text”: “<A text to be analyzed>”,
    “language”: “Ukrainian”,
    “input information field”: “text”,
    “possible intents”: [
        “quantity”,
        “place”,
        “way of doing”,
        “object”,
        “subject”,
        “action”,
        “location”,
        “direction”,
        “scene of action”,
        “conditions”,
        “instrument”,
        “collaborator”,
        “relation”,
        “cause”,
        “sequence”,
        “origin”
    ],
    “several intents”: true,
    “intents probability”: true,
    “show intent subject”: true,
    “max intents number”: 4,
    “intents arrange”: “by probability”,
    “output format”: “JSON”,
    “output representation template”: {
        “result”: [
            {
            “intent”: “intent name - string”,
            “type”: “narration, interrogation or imperative”,
            “probability”: “float value”,
            “subject”: “subject of the intent as a name group - string”,
            “object”: “object of the intent as a name or verb group - string”
            }
        ]
    }
}
```

Fig. 1.2. JSON schema of an example prompt focused on extracting intents and their connection to relevant entities.

Final responses provided by ChatGPT based on identified intents and selected contexts may vary depending on the intents themselves. However, they share a common structure, which consists of a list of dictionaries with "intent" (intent name) and "results" (text or list of texts or other data structures) fields, or simply "none" if there is no relevant information to provide.

The proposed method was tested using the first chapter of the Ukrainian version of the "White Paper on Physical and Rehabilitation Medicine in Europe" as the subject area, which served as the basis for building the context ontology. The responses obtained from the system were classified into the following categories:

- True Positive (TP): ChatGPT provided a response, and it was correct.
- True Negative (TN): ChatGPT did not provide a response, indicating either its acknowledgment of its lack of knowledge or the insufficient amount of information in the texts. This category also includes cases where ChatGPT returned a response of "None" when the corresponding information was indeed absent in the contexts.
- False Positive (FP): The system attempted to provide a response, but it was incorrect.
- False Negative (FN): ChatGPT did not provide a response, although the correct answer was present in the contexts.

It is important to note that during the testing phase, questions and phrases from unrelated subject areas containing entity names not present in the contextual ontology were excluded. In such cases, a negative result would be guaranteed since no relevant context would be selected, and further processing would not be conducted. Therefore, all queries participating in the testing were formulated to pass through all stages of the proposed approach.

Additionally, it should be emphasized that the proposed approach allows for the possibility of multiple responses to a single question, primarily due to the presence of multiple defined intents. In the evaluation process, all provided responses were considered, regardless of whether they were given for the same or different queries.

These considerations ensure a comprehensive assessment of the system's performance and its ability to handle various queries, considering the defined intents and extracting relevant information from selected contexts.

The testing results are presented in Table 1.1.

The testing of the proposed method yielded the following values for the standard evaluation metrics:

- Accuracy: 0.7059;
- Precision: 0.6534;

- Recall: 0.9444;
- F1 Score: 0.7724.

Table 1.1

**The proposed method testing results**

| | True | False |
|---|---|---|
| **Positive** | 17 | 9 |
| **Negative** | 7 | 1 |

These metrics provide a quantitative assessment of the system's performance in terms of its accuracy, precision, recall, and overall effectiveness. The accuracy metric represents the proportion of correct answers provided by the system compared to the total number of queries. The precision metric measures the system's ability to provide accurate responses among the answers it generates. The recall metric indicates the system's capability to retrieve all relevant answers from the available contexts. The F1 score combines precision and recall to provide a balanced measure of overall performance.

In addition to the standard metrics, we also considered additional criteria, namely Precision* and Recall* (Eq. 7, 8). These metrics differ from the standard Precision and Recall in that they treat both true positive and true negative results as true results, without distinguishing between them. These additional criteria provide a broader evaluation of the system's effectiveness in capturing true answers and identifying relevant information from the contexts. By considering both positive and negative results, we gain a more comprehensive understanding of the system's performance in terms of precision and recall.

$$Precission^{*} = \frac{TP + TN}{TP + TN + FP} = 0.7273 \qquad (1.7)$$

$$Recall^{*} = \frac{TP + TN}{TP + TN + FN} = 0.96 \qquad (1.8)$$

Thus, $F1^{*} = \frac{2 * \left(Precission * Recall\right)}{Precission + Recall} = 0.8276$.

The obtained metric values demonstrate the potential usability of the proposed method, although there is still room for improvement.

One of the main drawbacks identified in the current implementation is the high rate of false positive responses, resulting in a relatively low Precision value. This behavior can be attributed to ChatGPT's tendency to attempt to provide an answer even when there is insufficient information available in the given contexts. Additionally, the defined possible intents may not always align perfectly with the provided message, although such intents are often assigned a relatively low probability.

However, it is worth noting that many of these false positive answers were accompanied by true positive answers. In other words, while an incorrect answer was provided in some cases, a correct answer was given alongside it. These accompanied false positive answers could be viewed as supplementary information that may be tangentially related to the main answer.

Although the presence of false positives impacts the Precision value, the fact that they often coexist with true positives suggests that the system is capable of providing additional insights or related information. This observation highlights the potential value of considering the accompanied false positive responses in a practical context.

Addressing the issue of false positives and refining the alignment between possible intents and the message content are areas for further improvement to enhance the Precision of the system.

Let us consider an example. The initial phrase (in Ukrainian) is "На що повинна спиратися ФРМ?" (Eng. — "What should the physical and rehabilitation medicine (PRM) be based on?"). The system detects the following intents:

```
[
    {
        "intent": "subject",
        "type": "interrogation",
        "probability": 0.8,
        "subject": "ФРМ",
        "object": null
    },
    {
        "intent": "cause",
        "type": "narration",
        "probability": 0.6,
        "subject": "ФРМ",
        "object": "спиратися"
    },
    {
        "intent": "way of doing",
        "type": "interrogation",
        "probability": 0.4,
```

“subject”: null,
“object”: “спиратися”
}
]

And the following named entities were found to select the contexts:

[
{
“words”: [“ФРМ”],
“type”: “noun”,
“main word”: “ФРМ”
},
{
“words”: [“повинна”, “спиратися”],
“type”: “verb”,
“main word”: “спиратися”
}
]

We have the following intents that were defined: “subject” (interrogation), “cause” (narration), and “way of doing” (interrogation). All of these intents have relatively high probabilities and should be considered for obtaining the final answer set. However, only the second intent (cause/narration) resulted in the ChatGPT providing a completely correct answer. Surprisingly, this was not the intent with the highest probability. The other intents also led to comprehensive and concise answers, but they were not directly relevant to the initial question. The first intent provided information about physical and rehabilitation medicine (PRM) and the “White Book on Physical and Rehabilitation Medicine in Europe”, while the third intent focused on the purposes of the International Classification of Functioning, Disability and Health (ICF). Although this additional information might be interesting to the user, it is not directly related to the original question.

## 1.3. CONCLUSIONS

As a result of the conducted research, a productive triad has been developed, encompassing three key components: methodological foundations for using ontology-driven structured prompts in ChatGPT meta-learning, advanced information technologies, and a comprehensive service known as the RDWE OntoChatGPT system.

By utilizing ontologies and structured prompts, the RDWE Onto-ChatGPT system has demonstrated its potential to enhance productivity and expand ChatGPT's knowledge base in specific subject domains and languages. A formalized method has been developed, involving meta-learning for the creation and configuration of structured prompts and context ontologies, as well as a sequence of operations for intent detection, named entity identification, context selection, and formulation of the final prompt based on the information and conclusions provided in the contexts. The key feature of this method is its ability to provide ChatGPT with specific information using selected contexts, which can enhance its knowledge in specific subject areas and improve its data handling in different languages.

The use of structured prompts in JSON format enhances their reliability and facilitates obtaining relevant responses. By incorporating a meta-ontology, prompts become more flexible and formalized, elevating the meta-learning process to a higher level of efficiency.

Although the methodology has shown promising results, it still exhibits a tendency towards false-positive responses. However, these false-positive responses are often accompanied by true positive responses, and in cases where relevant information is lacking in the selected contexts, true negative responses are provided. These true negative responses can be considered as providing additional information about the research object.

Furthermore, it is important to emphasize that the proposed methodology can be applied not only to the specific ChatGPT model used in the research but also to other conversational AI systems based on LLM, such as Google's Bard, which relies on PaLM 2 LLM. The fundamental principles and methods of meta-learning, structured prompts, and information retrieval based on ontology can be adapted and utilized in combination with various LLM-based systems. This underscores the potential universality and scalability of the proposed approach across different platforms, enabling its broader application in natural language processing and conversational systems.

The proposed approach serves as a prototype for a more refined RDWE system. Enhancements are planned for structured prompts in JSON format to improve the system's productivity. Additional input data, such as emotions detected in the initial message, and selected contexts, will be included in the prompts. Prompt phrases for keys and values will also be refined along with the overall structure to increase their specificity and reduce the likelihood of detecting irrelevant intents, thereby reducing the number of false-positive responses. It is expected that these enhancements will improve the accuracy criterion of the RDWE system.

**Data access.**

The source code of the RDWE OntoChatGPT system implementation, the meta-ontology, the ontology of contexts, SPARQL queries to the meta-ontology, samples of structured JSON prompts for ChatGPT, test queries and results, are available in open access in the public GitHub repository [28].

**Acknowledgement**

The study was supported by the National Research Foundation of Ukraine (Grant № 2021.01/0136 "Development of the cloud-based platform for patient-centered telerehabilitation of oncology patients with mathematical-related modeling") and conducted at the V. M. Glushkov Institute of Cybernetics, National Academy of Sciences of Ukraine, Kyiv, Ukraine.

# 2. ONTOLOGY RELATED NATURAL TEXT SEMANTIC ANALYSIS APPROACHES

***Vladislav Kaverinskiy***

*The article presents a new technique and its software implementation to create a deeply semantically structured ontology using plain natural language text as input, without regular structure or any previous tagging and markup. The new approach is primarily aimed at highly inflectional languages, and is implemented for Ukrainian.*

## 2.1. INTRODUCTION — NLP AND NLU RELATION WITH KNOWLEDGE ENGINEERING

### *2.1.1. Concept and role of ontology in computer science*

Ontology [1, 2] is primarily a philosophical term denoting the philosophy of being, a branch concerned with "what is, i.e., the nature of reality, and the structural aspects of existence as such." The Internet Dictionary of Social Science Methods by SAGE (2006) describes ontology as a "concept related to the existence and interrelationships between various aspects of society, such as social actors, cultural norms, and social structures." Ontology is the investigation of reality and our beliefs about things [3]. It is the essence of the universe and the existence of truth, but it also defines what can be said about it. Bhaskar (2008) defines social ontology as a philosophical significance ascribed to the social world. It is assumed that people ponder whether there is a truth that exists independently of human conceptions and interpretations and whether there is everyday reality or several context-specific ones [4, 5].

In computer science, the term "ontology" refers to a form of structured representation of knowledge within a specific domain. An ontology is a formalized knowledge model that describes concepts, relationships, and attributes within a particular domain [6]. Ontologies can be constructed using various languages and standards, such as OWL (Web Ontology Language) and RDF (Resource Description Framework). Graph databases are employed for the storage and optimized processing of ontological data [7].

Typically, ontology in computer science constitutes a graph-based knowledge base in which defined predicates interconnect concepts. The most prevalent description format for ontology is the OWL (Web Ontology Language), which encompasses the following fundamental entities: classes, properties, and named entities (individuals). Classes exhibit a hierarchical tree-like structure of the class-subclass type. Each class may have one or more parent classes. The presence of subclassing is optional, and theoretically, the number of subclasses is unlimited. The root for all classes is the class "Thing". Named entities ("Named Individuals") essentially serve as instances of classes and denote specific objects and concepts within a given domain, in contrast to classes describing more abstract, generalized concepts.

Named entities can be linked through properties, i.e., object A can have property B with value C, where A and C are named entities. The presence of property values is optional, as sometimes the property itself suffices as a characteristic, for example: the object "crow" has the property "can fly."

Properties also have characteristics such as Domain and Range. Their values are classes. Domain specifies the set of classes, instances (named entities) of which may have the given property. Range defines the set of classes, instances of which may serve as values for the given property. Properties may have additional constraints; for instance, if a property is functional, one specific object may have only one value for that property.

In ontologies, aside from the hierarchical structure, classes may also be combined into unions and intersections, serving as distinct entities in the ontology. Horizontal relationships between classes can be inferred through the Domain/Range of specific properties. It is possible to explicitly specify the identity or distinctiveness of certain entities in the ontology. This brief description does not exhaustively cover all features of ontology structure but provides a fundamental understanding of such data structures and the OWL standard. A more detailed description can be found at the reference [8].

A distinctive feature of ontology, setting it apart from merely being one form of representing a graph database, lies in the ability of the ontological structure, through the use of descriptive logics, to infer conclusions and acquire new knowledge not explicitly represented within it. However, there is no impediment to utilizing ontology merely as a graph database. In such a case, it would be more appropriate to refer to it as a graph data structure of ontological type, built according to the OWL standard and represented in RDF/XML format. It is noteworthy that in the present work, ontologies are considered more in the context of being a form of a knowledge graph database.

The formal language for querying ontologies is SPARQL. Such a query is based on a set of conditions expressed in the form of RDF triples (subject — predicate — object), to which the selected results must conform. A more comprehensive description of the SPARQL language, including its syntax and capabilities, can be found in the official documentation at the following reference [9].

### *2.1.2. Definition and main tasks of ontological engineering*

The tasks involved in creating a correctly structured and comprehensible ontology, ensuring its most effective practical utilization, constitute a distinct field of research known as ontological engineering. The fusion of ontological approaches and machine learning methods has given rise to a new field referred to in literature as "ontology learning".

Ontological engineering represents a contemporary approach to the systematic organization of knowledge in various subject domains, particularly in computer science, cybernetics, and knowledge engineering. This approach is grounded in the creation and application of ontologies — formalized knowledge models that aid in organizing information and understanding the interconnections between different concepts within a subject domain. Ontological engineering enables the automated creation of ontologies and their use to enhance search, analysis, and knowledge integration [10].

Ontological engineering finds broad application in computer science and cybernetics. Ontologies contribute to improving the efficiency of data search and integration in diverse information systems. They enable the creation of semantically rich models, facilitating the automated processing and comprehension of data [11]. Moreover, ontological engineering assists in developing expert systems and other intelligent applications built upon formalized knowledge [12].

Ontological engineering continues to evolve actively. With the growth in the volume of available information and the increasing complexity of subject domains, the importance of systematizing and understanding knowledge becomes even more pronounced. Machine learning and other technologies contribute to the automated creation of ontologies and enhance data integration [13]. Ontological engineering holds the promise of becoming a key tool in the development of intelligent systems and sophisticated data analysis.

### *2.1.3. Automated creation of ontologies*

One of the crucial aspects of ontological engineering is the automated creation of ontologies. This process involves the application of methods and tools that assist in defining concepts, relationships, and attributes within a subject domain. An approach employed in this context is ontology learning, which utilises machine learning and textual information analysis to extract knowledge and construct ontologies [14]. Such an approach streamlines and expedites the ontology creation process, particularly in large and complex subject domains.

Ontological engineering represents a progressive methodology for the systematic organization of knowledge in subject domains, particularly in computer science, cybernetics, and knowledge engineering. The use of ontologies and graph databases contributes to the enhancement of data search and analysis, as well as the development of intelligent systems. The automated creation of ontologies, including ontology learning methods, renders the ontology creation process more accessible and efficient. Overall, ontological engineering opens up new possibilities for organizing and utilizing knowledge in various subject domains. With the automation and advancement of new technologies, this approach becomes even more relevant and promises a significant contribution to the development of information technologies and intelligent systems.

The automated creation of ontologies and graph databases based on natural language text analysis is a pertinent topic in the fields of computer science, cybernetics, and knowledge engineering. This approach is employed for the creation of formalized knowledge models and databases based on information extracted from textual sources. Let us consider key aspects of the automated creation of ontology and graph databases using natural language text analysis.

The creation of an ontology typically requires a substantial amount of manual effort and subject matter expertise. However, with the advancement of Natural Language Processing (NLP) and machine learning technologies, ontology creation can be automated. One approach to achieve this is ontology learning, which leverages text analysis for knowledge extraction and automatic ontology construction [14].

This process involves steps such as:

— text Information Extraction: Analyzing textual sources to identify concepts, relationships, and attributes.

— data Cleaning and Processing: Transforming textual information into structured data.

— ontology Construction: Developing a formalized knowledge model based on the extracted information.

— tools and libraries, such as Stanford NLP, OpenNLP, and others, are employed for automated ontology creation, providing capabilities for natural language text analysis and knowledge extraction.

Following ontology creation, graph databases can be utilised for the storage and optimized processing of data represented in graph form. Graph databases enable the preservation of relationships between concepts and objects, facilitating easy querying and data analysis [15].

#### *2.1.4. Ontology and graph databases*

Applying graph databases in conjunction with the created ontology can achieve several important objectives:

1. Information retrieval and analysis: Graph databases enable the execution of complex queries, including pathfinding between concepts and analysis of graph structure to discover significant relationships.
2. Data integration: Graph databases aid in consolidating data from various sources, which is crucial for data processing in global systems.
3. Data visualization: Graph databases provide the ability to visualize the graph structure, facilitating data understanding and analysis.

The automated creation of ontology and graph databases based on natural language text analysis finds widespread applications in various fields. For instance, in medicine, this approach can be used to create ontologies that help account for different medical terms and their interconnections. In the field of bioinformatics, automated ontology construction assists in integrating data about genes, proteins, and reactions into a unified model. This approach is also applied in e-commerce for analyzing customer behaviour and recommendation systems. In the financial sector, graph databases are utilized for detecting financial fraud and risk analysis.

Automated ontology creation and graph databases based on natural language text analysis represent a significant research direction in computer science, cybernetics, and knowledge engineering. The use of Natural Language Processing (NLP) technologies and graph databases allows the automation of the knowledge creation and processing process, which is valuable across various domains. The research findings in this direction are already being applied in different industries, and this approach has the potential for further development and refinement.

## 2.2. AUTOMATIC KNOWLEDGE BASES BUILDING ON NATURAL LANGUAGE TEXTS

### *2.2.1. Regularly structured text sets*

#### *2.2.1.1. Texts with regularly predefined structure as a base for ontology*

In certain cases, a text (or a set of texts) exhibits a regularly predefined structure. That is, in specific locations, it is guaranteed or with a high probability to contain information of a certain type presented in a defined format.

Thus, it is possible to devise a software instruction by which a machine can construct an ontology-based database from a set of similar texts.

Examples of such input material from our practical experience include, for instance, a collection of letters [23] or a set of PDF files of EBSCO medical and medical rehabilitation articles [24].

Let us regard an example of the ontology structure for letters:

The ontology includes the following main classes:

*Date* — dates of the letters;

*LetterType* — types of letters based on their content. It has descendant classes corresponding to the following types:

*Apology* — apology;

*Attachment* — enclosed items in the shipment;

*Congratulation* — letters contain congratulations;

*Invitation* — invitation;

*Letter* — regular, traditional letter;

*Narration* — contains long narrative texts;

*Telegram* — telegram;

*WithPoems* — letters contain poems.

*Name* — names;

*Patronymic* — patronymics;

*Surname* — surnames;

*Person* — authors of the letters;

*Place* — locations (usually cities) from which the letters were sent;

*Year* — years for which the letters were sent.

*TextLink* — links to the full texts of the letters.

The properties within the ontology are as follows:

*Authorship* — authorship, descendants of this property associate authors with the letters they wrote. In the domain block of each descendant of this property, there is one author, and the range block includes a list of links to all their letters.

*FullName* — descendants of this property combine names, patronymics, and surnames (or other components of the full name available for a specific case) with the corresponding class representing the author (a descendant of the Person class). The Person descendant is included in the domain block, and the components of the name are in the range block. This approach is necessary because in their query, the user may not always specify the full name of the letter author but may only provide the name and surname, or just the surname, or even the name and patronymic.

*LettersTypes* — the association of letters with their types. This property has a set of descendants, constrained by corresponding classes, descendants of the LetterType. Each such descendant property includes in its domain block the corresponding descendant class of LetterType. In the range block, there is a list of links to the texts of the respective letters.

*Locations* — descendants of this property combine links to letters and the places of their dispatch. The dispatch location (descendant of the Place class) is in the domain block, and the range block contains a list of letters that were sent from this location.

*SendingDate* — descendants of this property combine links to letters with the dates of these letters. The link to the letter is in the domain, and the dispatch (writing) date is in the range.

*YearOfDate* — descendants of this property serve to associate years (subclasses of Year) with dates (subclasses of Date). Each year enters the domain of a separate property, and the list of corresponding dates is in the range.

The schema of the main classes of the proposed ontology of letters is presented in Fig. 2.1. As observed, such an ontology is semantically structured. To operate, it necessitates a set of SPARQL queries. Each such query is directed towards obtaining specific concepts based on the provided input data. Determining the required query and identifying entities for substitution into it necessitates semantic analysis of the user's phrase. It is anticipated that a contextually semantically structured ontology will contribute to reducing the "noise" in results by minimizing weakly relevant data.

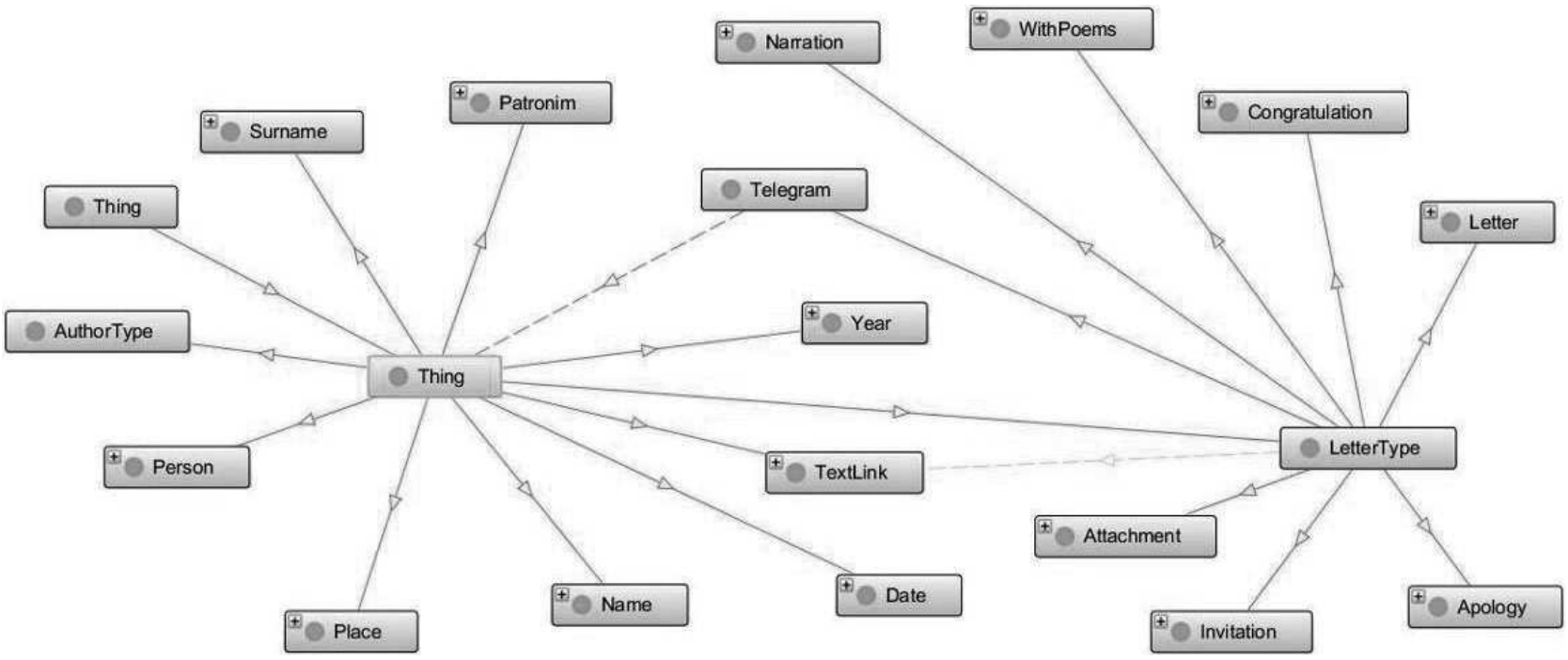

Fig. 2.1. Structural diagram of the basic classes of the ontology of letters

### *2.2.1.2. Automation of ontology creation based on regularly structured texts*

**Primary Processing:** Initial processing involves the input text, which should be presented as a text file. Preliminary cleaning of the file from “textual debris” is conducted, including the removal of page numbers, broken line breaks, reference notes, section headings (e.g., “letters for a certain year”), and headers. These processes are partially automated through the use of a text editor’s features or a relatively straightforward script that replaces specific patterns in the text.

**Automated Text Markup:** The principle of automated text markup is applied in this case as well. The process of creating annotated text is automated, with the programmatic implementation of text markup heavily dependent on the structural features of the analyzed texts. This approach is practical when dealing with a sufficiently large volume of texts. For instance, certain sections may be separated by a line with three asterisks and a space (“* * *”), which serves as a marker for tagging with <paragraph>. These tags, in turn, may contain other nested tags, with the presence of some being optional or allowing the possibility of no text within certain tags. For example, the <title> tag corresponds to a heading, but some texts in the analyzed set may not have headings.

**Example of Automated Text Annotation:**

In the annotation of a collection of letters, the following tags may be present within the <paragraph> container:

<link> — link to the text of the letter for ontology;

<date> — date of writing (dispatch) of the letter;

<place> — location of letter composition;

<author> — author of the letter.

The document’s location and characteristic surrounding words serve as markers during automated annotation. For instance, documents often commence with a line containing the author’s full name, preceded by the word “From”. Thus, by discarding this “From” and retaining the subsequent content in its original form, information for the <author> tag is obtained. Additionally, dates in documents are typically presented as: the day (in digits), the month name (in words), the year (in digits), and the abbreviation “yr.” at the end, without punctuation between components. However, in some documents from the input set, the day or even the month may be absent. In certain cases, a range of numbers is provided with a hyphen. Words such as “city,” “town,” “village,” “settlement,” “c.,” “v.,” and “s.” in short lines before or after the main document text act as markers for determining the geographical location. The word following them represents the name of the locality as part of the address. The automated identification of geographic localization features can be facilitated by creating lists of cities, towns, and other geographic names. Their presence in characteristic sections of the document also defines the content of the <place> tag. The process of automatically annotating texts, even in a large document set, takes only a few seconds, whereas manual annotation could consume many hours.

**Formation of OWL Ontology Based on Annotated Text:**

The annotated text is input into a program for ontology creation, generating an OWL file based on it. The fundamental principles of creating the OWL file are following: lists (more precisely, sets) of software objects are formed, which are then serialized into corresponding OWL components — classes and properties.

Let's provide examples typical for the ontology of letters:

**1) <author> Tag:** Information from the <author> tags is broken down into parts, surnames, and patronymics are identified based on characteristic features. Other parts (separated by spaces) are treated as names. Objects corresponding to these are created. The entire content of the <author> tag becomes a Person object.

**2) <link> Tag:** The content of the <link> tags has the string "_TextLink" appended, and objects containing links to the main texts are created based on them.

**3) Dates and Years:** Objects of type Date are formed from dates, to the name of which "Date_" is prepended — class names in OWL should not start with a digit in some interpretations. The date is written as is into the label field of the respective object. To create objects of type Year from the date string, split by spaces, the last component, which is a number, is taken. "Year_" is prepended to the name of the year object; in the label, the year is recorded as digits in string format. The content of the <place> tag is taken directly.

**4) Context Typing:** One of the most obvious and simplest ways to determine the type of the full context by its content is based on the occurrence of sets of characteristic words or textual structures. For example, poems within texts are identified as a sequence of short lines with a close (but not necessarily equal) number of characters. To be identified as narrative, contexts must have a sufficiently large volume and contain a high concentration of verbs in the third or first person forms of the present or past tense. Machine learning methods and transformer models designed for text classification may also be involved, determining moods, "tonality," positive/negative/

neutral sentiment, approval or criticism, and other classification criteria provided in such models.

**5) Combination of Created OWL Entities:** As all characteristic features of a text fragment (document) are combined in a single container, such as <paragraph>, it is straightforward to create corresponding property objects that link OWL class objects with each other. Subsequently, lists of OWL class and property objects are serialized and stored in a file. The population of the full-text database also occurs in an automated manner. This can be oriented towards document databases (such as MongoDB) or relational databases (PostgreSQL, MySQL, Oracle), or simply a set of text files with specified names and locations. In any scenario, each document in the collection must have a title corresponding to its reference in the OWL ontology.

#### *2.2.1.3. An example of automated construction of an OWL-ontology on medical rehabilitation based on EBSCO article files*

**Input Data.** Another example of an automated approach to ontology construction based on a set of texts with a specified regular structure is the development of a knowledge base on issues of medical rehabilitation using a collection of scientific articles in a related structure to medical protocols.

The knowledge base was automatically created from a dataset of EBSCO articles dedicated to medical rehabilitation, consisting of 1013 PDF files with a total volume of 192 MB. All articles are in English. The key concept behind the automated knowledge base creation is the coordinated and predefined file structure, which serves as a sort of instruction for the software system.

A general scheme of OWL ontology creation using of a documents with a regular structure set is presented in Fig. 2.2.

**Implementation of OWL Ontology Creation.** To implement the creation of a knowledge base in the form of an OWL ontology in RDF/XML format, specialized scripts were developed using the Python programming language. The process consists of two stages.

Automated creation of JSON representation of input article files: In the initial stage, the program extracts and analyzes text from PDF files, automatically identifying and structuring sections and topics in each file as JSON structures of a predefined format. Thus, a set of JSON files is generated, where each file corresponds to the original PDF file, representing its structured content.

Phrases, collocations, and concepts that serve as keys in the provided dictionary are located in headers of various levels in the texts. The search level is determined by the nesting level, and in the texts, headers of different levels are distinguished by text format and characteristic symbols, such as "’," " • ," "–." Additionally, markers of termination, such as a blank line, a combination of a period or a period with a comma and a line break, etc., are taken into account.

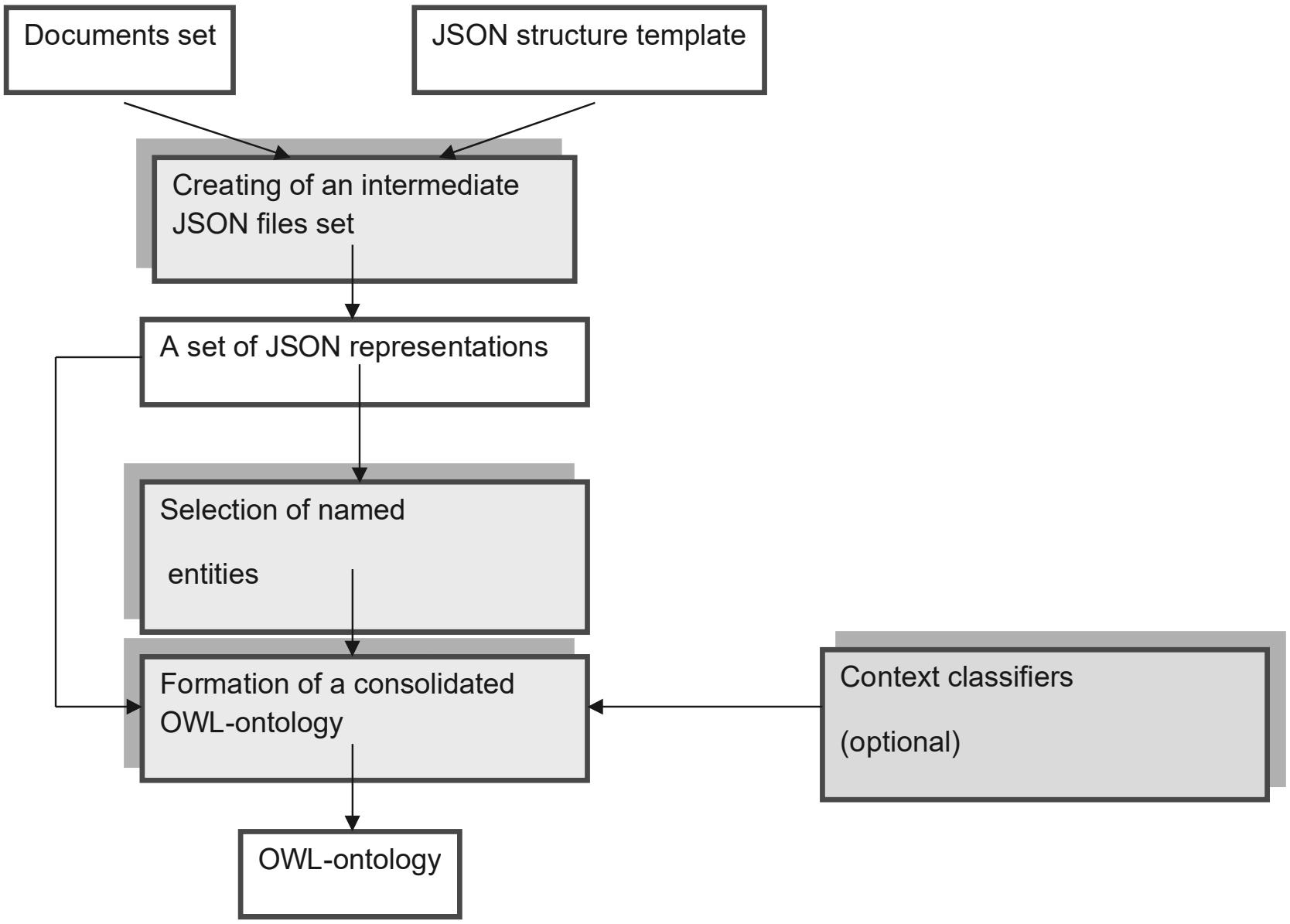

Fig. 2.2. A general scheme of OWL ontology creation using of a documents set with a regular structure

It is worth noting that in the analyzed files, key values are not always encountered in the exact form as presented in the template structure. Therefore, a synonym dictionary has been created, indicating the basic values of the dictionary keys with an indication of their possible variability. For instance:

"causes & risk factors": [
    "causes pathogenesis & risk factors",
    "causes pathophysiology & risk factors",
    "causes and risk factors",
    "causes pathophysiology and risk factors"
]

For each parsed PDF file, a corresponding structured JSON file of the specified form is created.

**Formation of OWL Ontology.** At the second stage, utilizing the obtained set of JSON structures, an OWL ontology is generated. The hierarchical structure of the JSON dictionary keys forms the foundation of the future OWL class system, while the corresponding contextual values become named entities within their respective classes. Each article file name is transformed into a named entity in the “Articles” class. The OWL property “Link to Article” establishes connections between contexts and the corresponding articles in which they appear. Named entities defined within contexts are also transformed into named entities in the “Word” class and linked to the respective contexts using the OWL property “Link to Context.” This structure allows specific contexts to be selected in the ontology through SPARQL queries.

The basic class schema of the ontology obtained from the set of medical articles described by this method is illustrated in Fig. 2.3. Due to its extensive size and complexity, this ontology is only partially presented in this illustration.

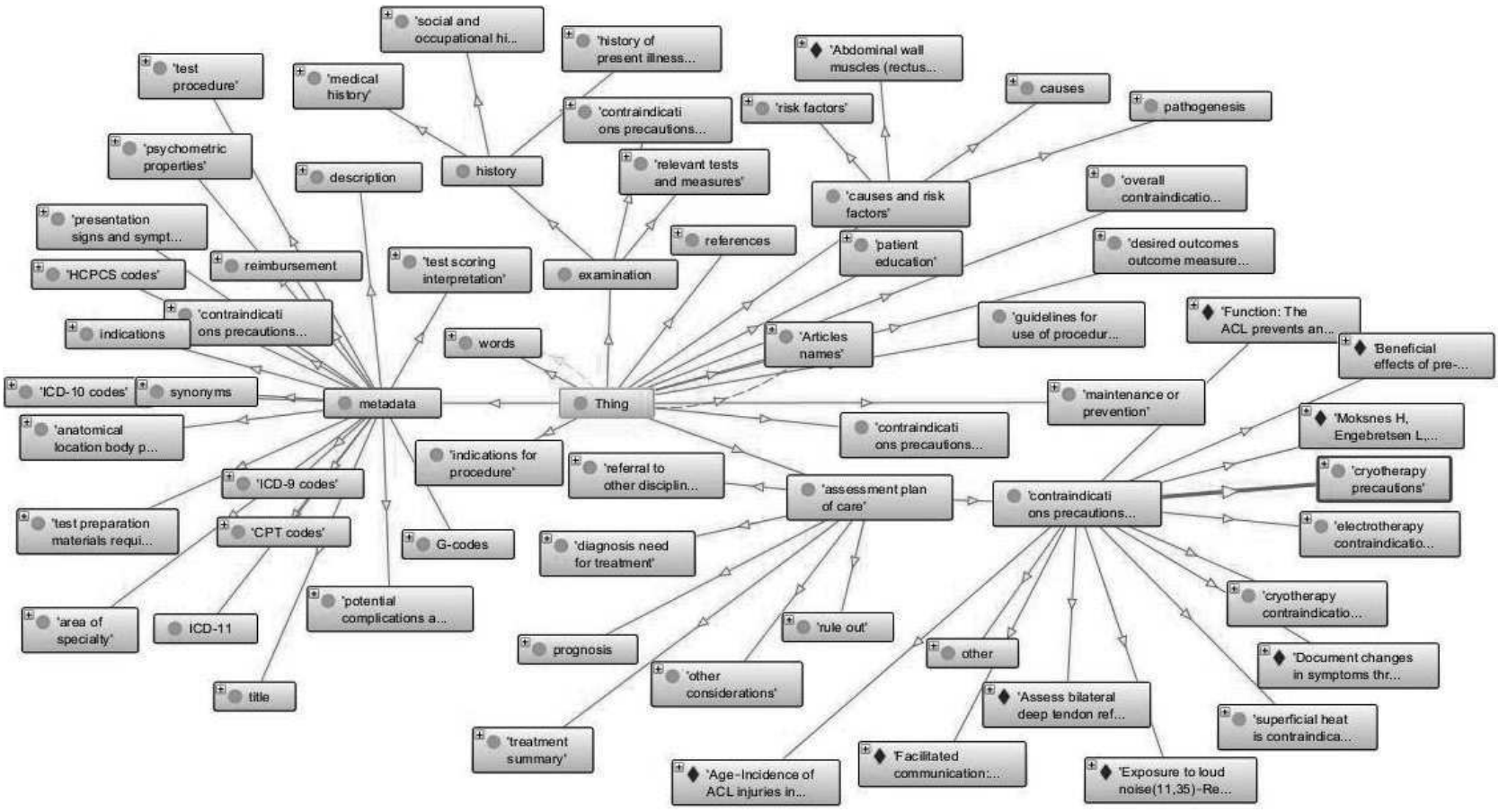

Fig. 2.3. Scheme (abbreviated) of ontology based on the materials of articles on medical rehabilitation

Due to the substantial size of the acquired knowledge base, it has been partitioned into 10 sections to facilitate parallel queries for practical use.

**General Principles of Working with the Ontology of the Specified Structure.** Working with the ontology can be carried out in the following manner. For instance, a doctor describes observed symptoms explicitly specifying the type of information desired, such as anticipated diagnoses, methods of diagnosing these diagnoses, treatment methods, etc. Named entities are identified in the input description, and contexts are selected from a specific section, such as symptom descriptions. Subsequently, using the "relate to article" property, the relevant article or set of articles and associated items related to the individuals (representing types of information) are retrieved. These texts from individuals (recorded, for example, in their "comment" or "label") are returned to the user or subjected to further analysis using a large language model.

### *2.2.2. Plain text syntactic and semantic analysis for knowledge graphs extraction*

#### *2.2.2.1. Semantic texts analysis in ontology engineering adapted to inflective languages*

The proposed approach is based on the rules of syntax-semantics analysis. It is known that in inflected languages, the combination of specific flexions (variable word endings) plays a crucial role in connecting words within a sentence. A developed linguistic system of various flexion combinations for different parts of speech allows expressing significant semantic information. Thus, the analysis of existing word form combinations enables obtaining not the entire but a substantial portion of semantic information.

In the considered approach, we have suggested approximately 90 semantic types, each of which can effectively have several (up to over a hundred) subtypes depending on additional characteristics such as gender, tense, number, or a specific preposition for each word in the examined pair. These additional subtypes are currently not utilized in ontology creation, but they may prove useful in the future for a more profound and accurate information structuring. Additionally, they directly originate from so-called "correlators," which are essentially part of the analyzer providing instructions to the program on how a particular semantic type can be expressed.

Transitioning to the examination of the issue of automated ontology construction based on natural language texts and working with them, the focus should be on semantic text analysis. Currently, there exists a variety of solutions for different types of text analysis, including ontology construction. In [16] close to two dozen such systems have been analysed, each with its own merits and shortcomings. IBM Analytics, Oracle BI, Ontos, and Onto-Text were considered the most sophisticated at that time. These systems implement intelligent procedures of linguistic-semantic and concep-tographic analysis in processing textual arrays. However, only a few of them offer the capability of automated ontology updates. Moreover, they are incapable of forming a comprehensive ontological interactive knowledge system that supports decision-making processes. Additionally, commercial systems do not disclose algorithms and approaches to text processing to developers and lack detailed explanations of the technologies used in implementing software modules. Another issue is the orientation of the majority of those systems primarily towards the English language, whereas the capabilities for working with other languages, particularly inflected ones such as Ukrainian, are inadequately represented.

Large language models, such as ChatGPT, are fundamentally capable of forming ontologies. However, they face several limitations. They are restricted in processing large volumes of input text and lack tools for ontology supplementation. The structure of such ontologies is, in many ways, constructed "at their discretion" and managing this process is not entirely convenient. In principle, it is possible to create a small graph structure of ontological type using GPT, but it would fall far short of practical needs.

Therefore, addressing the challenge of semantic research in an interdisciplinary information environment necessitates the development of tools ensuring a comprehensive ontological representation of semantics, facilitating the storage, processing, and access to its diverse objects and informational units.

Efforts towards developing software systems for ontological representation of text have been undertaken in works [17–20]. However, the implementations developed at the outset of the mentioned researches required significant refinement. The first aspect involves the format for describing and storing the ontology, which, for future usability, had to adhere to the OWL standard, such as the one presented in RDF/XML format. Secondly, the existing capabilities and quality of text parsing in the "Konspekt" program [17] needed refinement and improvement. Thus, our research in this direction has bifurcated into two branches. Firstly, there is an effort to enhance the existing "Kaonspekt" program. Secondly, the focus is on creating an alternative semantic analyzer, operating on a somewhat different scheme but producing results in a format similar to the "Konspekt" program. Additionally, it was necessary to implement a software module for representing this format in a graph database-friendly form, specifically in RDF/XML, aligning with the OWL standard.

#### *2.2.2.2. Semantic analyzer for knowledge graphs building from a natural language text*

Regarding the enhancement of the existing tool, it is pertinent to provide an overview of the "Konspekt" program. This syntactic-semantic analyzer processes sentences in the Ukrainian language, identifying name groups and constructing a graphical representation of semantic relationships within the sentence. Its algorithm is grounded in the principle that in inflected languages, a substantial number of semantic relations can be delineated through the combination of specific parts of speech in particular word forms, determined by flexions (variable parts of words) and prepositions. The system incorporates corresponding dictionaries of roots and flexions, as well as files of so-called determinators and correlators, defining the combinations of parts of speech, their flexions, and the presence of specific prepositions corresponding to semantic categories.

In the initial versions employed at the outset of the research, a limited number of semantic categories and their corresponding determinators and correlators were utilised, essentially constituting a demonstration version. Subsequently, a significantly expanded set of determinators and correlators was introduced to the system, capable of recognising around 80 semantic categories, each with up to a hundred subcategories. The typification and names of semantic categories were somewhat altered compared to the original version. Additionally, a series of corrections and minor modifications were implemented in the program's source code. These refinements contributed substantially to improving the quality of text parsing [25].

Perhaps the central and most demanding aspect involves the syntactic-semantic analysis of the input text. The corresponding software module encompasses a manually created set of so-called "correlators" and "determinators". "Determinators" represent combinations of possible cases and prepositions between them (if any), along with corresponding semantic subtypes for each variant. Additionally, it specifies whether the combination is reversible and identifies the main word within the pair. Semantic subtypes are denoted symbolically, for example, K1001, K4801, K6201, and so forth. Here is an example line from the file of "determinators" for the Ukrainian language:

ім під ям L I K6201K8644K8646

In this example, flexions for the first and second words, "ім" and "ям," are indicated. It is assumed that the preposition "під" stands between them. The symbol "L" signifies that the main word is on the left, while the symbol "I" indicates that the link is reversible. Possible semantic subtypes for this combination are denoted as K6201, K8644, K8646.

“Correlators” represent correspondences between each semantic subtype and possible combinations of parts of speech, including their order within the pair. The file also lists the names of semantic types (“macro types”). For each of these “macro types” there can be several (up to over a hundred) corresponding subtypes. Here is an example line from the “correlators” file for the Ukrainian language:

K3506 subtract_actions S4S1;S4S6;S4S13;S4S5;S4S3;S4S10;S4S11;S4S12;S4S18;S4S22;S4S25;S4S28

In this instance, “K3506” serves as the name of the semantic subtype. Following that is “subtract_actions” (“separability of an action”), which represents a verbal name of the corresponding semantic type. Subsequently, there is a sequence of possible pairs of parts of speech represented by corresponding symbols. For example, S1 stands for “noun,” and S4 represents “verb.” Pairs are separated by commas.

The program also includes a dictionary containing word roots, lemmas, and flexions. The dictionary provides correspondences between roots, lemmas, and sets of flexions. Dictionaries are stored in a special compact format and can be automatically generated using open language data.

The primary goal of the program is to utilise these data, along with the input text, to identify and categorise words, determine their roots and flexions, recognize relationships between words, and ascertain their semantic types.

Another outcome of such analysis is the identification of groups of related words in sentences. Formally, a group constitutes a fully connected graph. In practice, such groups can correspond to a simple sentence, a part of a complex sentence, or a participial construction.

However, despite the mentioned improvements a series of shortcomings persisted. Therefore, an attempt was made to create a new analyser module in the Python programming language, operating on substantially altered principles.

One of the primary tasks in text parsing is the extraction of name groups. In our analyser, this process is integrated with the construction of a graph representation of the syntactic-semantic analysis. In the “Konspekt” program [17], name groups were identified based on characteristic patterns of combining nouns and adjectives. In our case, we adopted a slightly different approach based on sequence analysis.

The general scheme for identifying name groups in a sentence is provided on Fig. 2.4. Initially, the subject group is located, indicated by a noun in the nominative case. Subsequently, name groups are formed around other existing nouns that did not enter the subject group or other name groups.

Words within the identified name groups are removed from the analysed sequence. Formally, in the program, individual nouns are also represented as name groups consisting of a single word. A more detailed scheme of the algorithm for forming name groups is presented in Fig. 2.5.

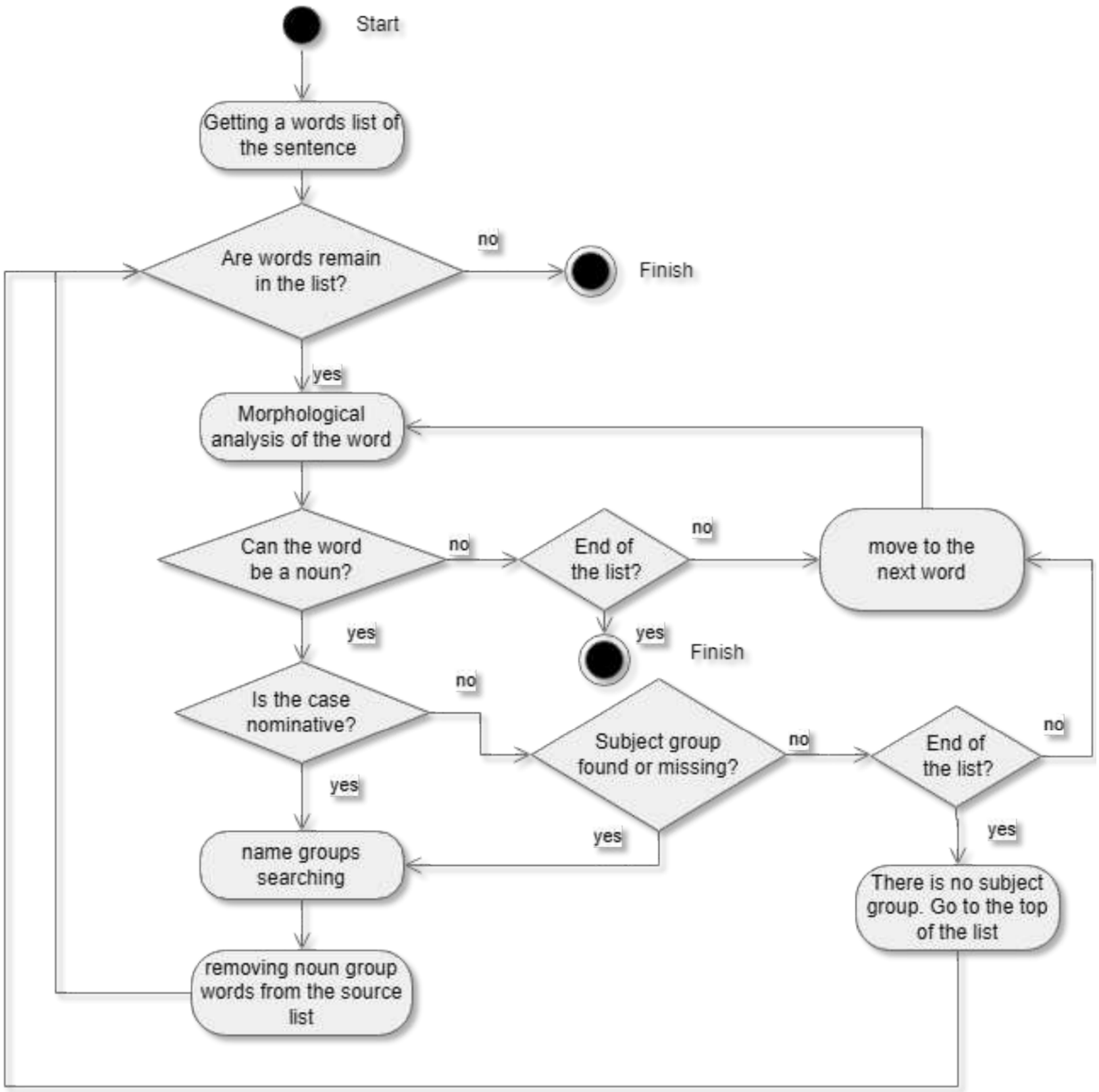

Fig. 2.4. The general scheme for determining name groups in a sentence

As mentioned, the key distinction lies in the scheme itself for forming name groups. The formation of a name group starts with the main noun. The sequence of words to the right and left of it is analysed based on their presence. Other nouns in oblique cases (usually genitive, less frequently others if not separated from the main group by verbs), adjectives, and participles related to the main noun or other nouns in the group, adverbs related to ad-jectives, can be included in the name group. For each word in the sequence, using the PyMorphy2 analyser, its variants of belonging to certain parts of speech in a specific word form are determined. If at least one variant of the parsing corresponds to the conditions for inclusion, the word is added to the group, and the focus shifts to the next word in the sequence, if present. If the conditions are not met, the analysis of the sequence either continues under the conditions of joining the word chain, or the analysis in this direction is terminated due to the lack of connection or the formation of a group.

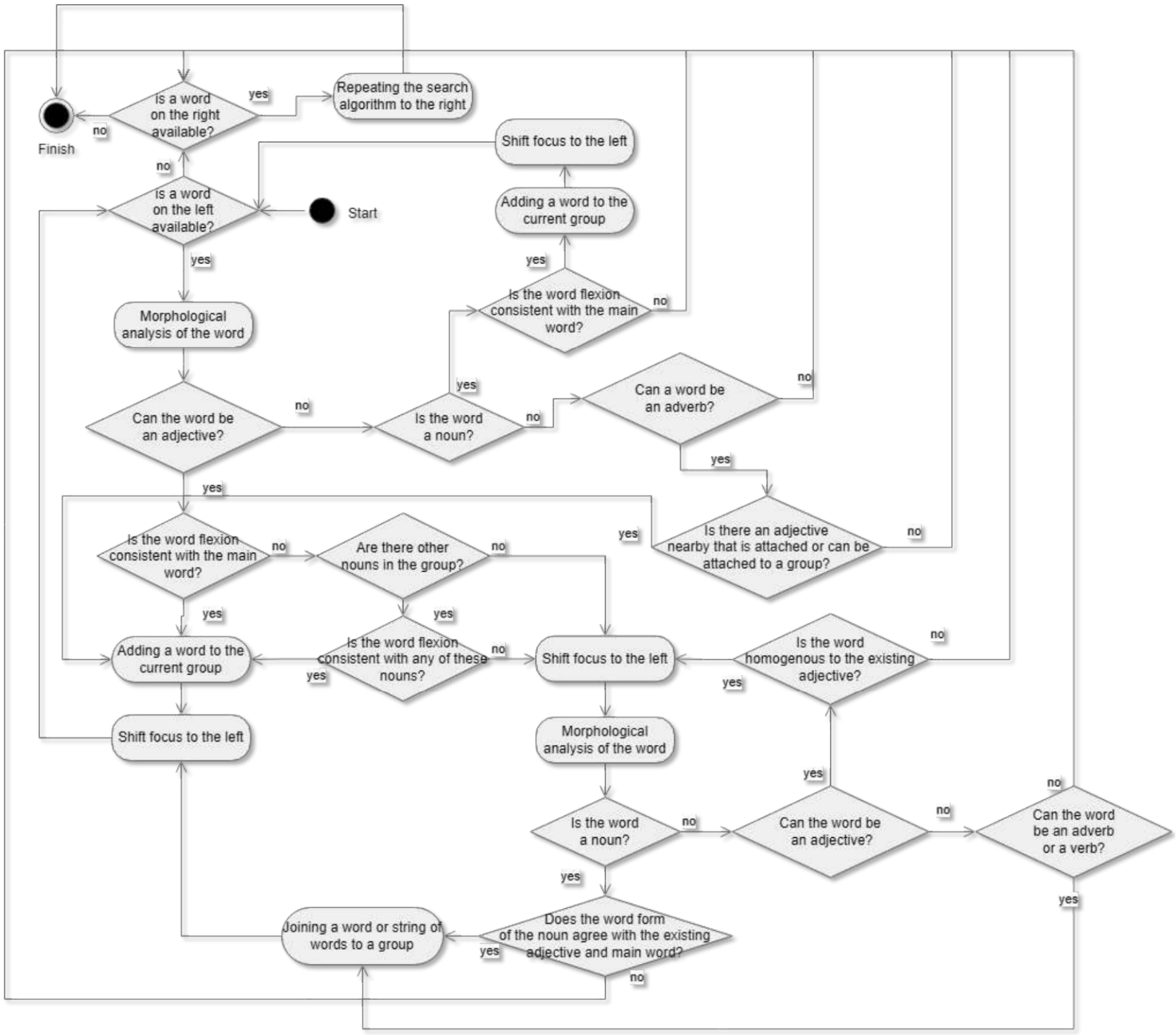

Fig. 2.5. The scheme of a name group formation

The continuation of the analysis in case of situational inconsistency of a word can be determined, for example, by the disagreement of an adjective in case with the main word of the group. However, it may be consistent with one of the subordinate nouns, which can be found in subsequent iterations. Similarly, an adverb can be added to the group in subsequent iterations if an adjective meeting the conditions for inclusion is found in the next step. The termination condition for group formation is finding a verb, numeral, adverbial participle, or an uncoordinated adjective that is not related to the main word or a word associated with it.

In this way, name groups of arbitrary length and complexity can be identified and formed.

Similar to the “Konspekt” program, name groups are decomposed into hierarchical subgroups. At the top of this hierarchy are the main nouns of the groups. Children groups are formed by adding words according to the available variants found in the text.

The extraction of name groups and their hierarchies constitutes a distinct outcome, and the corresponding structure is preserved in an XML file. However, a second crucial task is the construction of the syntax-semantics parsing tree for a sentence. The scheme of this process is presented in Fig. 2.6 This becomes more straightforward when name groups are already identified, effectively forming the subgraphs of such a tree.

In the presence of verbs and a subject group, subject-predicate pairs are initially established. In the absence of verbs, a check is performed for the existence of a nominal predicate group.

Next, verb groups are assembled, including adverbs and adverbial participles linked to verbs, as well as nouns (and name groups) in oblique cases — verification is carried out for the connection of their head words with verbs.

If there are unconnected words remaining in the analysed sequence, an analysis is conducted for the possibility of distant (disjointed) connections. This involves examining the potential correlation of these words with words in existing groups.

Subsequently, an analysis is performed for the possibility of merging connected groups based on the connectivity feature of the words they contain.

It is worth noting that the last two types of checks yield less probable binding results.

In a sentence one (simple sentence) or several connected groups can be identified. In such cases, the sentence is likely to be complex or contain clauses.

The presented schemes here are somewhat simplified. In reality, there is another feature of the developed method. Quite often, a homonymy of word forms could appear. However, in many cases, it is possible to determine the existing part of speech and its word form through the surrounding context. For this purpose, the developed program includes a set of rules that allow recognizing a significant number of such cases based on distinctive features.

Moreover, if such determination based purely on grammatical characteristics is impossible, parsing variants are selected in a way that maximizes the connectivity of the resulting parsing graph.

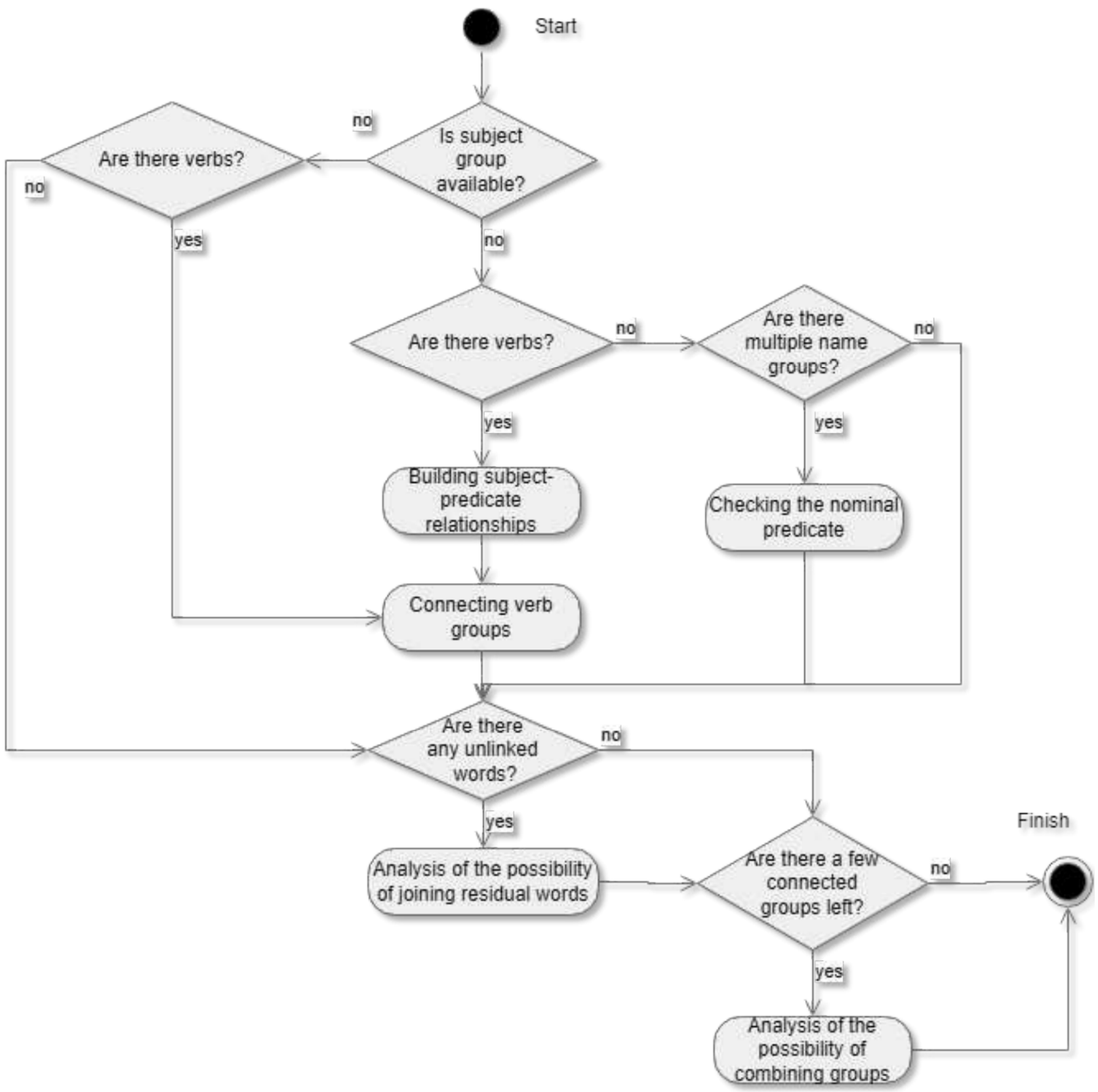

Fig. 2.6. Scheme of forming the syntactic-semantic graph of a sentence

### *2.2.2.3. Intermediate files of semantic analyzer results*

The output of both developed analysers consists of two XML files: allterms.xml and parce.xml. They are utilised for subsequent ontology creation in OWL.

Description of the allterms.xml file: The file allterms.xml represents a list of terms—nouns and name groups—contained in the analysed text along with their primary characteristics. It comprises two main sections: <export-terms> and <sentences>. The former encompasses the terms. Here is an example of term representation (for Ukrainian):

```
<term>
    <ttype>Noun_noun</ttype>
    <tname>тіло людини</tname>
    <wcount>2</wcount>
    <osn>тіл</osn>
    <osn>люд</osn>
    <sentpos>1/1</sentpos>
    <sentpos>1/2</sentpos>
    <reldown>2</reldown>
    <reldown>4</reldown>
</term>
```

The <ttype> tag denotes the sequence of parts of speech constituting the term. The <tname> tag represents the text of the term as it appears in the text. The <wcount> tag determines the number of words in the term. The <osn> tags are provided for each word, representing their stems. The <sent-pos> tags specify the positions of the term's words in the text in the format sentence number — (from 0) / word position in the sentence — (from 1). The <reldown> and <relup> tags are optional and indicate the relationship of the considered term with other terms in the file. The <reldown> points to the narrowing term—each word of which can be found in this term. The <relup> indicates the expanding term—contains all words from this term and some others. The <reldown> and <relup> tags aid in constructing the hierarchy of terms in the ontology.

The <sentences> part contains only the texts of all sentences from the analysed text within the <sent> tag each.

Description of the parce.xml file: The parce.xml file represents the syntax-semantics schema of each sentence in the text. Sentence structures are presented in container tags <sentence>. This container contains tags such as multiple <item> tags representing words and their characteristics, <sent_pos> indicating the sentence number in the text (starting from 1), and <sent> containing the text of the sentence. Here is an example of the <item> tag (for Ukrainian):

```
<item>
    <word>Книга</word>
    <osnova>кни</osnova>
    <lemma>книга</lemma>
```

```
    <kflex>a</kflex>
    <flex>ra</flex>
    <number>1</number>
    <pos>1</pos>
    <group_n>1</group_n>
    <speech>S1</speech>
    <relate>0</relate>
    <rel_type>K0</rel_type>
</item>
```

The <word> tag contains the word text as presented in the analysed text. The <osnova> tag represents the word's stem. The <lemma> tag provides the lemma—the base form of the word. The <kflex> and <flex> tags are flexions, where <kflex> is the grammatical ending, and <flex> is the variable part of the word, including cases of words that modify their stems. The <number> tag is the word's position in the sentence. The <group_n> tag indicates the word's membership in the associated group from the sentence. The <speech> tag contains the notation of the corresponding part of speech. The <relate> tag indicates the number of the word from which a semantic connection goes to the considered word. If the word has no inbound links, as in the above example, the content of this tag is set to zero. The <rel_type> tag represents the type (subtype) of the semantic connection, where the value K0 signifies the absence of a connection or its unknown type.

#### *2.2.2.4. Creation of OWL-ontology based on the results of text analysis*

The depicted general scheme of ontology creation from natural language text, as illustrated in Fig. 2.7, encompasses not only syntactic analysis and the determination of semantic relationship types but also the generation of an ontology OWL description in RDF/XML format. This format is essential for the subsequent effective processing of the presented information by software tools. Such tools may include knowledge base manipulation tools and SPARQL, such as Jena Fuseki, or graph databases like Neo4J, utilizing the Neosemantics module for handling OWL ontologies. In this context, the query language would not be the conventional SPARQL but Cypher.

In the construction of an ontology based on the materials of syntactic-semantic parsing, additional classifiers characterizing sentence structures, related groups within sentences, and even groups of sentences can be employed. This classification may take a formal nature (questions, imperatives, statements, main or subordinate clauses, types of subordinate clauses) or be based on the analysis of moods, expressed intentions, and other criteria.

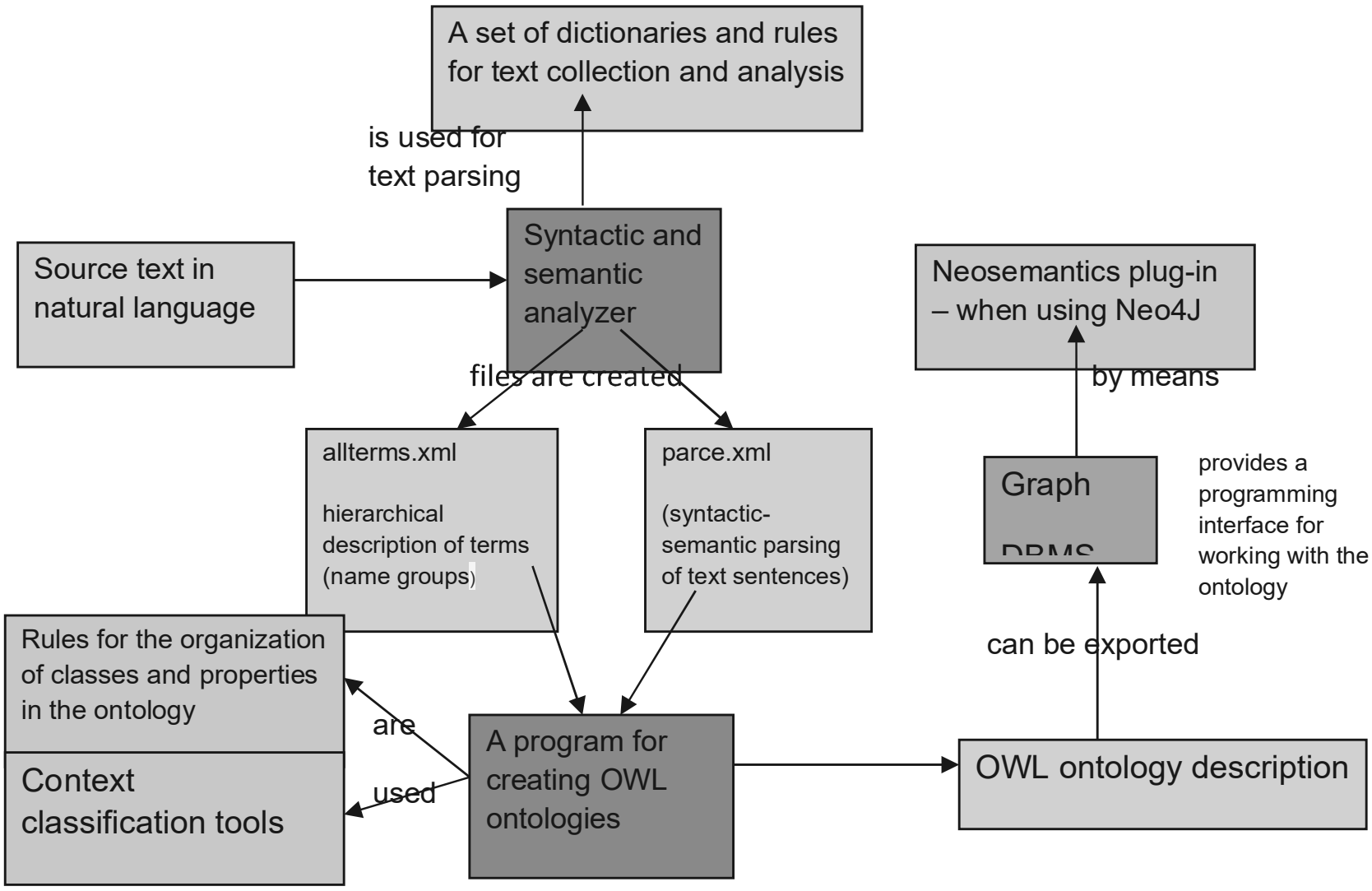

Fig. 2.7. General scheme of the procedure for automatically creating OWL-ontology based on text

## 2.3. SHORT NATURAL LANGUAGE MASSAGES ANALYSIS FOR AUTOMATIC SPARQL AND CYPHER QUERIES FORMATION

### *2.3.1. The proposed approach to formal queries creation based on natural language users' massages analysis*

Formal queries to the knowledge base are essential for working with ontology. In our work, we consider queries in SPARQL and the increasingly promising query language Cypher, used in the Neo4J graph database. It is important to note that, presently, the creation of Cypher queries based on natural language phrases is underexplored in other research, especially for the Ukrainian language. Therefore, this research direction is relatively novel and relevant.

Communication with the knowledge base involves the use of formal query languages. Consequently, when creating dialogue and reference systems with a natural language interface, there is a need for the automated generation of packages of formal queries based on natural language user queries. These queries aim to retrieve relevant information from the knowledge base expressed in natural language.

To address this challenge, we proposed an approach based on subjecting the user's natural language query to a series of checks. The results of these checks determine the set of semantic types expressed in the phrase and the corresponding concepts that specify them. The schema of a reasonably straightforward yet effective version of this method is provided in the Fig. 2.8.

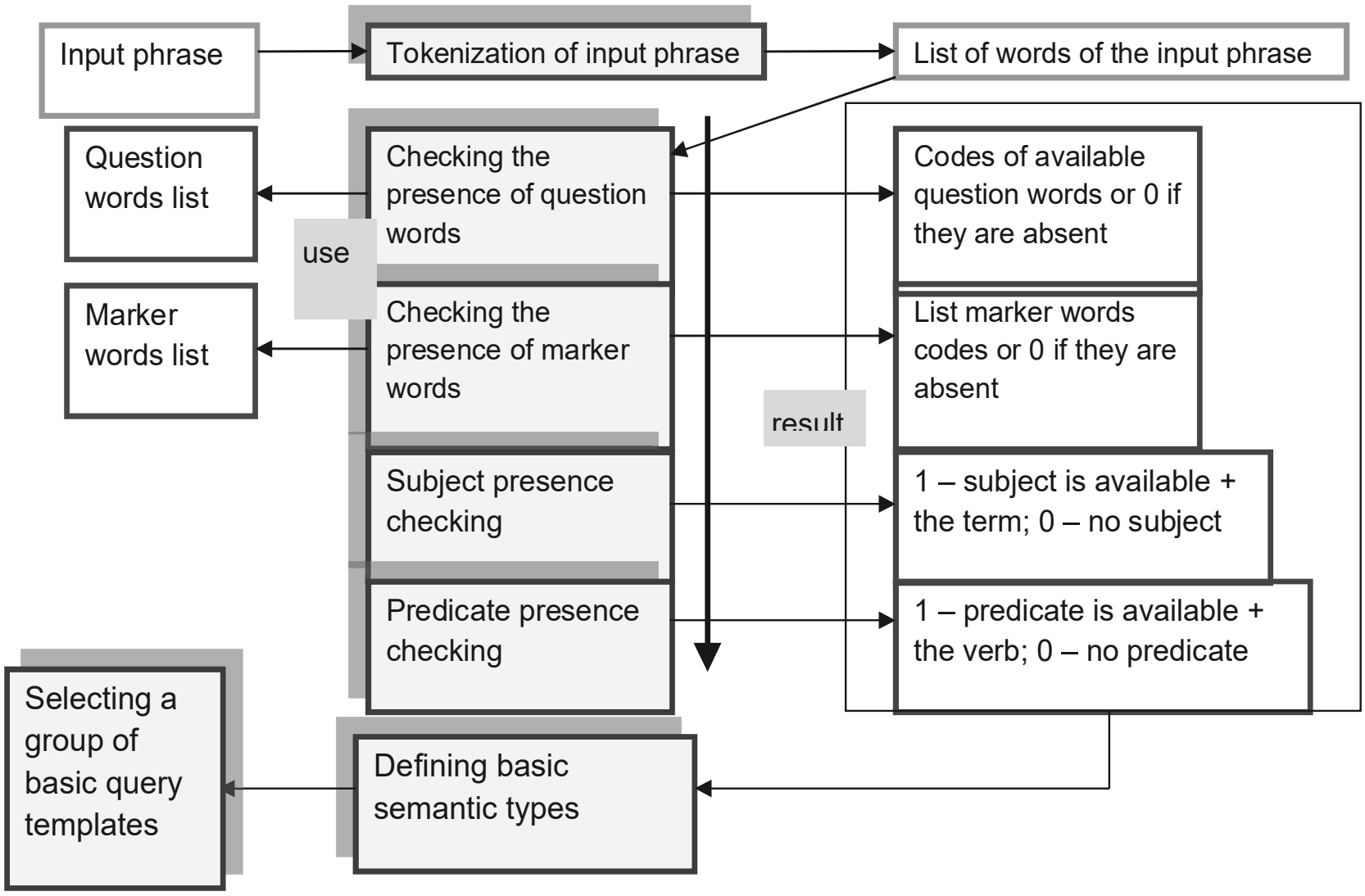

Fig. 2.8. A scheme for parsing the user's phrase to select a basic set of formal query templates

It is worth noting that for inflected languages such as Ukrainian, word order is less crucial, and the presence of specific words and their word forms is more important. Fig.2.8 illustrates the schema for determining a set of basic semantic types and corresponding primary query templates. For this purpose, the processed expression, tokenized to the level of sentence lists and corresponding words, undergoes four successive checks:

Checking for the presence of question words (a critical point for determining the type of requested information).

Checking for marker words, primarily verbs such as "located," "works," "stands," "sends," etc. These words are categorized into groups of synonyms with similar semantic colouring.

Checking for the presence or absence of the subject and the subject itself, if present.

The next important aspect after the subject is checking for the presence of a meaningful predicate, if it does not fall into the categories of words from the second check.

The result of this analysis is a set (or list of sets) of four values — result codes from the checks — as well as subjects and predicates if present. This information is sufficient for selecting a set of basic templates for formal queries. Even based on the results of such basic checks, more than a dozen basic templates can be created.

Further, additional checks follow to determine more significant semantic nuances. However, for these checks, there are no separate basic templates; otherwise, the number of templates would significantly increase, and the templates themselves would have considerable code duplication. Instead, based on the results of these additional checks, modifications (changes and additions) are made to the basic templates according to the corresponding instructions. This makes the method more flexible and simplifies both the process of automatic analysis and the construction of the corresponding software system.

### *2.3.2. XML templates for queries in the Cypher language*

Query templates are stored in the form of an XML file with a specific structure. Here, we provide examples of templates for the Cypher query language, utilised in the Neo4j graph database. Its application is more preferable due to the substantial size of the ontology created through entirely automatic syntactic-semantic analysis of the text. This preference arises from the high performance of the Neo4j database management system in handling extensive datasets.

Let us consider an example of one of the simplest of such templates:

```
<template>
    <verbose_name>Common information</verbose_name>
    <id>1</id>
    <type>base</type>
    <variables>
        <variable>
            <name>INPUT_VALUE_1</name>
            <destination>input</destination>
        </variable>
        <variable>
            <name>CONTEXT</name>
            <destination>output</destination>
        </variable>
```

```
    </variables>
    <match>
        (inp:Class)-[]-(n:Relationship),
        (n:Relationship)-[]-(x:Class),
        (n)-[:SPO]->(rel_group),
        (rel_group)-[:SPO]->(rel_sent),
        (rel_sent)-[:SPO]-(sent_super)
    </match>
    <where>
        inp.label = "INPUT_VALUE" and
        sent_super.name = "SentenceGroups"
    </where>
    <return>
        DISTINCT rel_sent.label as CONTEXT;
    </return>
</template>
```

The sections of the XML template, namely <match>, <where>, and <return>, correspond to specific sections of the formal query in the Cypher language [21]. Certain fragments of the content (text) in these sections serve as variable templates. Variables are described in the <variables> section, where each variable is defined by its name — <name> and its destination — <destination>. The destination can have values of either 'input' — indicating values substituted into the template, or 'output' — signifying variables not replaced during query formation by specific output values. Instead, they serve as references to the names and quantities of parameters whose values are obtained upon query execution. The <id> tag for the template identifier serves to match it with the result of the user phrase analysis, as well as the corresponding template for response formation. The <verbose_name> tag is included solely for human recognition of query templates during system development and maintenance. In the subsequent examples, query templates will be presented in a simplified form without XML tags. The <type> tag indicates the template type; currently, there are two types: 'base' — representing the primary template, and 'additional' — signifying an additional modifier template.

We should also dwell on the structure of additional templates. Here is an example of one of them:

```
<template>
    <verbose_name>An adjective related to the subject</verbose_name>
    <id>1</id>
    <type>additional</type>
```

```
    <variables>
        <variable>
            <name>INPUT_VALUE_ADJ</name>
            <destination>input</destination>
        </variable>
        <variable>
            <name>ADJ_PLUS</name>
            <destination>intermediate</destination>
        </variable>
        <variable>
            <name>INP_ADJ</name>
            <destination>intermediate</destination>
        </variable>
    </variables>
    <block_union>and</block_union>
    <next_item_union>or</next_item_union>
    <match>
        (inp:Class)-[]-(ADJ_PLUS:Relationship),
        (ADJ_PLUS:Relationship)-[]-(INP_ADJ:Class),
        (ADJ_PLUS)-[:SPO]->(rel_group)
    </match>
    <where>
        INP_ADJ. label = "INPUT_VALUE_ADJ"
    </where>
    <return></return>
</template>
```

This type of template also includes the <match>, <where>, and <return> blocks. The content of these sections is added to the corresponding blocks of the base template. Each of these blocks can be empty or absent. Unique features of additional templates include the <block_union> and <next_item_union> tags. The <block_union> tag contains the type of union for the entire formed <where> block with the base template. The <next_item_union> tag indicates the type of union for repeated elements of the <where> block if the corresponding input variable is represented as a list (array). For example, in the above template, the variable INPUT_VALUE_ADJ may correspond to a series of adjectives related to the subject. The parameter values for <block_union> and <next_item_union> can be "and" or "or." Additionally, intermediate variables constitute a third type (<destination>) specific to additional templates. These variables do not participate in data transmission to the query or in their direct retrieval. Their main feature is that when repeating the block during query formation, these variables are not fully duplicated but are added with the next sequential number, such as ADJ_PLUS_1, ADJ_PLUS_2, ADJ_PLUS_3, and so forth.

### *2.3.3. The method of automatically forming requests according to templates*

Let's delve deeper into the structure of formal queries and the method of their formation. The ontology's structure allows for targeted search of both contexts and individual concepts, considering the presence of these concepts in the context and their relatedness based on a specific semantic type criterion. In the proposed scheme, there is a basic query template aimed at obtaining information of a specific type in the specified form, along with additional modifier templates that optionally construct query strings in corresponding blocks of the main query, introducing additional conditions. Let's examine examples of some query templates.

Let's start with perhaps the simplest one mentioned earlier. This template is designed to obtain the context (sentence) in which the queried single concept (word) is not just present but forms a connection with other words. This ensures that the concept "organically" fits into the context.

In Cypher, queries are divided into three main blocks: MATCH, WHERE, and RETURN. The MATCH block specifies the pattern of relationships between nodes in a directed graph. The WHERE block imposes conditions on the properties (characteristics) of the nodes and/or relationships specified in the MATCH block. The RETURN block indicates what should be output as a result and under what name (alias). In this case, there is a specific class marked with the variable 'inp.' In the WHERE block, we imposed a condition that the label property of the inp node should be equal to the queried concept (hereafter, in query templates, INPUT_VALUE represents the text of the input concept). In the MATCH block, it is specified that inp is a node (enclosed in parentheses) of type Class. It is connected to another node 'n' that has a type Relationship (property in OWL). The type of the relationship is not defined (square brackets are empty), and the direction of the relationship is not indicated. This means it can be bound both to the DOMAIN and the RANGE. Specifying directions is unnecessary since it is known that such relationships are constructed from the property to the class. Additionally, we specified that this property should also be associated with a certain class 'x'. Then, we indicated that the property uniting these classes should relate to some sentence 'rel_sent.' The condition 'sent_super.name = "SentenceGroups"' guarantees that 'rel_sent' is indeed a sentence.

As a result, we request to output 'rel_sent.label,' which contains the sentence context under the alias CONTEXT.

Let's consider a somewhat more complex example. We want to inquire about the known characteristics (definitions) of the object INPUT_VALUE in the ontology. In other words, what can INPUT_VALUE be (or is)? The query will take the following form:

```
MATCH (inp:Class)-[]-(n:Relationship),
    (n:Relationship)-[]-(x:Class),
    (n)-[:SPO]->(prop_type_1),
    (n)-[:SPO]->(rel_group),
    (rel_group)-[:SPO]->(rel_sent),
    (rel_sent)-[:SPO]-(sent_super)
WHERE
    inp.name = "INPUT_VALUE" and
    (prop_type_1.label = "object property" or
    prop_type_1.label = "action property" or
    prop_type_1.label = "action separability" or
    prop_type_1.label = "impact level")
    and
    sent_super.name = "SentenceGroups"
RETURN DISTINCT x.label as result, rel_sent.label as context;
```

Compared to the previous example, in the MATCH block, one line has been added: (n)-[:SPO]->(prop_type_1). This provides information that the property 'n' must be a child in relation to 'prop_type_1' (relationship type). Here, we explicitly specify the direction of the relationship. In the WHERE block, we specify 'prop_type_1' through possible values of the label parameter. To make the template more universal, as we a priori do not know whether INPUT_VALUE is a noun or a verb, several options for the value of prop_type_1.label are provided through logical OR. With the introduction of additional hierarchy of semantic relations into the ontology, this construction can be simplified:

```
MATCH (inp:Class)-[]-(n:Relationship),
    (n:Relationship)-[]-(x:Class),
    (n)-[:SPO]->(prop_type_1),
    (n)-[:SPO]->(rel_group),
    (rel_group)-[:SPO]->(rel_sent),
    (rel_sent)-[:SPO]-(sent_super),
    (prop_type_1)-[:SPO]->(prop_type_category)
WHERE
    inp.name = "INPUT_VALUE" and
```

```
    prop_type_category.label = “property types”
    and
    sent_super.name = “SentenceGroups”
RETURN DISTINCT x.label as result, rel_sent.label as context;
```

As a result, we output the label for the nodes ’x’. This represents the characteristics (properties) of the object ’inp.’ Additionally, we inquire about the sentence text to understand the context in which this property is mentioned.

Similarly, one can inquire about the actions of the object. For this, it is only necessary to specify a different value for prop_type_1.label in the WHERE block, specifically: prop_type_1.label = “object-action”.

In the case of using a condition with multiple possible types of relationships (prop_type_1.label), the obtained type value can also be included in the result, aiding in response synthesis. Let’s provide an example where the location of an object is queried without specifying the type of localization (“Where is INPUT_VALUE located?”).

```
MATCH (inp:Class)-[]-(n:Relationship),
    (n:Relationship)-[]-(x:Class),
    (n)-[:SPO]->(prop_type_1),
    (n)-[:SPO]->(rel_group),
    (rel_group)-[:SPO]->(rel_sent),
    (rel_sent)-[:SPO]-(sent_super)
    (prop_type_1)-[:SPO]->(prop_type_category)
WHERE
    inp.label = “INPUT_VALUE” and
    prop_type_category.label = “localization types” and
    sent_super.name = “SentenceGroups”
RETURN DISTINCT x.label as result, rel_sent.label as context,
    prop_type_1.label as predicate;
```

The main distinction here is the presence of the prop_type_1.label as predicate expression in the RETURN block. This allows for returning the specific semantic type of the obtained result. It should be considered when generating response text.

In some cases, in addition to predicates in queries, lists of concept variants (characteristic verbs, nouns, adjectives) can be used. The main difference is that conditions are imposed on the ontology vertex associated with x.label. In other words, the queried object must be linked to a certain concept ’x’ with a relationship of a specific type, for example, “object-action,” and this ’action’ should be described by the text parameter label as one of those listed in the set. In the future, there are plans to introduce an a priori (independent of the analyzed text) classification of actions, features, and concepts into the ontology, which will simplify the query and eliminate the need for such lists — the concept 'x' simply needs to be a child in relation to, for example, verbs of a certain category.

Let's discuss template modifiers separately — fragments that are added to the main query templates. For instance, the input parameter is not a single word but a name group, i.e., connected nouns and adjectives. To associate adjectives with the input concept, the following lines should be added to the respective blocks:

```
In the MATCH section:
    (inp:Class)-[]-(adj_plus:Relationship),
    (adj_plus:Relationship)-[]-(inp_adj_1:Class),
    (adj_plus)-[:SPO]->(rel_group)
In the WHERE section:
    and
    inp_adj_1.label = "INPUT_VALUE_ADJ"
```

Similar blocks can be added for additional adjectives with variables like inp_adj_2, inp_adj_3, and so forth. Conditions for the presence of a noun in the genitive case can be added to the query using the following blocks:

```
In the MATCH section:
    (inp_noun_1:Class)-[]-(noun_plus:Relationship),
    (noun_plus)-[:SPO]->(rel_group)
In the WHERE section:
    and
    inp_noun_1.label = "INPUT_VALUE_NOUN"
```

Here, a condition is added only for the inclusion of this additional noun in the same group as the main queried concept. Conditions for the presence of associated adjectives with this noun can also be imposed:

```
In the MATCH section:
    (inp_noun_1:Class)-[]-(adj_plus_add:Relationship),
    (adj_plus_add:Relationship)-[]-(inp_adj_add:Class),
    (adj_plus_add)-[:SPO]->(rel_group)
In the WHERE section:
    and
    inp_adj_add.label = "INPUT_VALUE_ADJ_ADD"
```

In specific cases, it may be necessary to attach a predicate of negation to the query. To achieve this, conditions for inclusion in the group of the negation particle or another negation predicate are added to the query:

In the MATCH section:

```
(neg:Class)-[]-(neg_rel:Relationship),
(neg_rel)-[:SPO]->(rel_group)
```

In the WHERE section:

```
and
(neg.label = "no" or
neg.label = "not" or
neg.label = "forbidden" or
neg.label = "prohibited" or
neg.label = "unnecessary" or
neg.label = "impossible" or
neg.label = "needlessly")
```

### *2.3.4. Automatic generation of SPARQL queries to contextual ontology using the example of a knowledge base of EBSCO medical articles*

The system receives a textual message as input. Initially, the text is cleaned from disallowed characters, which, in this case, include Latin (English) alphabet letters, whitespace, period, hyphen, paragraph symbols, and line breaks. All other characters are considered disallowed and are removed for further processing. Tokenization of the text into individual words is performed using NLTK library tools, along with the identification of their parts of speech. The words are lemmatized — brought to their base form, and stop-words are removed. Stop-words include articles, conjunctions, prepositions, pronouns, question words, auxiliary and modal verbs, and particles. The list of stop-words has been extended to include common conversational phrases such as "give," "show," and "present," which, while prevalent in interactions with the help system, do not provide informative content in this context. Additionally, the text is purged of words not included in the list of concepts presented in the knowledge base. Consequently, the processed text represents a list of meaningful words brought to their base form.

Subsequently, the system determines the specific semantic category or set of categories expressed in the given message. Each of these categories corresponds to a separate SPARQL query. Currently, 26 such query categories have been implemented: "synonyms," "symptoms," "indications," "patient education," "description," "reimbursement," "test," "rule out," "treatment summary," "relations," "prognosis," "diagnosis need for treatment," "contraindications precautions," "causes," "pathogenesis," "g-code," "any_code," "any_icd," "icd 9," "icd 10," "icd 11," "hcpcs," "cpt," "articles," "references," and "contexts." However, there is room for expanding their number. The determination of a specific semantic category is based on the presence of certain marker words in the obtained list. The mapping of these categories to marker words is implemented in the form of a dictionary in the file marker_words.json. In this dictionary, each of the specified categories is associated with a list or set of lists of marker words. For example:

```
„icd 10“: [
    [„icd“, „10“, „code“],
    [„icd-10“, „code“],
    [“icd”, “10”],
    [“icd-10”]
],
“risk factors”: [
    [“higher”, “risk”, “factors”],
    [“risk”, “factor”],
    [“risk”]
]
```

The list of words undergoes verification across all available categories. During this process, it is noted which specific marker words from the list identify each particular category. If the lists of markers intersect, the longest one is selected, containing all words found in the analyzed list. Words that belong to a syntactic-semantic group (sentence or sentence part) and were not identified as markers for any of the categories are marked as query parameters. The semantic category determines the query type. If no existing special semantic category is assigned to a syntactic-semantic group, it is categorized as “contexts,” and the words themselves become parameters for the corresponding query.

The output returns lists of identified special semantic categories and their corresponding words — query parameters.

A dedicated module within the software system is responsible for the generation of SPARQL queries. Each identified special semantic category corresponds to its own SPARQL query template. The templates are stored in the form of a JSON dictionary in a file. Here is an example of such a template:

```
“description”: {
“verbose”: “description for the case of”,
“parts”: [{
“type”: “constant”,
“n”: 1,
```

```
“body”: [“PREFIX rdf: <http://www.w3.org/1999/02/22-rdf-syntax-ns#>”,
          “PREFIX rdfs: <http://www.w3.org/2000/01/rdf-schema#>”,
          “PREFIX owl: <http://www.w3.org/2002/07/owl#>”,
          “PREFIX xsd: <http://www.w3.org/2001/XMLSchema#>”,
          “PREFIX name: <http://www.semanticweb.org/ContextOntology#>”,
          “SELECT DISTINCT ?topic ?context_text ?article_name ?scope”,
          “WHERE {“
    ]},
{
     “type”: “listed input”,
     “n”: 2,
     “body”: [“?word_>_order_< rdfs:label >_input_<@en.”,
          “?word_>_order_< name:relate_to_context ?context.”
          ]},
{
“type”: “constant”,
“n”: 3,
“body”: [
     “?context rdf:type ?context_class.”,
     “?context name:relate_to_article ?article.”,
     “?article rdfs:label ?article_name.”,
     “?title name:relate_to_article ?article.”,
     “?title rdf:type name:cl_title.”,
     “?title rdfs:label ?scope.”,
     “?out_context name:relate_to_article ?article.”,
     “?out_context rdf:type ?out_context_class.”,
     “?out_context rdfs:comment ?context_text.”,
     “?out_context_class rdfs:label ?topic.”,
     “FILTER ((?out_context_class = name:cl_description) &&”,
          “(?context_class = name:cl_title ||”,
          “?context_class = name:cl_description ||”,
          “?context_class = name:cl_synonyms))”,
          “.} ORDER BY ?scope ?topic”]}],
“outputs”: {
“scope”: {“sortage”: “primary”, “verbose”: “Scope: “},
“topic”: {“sortage”: {“by each”: “scope”}, “verbose”: “Topic: “},
“context_text”: {“sortage”: {“by each”: “topic”}, “verbose”: “”},
“article_name”: {“sortage”: {“group”: “scope”},”verbose”: “Related articles:”}}}
```

In this example, the keys in the dictionary represent the names of the corresponding special semantic categories, here, “description” (description of the specified phenomenon). The values are dictionaries with the following keys:

"verbose" — a phrase fragment preceding the response.

"parts" — a list of sections forming the SPARQL query (the main section).

"outputs" — instructions for structuring the response data obtained during the query execution.

Let's examine the "parts" section in more detail. The value for the key "parts" is a list, each element of which corresponds to a query section. The template for a query section is a dictionary with the following keys:

"type" — the type of the section. The following types are anticipated: "constant" — the section is inserted into the query without modifications; "listed input" — for each of the provided parameters (words), the section is inserted into the query. In this case, the part >input< is replaced with the value of the respective input parameter of the query, and the part >order< is replaced with an incremental integer value (1, 2, 3, ... n), converted to a string type to avoid repetition of the same query variable; "single input" — instead of the part >input<, one parameter is inserted only once, and for the next parameter, if any, a completely new query is generated.

"n" — the order of the section template in query formation to prevent mixing of sections.

"body" — a list of query lines to be formed. The lines may contain placeholders for input parameters >input< and incremental values >order<.

The functioning of the module essentially involves selecting the appropriate templates from the dictionary and concatenating the query sections in the specified order, replacing the corresponding placeholders with input parameters when necessary.

The "outputs" section, like "verbose," is not directly used by the program module but is passed along with the query and is necessary for further response formation.

The execution of SPARQL queries, as well as the storage of the knowledge base, is performed by the Jena Fuseki database management system. Interaction with it is facilitated by the process_queries.py module using the SPARQLWrapper library tools. Since the ontology is distributed — divided into parts, each query from the received package is executed separately for each part. Executing SPARQL queries is a relatively slow process. To expedite the process, queries to all ontology parts are executed concurrently in multiple threads. Substantive responses may be obtained from one or several ontology parts. The tables obtained from several ontology parts are merged.

## 2.4. USAGE OF LARGE LANGUAGE MODELS FOR NATURAL LANGUAGE TEXTS GENERATION BASING ON DATA EXTRACTED FROM ONTOLOGY

### *2.4.1. Natural language phrases building basing on semantic structures*

The automatically generated ontology, derived from natural language text, comprises entities and semantic relationships that connect them. These relationships, specified by entities, are associated with groups of expressions, which, in turn, are linked to sentences in the input text. Consequently, with such a semantic representation, it is possible to formulate a natural language sentence. For this task, the template-based approach, as described in [22], can be employed. However, with the advancement of approaches based on the use of deep learning neural networks, epitomized by large language transformer models such as ChatGPT, the consideration arises to explore them as tools for synthesizing natural language sentences based on semantic structures. An attempt in this direction was made in the scope of our work.

As a test ontology, we utilised a knowledge base created from the text "Composition of a Computing System." Since the knowledge base was managed by the graph database Neo4J, the query language used for it is Cypher.

Below is the text of a test query to the specified ontology for obtaining the text of a specific sentence (for result comparison) based on its ID. It also retrieves the corresponding set of semantic categories and the related pairs of concepts (main and dependent) for this sentence. The results of such a query served as input data for the task of reverse synthesis of a natural language sentence:

```
MATCH (inp:Relationship)-[:SPO]->(inp_type:Relationship),
    (inp:Relationship)<-[:SPO]-(linked_group:Relationship),
    (linked_group:Relationship)-[:SPO]->(linked_group_type:Relationship),
    (linked_group:Relationship)<-[:SPO]-(certain_words_link:Relationship),
    (certain_words_link:Relationship)-[:SPO]->(sem_type:Relationship),
    (sem_type:Relationship)-[:SPO]->(w_link_type:Relationship),
    (certain_words_link:Relationship)-[:DOMAIN]->(main_entity:Class),
    (certain_words_link:Relationship)-[:RANGE]->(dependent_entity:Class)
WHERE
    inp_type.name = "SentenceGroups" and
    linked_group_type.name = "Groups" and
    w_link_type.name = "WordsLink" and
```

ID(inp) = specify sentence ID
RETURN DISTINCT ID(inp) as id, inp.label as text, main_entity.label as main_entity, dependent_entity.label as dependent_entity, sem_type.label as sem_type;

In this query, the specified ontology is queried for a sentence with a particular ID. The result includes the ID, text, main entity, dependent entity, and semantic type of the sentence, along with the associated semantic categories.

The semantic structure obtained from the ontology appears to be sufficient for constructing a coherent natural language sentence of corresponding content.

To initiate the synthesis task in a large language model such as ChatGPT, it is necessary to provide a relevant instruction or prompt. As mentioned earlier, it is preferable to use the English language for such instructions. The instruction itself is structured in JSON format. The corresponding text of the prompt is provided below:

{
"Intriduction": «You are an expert in knowledge engineering and ontologies as well as in meaningful text generation in inflect languages. You will be provided with data obtained from some ontology through a query. The ontology was made automatically basing on the results of semantic analysis of a natural language text. The results are pairs of lemmatized words ("main entity" and "dependent entity") accompanied with a name of syntactic-semantic relationship that linked them in the certain sentence.",
"Action to perform": «Assuming that all the data you will be provided belong to one sentence you are to make a try to restore the original sentence using such a prompt. Language of the ontology, input and output data is Ukrainian.",
"Restrictions": «Do not put the semantic relationships as a phrase as it given in the sentence you generate, it will be definitely wrong. It is just a prompt for syntactic linking. Remember that the provided words are lemmatized, so you are to put them in a correct form according to other entities of the sentence and the given syntactic-semantic relationships of the prompt.",
"Additional data to provide": «Also provide an estimated value of probability that the generated sentence corresponds the intent of the prompt given.",
"The essence of the syntactic-semantic relationship names and meaning explanation":
{
"object property: «the dependent entity express a property or some characteristic, or quality of the main entity. When the response sentence generation you should use the dependent entity as an adjective with the main entity which is noun",
"action property": "the dependent entity express a property or some characteristic, or quality of the main entity which is an action. When the

```
        response sentence generation you should use the dependent entity as an
        adverb with the main entity which is verb”,
        “quality change”: “the dependent entity express that the main entity may
        be subjected to some quality changes, which may follow from the other
        context”,
        “destination”: «the dependent entity express the destination of the main
        entity”,
        “object”: «the object (noun) affected throw the action expressed by the
        main entity”,
        “object / action”: «the main entity performs an action expressed by the
        dependent entity”,
        “preposition binding”: «merely shows that the main entity here in the con-
        text of the provided sentence is to be used with the preposition which is the
        dependent entity. This means that you should use this preposition with the
        main entity when the response sentence generation”,
        “possession”: “the dependent entity or somewhat relates to the main entity.
        When generation this usually should be expresses using genitive case”,
        “equality”: “the different name of the entity or an equivalent entity”,
        “objective entry”: “the main entity is a part or member of the dependent
        entity”,
        “state”: “a state or a constant characteristic of the main entity if it is noun
        or an entity linked to in if it is a verb”
    },
    “Input data”: [
        {
            “main entity”: “some word 1”,
            “dependent entity”: “some word 2”,
            “semantic relationship”: «semantic category 1”
        },
        ........................
        ........................
        {
            “main entity”: “some word n”,
            “dependent entity”: “some word n+1”,
            “semantic relationship”: “semantic category n”
        }
    ]
}
```

In the provided instruction, the “Introduction” section establishes initial settings for the large language model concerning its subsequent behavior and offers basic clarification of the input data.

The “Action to Perform” section formulates the direct task to be executed.

The "Restrictions" section provides additional instructions regarding the formed output text and attempts to eliminate ambiguity in the interpretation of the instruction.

The "Additional Data to Provide" section, in this case, serves to instruct the model to conduct its own assessment of the task's performance quality.

In the "The Essence of the Syntactic-Semantic Relationship Names and Meaning Explanation" section, a dictionary is presented, offering explanations for the interpretation of specific types of semantic relationships and how to use them in constructing the resulting sentence. Given the ontology's substantial number of semantic categories and the character limit for ChatGPT's input messages, the practical volume of such a dictionary can be constrained to the semantic categories present in the given sentence.

Direct pairs of entities and their corresponding semantic relationships are enumerated in the list of dictionaries under "Input Data."

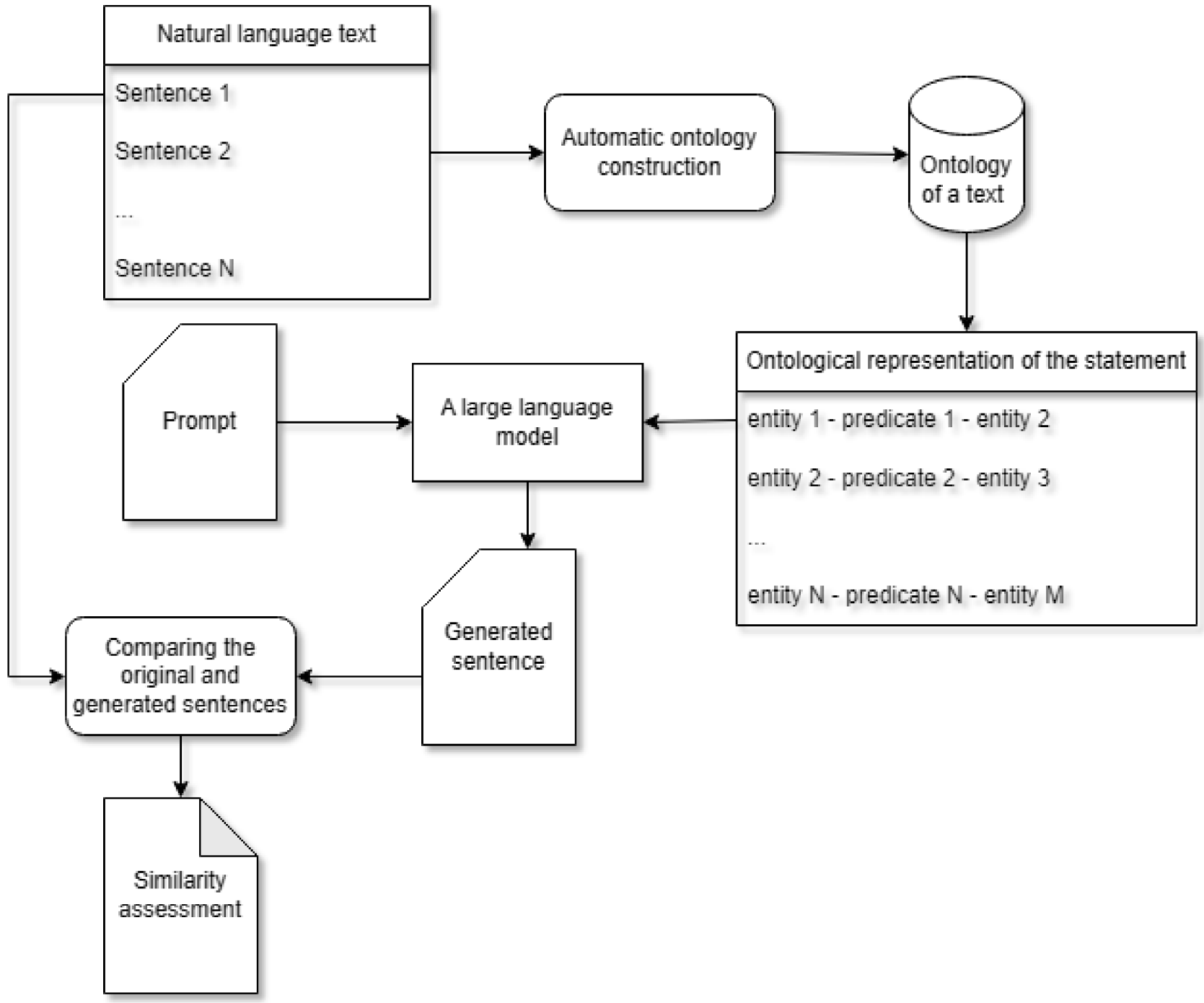

Fig. 2.9. The general scheme of generating natural language sentences based on an ontological representation using a large language model

The output yields a formulated natural language sentence and the model's own assessment, provided by ChatGPT, of the likelihood that the sentence was constructed accurately and corresponds to the original (whose appearance the model is unaware of).

The general scheme of the phrases formation process grounding of their ontological representation is given in Fig. 2.9.

#### *2.4.2. An experiment on the reverse synthesis of natural language sentences based on their ontological representation with the involvement of a large language model*

The essence of the experiment was as follows: from a test ontology created based on the text "Composition of a Computing System," individual sentences and their corresponding entity pairs with semantic categories linking them within the given sentence were extracted using a Cypher language query. Subsequently, employing a specialised prompt instruction, the extensive language model (ChatGPT) was tasked with generating a grammatically correct sentence in Ukrainian based on a set of entity pairs under the specified semantic relationships.

The prompt instruction elucidates to the model the existing semantic categories, particularly in the context of sentence creation. Importantly, the original sentence was not provided as input to the system.

As a result, the generated sentence and the model's self-assessment of the likelihood that the sentence was faithfully reproduced were returned.

For testing purposes, 10 sentences from the specified text were utilised.

The cosine similarity index was used to compare the ontological representation of the sentence formed on the basis of the original sentence. Cosine similarity is a measure of similarity between two pre-Hilbert space vectors used to measure the cosine of the angle between them. So, if there are two feature vectors (A and B), then the cosine similarity, cos(θ), can be represented using the scalar product and norm (2.1):

$$similarity = \cos(\theta) = \frac{A \cdot B}{\|A\|\|B\|} = \frac{\sum_{i=1}^{n} A_i \times B_i}{\sqrt{\sum_{i=1}^{n} (A_i)^2} \times \sqrt{\sum_{i=1}^{n} (B_i)^2}} \qquad (2.1)$$

In the context of information retrieval, the cosine similarity between two documents varies from 0 to 1. This is due to term frequency (tf-idf weights) not being negative, and the angle between two term frequency vectors cannot exceed 90°. Cosine similarity serves as an effective evaluative measure, particularly for sparse vectors, as it considers only non-zero dimensions.

The "soft" cosine measure accounts for similarity between pairs of features. Traditional cosine similarity treats features in the vector model as independent or entirely separate, while the "soft" cosine measure acknowledges the similarity of features in the vector model. This allows for the generalization of the cosine similarity concept and the concept of object similarity in the vector space.

In the field of natural language processing, the similarity between objects is quite intuitive. Features such as words, N-grams, or syntactic N-grams may exhibit substantial similarity, although they may formally be considered different features in the vector model. For N-grams or syntactic N-grams, Levenshtein distance can be applied (in addition to words).

To compute the "soft" cosine measure, a similarity matrix (denoted as s) between features is introduced. It is calculated using Levenshtein distance or other similarity measures, such as various WordNet similarity measures. Subsequently, multiplication is performed using this matrix.

If there are two N-dimensional vectors, a and b, the "soft" cosine measure is computed as follows (2.2):

$$soft_cosine_1(a,b) = \frac{\sum_{i,j}^{N} s_{ij} \cdot a_i \cdot b_j}{\sqrt{\sum_{i,j}^{N} s_{ij} \cdot a_i \cdot a_j} \cdot \sqrt{\sum_{i,j}^{N} s_{ij} \cdot b_i \cdot b_j}} \tag{2.2}$$

Here, $s_{ij}$ represents the similarity between feature i and feature j. In the absence of similarity between features (sii = 1, sij = 0 for i ≠ j), Equation (2.2) is equivalent to the commonly accepted formula for cosine similarity.

Since performing mathematical computations directly on strings is strictly impossible, and the calculation of a measure such as cosine similarity requires the existence of vectors, natural language texts undergo vectorization for processing and analysis. In the context of Natural Language Processing (NLP), sentences are represented as vectors in a multidimensional space, employing methods such as word embeddings. To obtain vector representations of sentences in our case, the Python library spaCy was utilised, incorporating language models uk_core_news_lg for the Ukrainian language and xx_ent_wiki_sm, which is multilingual. The TF-IDF method was also employed. Implemented methods within spaCy were engaged for computing cosine similarity values.

The values of quantitative assessments, characterizing the proximity of sentences generated by a large language model to the original, are presented in Table 2.1. The table compares the cosine similarity values under different methods of vector representation for the analysed sentences (original and generated). Additionally, a subjective assessment of the likelihood of accurate reproduction from ChatGPT is provided. It is important to note that this assessment cannot be considered a fully objective measure but rather serves as a guide and an evaluation of the model's self-critique within the GPT framework.

Table 2.1

**Quantitative Assessments of the Quality of Reverse Synthesis of Sentence s from Ontological Representation**

| Assessment of the probability of correct reproduction of the phrase by ChatGPT | | Cosine similarity | | | | | |
|---|---|---|---|---|---|---|---|
| | | Model xx_ent_ wiki_sm | | Model uk_core_ news_lg | | Model tf-idf | |
| Mean value ± confidence interval | Variation interval | Mean value ± confidence interval | Variation interval | Mean value ± confidence interval | Variation interval | Mean value ± confidence interval | Variation interval |
| 0.845 ±0.037 | 0.75–0.90 | 0.8716 ±0.0335 | 0.8193–0.9722 | 0.8108 ±0.1224 | 0.4067–0.9653 | 0.2927 ±0.1718 | 0.0607–0.7745 |

From the presented results, it is evident that the quantitative assessment of cosine similarity is highly dependent on the method of vector representation applied to the analysed texts.

Immediate observations reveal that the language models xx_ent_wiki_sm and uk_core_news_lg lead to quite high cosine similarity values (0.8716 and 0.8108, respectively). Meanwhile, a simpler vectorization method based on tf-idf yields significantly lower mean values and a larger range of variation. Let's analyse this behaviour.

The xx_ent_wiki_sm model (multilingual) demonstrates a narrow range of variation and a relatively high mean cosine similarity value. The reduction in the mean score when using the uk_core_news_lg model (for the Ukrainian language) is attributed to a greater downward variation. However, the maximum obtained values for these two models are quite close. In simpler terms, the application of the uk_core_news_lg model in certain cases results in a significantly lower cosine similarity score.

Comparing the cosine similarity scores obtained from the vectorization models xx_ent_wiki_sm and uk_core_news_lg is illustrated in Fig. 2.10 (a), revealing an absence of any significant correlation between the obtained values. The R-squared value is merely 0.0006. These models perceive natural language text somewhat differently.

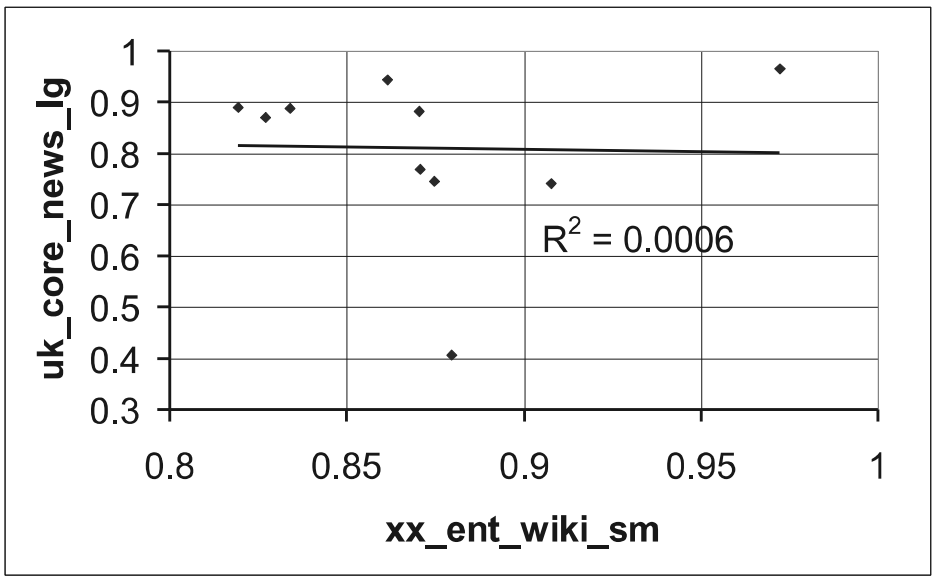

a)

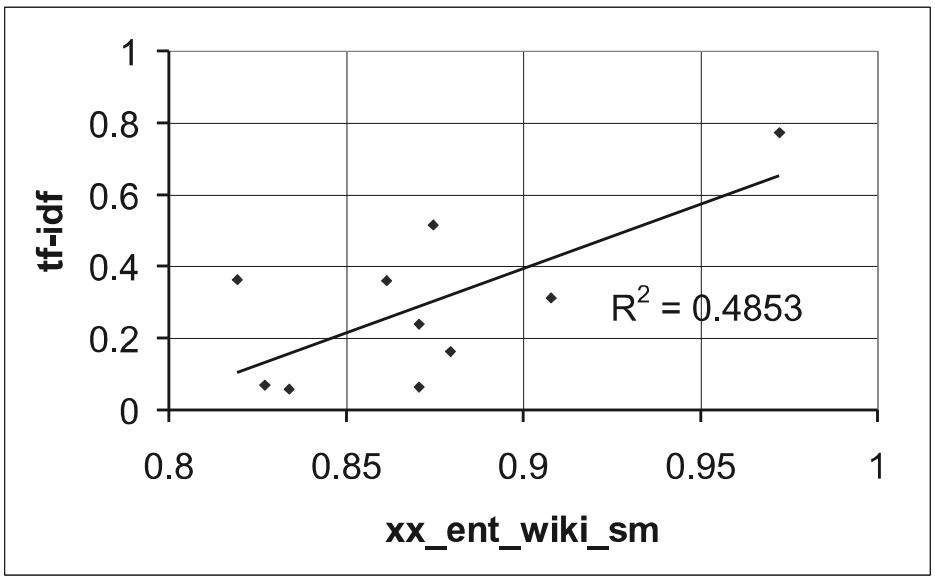

b)

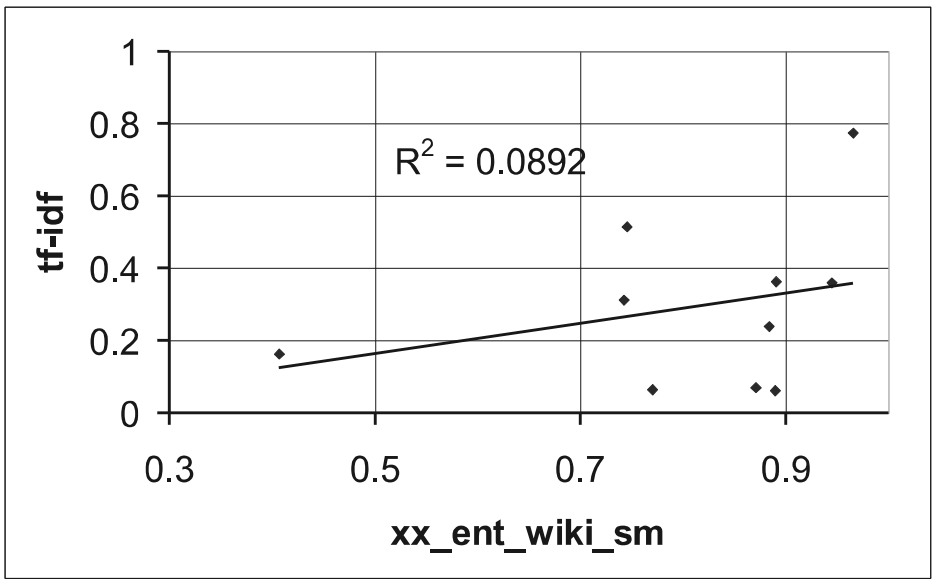

c)

Fig. 2.10. Comparison of cosine similarity values of original and formed sentences when vectorizing the text according to different models:

a) xx_ent_wiki_sm / uk_core_news_lg;
b) xx_ent_wiki_sm / tf-idf;
c) uk_core_news_lg / tf-idf

Analysis of directly generated sentences revealed that when vectorizing with the uk_core_news_lg model, reduced cosine similarity scores occur in cases of generating complex, branching sentences, while the original sentence is considerably simpler, albeit semantically close. For instance, the original sentence was: "The hard disk (Winchester) is also part of the system unit." From its ontological representation, the large language model generated the following sentence: "The disk, known as Winchester, is hard and is part of a block that is the system." It is evident that the second (generated) sentence appears somewhat unnatural, filled with unnecessary entities and convolutions, although it is close in meaning to the original. The xx_ent_wiki_sm model seems less sensitive to such manifestations and provides a cosine similarity score of 0.8795 (close to the sample mean). Simultaneously, the uk_core_news_lg model yields a score of 0.4067, significantly lower. This trend is observable multiple times, albeit with a smaller difference — the uk_core_news_lg model is more sensitive to the formation of formally twisted redundant phrases, whereas xx_ent_wiki_sm is more inclined towards content analysis rather than form.

However, as seen from the graph, there is no clear correlation between the models. There are instances where uk_core_news_lg gives a higher cosine similarity score, and xx_ent_wiki_sm, accordingly, a lower one. A specific analysis suggests that in such cases, while preserving a lexically and syntactically close structure to the original, the content was somewhat distorted. An example of this can be the following case. Original phrase: "The main device of the computer is the central processor."; generated phrase: "The device that belongs to the computer is an object where the central processor is the main device." Here, we observe a distortion in the stylistics of the output phrase. Additionally, there is a distortion in the content. In the generated sentence, it implies that the computer has some device, where the central processor is the main device, which does not fully correspond to the original sentence.

At the same time, there is a noticeable correlation in the cosine similarity scores between the xx_ent_wiki_sm model and the scores obtained for the tf-idf method. The highest scores for these methods were obtained when the sentences practically matched: "The content of this memory is stored only when the power is on." and "The content of the memory is stored only when the power is on." Regarding the lowest scores, the tf-idf method also proves to be more sensitive to deformed sentences, hence the scores will be numerically lower. However, the content correlation is preserved. Meanwhile, the uk_core_news_lg model weakly correlates with the tf-idf method.

Thus, if the representation form is of little importance and emphasis is on content, vectorization using the xx_ent_wiki_sm model can be employed. Meanwhile, the uk_core_news_lg model is sensitive to both content and rearrangement of form, making it suitable for obtaining a more rigid and sensitive comparison based on cosine similarity. The tf-idf method is highly sensitive to rearrangement of form but less adept at discerning content similarity. It is better suited when only the formal similarity of texts is crucial.

Overall, comparing the scores obtained from different methods and visually reviewing the experiment results allows summarizing that the proposed approach of generating natural sentences in the Ukrainian language based on their ontological representation using a large language model is capable of conveying the general meaning and sense of the original phrase. However, often, though not in all cases, the generated phrase may appear somewhat unnatural, containing excess entities and expressions. In some instances, errors may be associated with the model's imperfect understanding of professional terminology and slang. An example of such a situation is the following case: the original phrase — "The printer is designed to create hard copies of documents."; the generated phrase — "The printer, designed to create copies that are documents and have a hard state." We observe that the model lacks rearrangement for the concept of "hard copy" and attempts to construct a phrase based on its paradigm of common word combinations.

These results indicate that while large language models can be applied for text generation based on ontological representation and convey the general meaning, such phrases often fall short of the ideal in terms of form (and sometimes nuances of meaning). This underscores the relevance of text generation systems based on ontological representations (including results of queries to the ontology) built on rules and robust templates, as demonstrated in [22]. Such approaches, with well-established and detailed rule and template systems, can generate significantly higher-quality natural phrases based on semantic representations than large language models can. However, large language models can be successfully applied in tasks of semantic analysis, enhancing the relevance of identified intents and the material obtained from the knowledge base for response synthesis. Moreover, large language models excel in generating text responses based on a set of contexts and a list of relevant intents.

# 3. COMPLEX AUTOMATED COMPETITIVENESS DIAGNOSTIC SYSTEM OF FUTURE IT-SPECIALISTS

***Liliya Ivanova***

*The article presents an automated computer system for diagnosing the competitiveness of future specialists in information technology, which consists of a set of tests for determining the psychological, creative and personal qualities of an education seeker. It was determined that the application of a complex approach to system diagnostics using multi-directional tests allows obtaining a systematic assessment of the components of the competitiveness of future specialists in dynamics. The use of an automated computer system for the diagnosis of the student's personality for the assessment and attestation of the psychological state of students of educational institutions of various accreditation levels, the identification of personal and professional characteristics, which will allow quantifying the level of competitiveness of future information technology specialists and greatly simplifies the recording and processing of the respondent's answers while simultaneously reducing the probability of errors at a certain stage of diagnosis. The functions of an automated computer system for diagnosing the competitiveness of future information technology specialists are considered. The structure of the blocks of the automated computer system for the diagnosis of competitiveness is presented. The results of an experimental study of the competitiveness of students of the Odesa Technical Vocational College using an automated computer system for diagnosing the competitiveness of future information technology specialists are presented. It is proposed to use an automated system for diagnosing and evaluating the creative, psychological and personal components of students' competitiveness for the work of structural divisions of the educational, psychological and educational direction in educational institutions of different levels of accreditation to determine the level of formation of key personal qualities of education seekers.*

## INTRODUCTION

In the conditions of the multifaceted development of our country, an irreversible process of modernization of the entire education system is taking place, the main goal of which is the training of a competitive specialist of the appropriate level and profile, who is fluent in his specialty and oriented in related fields of activity, capable of effective activity in his specialty at the level of world standards, ready to continuous professional development. Today, competitiveness is considered as a sign of a modern specialist.

The problems of the national education system have recently been constantly under the watchful eye of all sections of the public. Ukrainian higher education loses to the Western education system in terms of providing students with knowledge and forming practical skills: if everything is relatively safe with theoretical foundations in Ukrainian higher educational institutions, the practical knowledge of graduates and tomorrow's specialists is clearly insufficient.

The ideas of general and personal development, formulated in the context of psychological and pedagogical concepts of developmental and personally oriented education, are a genetic prototype of modern ideas about competences, which are considered as transversal, extra-, extra- and meta-subject education, integrating both traditional knowledge and various kind of generalized intellectual, communicative, creative, methodological, worldview and other skills.

The strategic tasks of education modernization and the specifics of professional training of future IT specialists in accordance with the requirements of the innovative labor market are reflected in the state fundamental laws and by-laws: Laws of Ukraine "On Education" (2017), "On Higher Education" (2014), "On scientific and scientific and technical activities" (2016), Concept of development of the digital economy and society of Ukraine for 2018–2020 (2018), Strategy for the development of the information society in Ukraine for 2013–2020 (2013), Decree of the President of Ukraine "On the goals of sustainable development of Ukraine for the period until 2030" (2019) and other documents.

A large number of studies are devoted to the study of trends in the development of higher education in the conditions of a change in the educational paradigm and the formation of professional competence of a specialist (H. Artyushin, L. Baranovska, V. Bobrytska, N. Bulgakova, I. Zarubinska, O. Kovtun, A. Kokareva, O. Kotykova, N. Ladogubets, V. Lugovoi, E. Luzik, L. Pomytkina, T. Sayenko, V. Semichenko, etc.); definition and development of theoretical and methodical foundations of continuous professional education (O. Volosovets, S. Honcharenko, L. Huberskyi, V. Kremen, L. Lukyanova, N. Nychkalo, S. Sysoeva, N. Ridey, etc.), study the personality of the future competent specialist and the issue of professional training of an IT professional (N. Bulgakova, M. Zgurovskyi, G. Kozlakova, V. Kruglik, P. Luzan, E. Luzik, V. Osadchyi, T. Sayenko, A. Eckerdal, M. Caspersen, C. Masuck, J. Miller, N. Truong, etc.).

"IT specialist" includes a number of different professions, the responsibilities of which are very different and depend on the specific position. Conventionally, all information technology specialists can be divided into those who work with "hardware" and those who deal with "software". The digital environment is developing rapidly, so the list of IT professions is constantly updated with new specialties. Also, various technical devices and information systems are constantly being improved and updated, which requires a high level of knowledge and skills of IT specialists, their mobility, constant self-development and self-improvement.

Intensive development of information technologies is predicted, therefore the demand for IT specialists is constantly growing. In order to achieve success and earn good money, you need to master the most relevant areas and keep your finger on the pulse of the latest trends. And for this, IT specialists need to be psychologically ready to implement new technologies, to perceive and master innovations, to demonstrate flexibility and a creative approach. And that is why the formation of professional competence of IT specialists is the main condition for effective professional activity in a high-tech innovative society.

According to the Exploring Ukraine IT Outsourcing Industry study, Ukrainian educational institutions annually graduate about 16,000 specialists in the field of information technologies [1]. However, supply and demand in the information technology market remains unbalanced. And every month, Ukrainian IT companies publish about three thousand new vacancies. Unfortunately, nowadays it is more and more common to talk about the inconsistency of the existing system of training specialists with the real requirements of the market and, in particular, in the field of information technologies. A. Yavorskyi, vice president of the GlobalLogic company, claims that "the number of young professionals capable of working in IT is currently sufficient, but their quality plays an equally important role. Unfortunately, the modern education system in Ukraine does not keep up with trends in the field of IT technologies." Y. Lyubinets, chairman of the board of directors of Soft Serve, agrees with this opinion. The Managing Director of Luxoft Ukraine O. Alkhimovych notes that educational institutions do not give young professionals an understanding of business requirements due to outdated training programs. She noted that "the market needs specialists who work on the edge of innovative solutions, who know how to think, create, create complex creative solutions" [2].

On the "Professional Consulting Portal" [3], the professional profile of an information technology specialist is presented. In essence, this is a document that provides a comprehensive, systematized and comprehensive description of the objective characteristics of the profession and the requirements for individual psychological characteristics of a person. The list of personal qualities that ensure successful performance of professional activities in the IT industry includes the following: responsibility, attentiveness, patience, diligence, systematic work, accuracy, logical thinking, tenacity, perseverance, flexibility and dynamic thinking, the ability to make decisions independently, purposefulness, technical abilities, mathematical abilities, analytical thinking.

The professional activity of a specialist in the field of information technologies (in primary positions) consists in: implementation of general functions of planning and management by implementing partly administrative and mainly operator work procedures; collecting, systematizing, and accumulating primary information both for the performance of his job duties and for the needs of the structural unit where he works; development and research of predictive mathematical models of organizational-technical and socio-economic objects and systems for the purpose of their analysis and improvement; making operational decisions within their competence; functional and informational preparation of draft solutions; management of subordinates whose competence is not higher than that of technical employees or junior specialists; carrying out an assessment of the potential of computer hardware and software; organization and control of the functioning of the information system in these operations of the technological process of information processing; development of software for professional activity tasks using high-level programming languages and database management systems [5].

In fact, during personnel selection, the applicant's compliance with the requirements is checked, which are divided into two categories: "hard" and "soft" requirements.

"Hard requirements" (professional qualities or "hard skills") are professional abilities, skills and knowledge that are necessary for the performance of professional tasks. They can be measured and evaluated objectively. They are related to knowledge of fundamental and special disciplines, acquisition of practical training, etc. For an IT specialist, these can be: deep knowledge of programming languages; ability to create algorithms; knowledge of operating systems, software architecture; software and hardware coding and testing; software debugging; knowledge of software development process methodologies; knowledge of English; ability to create technical documentation, etc. But they depend significantly on the company's field of activity and may differ.

“Soft” requirements (personal qualities or “soft skills”) combine a number of psychological characteristics, properties and skills that can be grouped. These qualities are difficult to measure, and therefore their evaluation is subjective. Most employers consider them as important as professional knowledge and skills. Professional skills and abilities become obsolete, and “soft skills” are always relevant. According to G. Babii, personal qualities are a complement to professional qualities, and professional qualities determine those personal qualities that a specialist must have or develop in himself for professional growth [6].

Basically, soft skills are skills, abilities and characteristics that allow you to be successful in professional activities. These include: leadership qualities and the ability to work in a team, the ability to teach and negotiate, the ability to set and achieve goals, time management, goal orientation, presentation skills, effective communication skills, stress resistance, creativity, a creative approach to solving tasks and analytical abilities, etc. There is no fixed list, just like the classification of soft skills. Because it is clear that different types of soft skills are prioritized for different types of activities.

Information technologies are becoming one of the most important tools for forming needs, interests, views and value attitudes, influencing the worldview of a person; act as a mechanism of education and training and, in general, are a means of forming professionally important qualities of a specialist. Professional training of highly qualified specialists in the IT field, employed in industry, business, scientific centers, becomes a strategically important task that requires the maximum use of achievements of scientific and technical progress, a comprehensive approach to planning the educational and scientific-methodical process, bringing methods, means and forms learning according to the demands of modern life.

One of the global tasks of training an IT specialist is the formation of abilities and skills to navigate in a huge flow of information, to quickly reorganize one’s activities in accordance with modern requirements in the conditions of informatization, to master new technologies and knowledge.

The field of information technologies and telecommunications is characterized by rapid development, and in order to maintain the gained pace, specialists who would help it in this are becoming more and more necessary. At the same time, specialists should meet certain requirements, without which they will not be able to successfully cope with their professional tasks. Such specialists must first of all be inclined to mathematics, informatics and working with technology, including computers. They need to have an analytical mind, a good memory and the ability to work with a large amount of information. Also, irreplaceable qualities for all employees in this field, regardless of position, are responsibility, organization, stress resistance, and the ability to study independently from specialized literature.

It is worth noting that the image of an IT specialist who worked, for example, ten years ago, is significantly different from today. Now he is not a silent, focused person who does not take his eyes off the computer all day, but a sociable employee, ready to work in a team and have a direct dialogue with customers. When creating a product, IT and telecommunications workers focus on its future consumers, so they know the interests and needs of their potential customers well.

It is important not only to have all these qualities, but also to present them correctly, for example, during employment. The main business card here is a resume, in which you should definitely indicate your advantages and abilities. "LinkedIn" — a social network for establishing business connections — studied the resumes of IT specialists and made a list of the most frequently used qualities:

- analytical mindset — 6.5 % of applicants;
- ability to work in a team — 4.3 % of applicants;
- hard work — 2.5 % of applicants;
- the ability to work with a large amount of information — 1.9 % of applicants.

Holders of such qualities really have many chances to find a job in their specialty, especially since the demand for workers in the field of information technology and telecommunications significantly exceeds the supply. Of course, in order to be competent and in demand, they will need not only a psychological inclination to the profession. They cannot do without a good education, professional knowledge and skills, constant improvement of their skills. And knowledge of the English language, besides, will help to find a job in an already known or very promising foreign company.

The analysis of the specifics of the tasks in the field of information technologies and the question of the professional qualities of IT specialists were at different times dealt with by psychologists and educators from abroad — F. Brooks, G. Weinberg, N. Wirth, E. Dijkstra, S. McConnell, M. Smulson, B. Schneiderman and others, who claim that professional IT specialists have their own psychological and human traits, qualities and determine the abilities and features of thinking that should be characteristic of them.

The analysis of the practical activities of modern specialists in the IT field makes it possible to identify a kind of "standard of professional competence of IT specialists", which can be presented as a set of knowledge and skills that an information technology specialist must possess for successful professional activity:

1) highly specialized competencies that allow creative use of computer programs, various “software packages”, utilities and gadgets in the process of solving emerging problems;

2) knowledge of a wide range of programming languages and the ability to use them in the process of joint organizational activities when solving set tasks;

3) the ability to enter a special mode of information activity, which involves focusing consciousness on symbolic information and transforming the received information into organizationally significant information / knowledge;

4) the ability to integrate into a single field of the organization the information produced during the operation of computer networks with the information circulating through other channels of intra-organizational information /

Despite the importance of the qualification factor of IT specialists, their internal group status and internal organizational perception are largely determined by such important individual parameters as the type of temperament and personality traits. In the studies of Yu. Babayeva and A. Voyskunsky, the following two groups of professionally important qualities are given: personal and communicative [3]. According to O. O. Gurska, the set of the most expected personal qualities that modern IT specialists should display should include emotional stability, punctuality, accuracy, thrift, high efficiency, the presence of extreme attentiveness and logic in thinking. The factor of passion for work and high professional motivation is also of great importance, because without a sustained interest in everyday professional activities, it is impossible to count on internal positive motivation, which, according to the point of view of researchers F. Brooks and M. Smulson, is closely related to a feeling of joy, with a feeling of absolute freedom of creativity, which IT specialists are able to feel in the course of their work.

Summarizing the theoretical and practical studies discussed above, it can be stated that IT specialists must have a set of mandatory abilities based on a combination of certain psychological characteristics, namely:

1) the ability to carefully consider one’s professional actions, their expediency and safety in view of the high risk of irreversible consequences of professional mistakes, and also to take responsibility for their results;

2) the ability to long-term and high-level intellectual concentration of activity, in which various manifestations of psycho-emotional and phys-iological discomfort are possible due to monotony, low and monotonous physical activity, which is based on a combination of a high level of intellectual abilities, emotional-volitional self-regulation and a balanced type of temperament;

3) the ability to convey one's thoughts in a language understandable to ordinary users-colleagues (who are not specialists in the IT field) when explaining the rules of working with computer / software equipment and operating systems, which is based on a combination of a high level of development of social intelligence and basic communication skills;

4) the ability to react in an emotionally balanced manner to possible mistakes of colleagues and direct supervisors that arise in the process of their work with office computer programs, networks, and in an accessible, correct form to explain everything they need to know in order to avoid similar mistakes in the future, which is based on a combination of a high level of development of emotional self-control and appropriate communication skills.

Today there are many different areas of professional activity of IT specialists, which have their own characteristics and require deep knowledge in the field of information technologies: programmer; system architect; specialist in information systems; system analyst; system administration specialist; information technology manager; manager of sales of solutions and complex technical systems; specialist in information resources; database administrator.

The conducted analysis showed that today employers need specialists with work experience and an available range of professional knowledge, with developed personal skills that significantly increase work efficiency. This position is confirmed by the results of studying resumes and vacancies on specialized resources and websites of leading employers in the IT industry.

**Formulation of the problem**. The study of competitiveness as an integral property of the individual, the dynamics of its development in the process of professional formation, the possibility of discovering new reserves of the individual, is gaining relevance and requires in-depth study.

The study of competitiveness as an indicator of the quality of training of specialists is devoted to the works of R. Fathutdinov, D. Chernilevsky, S. Shirobokov, D. Bogyna, N. Hlevatska, O. Grishnova, M. Krymova, O. Krymova, L. Lisohor, M. Semikina, S. Sotnikova, N. Shulgy. In foreign pedagogy, the problem of specialist competitiveness became the subject of research by R. Kvasnytsia, V. Landsheer, M. Lennon, P. Mercer, M. Robinson, and others. In the studies of O. Dushkina, M. Knyazeva, M. Mashnikov, V. Oganesov, and M. Semenova, problems of the development of the personality of a specialist were considered. The structure and characteristics of competitive personality qualities are considered in the works of I. Drach, G. Dmytrenko, V. Andreev, N. Borisov, E. Klimov, A. Markova, and L. Mitina. L. Mitina's research presents the psychological aspect of the development of a competitive personality.

It should be noted that in various countries there is an ongoing dialogue between employers and educational institutions regarding the definition of personal and professional skills that can become key for a specialist. This issue is also being studied in Ukraine. The World Bank's "Skills for Modern Ukraine" study (2019) showed that there is a significant demand in Ukraine for developed cognitive skills, self-organization, resilience, teamwork and willingness to learn. This list has not undergone significant changes in recent years. Thus, according to a study conducted by the CSR Development Center during July-August 2021, the following skills are in greatest demand among Ukrainian employers: the ability to work in a team, communication skills, analytical thinking, the ability to learn quickly, flexibility, responsibility, initiative, competent written and oral language, emotional intelligence. At the same time, in the perspective of 2030, according to the same respondents, the greatest demand will be for critical thinking and a project approach to solving problems. An important place in the system of requirements for the training of a future specialist is occupied by appropriate psychological characteristics, personal and professional qualities that could ensure a high level of competitiveness in the modern labor market.

The problems and features of training future specialists in information technologies were considered in their studies by many domestic and foreign research scientists, in particular A. Vlasyuk, T. Hura, L. Grishko, L. Dobrovska, L. Zubyk, O. Kaverina, T. Kovalyuk, L. Kurzayeva, I. Mendzebrovskyi, T. Morozova, O. Pavlov, V. Sedov, D. Shchedrolosiev and others. The question of compliance of the existing standards of training of IT specialists with the requirements of employers was also investigated (O. Pavlov, T. Kovalyuk, P. Pavlenko, S. Popershnyak, V. Osadchiy).

In the works of T. Gur, L. Dobrovska, O. Kaverina, T. Kovalyuk, V. Medved, D. Mustafina, attention was paid to the problem of forming the competitiveness of graduates of technical universities, software specialists, and software engineers.

The basis of the new standards of higher and professional higher education of Ukraine in the field of knowledge 12 "Information technologies" are Computer Science Curricula documents, which regulate the process of training IT bachelors in the USA and other countries [7]. In turn, the developers of Computer Science Curricula focused on the recommendations of world leaders in information technology, namely the Institute of Electrical and Electronics Engineers (IEEE) and the Association of Computer Engineering (ASM).

Determining the personal qualities and characteristics necessary for a future IT specialist to effectively solve production problems, taking into account the requirements of all interested parties — employers, state and private institutions, educational institutions — requires a comprehensive approach of system diagnostics using multi-directional tests, which will allow obtaining a system assessment components of the competitiveness of future specialists in dynamics.

Diagnostics should be based on clear and precise ideas about the essence of this process. Psychologists, unanimously recognizing that the phenomenon of personality itself exists and is one of the basic objects of research in psychological science, also unanimously believe that the problem of objectively defining the essence of personality and its interpretation is one of the most difficult.

The diagnosis of the formation of competitiveness can be implemented using a set of complementary methods and techniques, namely:

– surveys and interviews during which the opinions of students, teachers and specialists on the problem of competitiveness and determination of the list of priority professionally important and personal qualities of a competitive specialist in information technologies were revealed;

– observation, the purpose of which was purposeful and systematic perception of the actions and behavior of future competitive IT specialists;

– self-assessment, in the process of which the level of formation of personal qualities of a competitive information technology specialist was determined;

– testing that revealed the level of professional knowledge, abilities and skills of the future IT specialist.

The principle of diagnostic efficiency is provided by the use of a number of methods that allow assessing the level of formation of the components of students' competitiveness, for example, expert evaluation, testing, individual and group interviews, and others. The selected diagnostic device allows you to measure the level of formation of the components of the competitiveness of future IT specialists and quickly process the results.

The principle of objectivity — diagnosis should be based on scientifically based criteria and indicators of the competitiveness of future IT specialists.

The developed evaluation tests were composed in accordance with the evaluation object and were based on the theory of pedagogical testing.

Levels of formation of the components of the competitiveness of future IT specialists by components [8]:

1. Motivational and value component

*Low*. The student has a weakly expressed need to achieve success, gain knowledge for future professional activity, expand his horizons, self-education, updating his own experience, the student needs constant encouragement and control. The student does not show activity, does not strive for successful activities, does not know how to build life and professional priorities.

*Average*. The student has a pronounced need to achieve success, gain knowledge for future professional activity, expand his horizons, self-education, updating his own experience, but the student needs a certain encouragement. The student is active and strives for successful activities, knows how to build life and professional priorities.

*High*. The student has a clearly expressed need to achieve success, gain knowledge for future professional activities, expand one's horizons, self-education, update one's own experience. The student is active and strives for successful activities, knows how to build life and professional priorities.

2. *Cognitive component*

*Low*. The student has a low level of formation of professional knowledge.

*Average*. The student has an average level of formation of professional knowledge.

*High*. The student has a high level of formation of professional knowledge.

3. Active component

*Low*. Low level of generally recognized practical professional skills and abilities.

*Average*. Sufficient level of generally recognized practical professional skills and abilities.

*High*. High level of generally recognized practical professional skills and abilities.

4. *Personal reflective component*

*Low*. The student does not show persistence and initiative; however, can be attentive, diligent, disciplined, rarely shows independence and autonomy in decisions; does not know how to adequately evaluate his own achievements and predict the consequences of future activities, takes criticism with difficulty and does not always respond to criticism with restraint.

Does not know how to organize and plan his own employment; does not show creativity, acts according to an algorithm; not communicative, does not know how to think logically and justify his own judgments, analyze information; not mobile, passive; has a limited range of ideas; hard to adapt to the new; needs constant control, rarely takes responsibility, withdrawn, has a low rate of development of technical abilities. *Average*. The student shows persistence and initiative in achieving the goal, depending on the situation; attentive, diligent, disciplined, can show independence and autonomy in decisions; is able to adequately assess one's own achievements and opportunities, predict the consequences of future activities, reacts to criticism with restraint. Able to organize himself and others for successful activities and plan his own employment; sociable, open to communication, likes to be the center of attention; shows creativity and unconventionality when solving professional tasks; knows how to think logically, but cannot always substantiate his own judgments and analyze information; sufficiently mobile, has a broad outlook that goes beyond the specialty, but has a limited range of ideas; able to work in a team, can defend his own views, but sometimes experiences difficulties in adapting to new things; has an average indicator of the development of technical abilities.

*High*. The student shows persistence and initiative in achieving the goal, is attentive, diligent, shows independence and autonomy in decisions, is disciplined; knows how to adequately evaluate own achievements and opportunities, predict the consequences of future activities, adequately responds to criticism. Able to organize himself and others for successful activities and plan his own employment; sociable, open to communication, likes to be the center of attention; shows creativity and unconventionality when solving professional tasks; is able to think logically and justify his own judgments, analyze information; mobile, has a broad outlook that goes beyond the specialty; able to work in a team, can defend his own views; does not experience difficulties in adapting to the new; has a high rate of development of technical abilities.

**The purpose of the article** is to present an automated complex system for diagnosing the student's personality, consisting of a set of tests for determining the psychological, creative and personal qualities of the student; for assessment and attestation of the psychological state of education seekers, identification of personal and professional characteristics, which will allow to quantitatively determine the level of the personal component of the competitiveness of future information technology specialists.

## 3.1. AUTOMATED COMPETITIVENESS DIAGNOSTIC SYSTEM OF FUTURE INFORMATION TECHNOLOGY SPECIALISTS

For diagnosis and evaluation of competitiveness, an automated system of diagnosis is proposed, which is a computer system of personality diagnosis, consisting of a set of tests for determining the psychological, creative and personal qualities of a future specialist. The system can be used to assess and certify the psychological state of school students, identify personal and professional characteristics. The use of modern computer technology provides new opportunities for diagnosing an individual or a group. The recording and processing of the respondent's answers is greatly simplified, while at the same time reducing the probability of errors at this stage of diagnosis (which are practically inevitable during manual processing). A significant advantage of computer tools for psychological and professional diagnostics is the speed of conversion of the received primary data into standard values.

Task:

1) present an automated computer system for diagnosing the competitiveness of future information technology specialists, consisting of a set of tests to determine the psychological, creative and personal qualities of the listener.

2) experimenting verify its effectiveness.

Functions of the automated system for diagnosing the competitiveness of a future IT specialist: diagnostics, collection and analysis of test results, saving the results in the database, forming a report according to the diagnostic results in the format of an Excel table.

The automated system for diagnosing the competitiveness of future IT specialists is a desktop application running on the Windows operating system, which is implemented in the C# programming language with support for Windows Forms technology, which is one of the most popular technologies for developing Windows windowed applications.

The comprehensive automated system for diagnosing the competitiveness of future IT specialists is a computer program with 16 available tests. According to the principle of passing the tests can be divided into two categories: "question-answer" and "identification of priorities". According to the content of the questions, the tests are divided into: text and text with the use of graphics. The complex of diagnostic tools is determined on the basis of the analysis of scientific works and dissertation studies of the psychological and pedagogical direction on the study of the phenomenon of "competitiveness" [4].

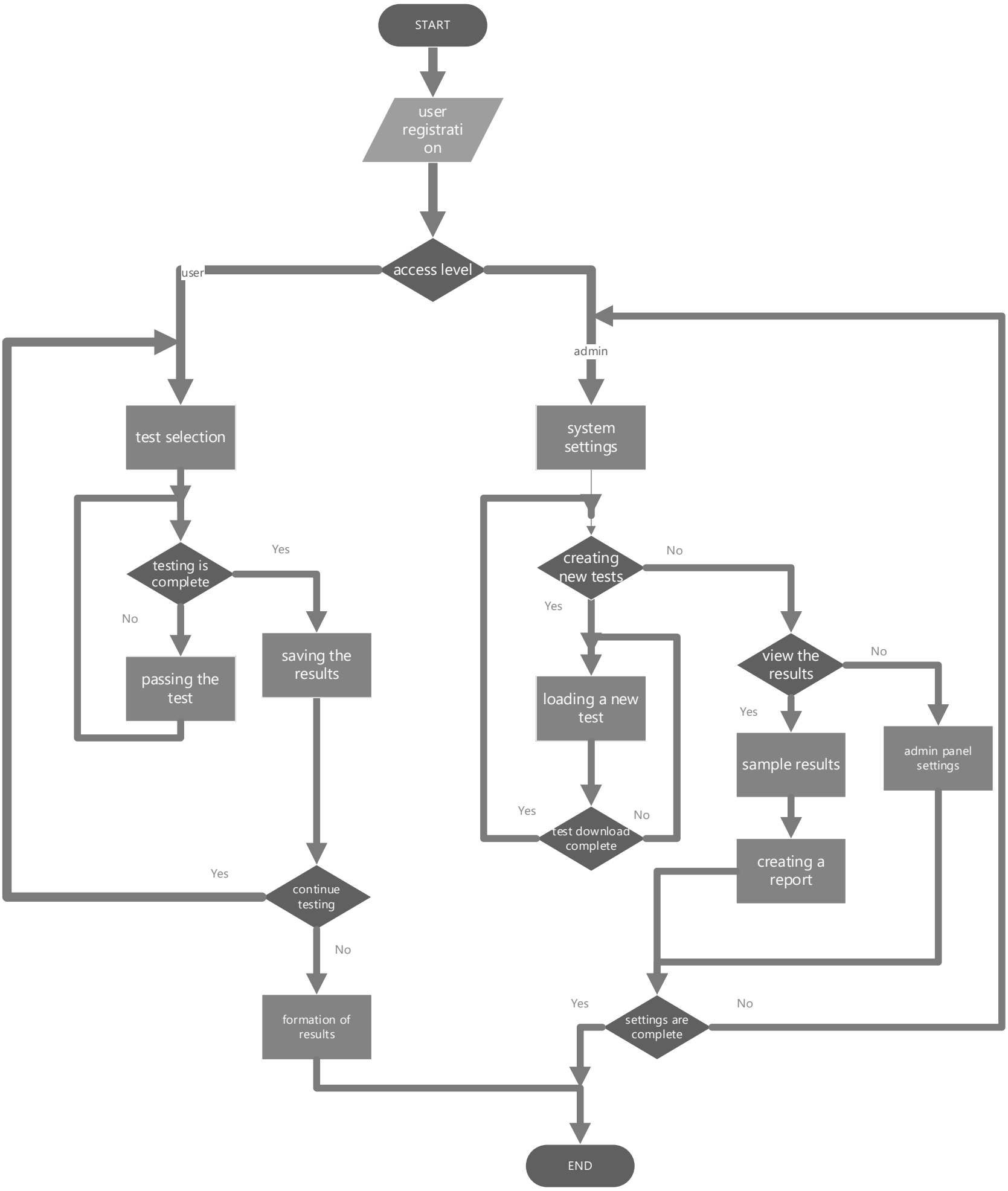

Fig. 3.1. Algorithm of the program

In general, the structure of the program “Computer diagnostics of personality” can be divided into three main blocks: registration block, testing block, administration block.

The registration block is the first window of the program and consists of several test fields in which the person being tested enters basic data, such as last name, first name, patronymic, group and gender. The program al-gorithm checks the data entered by the student. For example, in the last name, first name, patronymic and group fields, it is not allowed to enter any symbols and letters, except for the Cyrillic alphabet and the "-" sign, and the algorithm takes this into account.

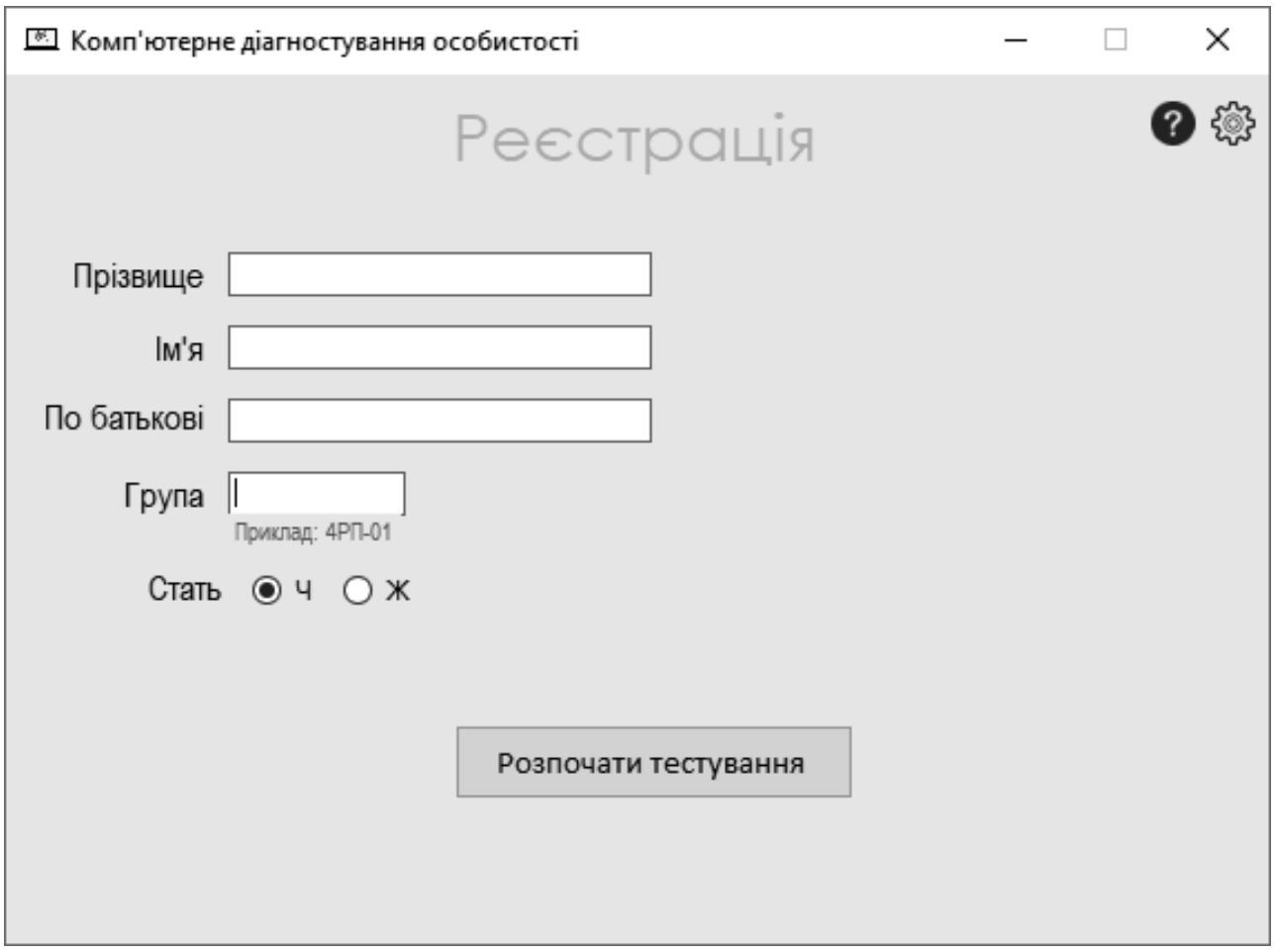

Fig. 3.2. Registration window

After clicking on the "Start testing" button, the program, according to the algorithm, checks whether all the fields with the surname, first name, patronymic and group are filled in, if at least one field is empty (not filled), the program displays a window with the error text "Fill all fields!". If all the fields are filled in, the registration window is closed, and work with the unit is considered completed, and the test selection window, which is related to the test unit, opens.

The test block consists of eighteen windows, one for test selection, one window for displaying test results, and sixteen windows that directly display each of the sixteen tests.

First, the test selection window is displayed, on which there are sixteen buttons. When you click on one of the buttons, a window with the corresponding test number opens. All tests are divided into two types: question-answer and prioritization.

In tests with a "question-answer" structure, the window displays the question and a list of possible answers next to it. In the list, the user has the opportunity to choose one of the answers.

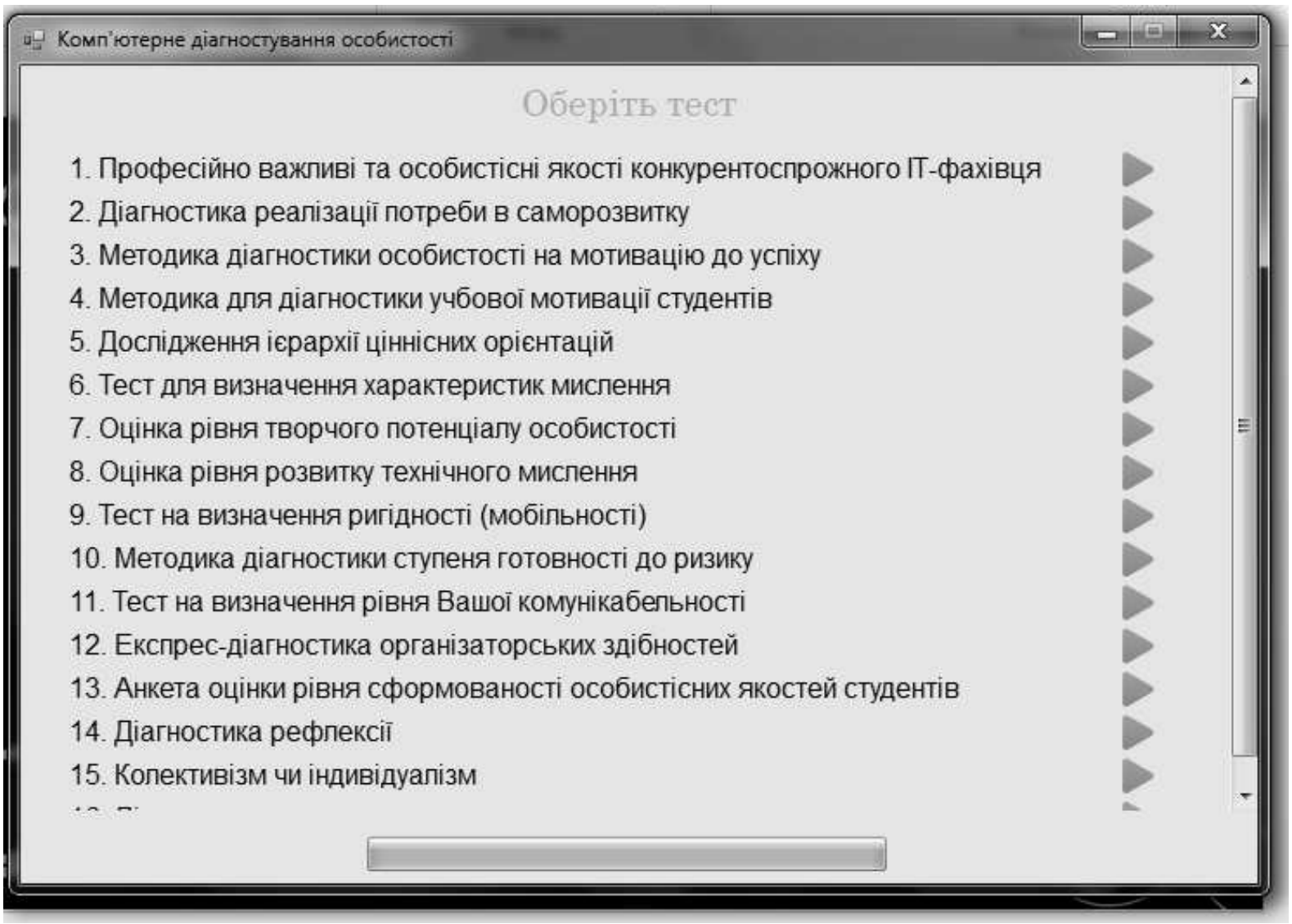

Fig. 3.3. Test selection window

***Professional and personal qualities of a competitive IT specialist are important***

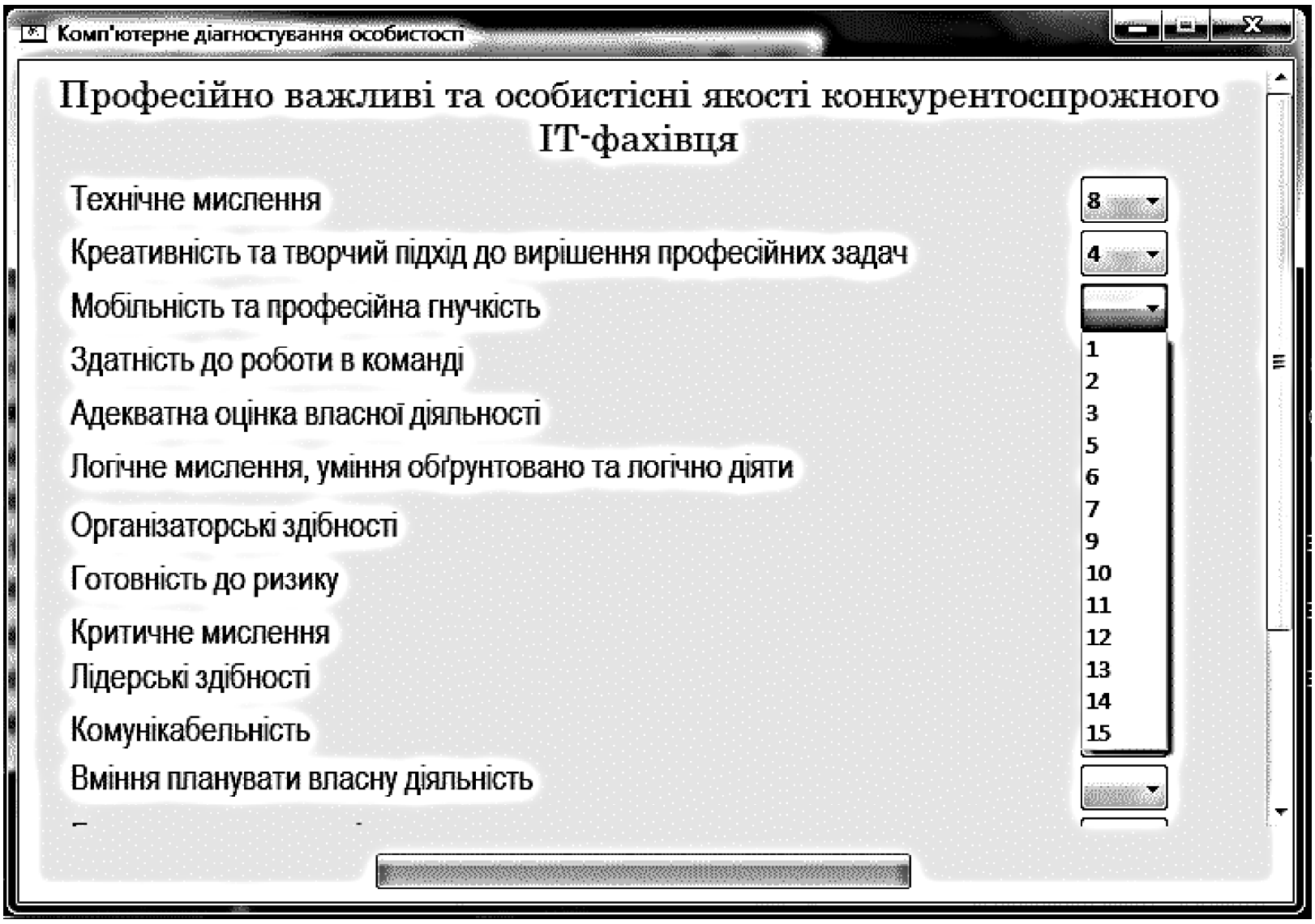

*Diagnosis of the realization of the need for self-development*

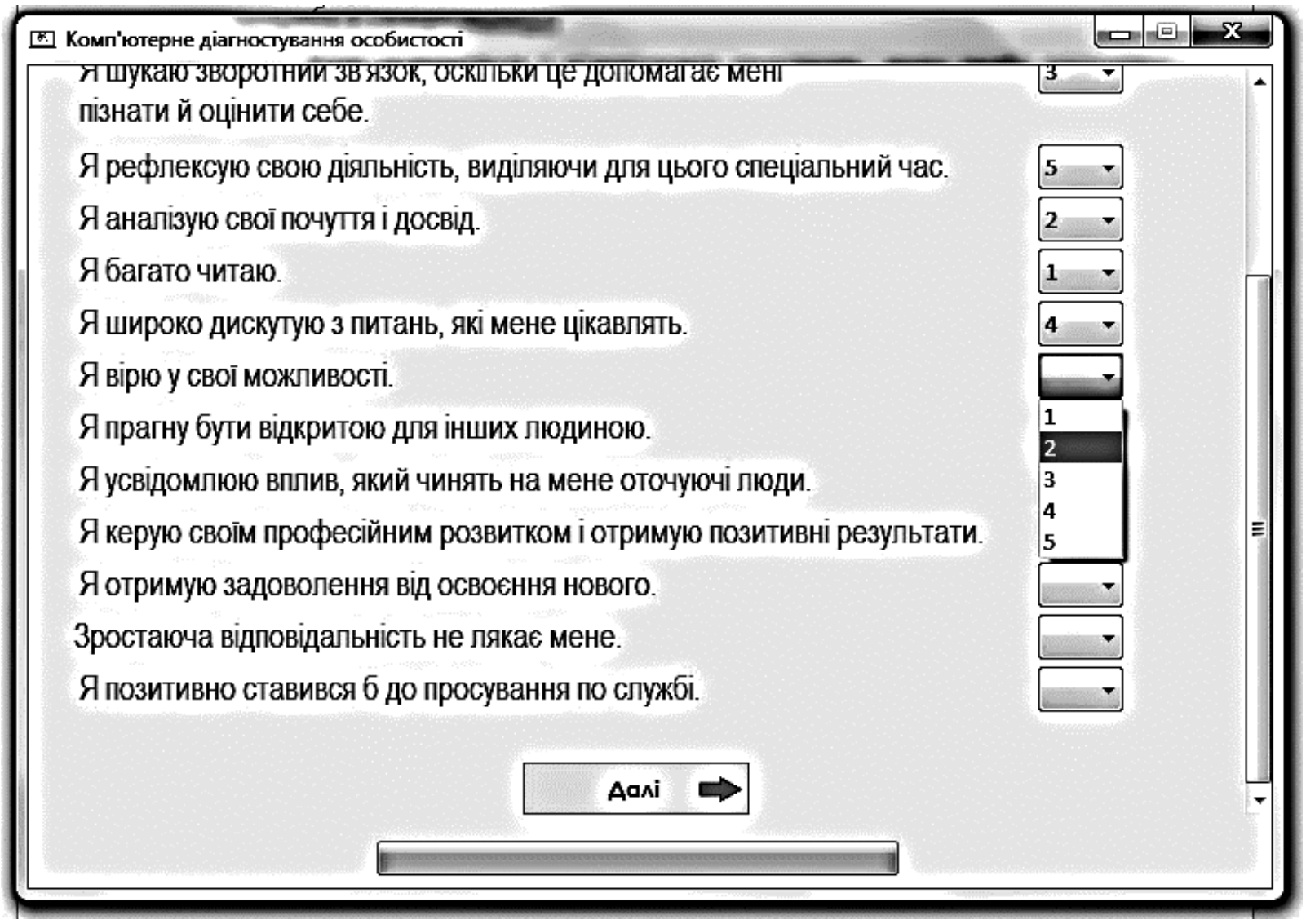

*Diagnostics of the degree of critical thinking*

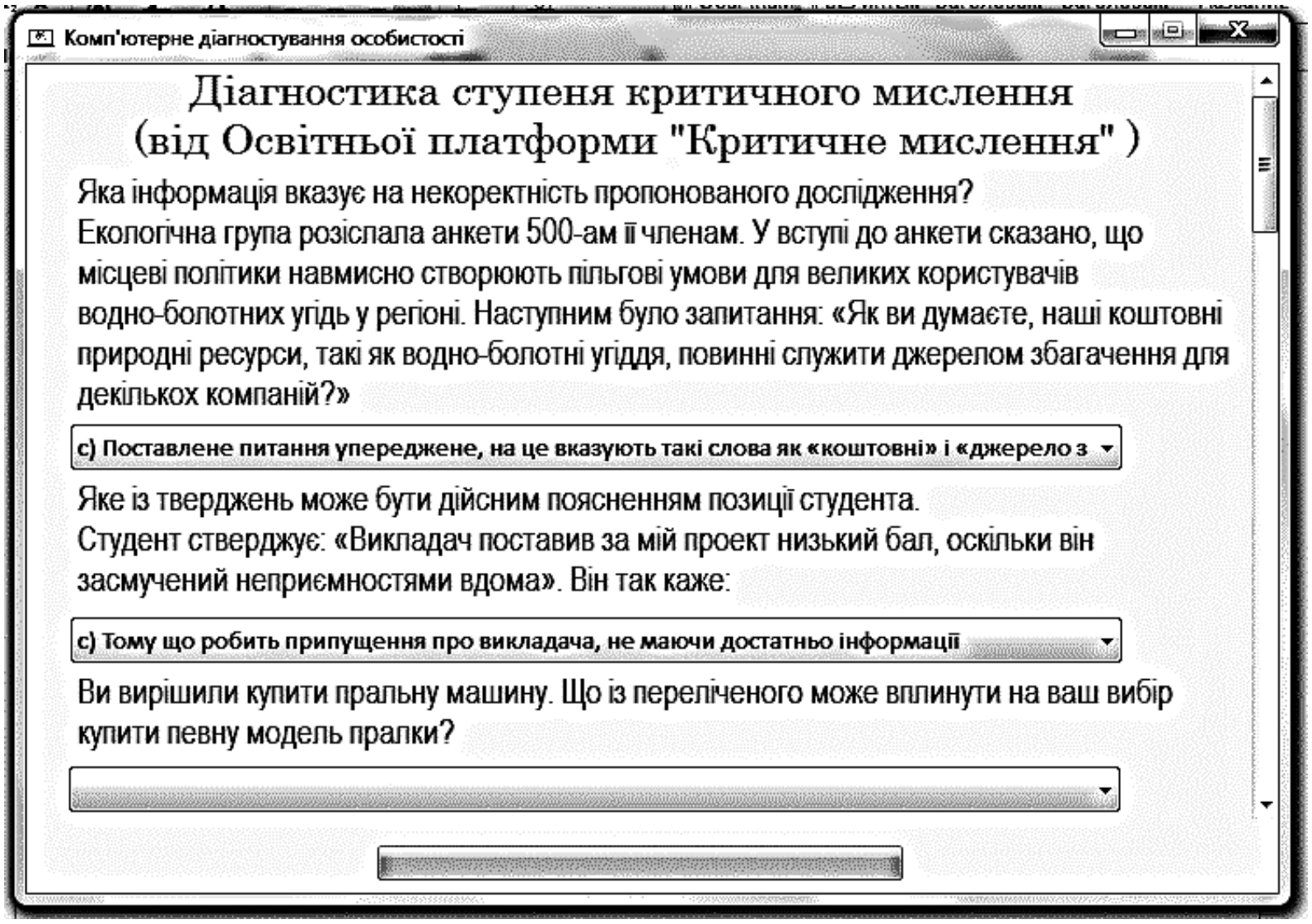

*Collectivism or individualism*

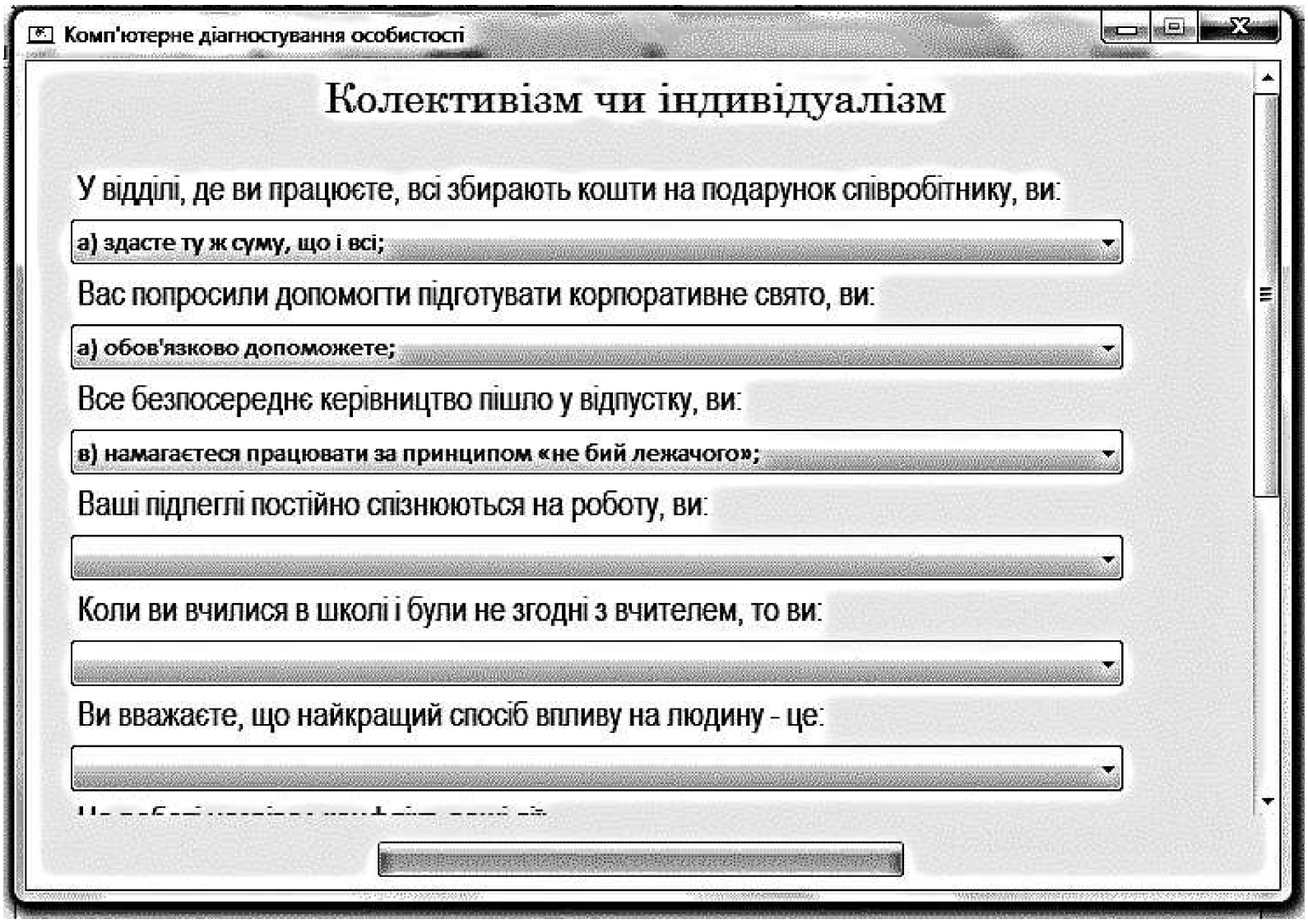

*Diagnosis of reflection*

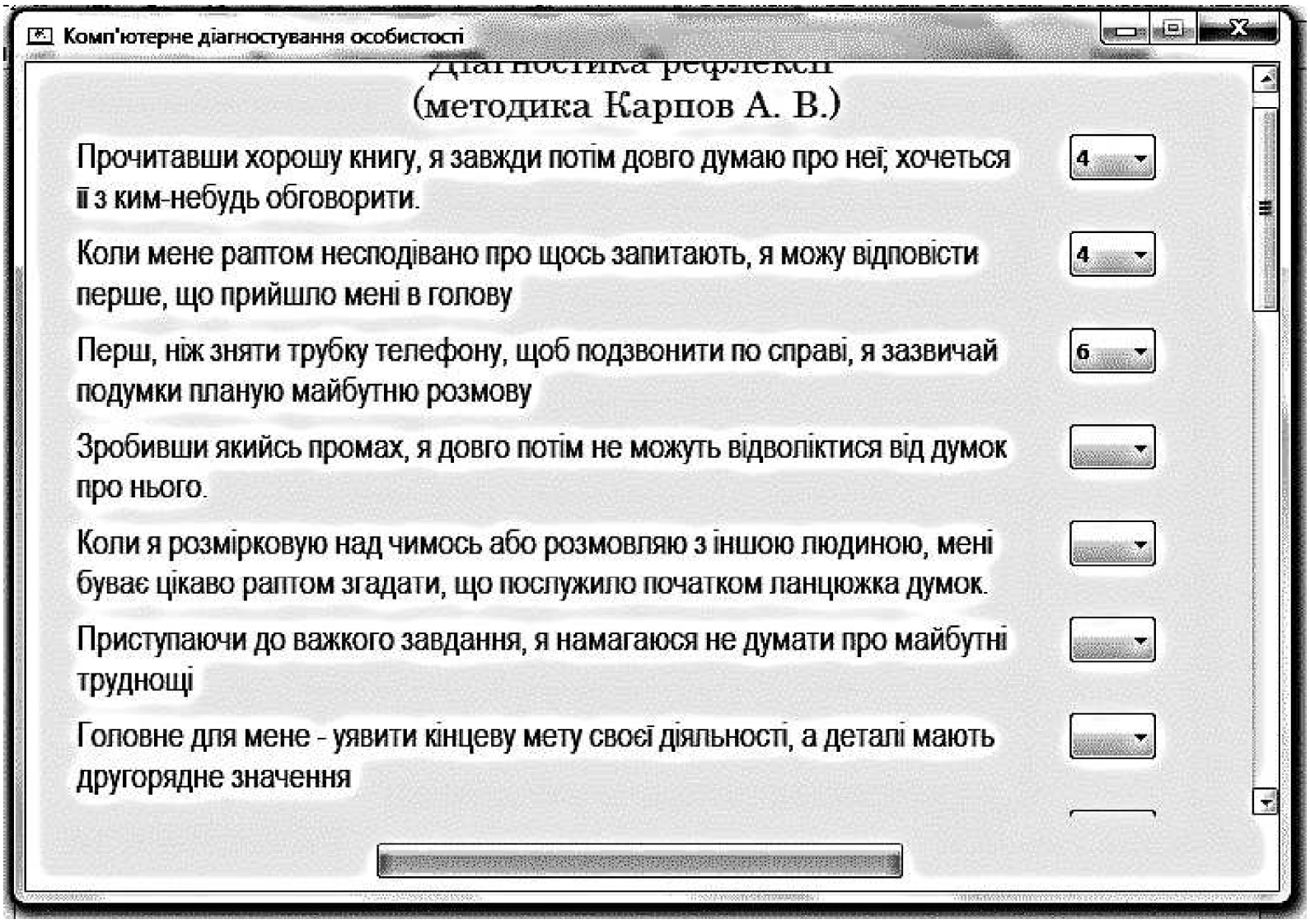

*Questionnaire of the level of formation of personal qualities of students*

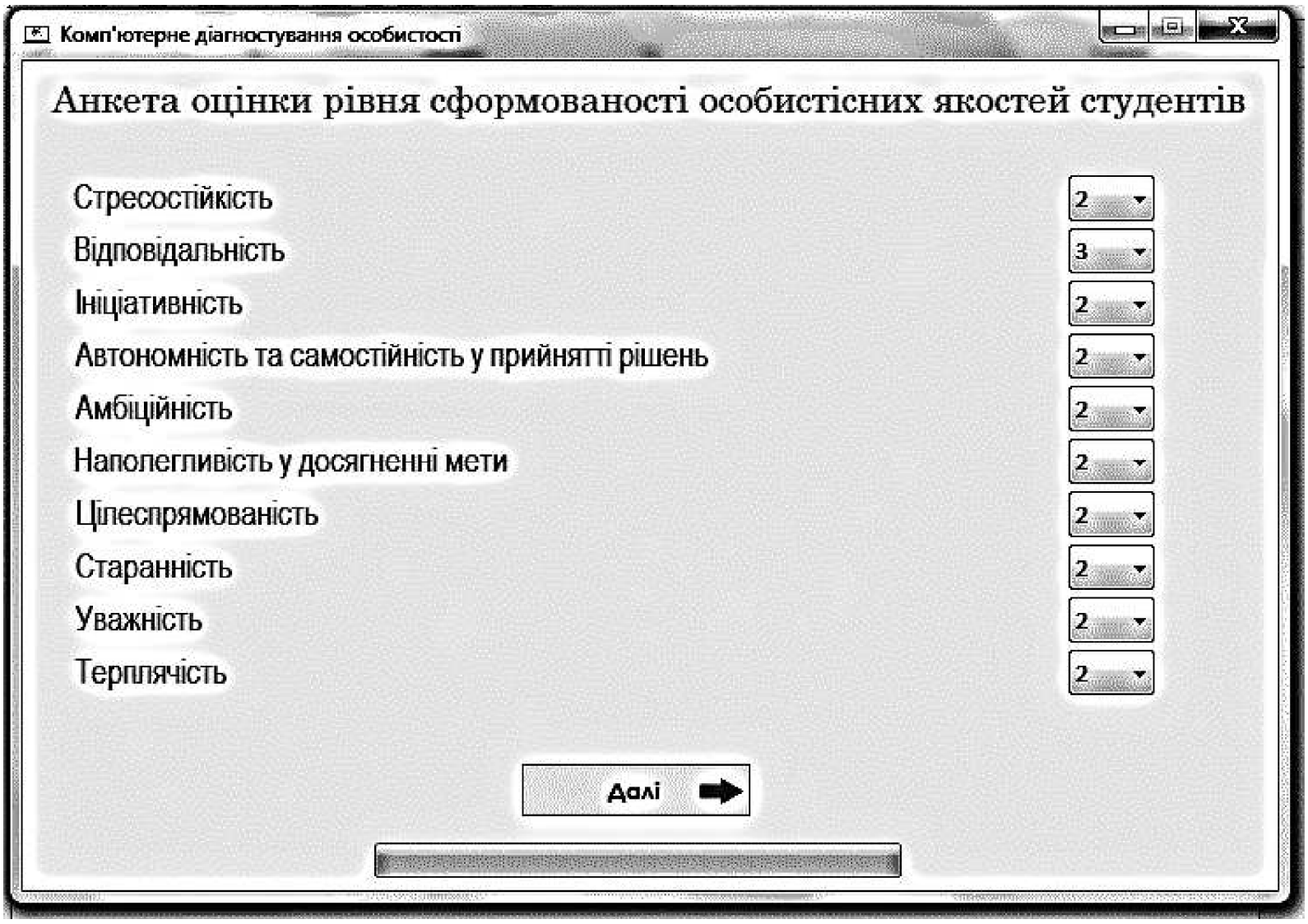

*Express diagnostics of organizational abilities*

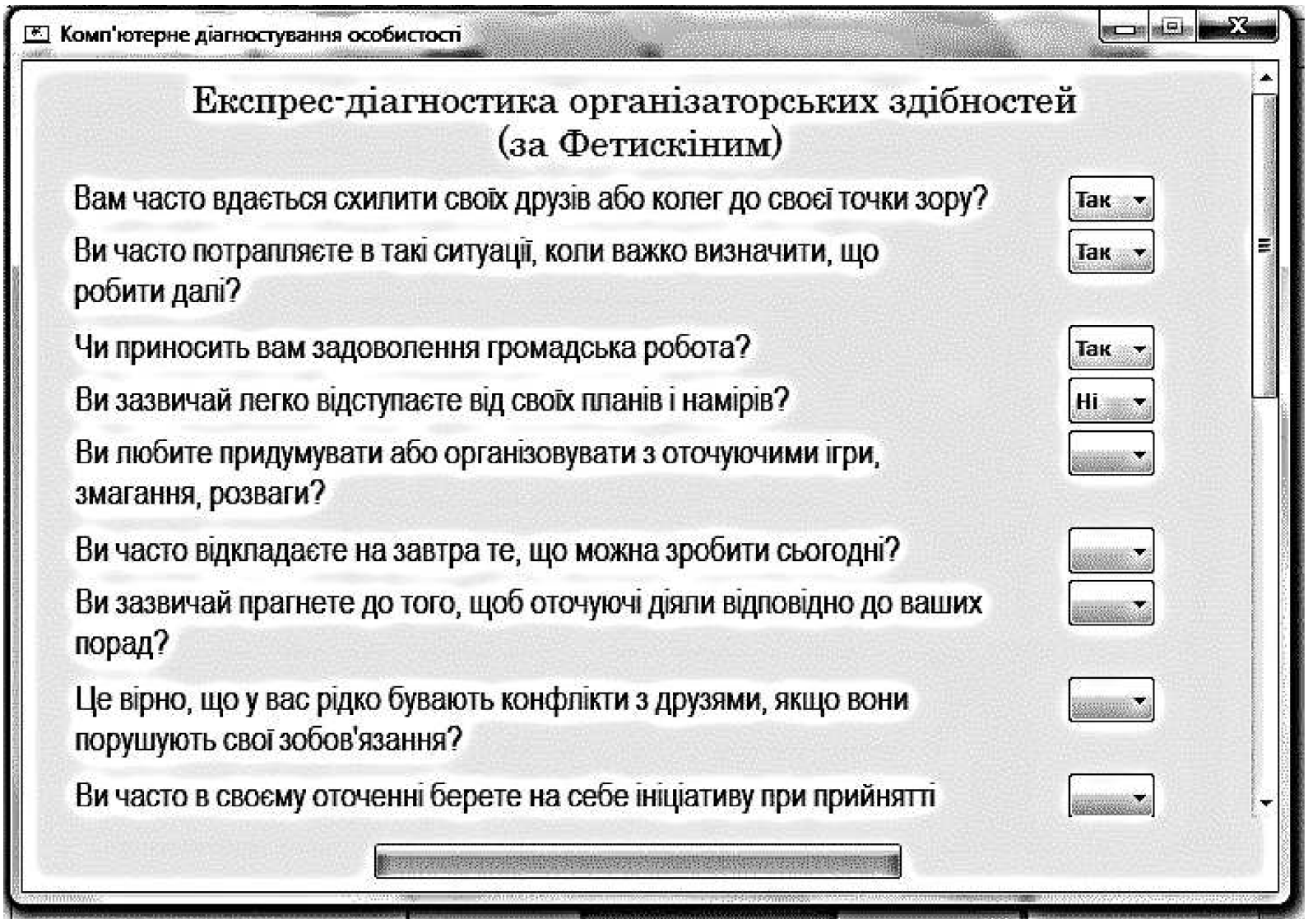

*A test to determine the level of sociability*

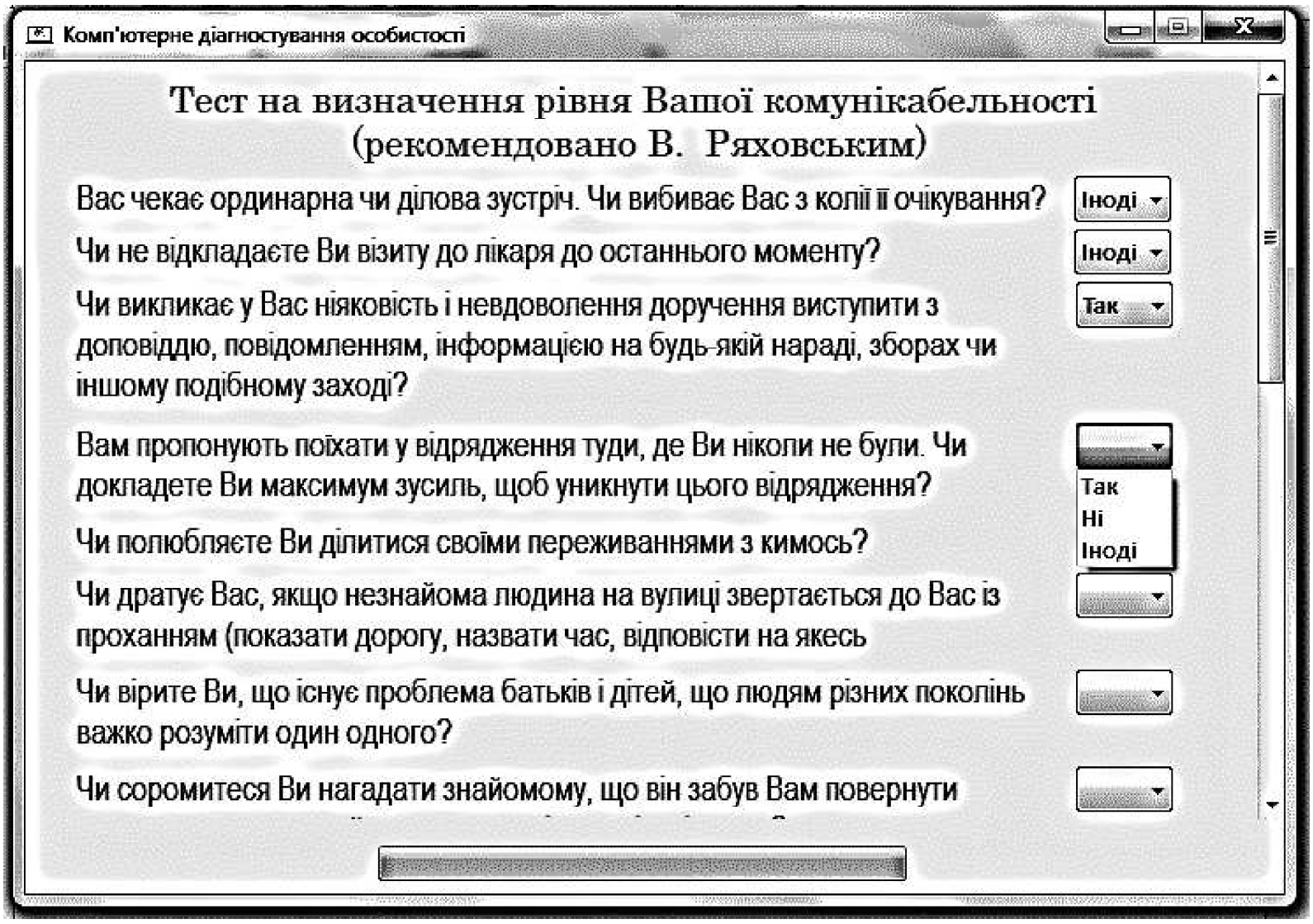

*Diagnosis of the degree of readiness for risk*

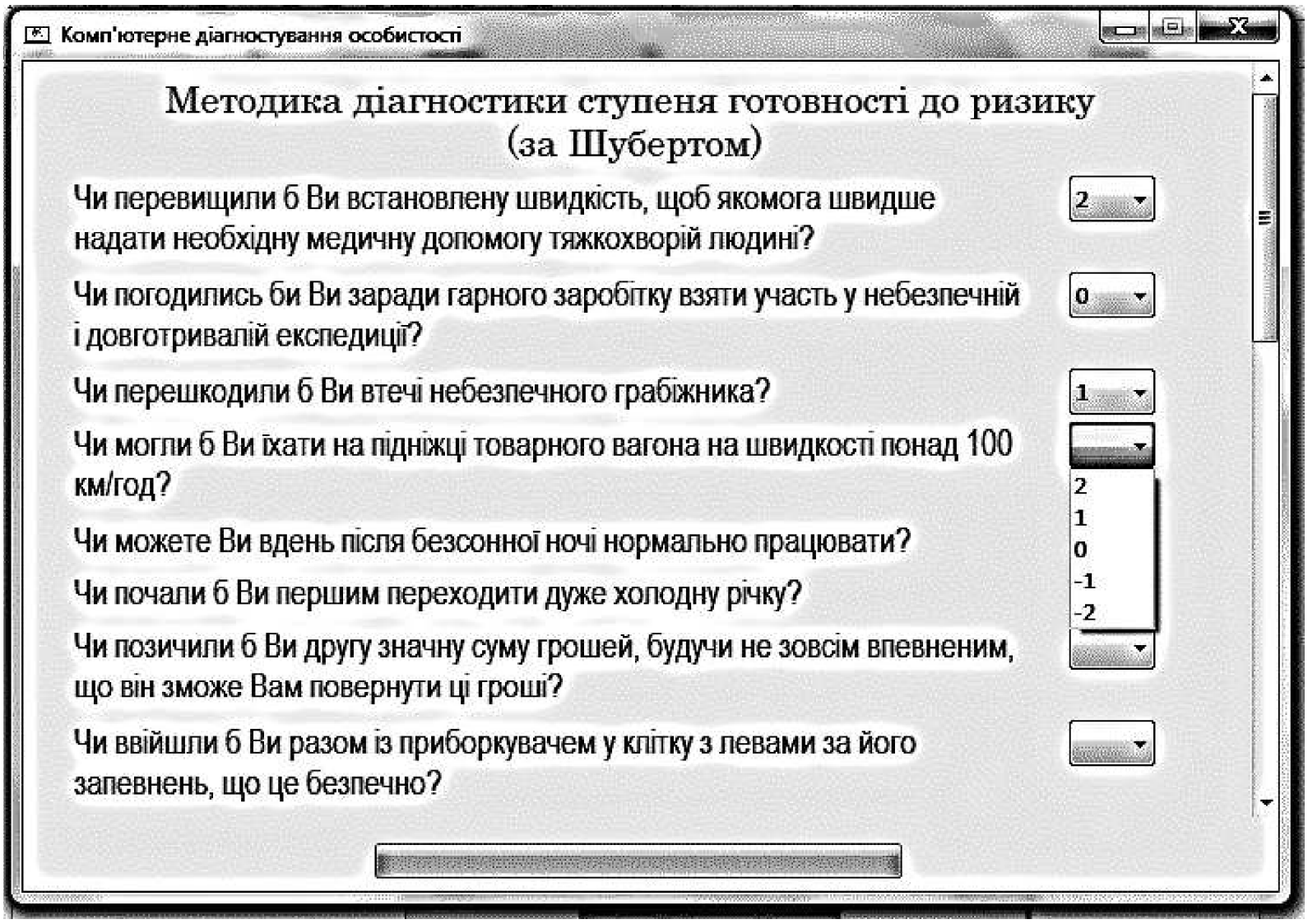

*Rigidity determination test*

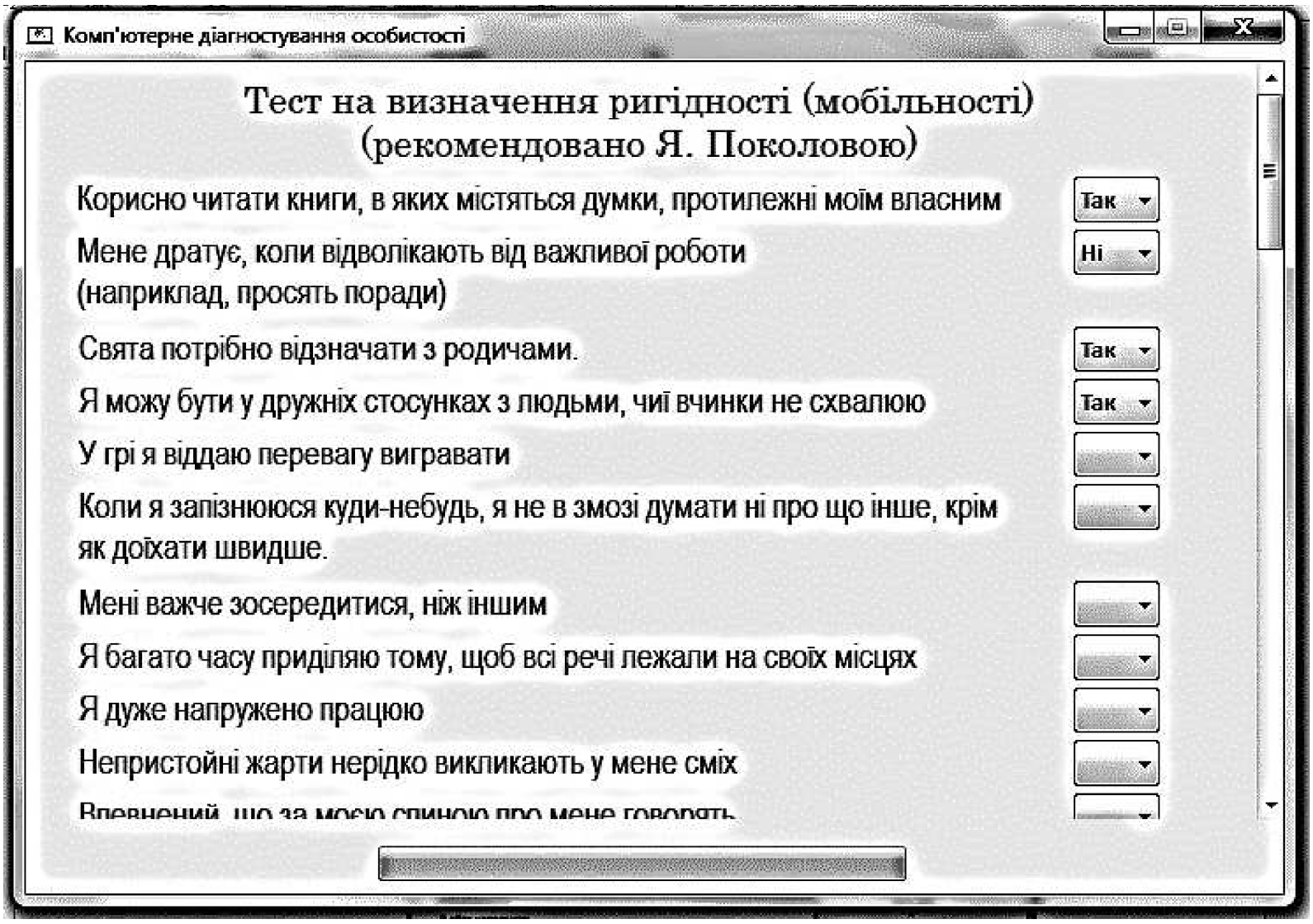

*Assessment of the level of development of technical thinking*

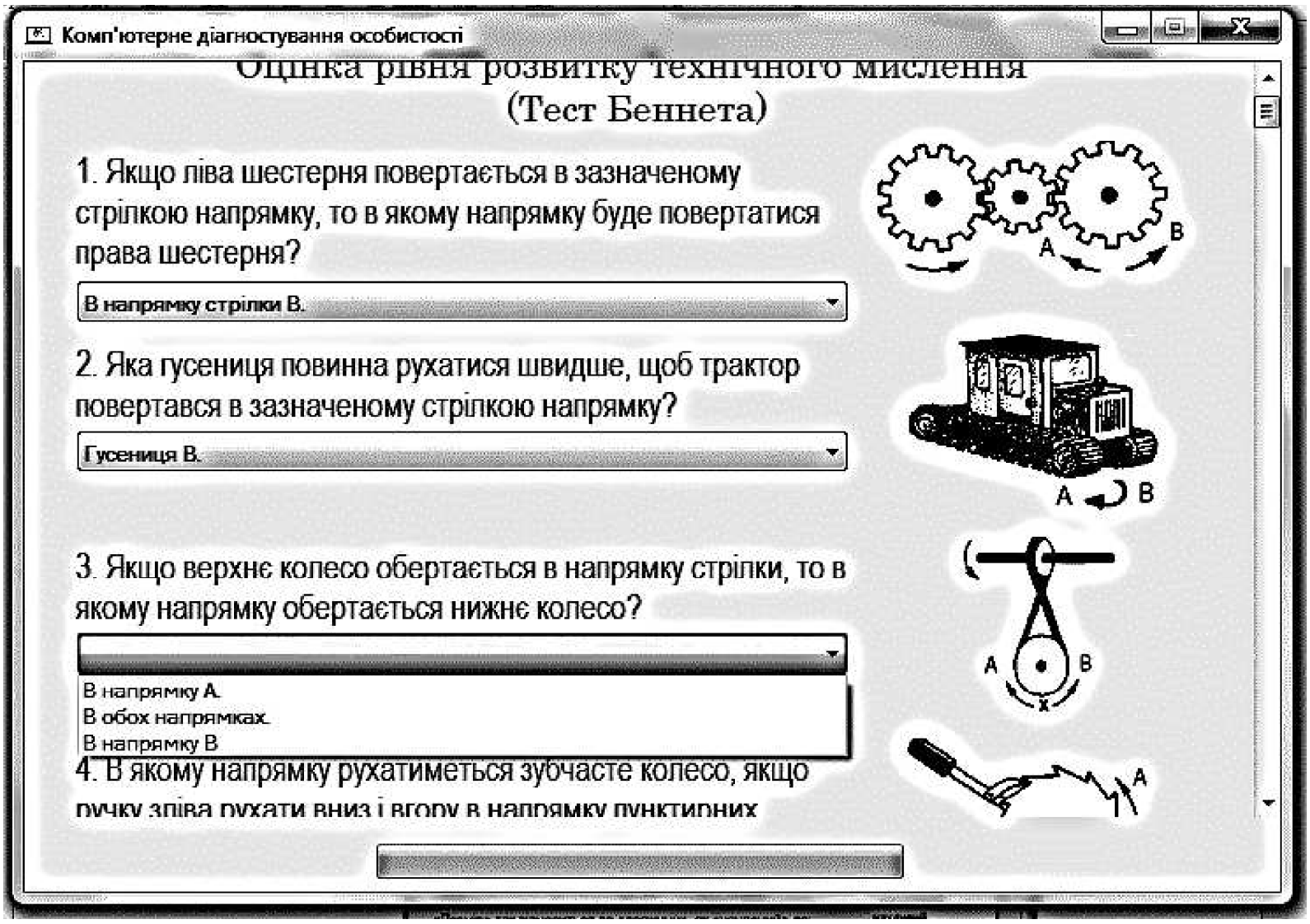

*Assessment of the level of creative potential of an individual*

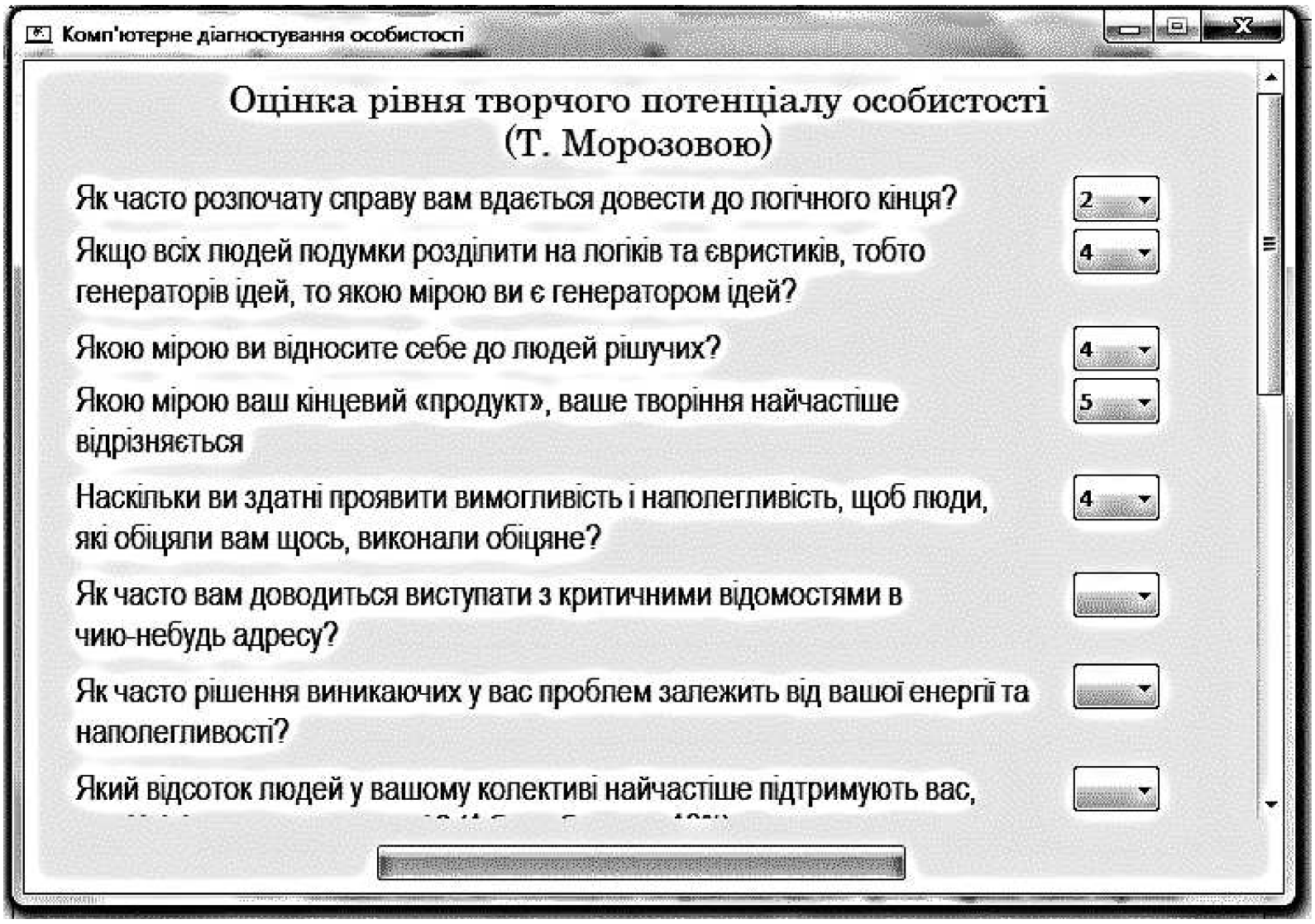

*A test to determine the characteristics of thinking*

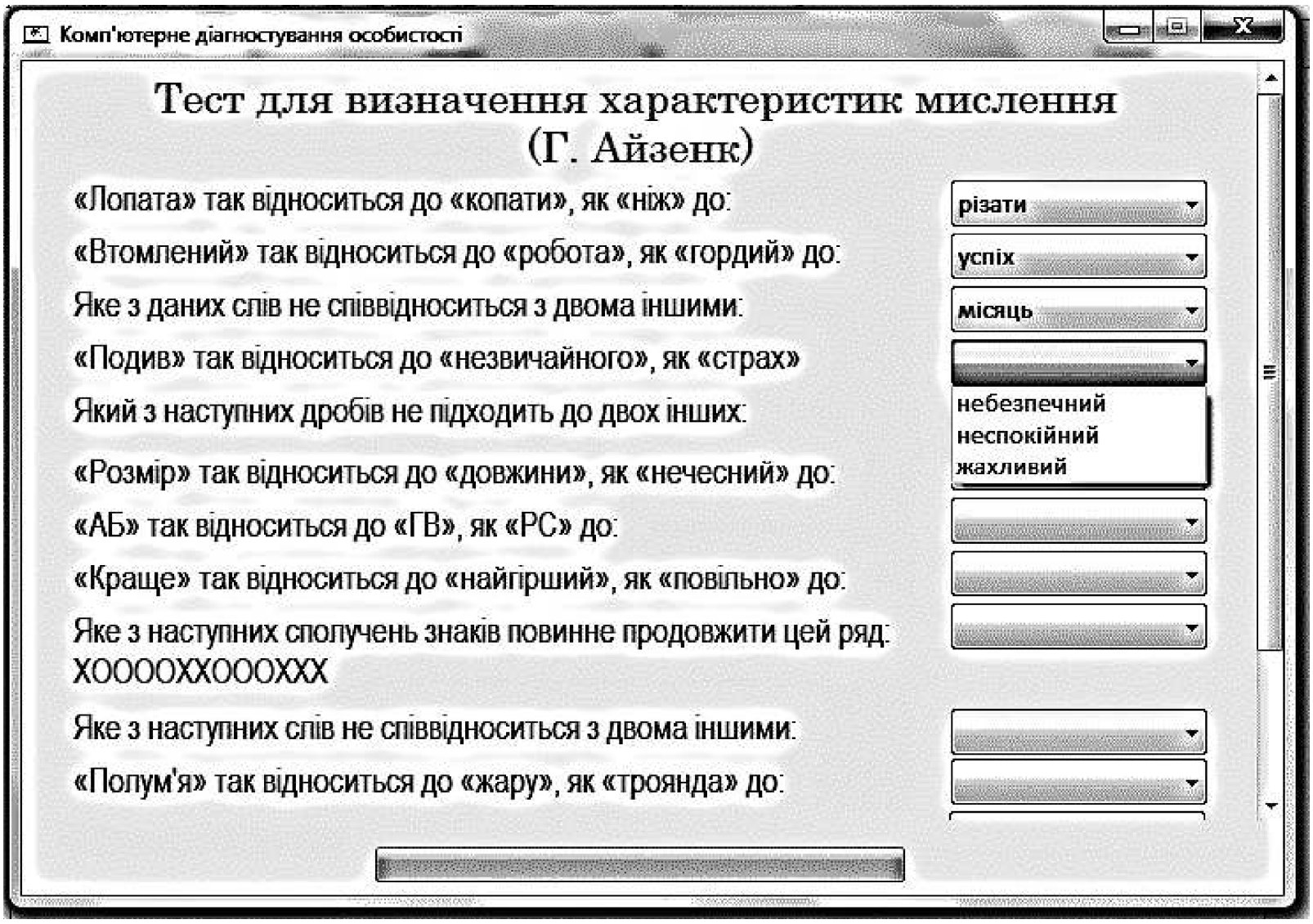

*A study of the hierarchy of value orientations*

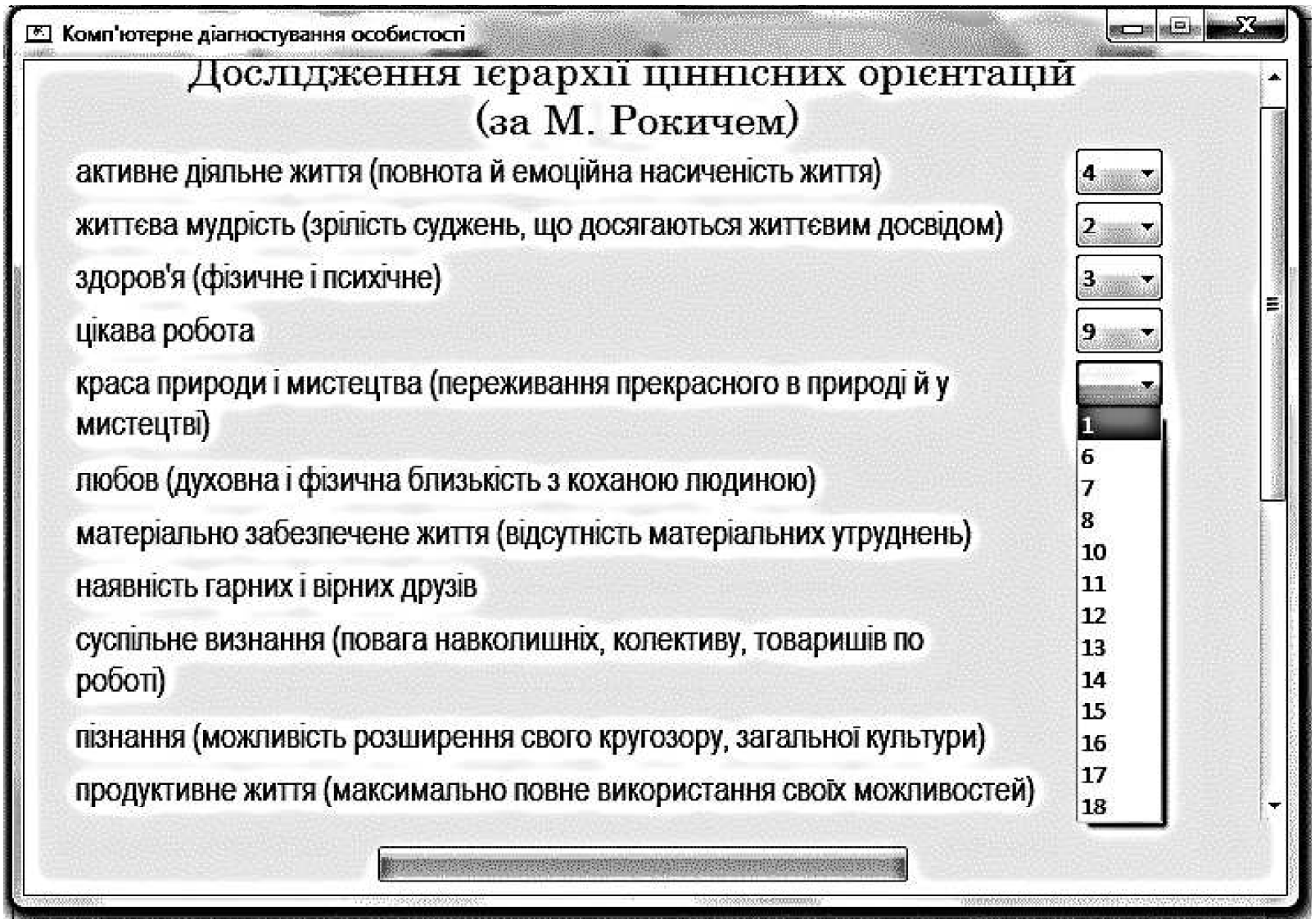

Fig. 3.4. Diagnostic tests

In tests with a "prioritization" structure, a statement is displayed in the window and next to it — a list with the place where this statement should be placed. When choosing a position of an assertion in the priority list, this position becomes unavailable for other assertions.

The algorithm of the program provides for checking whether the answer is selected from all the lists. If the answer to even one question is not selected, the program displays an error window with the text "Fill in all the answers!", otherwise the process of processing the test results begins.

The process of processing the results is divided into several stages. The first stage is the direct calculation of the number of points for the past test. The scoring scheme for each test is different. In tests where there is only one correct answer, points are awarded for choosing this answer. In addition, there are tests where there is no correct answer, and the student must simply evaluate this or that statement, in which case points are awarded according to the student's evaluation.

The second stage of processing the results is determining the level of the test result. The results of each test are reduced to a three-level scale, depending on the number of points scored (high, medium and low level), the higher the level, the higher the student has developed the quality that the test was aimed at identifying. In tests with a “prioritization” structure, points are not calculated, the process of processing the results consists only in saving the selected place in the list of priorities for each of the statements.

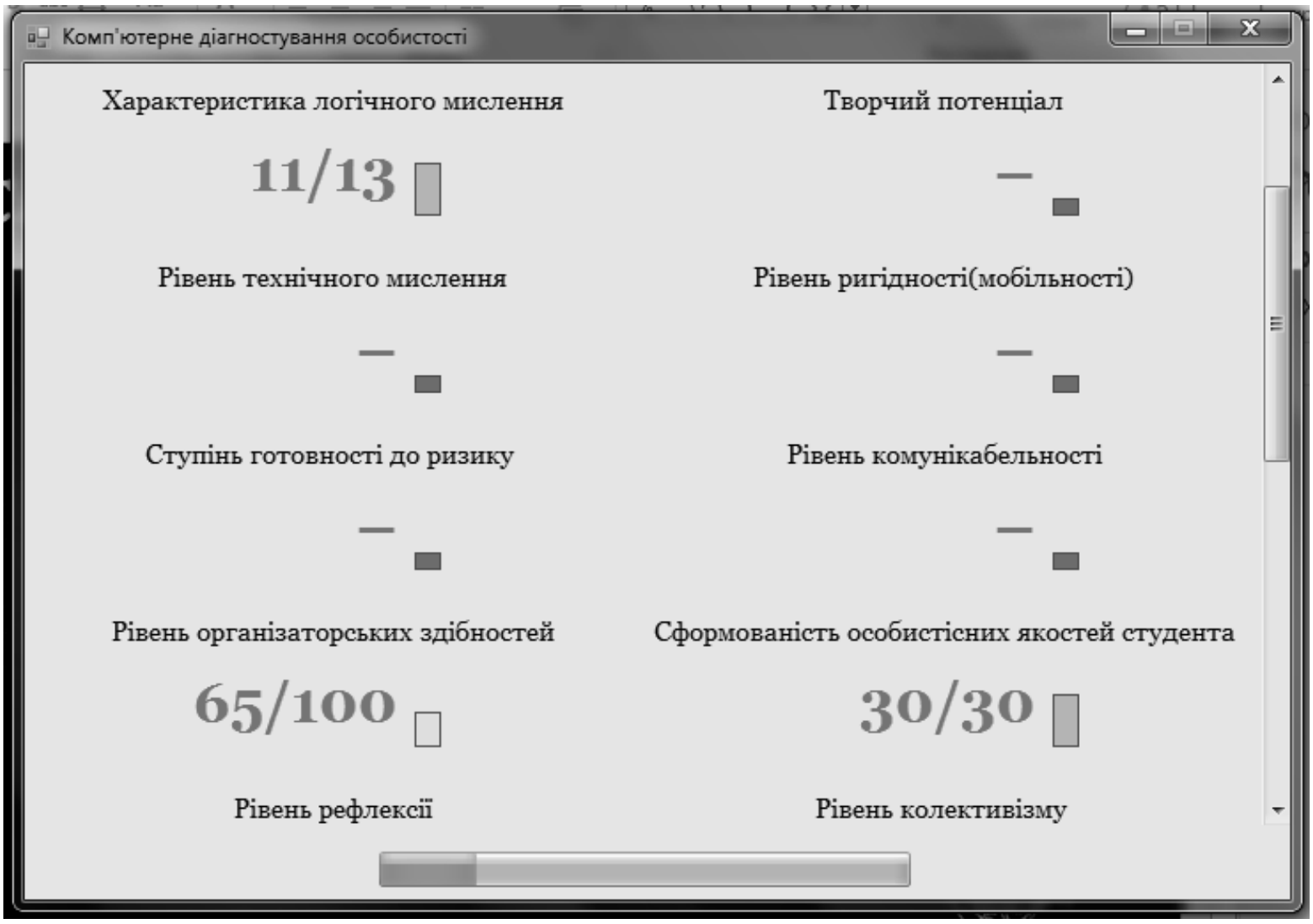

Fig. 3.5. Test results window

The conservation stage is the third stage. During the third stage, the results are stored in the database. This process consists of connecting to the database, forming the query text to the database, executing the command with the query text to the database (step of saving the results directly), and closing the connection to the database. If an error occurred at one of the stages (for example, there is no Internet connection or the administrator entered incorrect data for connecting to the database), the user will see an error with the text: “Error connecting to the database. Check your network connection or contact your administrator’, the results are not saved to the database, but may be saved later when connecting to the database. The stage of saving the results to the database takes place in parallel with the execution of the program, so the student has the opportunity to freely use the program simultaneously with the process of saving the results.

After passing the test, the test selection window opens again to the user. At the same time, the passed test will be marked as completed. It is not pos-sible to pass the same test in one session. The program provides an opportunity to view the results after the test, a test results window will open where the user can view his results for all the tests he passed (the number of points scored and their expression on a three-level scale). The user cannot see the results of the tests with the “prioritization” structure, because he does not have these results by definition. The Back button on the results panel returns the user to the test selection panel. In addition, after passing another test, the value of the progress scale, which reflects the degree of passing the diagnosis, increases.

If, due to connection problems, the program failed to save all diagnostic results, when it is closed, according to the algorithm, a warning will be displayed with the text: “Not all results were saved, are you sure you want to close the application?” with two buttons for answering “Yes” and “No”. If the user clicks the “No” button, the script will try again to save all unsaved diagnostic results. If the user selects the “Yes” option, the program will complete its execution.

Administration block. This block is a technical part of the program that is not accessible to the normal user (the one being tested), access to this part of the program is only available to the administrator (the person who conducts the diagnosis), based on this, to ensure security to access the administration panel, you must enter password. The first thing that a person who opens the administrator panel sees is a field for entering a password, a button for confirming the password entry, and a button for changing the password entry mode (hide the characters that are entered or display them). If the user entered the password correctly, the elements for entering the password will disappear, and the main part of the administrator panel will appear, if the password was incorrect — access will be denied until the user enters the correct password.

The admin panel is divided into two parts: the settings block, the results selection block.

The settings block consists of several text fields that are used to enter such information as the database server address (host), database user name, database name, database access password, and administrator panel access password. In the program, a password for accessing the administrator panel is set by default, the administrator can optionally change this password by writing a new password in the text field “password for accessing the administrator panel”.

In addition, this block has two functions “Save settings” and “Check connection to the database”. The “Save settings” button saves the entered data for connecting to the database and the password for accessing the administrator panel in a special service configuration file, which is located in the same folder as the program itself. The button “Check the connection to the database” starts the script for checking access to the database according to the entered settings, if the connection was successful, the program creates empty tables in the database in which the results of the diagnostics will be stored and displays a window with the message “The connection to the database was successful! “. If it was not possible to connect to the database, the program will display an error window with the text: “Error connecting to the database. Check your network connection or contact your administrator.”

The block of selection of results consists of one list and three text fields and serves to assign parameters for forming diagnostic results. In the list, the administrator selects the number of the test, the results of which he wants to receive, in addition, the student’s name can be entered in the text fields, if the administrator wants to see the results for an individual student, group number, if the administrator wants to see the results for an individual group, and gender (all parameters can be combined among ourselves). After pressing the “Send request” button, a request to the database is generated according to the parameters set, if the data corresponding to the specified parameters exists, the process of generating results begins.

The process of generating the results begins with the display of a window for choosing the storage location and the name of the file in which the results will be recorded. If the file in which the user wants to save the result file already exists, a warning will be displayed about this, if the user agrees — the file will be overwritten. Next, a new Excel book is created, in which data about students and their diagnostic results are recorded, according to the received data according to the specified parameters. In addition, the file records the number of students who received a high, medium and low level from the selected test, as well as statistics separately for boys and girls. According to these statistics, a histogram is built, which has nine columns.

After generating the results, the file is saved and becomes available for reading and editing by the user.

## 3.2. RESULTS OF THE AUTOMATED COMPETITIVENESS DIAGNOSTIC SYSTEM OF FUTURE INFORMATION TECHNOLOGY SPECIALISTS

The automated diagnostic system was used in the course of research and experimental work on the topic "Formation of the competitiveness of future information technology specialists in technical colleges" [3].

Students of Odessa Technical College participated in the experimental study. The control group (CG) of students studied according to the traditional system, for the experimental group (EG) pedagogical methods were applied, which contribute to the formation of the competitiveness of future information technology specialists in the process of professional training [4].

After the completion of the formative experiment, the students of the control and experimental groups were diagnosed with the help of an automated system for diagnosing the competitiveness of future specialists in information technologies according to two criteria that are structurally determined by the psychological, creative and personal qualities of the future specialist, namely motivational-value, cognitive, personal- reflective, professional and activity criteria.

The final analysis was carried out on the basis of the methods used at the diagnostic stage of the experimental work. After the mathematical processing of the collected empirical data, the results of the formation of the main indicators of the competitiveness of future specialists in information technologies according to the criteria look as follows.

The results of diagnostics based on motivational-value, cognitive, personal-reflective, professional-activity criteria after the diagnosis is completed are presented in tables 3.1–3.5.

According to the results of the automated system for diagnosing the competitiveness of future information technology specialists, as evidenced by the analysis of the overall result, there is an improvement in indicators in the experimental group, which characterizes the number of students with high and medium levels, and a reduction in the number of students with low levels. We can state an increase in the level of competitiveness of future information technology specialists based on motivational-value and personal-reflexive criteria, both in the experimental and control groups. However, small results indicate that it is necessary to further introduce new learning technologies into the educational process, to search for effective ways to increase the competitiveness of future specialists.

Table 3.1

**The results of the diagnosis of the formation of the competitiveness of future IT specialists according to the motivational and value criterion after the completion of the formative experiment**

| Indexes | %, at the ascertainment stage | | | | | | %, at the formative stage | | | | | |
|---|---|---|---|---|---|---|---|---|---|---|---|---|
| | KG | | | EG | | | KG | | | EG | | |
| | H | A | L | H | A | L | H | A | L | H | A | L |
| the presence of motivation for development, self-improvement | 11.32 | 43.4 | 45,28 | 10,19 | 43.51 | 46.3 | 16.98 | 48,11 | 34.91 | 16.2 | 58.32 | 25,48 |
| the level of education motivation | 12.53 | 28.3 | 59,16 | 15.08 | 29.1 | 55.82 | 20.75 | 35.85 | 43.4 | 24.84 | 42.12 | 33.04 |
| formation of value orientations | 14,15 | 65.01 | 20.84 | 15.75 | 65.75 | 18.5 | 19.81 | 67.92 | 12,27 | 27.0 | 69.12 | 3.88 |
| Together: | 12.67 | 45,57 | 41.76 | 13.67 | 46,12 | 40,21 | 19,18 | 50,62 | 30,19 | 22.68 | 56,52 | 20.8 |

Table 3.2

**The results of the diagnosis of the formation of the competitiveness of future IT specialists in terms of *cognitive* criterion after completion of the formative experiment**

| Indexes | %, at the ascertainment stage | | | | | | %, at the formative stage | | | | | |
|---|---|---|---|---|---|---|---|---|---|---|---|---|
| | KG | | | EG | | | KG | | | EG | | |
| | H | A | L | H | A | L | H | A | L | H | A | L |
| Possession of professional knowledge | 13,21 | 38,68 | 48,11 | 12.04 | 38.89 | 49.07 | 18.87 | 43.4 | 37.73 | 20.75 | 48.15 | 31.1 |
| Possession of professional skills and abilities | 38,68 | 37.74 | 23.58 | 38.89 | 38.89 | 22,22 | 45.3 | 42,45 | 12.25 | 48.15 | 49.07 | 2.78 |
| Together | 25.94 | 38,21 | 35.85 | 25,46 | 38.89 | 35.65 | 32.1 | 42.92 | 25.0 | 34,45 | 48,61 | 16.94 |

Table 3.3

**The results of diagnostics of the formation of competitiveness of future IT specialists according to *the personal-reflexive* criterion after completion of the formative experiment**

| Indexes | %, at the ascertainment stage | | | | | | %, at the formative stage | | | | | |
|---|---|---|---|---|---|---|---|---|---|---|---|---|
| | KG | | | EG | | | KG | | | EG | | |
| | H | A | L | H | A | L | H | A | L | H | A | L |
| The level of formation of personal qualities | 11.32 | 44,34 | 44,34 | 12.04 | 44,44 | 43.52 | 21.7 | 53.77 | 24.53 | 27.78 | 66,67 | 5.55 |
| Reflection of activity | 14,15 | 37.74 | 48,11 | 13.89 | 38.89 | 47.22 | 21.7 | 50.94 | 27,36 | 19.45 | 57.4 | 23.15 |
| Together | 12.74 | 41.04 | 46.23 | 12.96 | 41.67 | 45,37 | 21.7 | 52.36 | 25.94 | 23.62 | 62.0 | 14.35 |

Table 3.4

**The results of the diagnosis of the formation of the competitiveness of future IT specialists according to *the professional activity* criterion after the completion of the formative experiment**

| Indexes | %, at the ascertainment stage | | | | | | %, at the formative stage | | | | | |
|---|---|---|---|---|---|---|---|---|---|---|---|---|
| | KG | | | EG | | | KG | | | EG | | |
| | H | A | L | H | A | L | H | A | L | H | A | L |
| The level of development of the creative potential of the individual | 10.38 | 23.58 | 66.0 | 11,11 | 25.00 | 63.89 | 16.04 | 47.17 | 36.8 | 15.74 | 39.81 | 44,45 |
| The level of development of technical thinking | 8.49 | 19.81 | 71.7 | 9.26 | 21.30 | 69.44 | 13.2 | 29,25 | 57.55 | 18.52 | 32,41 | 49.07 |
| Professional mobility and flexibility | 9.43 | 20.75 | 69.8 | 9.26 | 22,22 | 68.52 | 14,15 | 35.84 | 50.01 | 20.37 | 32,41 | 47.22 |
| Ability to work in a team | 16.04 | 26.42 | 57.5 | 16.67 | 26.85 | 56,48 | 24.53 | 47.17 | 28.3 | 32,41 | 37.96 | 29.63 |
| The level of development of communication skills | 13,21 | 23.58 | 63.2 | 13.89 | 23.15 | 62.96 | 24.53 | 39.62 | 35.85 | 32,41 | 39.81 | 27.78 |
| The level of development of organizational skills | 14,15 | 29,25 | 56.6 | 12.96 | 29.63 | 57.41 | 24.53 | 38,68 | 36.8 | 20.37 | 37.04 | 42.6 |
| Critical thinking | 10.38 | 24.53 | 65, 1 | 10,19 | 23.15 | 66,67 | 13.2 | 34.9 | 51.9 | 15.74 | 30.56 | 53.7 |
| Together | 11.73 | 2 4, 0 | 64, 3 | 11.90 | 24.47 | 63,62 | 18.6 | 38.95 | 42,46 | 22,22 | 35.71 | 42.07 |

Table 3.5

**The results of the diagnosis of the formation of the competitiveness of future specialists in information technologies according to the specified criteria after the completion of the formative experiment**

| Criteria | %, on the affirmative stage | | | | | | %, on the formative stage | | | | | |
|---|---|---|---|---|---|---|---|---|---|---|---|---|
| | KG | | | EG | | | KG | | | EG | | |
| | H | A | L | H | A | L | H | A | L | H | A | L |
| motivational and valuable | 12.67 | 45,57 | 41.76 | 13.67 | 46,12 | 40,21 | 19,18 | 50,62 | 30,19 | 22.68 | 56,52 | 20.8 |
| cognitive | 25.94 | 38,21 | 35.85 | 25,46 | 38.89 | 35.65 | 32.1 | 42.92 | 25.0 | 34,45 | 48,61 | 16.94 |
| professional activity | 11.73 | 2 4, 0 | 64, 3 | 11.90 | 24.47 | 63,62 | 18.6 | 38.95 | 42,46 | 22,22 | 35.71 | 42.07 |
| personally reflective | 12.74 | 41.04 | 46.23 | 12.96 | 41.67 | 45,37 | 21.7 | 52.36 | 25.94 | 23.62 | 62.0 | 14.35 |
| Together | 15.77 | 37.2 | 47.0 | 16.0 | 37.8 | 46.2 | 22.9 | 46.2 | 30.9 | 25.7 | 50,71 | 23.54 |

## 3.3. CONCLUSIONS

An automated system for diagnosing the competitiveness of future information technology specialists can be recommended for the work of structural divisions of the psychological and educational direction in institutions of professional pre-higher and higher education to determine the level of formation of key personal qualities of education seekers.

# 4. PERSPECTIVE-CORRECT FORMATION OF GRAPHIC IMAGES

***Oleksandr Romanyuk, Oksana Romaniuk***

*The article presents advancements in color reproduction techniques, enhancing the realism of graphic scenes. It introduces novel methods for accurate shading of three-dimensional objects, improves the Barenbrug perspective-correct texturing method, and proposes quadratic approximation for perspective-correct texturing, reducing computation time and error.*

## 4.1. FEATURES OF FORMING REALISTIC THREE-DIMENSIONAL GRAPHIC IMAGES

The scope of computer graphics, which is the main means of communication between humans and computers, is constantly expanding, as graphic images are the most visual and adequately reproduce real objects and processes. The main task of modern computer graphics is to synthesize realistic three-dimensional images that reproduce real-world objects to the maximum extent possible. At the same time, it is important to achieve a graphic scene generation performance acceptable for a given application. Image synthesis involves the execution of a certain sequence of specialized stages, which together constitute a 3D graphics pipeline [1, 2].

The process of creating a three-dimensional image can be divided into the following main stages: the stage of describing a three-dimensional scene, the stage of geometric transformations, the stage of rendering, or visualization, and the stage of displaying the image on a monitor or printer. Each stage, in turn, contains several stages.

At the stage of describing a three-dimensional scene, the constituent objects of the scene, their main characteristics, state and relative position, and the strategy for further actions are determined. In the object space, three-dimensional objects, materials, light sources, virtual cameras, and additional tools for modeling special effects, including atmospheric phenomena, are operated with.

Synthesis of graphic images is a complex task, so in most cases, a graphic scene is decomposed into its components. Surfaces, including curved ones, are approximated by a polygonal mesh, which in most cases includes triangles. At this stage of computer graphics, the representation of surfaces by triangles is one of the ways to create real-time dynamic images and processes in an interactive mode.

At the stage of geometric transformations, in addition to tessellation, affine transformations are also performed, such as rotation, scaling, and displacement. After that, the triangles are merged into a solid wireframe model. To take into account the curvature of the surfaces, vectors of normals to the vertices of the constituent triangles are determined, for which geometric parameters are predefined.

When generating graphic images, it is necessary to take into account a number of cameras, which implies the inclusion of the corresponding stage in the graphic pipeline. The position of the cameras makes it possible to exclude invisible objects and surfaces, as well as objects that are located outside the camera's field of view, and, as a result, reduce computational costs.

At the final stage of the geometric transformation stage, a projection of the three-dimensional scene onto the visualization plane is formed, i.e. the output is a set of polygons in screen coordinates.

Projection is the mapping of points specified in a coordinate system with dimension $E$ to points in a system with a smaller dimension. In computer graphics, we mainly consider projections of three-dimensional images onto a two-dimensional picture plane. Flat geometric projections are divided into two main classes: central and parallel. The difference between them is determined by the ratio between the center of the projection and the projection plane. If the distance between them is finite, then the projection will be central, but if it is infinite, then the projection will be parallel. In real space, the reflection of rays from objects is perceived at the point of the observer's location, i.e., according to the principle of central projection [1]. Correct color reproduction takes place provided that the components of the color intensities of the corresponding surface points in the world (object) and screen coordinate systems coincide.

Existing shading methods do not provide correct color reproduction when shading three-dimensional graphic objects, since they do not take into account the $z$-coordinate in perspective projection. In this chapter, we develop the theoretical foundations for correct color reproduction in perspective projection.

At the final visualization (rendering) stage, pixels of the image of a three-dimensional scene are formed on the screen, which involves calculating not only their addresses but also color intensities. The rendering stage is the most time-consuming, since all operations are performed at the pixel and subpixel levels, unlike the geometric transformation stage, where only the vertices of polygons were processed.

At the rasterization stage, for a selected polygon that falls within the camera space, all its internal points are searched, which involves determining their addresses in the on-screen coordinate system. For the selected point, normalized vectors to the object surface, light source, and observer are calculated, as well as auxiliary vectors depending on the lighting model selected. These vectors are then used to determine the color intensities of the image points. The intensity value of the original light source is used or the color intensity value is sampled from the texture. If necessary, texture filtering or anti-aliasing is performed to eliminate artifacts.

When forming images of relief surfaces, the normal vector to the surface is perturbed by one of the known methods.

Additional procedures can be applied to the final image to simulate various natural phenomena.

At the post-processing stage, pixel visibility is checked once again and the finished image is displayed on a screen or printer.

In the course of their evolution, GPUs have gone from graphics pipelines with hard logic to programmable computing environments. Over the course of several generations, individual stages of pipelines were gradually replaced by programmable devices called shaders [1] A shader is a program that performs a certain stage of the graphics pipeline to determine the final parameters of an object or image. There are three types of shaders: vertex, pixel, and geometric, which process polygon vertices, pixels, and polygons, respectively. Vertex shaders determine the order of transformation of polygon vertices. Pixel shaders are used to dynamically change the properties of individual pixels. They calculate the color of a single pixel based on the input data received from the vertex raider and the specified lighting parameters. Geometric shaders were first used in NVidia 8 series video cards. They process primitives and can create new vertices and generate new primitives without using CPU resources.

A special place among the numerous stages of the graphics pipeline is occupied by the lighting stage, the texture mapping stage, and the painting stage. These three stages are directly responsible for the realism of objects.

Creating realistic images in computer graphics involves accurately conveying the external characteristics of natural objects in their computer model. The features of a realistic image are depth rendering, which allows us to determine the distance between objects, lighting of objects, as well as to evaluate the shape, color, and material of objects, shadow casting, and other optical effects.

One of the ways to increase the realism of an image is to display fine details, irregularities, and relief on the surfaces of objects that make up a graphic scene. This can be achieved by changing the geometry of the object [3]. The geometric model of the object is divided into a large number of low-level polygons, the vertices of which are shifted to the appropriate distance and in the appropriate direction. The larger the number of polygons and the smaller their size, the more realistic the object looks. This approach allows to display the relief and irregularities on the surface quite realistically, but is characterized by high computational complexity and low performance. It is difficult to use it to display such fine details as sand, skin pores, wood structure, etc.

An alternative to changing the geometry of an object is texture overlay [1, 2], which is widely used in graphics systems. A texture is a raster image that is superimposed on the surface of a polygon, which make up 3D models, to give it color or the illusion of relief [4]. Textures allow to simulate a variety of materials, complex surface structures (porous, with cracks, etc.) that are difficult to implement with a set of polygons. The quality of the texture surface is determined by the texture pixels — the number of pixels per minimum texture unit. Since a texture is an image, its resolution and format play a crucial role, which ultimately affects the quality of the synthesized graphic image. Therefore, in this chapter, we propose methods to improve the efficiency of perspective-correct texturing.

## 4.2. PERSPECTIVE-CORRECT COLOR REPRODUCTION WHEN INTERPOLATING COLOR INTENSITIES

In the classical implementation of Gouraud shading [1–4], the $z$-coordinate is taken into account only at the vertices of the triangle when determining their color intensities. Subsequently, the $z$-coordinate is used only to remove invisible surfaces. Thus, the perspective of the object is not taken into account when Gouraud coloring.

Fig. 4.1 shows a line segment $A'B'$, projected onto an screen plane. When using linear interpolation to calculate the color intensities along the edge $A'B'$, which is located in the world coordinate system, the color intensity at point $Q'$ $(A'Q' = B'Q')$ is equal to $I_{Q'} = (I_{A'} + I_{B'})/2$. Since $BQ \neq QA$, this regularity at the point $Q$ will be broken (Fig. 4.2).

The example above shows that there is a violation of the regularity of color change between a real object and its image on the screen.

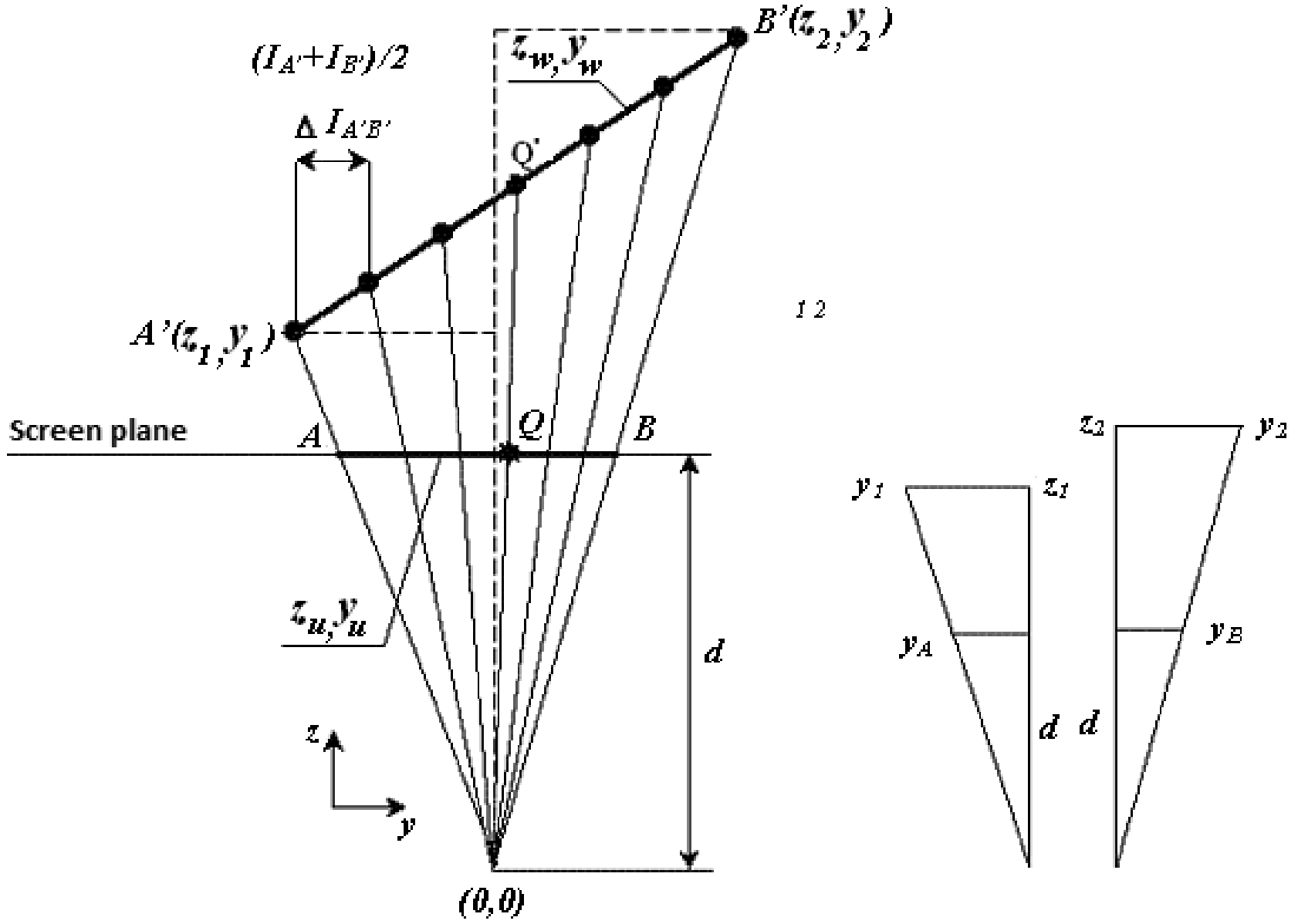

Fig. 4.1. Perspective vector projection on the screen plane

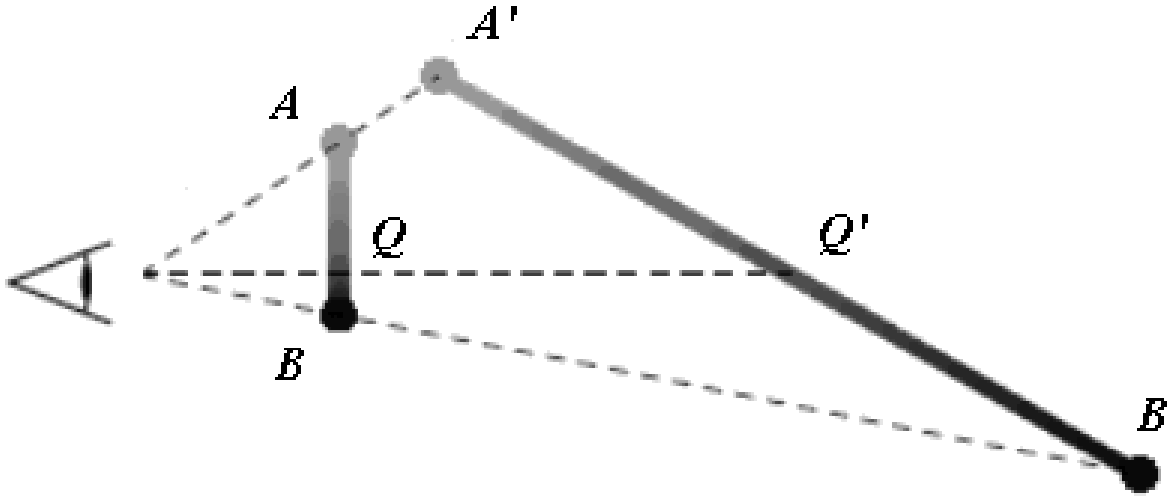

Fig. 4.2. Color mismatch for segments in the screen and world coordinate systems

This subsection discusses the issues of correct coloring of three-dimensional objects using the Gouraud method. According to the Gouraud method [1–4], the intensity of a color along a segment $AB$ in the world coordinate system is determined by the formula:

$$I_w = I_A + w \cdot (I_B - I_A)\,.$$

In the screen coordinate system:

$$I_v = I_A + v \cdot (I_B - I_A).$$

Since $v \neq w$, proportionality is violated when reproducing colors in all cases when $AB$ and $A'B'$ are not parallel (Fig. 4.2). Let's find the relationship between the variables *v* and *w*.

Let the observation point be located at a distance $d$ from the screen plane at the origin of the *yoz* coordinate system (Fig. 4.1).

For the segment $A'B'$, write the parametric equations

$$z_w = z_1 + w \cdot (z_2 - z_1),\; y_w = y_1 + w \cdot (y_2 - y_1).$$

For the segment $AB$ $y_u = y_A + u \cdot (y_A - y_B)$.

Given the similarities of the corresponding triangles (Fig. 4.1), we can write

$$\frac{d}{z_1} = \frac{y_A}{y_1};\quad \frac{d}{z_2} = \frac{y_B}{y_2};\quad \frac{d}{z_w} = \frac{y_u}{y_w}.$$

From the last expressions we find

$$y_1 = \frac{z_1 \cdot y_A}{d};\quad y_2 = \frac{z_2 \cdot y_B}{d};\quad z_w = \frac{y_w}{y_u} d.$$

Taking into account that $y_w = y_1 + w \cdot (y_2 - y_1)$ and the values for $y_1,\ y_2$

$$z_w = \frac{(y_1 + w \cdot (y_2 - y_{1})) \cdot d}{y_u} = \frac{(\frac{z_1 \cdot y_A}{d} + w \cdot (\frac{z_2 \cdot y_B}{d} - \frac{z_1 \cdot y_A}{d})) \cdot d}{y_A + u \cdot (y_B - y_A)} =$$

$$= \frac{(z_1 \cdot y_A + (z_2 \cdot y_B - z_1 \cdot y_A)) \cdot w}{y_A + u \cdot (y_B - y_A)}.$$

The right-hand side of the last expression is equivalent to the equivalent value, which is equal to $z_1 + w \cdot (z_2 - z_1)$. We get the expression

$$\frac{(z_1 \cdot y_A + (z_2 \cdot y_B - z_1 \cdot y_A)) \cdot w}{y_A + u \cdot (y_B - y_A)} = z_1 + w \cdot (z_2 - z_1).$$

After the transformations, we find that

$$w = \frac{u \cdot z_1}{z_2 - u \cdot (z_2 - z_1)}.$$

The last expression relates the parametric variables $w$, $u$ using the $z$ -coordinates of the edge $A'B'$ in the world system.

According to the Gouraud method, the color intensity along an edge is determined by the formula: $I_w = I_A + w \cdot (I_B - I_A)$. Substituting the $w$ values into the above expression, we obtain

$$I_w = I_A + \frac{u \cdot z_1}{z_2 - u \cdot (z_2 - z_1)}(I_B - I_A) .$$

Let's estimate the error that occurs when the perspective correction of color intensity is not taken into account.

$$\Delta I = I_A + (I_B - I_A) \cdot \frac{u \cdot z_1}{z_2 - u \cdot (z_2 - z_1)} - I_A - (I_B - I_A) \cdot u =$$

$$= (I_B - I_A) \cdot (u - \frac{u \cdot z_1}{z_2 - u(z_2 - z_1)}) = (I_B - I_A) \cdot u \cdot (1 - \frac{1}{\frac{z_2}{z_1} + u(1 - \frac{z_2}{z_1})}). \quad (4.1)$$

Fig. 4.3 shows a graph of the change $\Delta = u \cdot (1 - \frac{1}{\frac{z_2}{z_1} + u \cdot (1 - \frac{z_2}{z_1})})$ in the multiplier for different ratios $z_2 / z_1$.

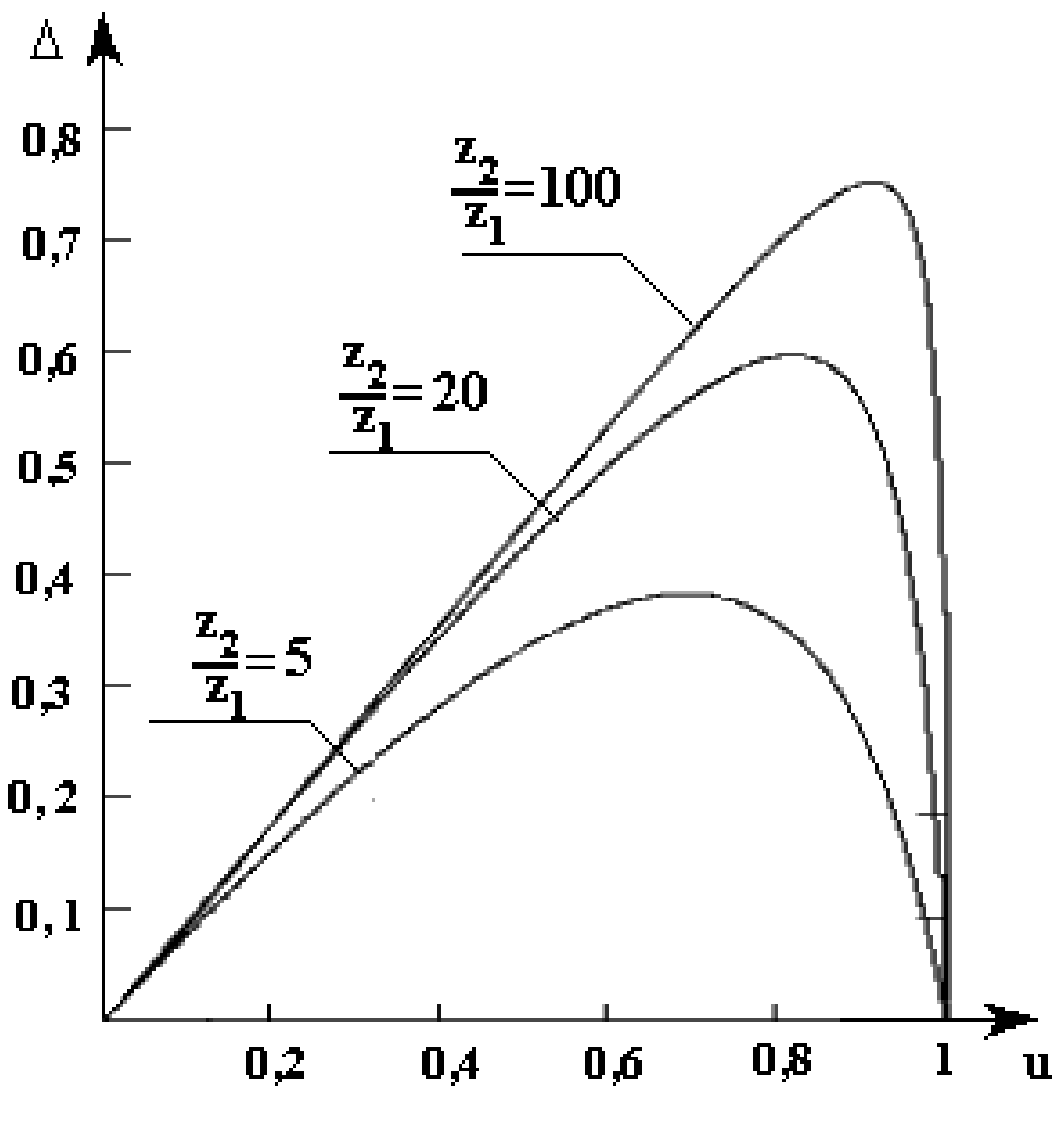

Fig. 4.3. Dependence $\Delta$ on $u$

When $z_1 > z_2$ the graph of dependence is similar to the one above with the difference that $\Delta < 0$, and the maximum value occurs at $1 - u$. Let determine the maximum values $\Delta$ for different ratios $z_2 / z_1$. To do this, find the derivative of $\Delta$ and set it to zero. We get two roots

$$\frac{\sqrt{\frac{z_2}{z_1}-\frac{z_2}{z_1}}}{1-\frac{z_2}{z_1}}, \quad \frac{-\sqrt{\frac{z_2}{z_1}-\frac{z_2}{z_1}}}{1-\frac{z_2}{z_1}}.$$

Researchers have shown that the first root determines the argument at which the function has a maximum value for $z_2 \geq z_1$, and the second root determines the maximum value for $z_1 > z_2$, and the maximum values of $\Delta$ coincide, so we will give only one expression for the first root.

$$\max\Delta = \frac{\sqrt{\frac{z_2}{z_1}-\frac{z_2}{z_1}}}{1-\frac{z_2}{z_1}} \cdot \left(1-\frac{1}{\frac{z_2}{z_1}+\left(1-\frac{z_2}{z_1}\right)\cdot\frac{\sqrt{\frac{z_2}{z_1}-\frac{z_2}{z_1}}}{1-\frac{z_2}{z_1}}}\right) = \frac{\left(\frac{z_2}{z_1}-\sqrt{\frac{z_2}{z_1}}\right)\left(\sqrt{\frac{z_2}{z_1}}-1\right)}{\sqrt{\frac{z_2}{z_1}-1}}.$$

The graph of the maximum value $\Delta$ versus the ratio $z_2/z_1$ is shown in Fig. 4.4 for different ratios $z_2/z_1$.

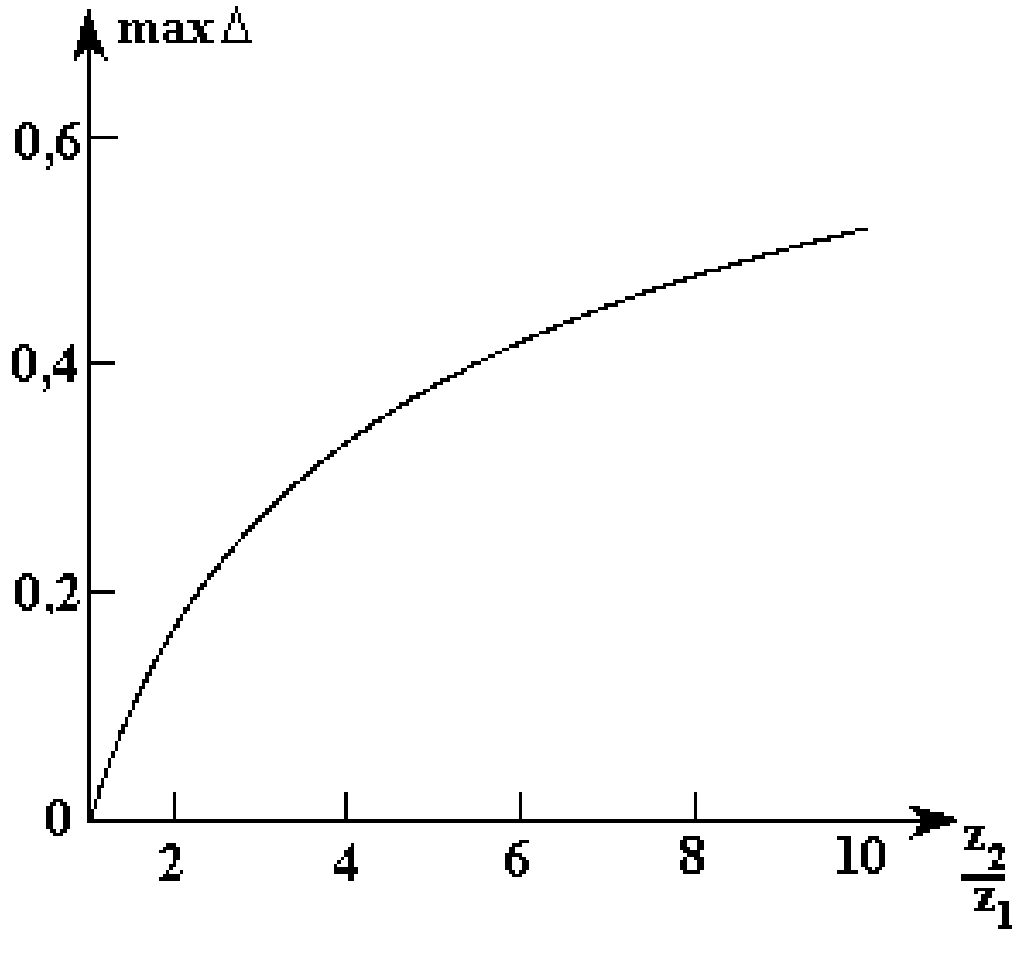

Fig. 4.4. Graph of dependence of the maximum value $\Delta$ on the ratio $z_2/z_1$

The graph shows that not taking into account the perspective of a graphic object leads to visual differences between images. It should be noted that in expression (4.1), the maximum value of the difference $I_B - I_A$ is equal to $I_B$ and occurs when the intensity at a point $B$ is equal to the maximum possible.

## 4.3. DETERMINATION OF VECTORS TO TAKE INTO ACCOUNT THE PERSPECTIVE OF A THREE-DIMENSIONAL SCENE

In the classical implementation of Phong shading [1–4], there are no geometric transformations that are inherent in perspective projections, which certainly affects the realism of reproducing graphic scenes. If Gouraud shading requires perspective correction of the color intensity of points, then when shading by Phong, it is necessary to correct the normal vectors and subsequently determine the color of the pixels. When correcting the normal vectors, it is necessary to take into account all the stages of geometric transformations established by the OpenGL and DirectX 10 standards.

We will consider the right-handed coordinate system [1], which is used in most computer graphics programs. Fig. 4.5 shows the features of perspective projection in three-dimensional space.

The volume of observation is set between the near and far cutoff planes, which are perpendicular to the axis $Z_w$. The display of the scene projection includes only those objects that are inside the cut-off pyramid. According to the OpenGL standard [1], the observation volume is displayed in a symmetric normalized cube (Fig. 4.6 — left image).

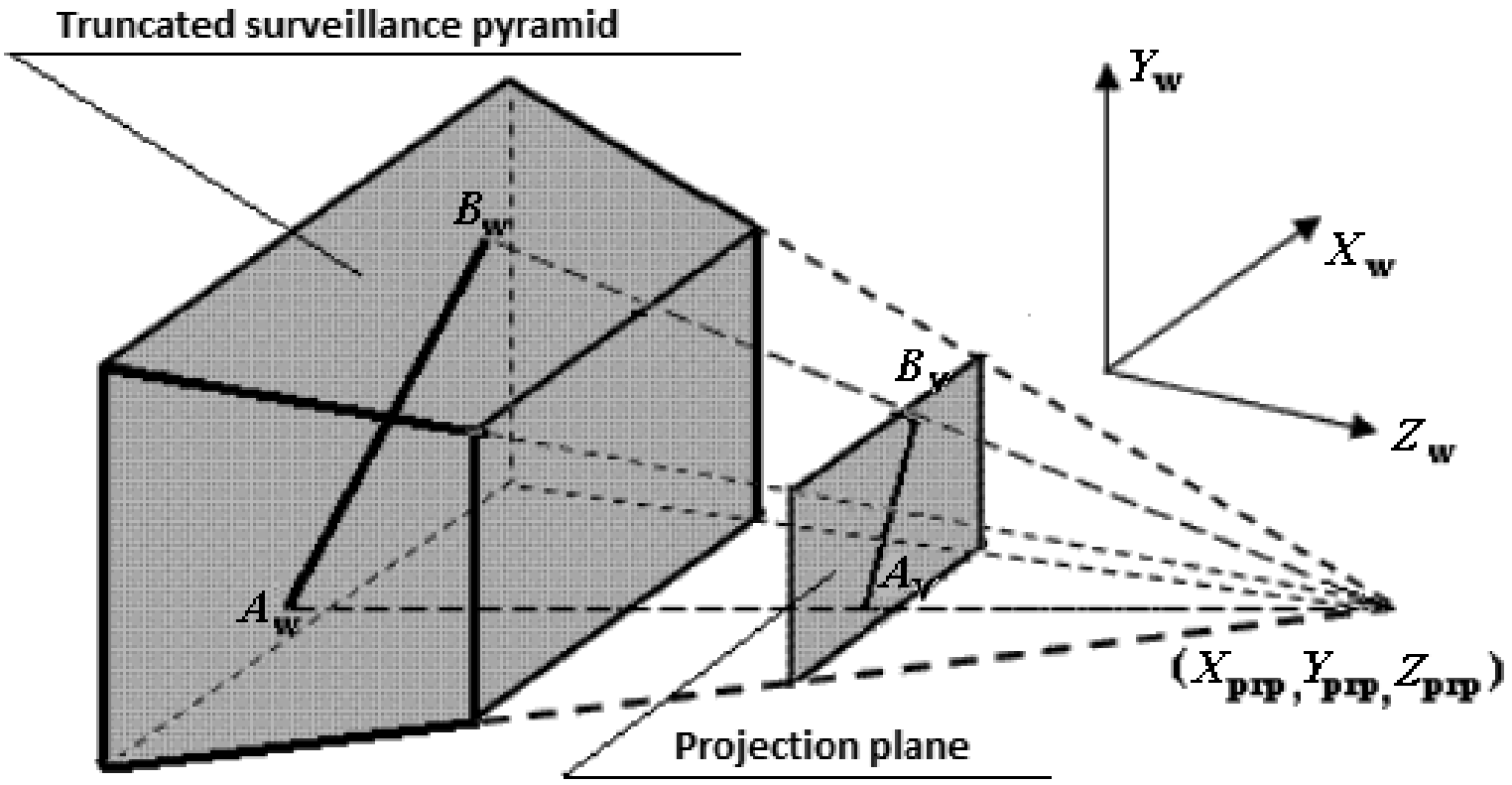

Fig. 4.5. Pyramidal observation volume of the perspective projection

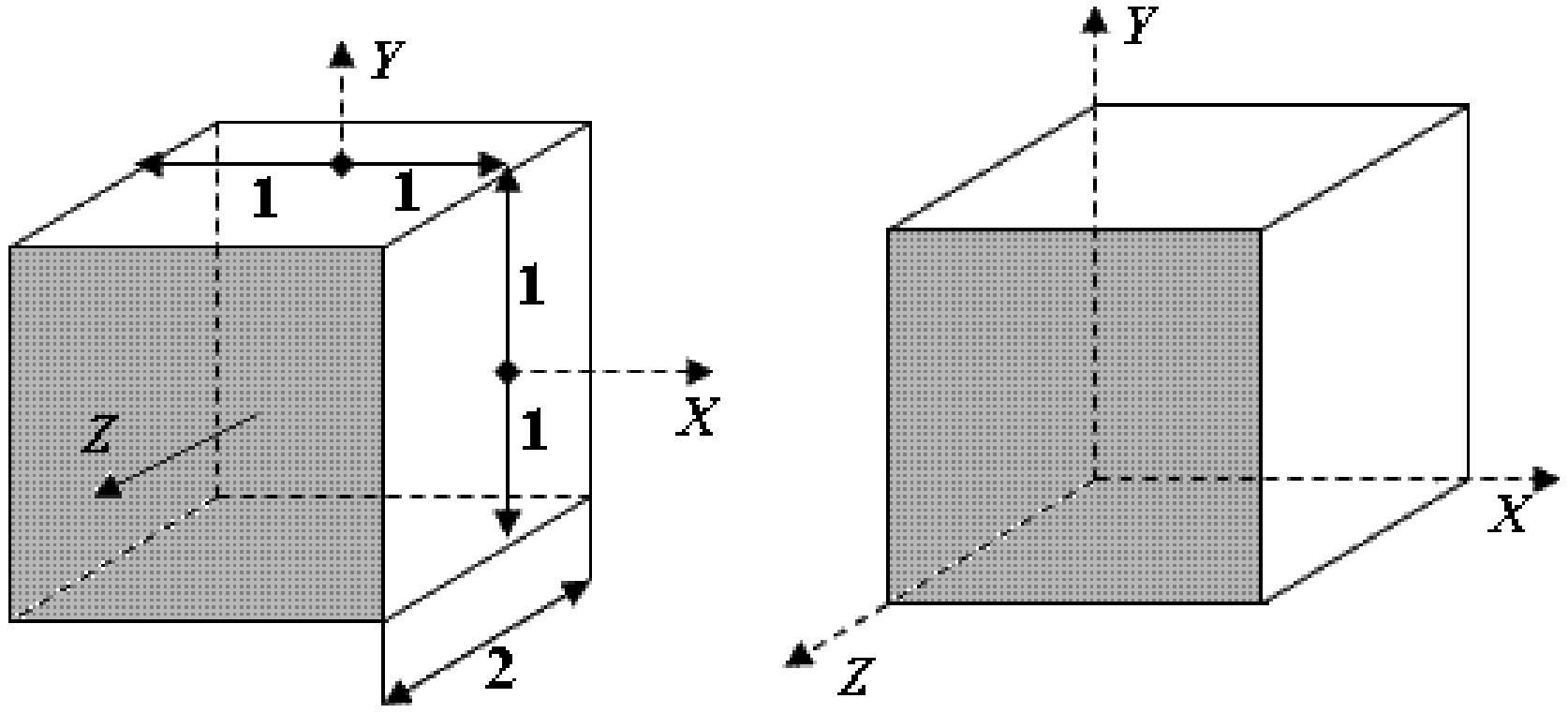

Fig. 4.6. Offset of the origin of the coordinate system of the normalized observation volume

Let a line segment $A_w B_w$ be given in the world coordinate space (Fig. 4.5). In the plane of observation, it corresponds to the segment $A_v B_v$.

The normalized coordinates $X_{norm}, Y_{norm}, Z_{norm}$ of the point $X_w, Y_w, Z_w$ in the world coordinate system are found by the homogeneous coordinates using the formula [2]

$$X_{norm} = \frac{-Z_n \cdot S_x \cdot X_w + S_x (X_m - X_n)/2}{-Z_w},$$

$$Y_{norm} = \frac{-Z_n \cdot S_y \cdot Y_w + S_y (Y_m - Y_n)/2}{-Z_w},$$

$$Z_{norm} = \frac{S_z \cdot Z_w + t_z}{-Z_w},$$

where $S_x = \frac{2}{X_m - X_n}$, $S_x = \frac{2}{Y_m - Y_n}$, $S_z = \frac{Z_m - Z_n}{Z_m - Z_n}$, $t_z = \frac{2Z_m \cdot Z_n}{Z_m - Z_n}$.

The contents of the normalized observation volume must be converted to screen coordinates. For simplicity, let's move the origin of the coordinate system of the normalized cube from its center to the leftmost corner, that is, shift all coordinates by one (Fig.4.6). Then the transformation matrix for the coordinates $(X_n + 1)$, $(Y_n + 1)$, $(Z_n + 1)$ will have the following form [3]

$$\begin{bmatrix} \frac{X_{vm}-X_{vn}}{2} & 0 & 0 & SM_{vx} \\ 0 & \frac{Y_{vm}-Y_{vn}}{2} & 0 & SM_{vy} \\ 0 & 0 & \frac{Z_{vm}-Z_{vn}}{2} & SM_{vz} \\ 0 & 0 & 0 & 1 \end{bmatrix}.$$

In most cases, the zero depth value is correlated to the screen area, for which $SM_{vz}=0$. The above transformations can be written in matrix form

$$\begin{bmatrix} X_v \\ Y_v \\ Z_v \\ 1 \end{bmatrix} = \begin{bmatrix} \frac{X_{vm}-X_{vn}}{2} & 0 & 0 & SM_{vx} \\ 0 & \frac{Y_{vm}-Y_{vn}}{2} & 0 & SM_{vy} \\ 0 & 0 & \frac{Y_{vm}-Y_{vn}}{2} & 0 \\ 0 & 0 & 0 & 1 \end{bmatrix} \times$$

$$\times \left[ \frac{-1}{Z_w} \times \begin{bmatrix} \frac{2Z_n}{X_m-X_n} & 0 & \frac{X_m+X_n}{X_m-X_n} & 0 \\ 0 & \frac{2Z_n}{Y_m-Y_m} & \frac{Y_m+Y_n}{Y_m-Y_n} & 0 \\ 0 & 0 & \frac{Z_n+Z_m}{Z_n-Z_m} & \frac{2Z_nZ_m}{Z_n-Z_m} \\ 0 & 0 & -1 & 0 \end{bmatrix} \times \begin{bmatrix} X_w \\ Y_w \\ Z_w \\ 1 \end{bmatrix} + \begin{bmatrix} 1 \\ 1 \\ 1 \\ 0 \end{bmatrix} \right]$$

From the latter system, we find that

$$\begin{bmatrix} X_v \\ Y_v \\ Z_v \end{bmatrix} = \begin{bmatrix} \frac{(X_{vm}-X_{vn})\cdot(X_n\cdot Z_w+Z_n\cdot X_w)-SM_{vx}\cdot(X_m-X_n)\cdot Z_w}{Z_w(X_n-X_m)} \\ \frac{(Y_{vm}-Y_{vn})\cdot(Y_n\cdot Z_w+Z_n\cdot Y_w)-SM_{vy}\cdot(Y_m-Y_n)\cdot Z_w}{Z_w(Y_n-Y_m)} \\ Z_n\cdot(Z_{vm}-Z_{vn})\frac{(Z_w-Z_m)}{Z_w(Z_n-Z_m)} \end{bmatrix}.$$

The $v$-index determines whether the point belongs to the on-screen coordinate system, and the $w$-index — to the world coordinate system. From the resulting system of equations, it is easy to find the $X_w, Y_w, Z_w$ point's coordinates in the world coordinate system.

$$\begin{bmatrix} X_{Aw} \\ Y_{Aw} \\ Z_{Aw} \end{bmatrix} = \begin{bmatrix} \dfrac{(SM_{vx} - X_{Av}) \cdot (X_m - X_n) - X_n \cdot (X_{vm} - X_{vn})}{(X_{vm} - X_{vn}) Z_n} \\ \dfrac{(SM_{vy} - Y_{Av}) \cdot (Y_m - Y_n) - Y_n \cdot (Y_{vm} - Y_{vn})}{(Y_{vm} - Y_{vn}) Z_n} \\ \dfrac{(SM_{vz} - Z_{Av}) \cdot (Z_m - Z_n) - Z_n \cdot (Z_{vm} - Z_{vn})}{(Z_{vm} - Z_{vn}) Z_n} \end{bmatrix}. \quad (4.3)$$

The system of equations (4.3) is similar for the points $B$ and $C$, and its denominators do not depend on the location of the points in the world system and are not equal to zero, provided that there is indeed a display window in the screen coordinate system. In the world coordinate system, for the current point $C_w$ belonging to the segment $A_w B_w$, we can write

$$\begin{bmatrix} X_{Cw} \\ Y_{Cw} \\ Z_{Cw} \end{bmatrix} = \begin{bmatrix} X_{Aw} \\ Y_{Aw} \\ Z_{Aw} \end{bmatrix} + t_w \cdot \begin{bmatrix} X_{Bw} - X_{Aw} \\ Y_{Bw} - Y_{Aw} \\ Z_{Bw} - Z_{Aw} \end{bmatrix}. \quad (4.4)$$

Similarly, for the screen coordinate system for the point $C_v$ of the segment $A_v B_v$

$$\begin{bmatrix} X_{Cv} \\ Y_{Cv} \end{bmatrix} = \begin{bmatrix} X_{Av} \\ Y_{Bv} \end{bmatrix} + t_v \cdot \begin{bmatrix} X_{Bv} - X_{Av} \\ Y_{Bv} - Y_{Av} \end{bmatrix}. \quad (4.5)$$

Let's write the first equation of the system (4.4) taking into account the first equation of the system (4.3).

$$\begin{aligned} &(SM_{vx} - X_{Cv}) \cdot (X_m - X_n) - X_n \cdot (X_{vm} - X_{vn}) \cdot Z_{Cw} = \\ &= (SM_{vx} - X_{Av}) \cdot (X_m - X_n) - X_n \cdot (X_{vm} - X_{vn}) \cdot Z_{Aw} + \\ &+ ((SM_{vx} - X_{Bv}) \cdot (X_m - X_n) - X_n \cdot (X_{vm} - X_{vn}) \cdot Z_{Bw} - \\ &-(SM_{vx} - X_{Av}) \cdot (X_m - X_n) - X_n \cdot (X_{vm} - X_{vn}) \cdot Z_{Aw}) \cdot t_w. \end{aligned}$$

In the last equation, instead of $X_{cv}$ and $Z_{cw}$ in accordance with the first formula of system (4.5), substitute the value of $X_{Av} + (X_{Bv} - X_{Av}) \cdot t_v$. We obtain

$$\begin{gathered}\left[(SM_{vx} - X_{Av} - (X_{Bv} - X_{Av})\cdot t_v)\cdot(X_m - X_n) - X_n\cdot(X_{vm} - X_{vn})\right]\times \\ \times(Z_{Aw} + (Z_{Bw} - Z_{Aw})\cdot t_w) = \\ \left[(SM_{vx} - X_{Av})\cdot(X_m - X_n) - X_n\cdot(X_{vm} - X_{vn})\right]Z_{Aw} + \\ \left[\ \left[(SM_{vx} - X_{Bv})\cdot(X_m - X_n) - X_n\cdot(X_{vm} - X_{vn})\right]Z_{Bw} - \right. \\ \left. \left[(SM_{vx} - X_{Av})\cdot(X_m - X_n) - X_n\cdot(X_{vm} - X_{vn})\right]Z_w\ \right]\cdot t_w\end{gathered}$$

Opening the brackets and performing the equivalent transformations, we find that

$$t_w = \frac{Z_{Aw}\cdot t_v}{Z_{Bw} - t_v(\cdot Z_{Bw} - Z_{Aw})}. \tag{4.6}$$

Dividing the numerator and denominator of the fraction by $Z_{Bw}$, we obtain

$$t_w = \frac{t_v}{\frac{Z_{Bw}}{Z_{Aw}} - t_v(1 - \frac{Z_{Bw}}{Z_{Aw}})}. \tag{4.7}$$

If we know the vectors of the normals $\vec{N}_l$ and $\vec{N}_p$, respectively, at the left and right points of the triangle rasterization line, then the intermediate vector $\vec{N}_s$ can be found by the formula

$$\vec{N}_s = \vec{N}_l + t_w\cdot(\vec{N}_p - \vec{N}_l).$$

Failure to take into account the depth of the object when calculating the vectors leads to an error in determining its orthogonal components, which can be determined by formula (4.2) by replacing the color intensity value with the value of the orthogonal component.

For perspective-correct color reproduction in Phong shading, it is necessary to use nonlinear interpolation of normal vectors using the variable $t_w$. Unfortunately, the calculation of $t_w$ by formula (4.6) involves performing a division operation for each current value of $t_v$. Let us consider the possibility of approximation $t_w$ to simplify the hardware implementation. Since the dependence of $t_w$ is nonlinear, the use of linear interpolation over the entire variable interval of $t_v$ is excluded. Let's approximate $t_w$ with a second-degree polynomial $a\cdot t_v^2 + b\cdot t_v + c.$ Let's find the unknowns $a, b, c$. To do this, let's create a system of equations using three points $t_v = 0,\ t_v = 1,\ t_v = 1/2.$

$$\begin{cases} c = 0, \\ a + b + c = 1, \\ \frac{1}{4} \cdot a + \frac{1}{2} \cdot b + c = \frac{Z_{Aw}}{Z_{Bw} + Z_{Aw}}. \end{cases}$$

The system has the following solution

$$a = \frac{2 \cdot (Z_{Bw} - Z_{Aw})}{(Z_{Bw} + Z_{Aw})},\ b = \frac{(3 \cdot Z_{Aw} - Z_{Bw})}{(Z_{Bw} + Z_{Aw})},\ c = 0.$$

If $\hbar = \frac{Z_{Bw}}{Z_{Aw}}$, then $a = \frac{2 \cdot (\hbar - 1)}{(\hbar + 1)}$, $b = \frac{(3 - \hbar)}{(\hbar + 1)}$.

The quadratic approximation gives satisfactory results only for $\hbar \leq 3$. Fig. 4.7 shows a graph of the change in the absolute approximation error from $t_v$ and $\hbar$.

A higher approximation accuracy can be achieved by using piecewise quadratic interpolation on two intervals of change $t_v$. For $0 \leq t_v \leq 0{,}5$

$$a = \frac{8 \cdot Z_{Aw} \cdot (Z_{Aw} - Z_{Aw})}{(Z_{Bw} + Z_{Aw})(3 \cdot Z_{Bw} + Z_{Aw})},\ b = \frac{(3 \cdot Z_{Aw} + Z_{Bw})}{(Z_{Bw} + Z_{Aw})(3 \cdot Z_{Bw} + Z_{Aw})},\ c = 0.$$

For $0{,}5 < t_v \leq 1$

$$a = \frac{-8 \cdot Z_{Bw} \cdot (Z_{Aw} - Z_{Bw})}{(Z_{Bw} + Z_{Aw})(3 \cdot Z_{Bw} + Z_{Aw})},\ b = \frac{2 \cdot (9 \cdot Z_{Aw} - 5 \cdot Z_{Bw}) \cdot Z_{Bw}}{(Z_{Bw} + Z_{Aw})(3 \cdot Z_{Bw} + Z_{Aw})},$$

$$c = \frac{3 \cdot (Z_{Aw} - Z_{Bw})^2}{(Z_{Bw} + Z_{Aw})(3 \cdot Z_{Bw} + Z_{Aw})}.$$

The analysis showed that in this case, with $\hbar = 2, 3, 4, 5$ the maximum modulus of the relative error does not exceed 1 %, 4 %, 8 %, 13 %, respectively. As for three-dimensional objects, $\hbar$, usually does not exceed 3.

Let's consider for approximation using of a third-degree polynomial of the form $a \cdot t_v^3 + b \cdot t_v^2 + ct + d$. To find the unknowns, we will create a system of four equations. To do this, we equate the values of the polynomial and $t_w$ (see Form 4.6) at points $t_v = 0,\ 1/3,\ 2/3,\ 1$. We find that

$$a = \frac{9\cdot(Z_{Bw}-Z_{Aw})^2}{(2\cdot Z_{Bw}+Z_{Aw})\cdot(Z_{Bw}+2\cdot Z_{Aw})},\quad b = \frac{-9\cdot(Z_{Bw}-Z_{Aw})(Z_{Bw}-2\cdot Z_{Aw})}{(2\cdot Z_{Bw}+Z_{Aw})\cdot(Z_{Bw}+2\cdot Z_{Aw})},$$

$$c = \frac{(2\cdot Z^2{}_{Bw}-4\cdot Z_{Aw}\cdot Z_{Bw}+11\cdot Z_{Aw})}{(2\cdot Z_{Bw}+Z_{Aw})\cdot(Z_{Bw}+2\cdot Z_{Aw})}.$$

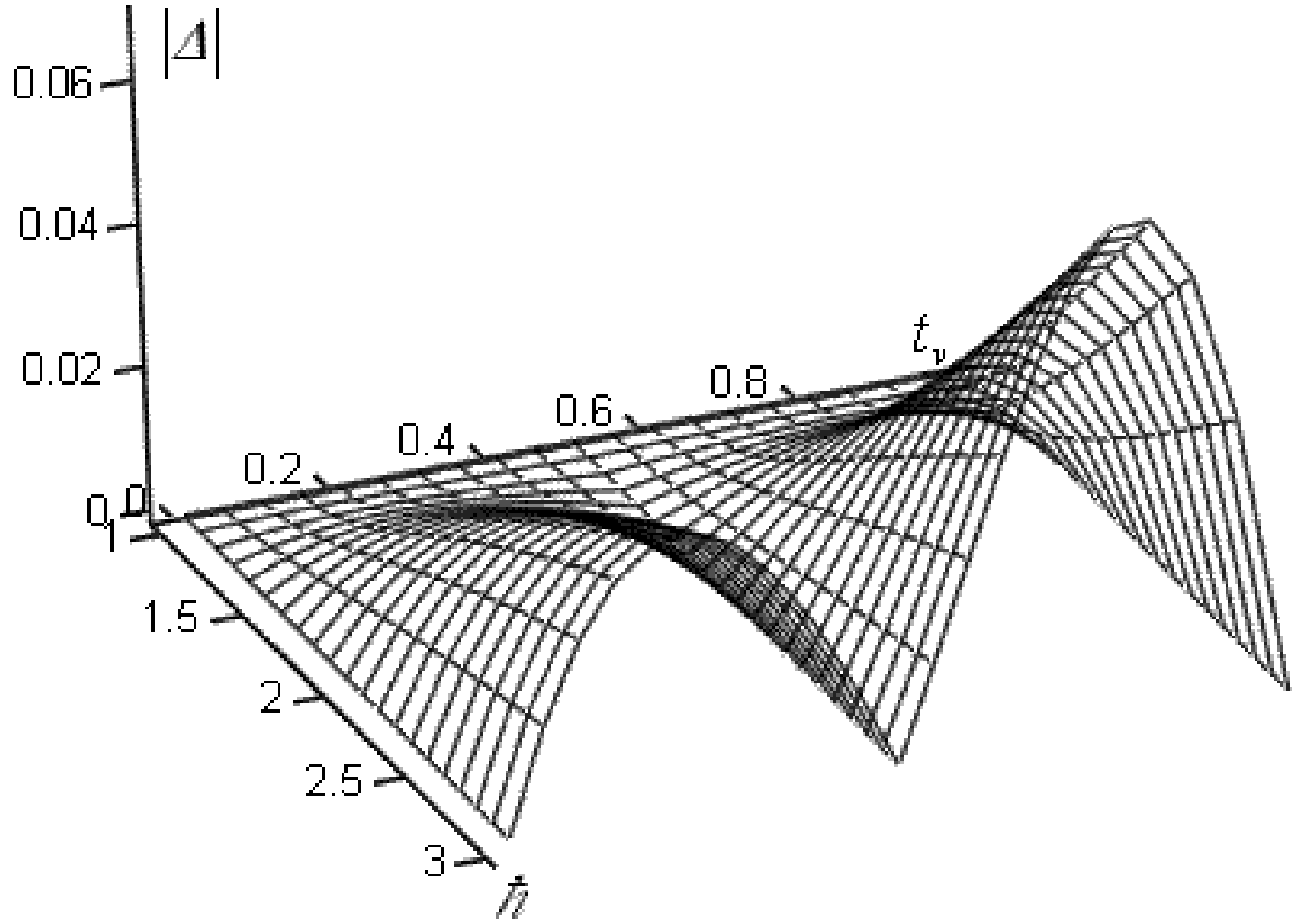

Fig. 4.7. Dependence of the absolute approximation error modulus on $t_v$ and $\hbar$

The analysis showed that the use of cubic interpolation results in higher accuracy compared to piecewise quadratic interpolation. For example, at $\hbar = 2, 3, 4, 5$ the maximum modulus of the relative error does not exceed 0.64 %, 2.9 %, 6.3 %, and 10.6 %, respectively.

The angular interpolation (Fig.4.8) of unit vectors of normals between the initial $\vec{N}_a$ and final $\vec{N}_b$ vectors is performed according to the expression [2]

$$\vec{N}(w) = \vec{N}_a \frac{\sin((1-w)\psi)}{\sin\psi} + \vec{N}_b \frac{\sin(w\psi)}{\sin\psi},$$

where $w \in [0, 1]$ is a parametric variable that determines the position of the streaming vector $\vec{N}(w)$ with respect to the vectors $\vec{N}_a$ and $\vec{N}_b$, and $\Psi$ is the angle between the normal vectors $\vec{N}_a$ and $\vec{N}_b$.

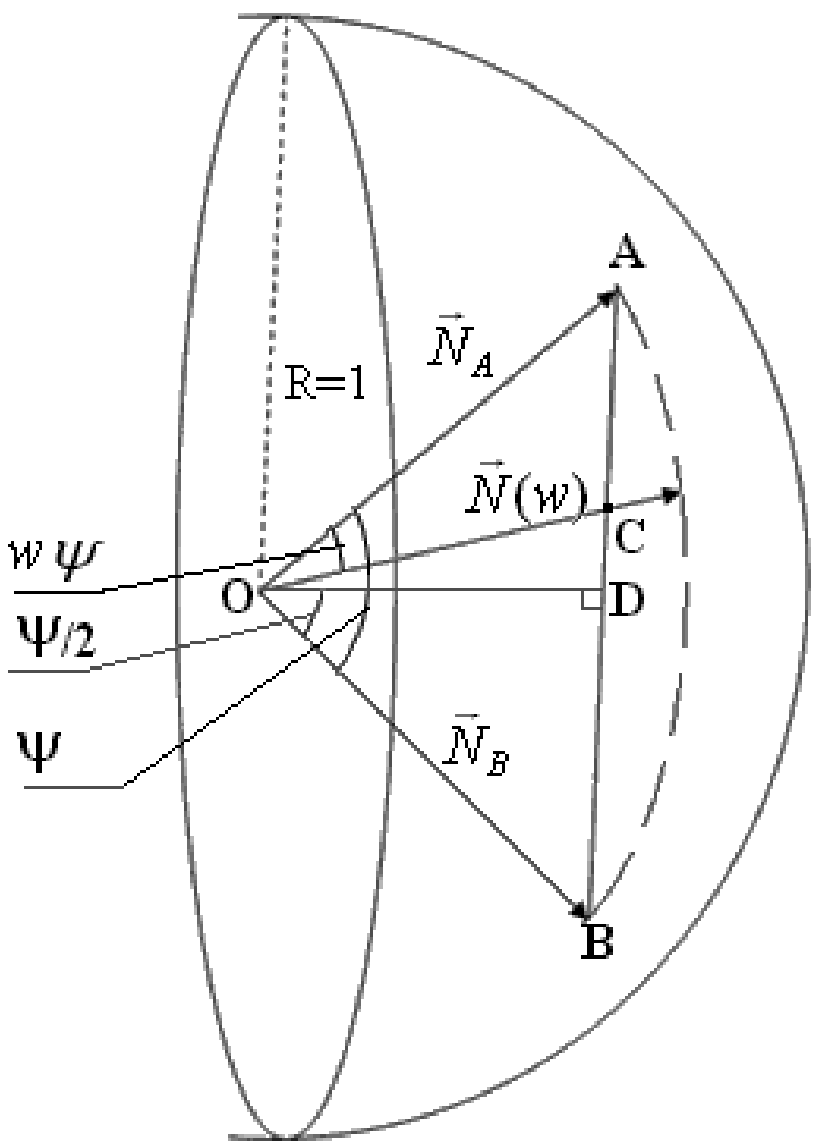

Fig. 4.8. Determination of normal vectors in spherical-angular interpolation

Since the vectors $\vec{N}_A$, $\vec{N}_B$ are unit (Fig. 4.8), the triangle *OAB* is isosceles. From the triangle *OBD* (*OD* $\perp$ *DB*) we find that $DB = \left|\vec{N}_B\right| \cdot \sin\frac{\psi}{2}$. Given that $\left|\vec{N}_B\right| = 1$, then

$$AB = 2 \cdot DB = 2 \cdot \sin\frac{\psi}{2}.$$

Consider the triangle *OAC*.

$$\angle OAC = \frac{180 - \psi}{2}. \ \angle OCA = 180^0 - \angle OAC - w \cdot \psi = \frac{180^0 + \psi}{2} - w \cdot \psi.$$

From the triangle *AOC*, by the sine theorem [3], we find that

$$AC = \left|\vec{N}_A\right| \frac{\sin(w \cdot \psi)}{\sin(\frac{180^0 + \psi}{2} - w \cdot \psi)} = \frac{\sin(w \cdot \psi)}{\sin(\frac{180^0 + \psi}{2} - w \cdot \psi)} = \frac{\sin(w \cdot \psi)}{\cos(w \cdot \psi - \frac{\psi}{2})}.$$

For the world coordinate system, the current value of the parametric variable $d$, which varies from 0 to 1, can be easily found through the ratio of the segment *AC* to *AB*.

$$d=\frac{\sin(w\cdot\psi)}{2\sin\frac{\psi}{2}\cdot\cos(w\cdot\psi-\frac{\psi}{2})}=\frac{\sin(w\cdot\psi)}{\sin(\frac{\psi}{2}+w\cdot\psi-\frac{\psi}{2})+\sin(\frac{\psi}{2}-w\cdot\psi+\frac{\psi}{2})}=$$

$$=\frac{\sin(w\cdot\psi)}{\sin(w\cdot\psi)+\sin(\psi-w\cdot\psi)}=\frac{1}{1+\sin\psi\frac{\cos(w\cdot\psi)}{\sin(w\cdot\psi)}-\cos\psi}.$$

It was proved [3] that the following relationship exists between a parametric variable $v$ in the on-screen coordinate system and a variable $d$ in the world coordinate system

$$d=\frac{1}{1+(1-\frac{1}{v})\cdot\frac{z_2}{z_1}},$$

where $z_2$, $z_1$ — Z-coordinates, respectively, of the start and end points of a line segment in the world coordinate system. For the screen coordinate system, the following relation is true

$$v=\frac{1}{1+\sin\psi\cdot\frac{\cos(\eta\cdot\psi)}{\sin(\eta\cdot\psi)}-\cos\psi},$$

where $\eta$ varies from zero to one.

Substituting the value $v$ into the formula for $d$, we obtain

$$d=\frac{1}{1+\frac{z_2}{z_1}\left(\sin\psi\frac{\cos(\eta\cdot\psi)}{\sin(\eta\cdot\psi)}-\cos\psi\right)}.$$

Thus, to find the vectors of normals in the world coordinate system, using the value of the parametric variable $\eta$ in the on-screen coordinate system, use the formula

$$\vec{N}(w)=\vec{N}_a\cos\left(\frac{1}{1+\frac{z_2}{z_1}\left(\sin\psi\frac{\cos(\eta\cdot\psi)}{\sin(\eta\cdot\psi)}-\cos\psi\right)}\psi\right)+$$

$$+\vec{N}_k\sin\left(\frac{1}{1+\frac{z_2}{z_1}\left(\sin\psi\frac{\cos(\eta\cdot\psi)}{\sin(\eta\cdot\psi)}-\cos\psi\right)}\psi\right).$$

For the world coordinate system

$$d=\frac{1}{1+\sin\psi\dfrac{\cos(w\cdot\psi)}{\sin(w\cdot\psi)}-\cos\psi}.$$

When perspective-correct shading

$$d=\frac{1}{1+\dfrac{z_2}{z_1}\left(\sin\psi\dfrac{\cos(\eta\cdot\psi)}{\sin(\eta\cdot\psi)}-\cos\psi\right)}.$$

Equating the right-hand sides of the above expressions, we find

$$\frac{1}{1+\sin\psi\dfrac{\cos(w\cdot\psi)}{\sin(w\cdot\psi)}-\cos\psi}=\frac{1}{1+\dfrac{z_2}{z_1}\left(\sin\psi\dfrac{\cos(\eta\cdot\psi)}{\sin(\eta\cdot\psi)}-\cos\psi\right)}.$$

From the last equation we find

$$1+\sin\psi\cdot\frac{\cos(w\cdot\psi)}{\sin(w\cdot\psi)}-\cos\psi=1+\frac{z_2}{z_1}\left(\sin\psi\cdot\frac{\cos(\eta\cdot\psi)}{\sin(\eta\cdot\psi)}-\cos\psi\right).$$

After simplification, we obtain

$$\sin\psi\cdot\frac{\cos(w\cdot\psi)}{\sin(w\cdot\psi)}-\cos\psi=\frac{z_2}{z_1}\cdot\left(\sin\psi\cdot\frac{\cos(\eta\cdot\psi)}{\sin(\eta\cdot\psi)}-\cos\psi\right).$$

We introduce the notation $a=\sin(w\cdot\psi)$. It is clear that $\sqrt{1-a^2}=\cos(w\cdot\psi)$

$$\sin\psi\frac{\sqrt{1-a^2}}{a}-\cos\psi=\frac{z_2}{z_1}\left(\sin\psi\frac{\cos(\eta\cdot\psi)}{\sin(\eta\cdot\psi)}-\cos\psi\right).$$

From the last equation we find

$$a=\frac{z1\cdot\sin(\eta\cdot\psi)\cdot\sin(\psi)}{\sqrt{(z1\cdot\sin(\eta\cdot\psi)\cdot\sin(\psi))^2+(z1\cdot\sin(\eta\cdot\psi)\cdot\cos(\psi)-z2\cdot\sin(\psi\cdot(i-1))^2}}.$$

The above ratio involves the calculation of two functions at once, sine and cosine, which complicates the calculation and requires the storage of tabular data for two functions. Dividing the denominator and numerator of the fraction by $z1\cdot\sin(\eta\psi)\cdot\sin(\psi)$ and performing the equivalent transformations, we find

$$\sin(w\cdot\psi)=\frac{1}{\sqrt{1+\left[\operatorname{ctg}\psi+\dfrac{z2}{z1}(\operatorname{ctg}(\eta\cdot\psi)-\operatorname{ctg}(\psi))\right]^2}}.$$

The above expression is valid for all cases except when $z1 \cdot \sin(\eta \cdot \psi) \cdot \sin(\psi) = 0$. This is the case at the starting point of the triangle rasterization line, where the normal vector is given and does not need to be calculated, so the proposed division can be performed. When the vectors $\vec{N}_A$, $\vec{N}_B$ coincide, $\psi = 0$, there is no need to perform interpolation.

Similarly, we find

$$\cos(w \cdot \psi) = \frac{\operatorname{ctg}\psi + \frac{z2}{z1}(\operatorname{ctg}(\eta \cdot \psi) - \operatorname{ctg}(\psi))}{\sqrt{1 + \left[\operatorname{ctg}\psi + \frac{z2}{z1}(\operatorname{ctg}(\eta \cdot \psi) - \operatorname{ctg}(\psi))\right]^2}}.$$

Finally, we can write down that

$$\vec{N}(w) = \vec{N}_a \cdot \cos(w \cdot \psi) + \vec{N}_k \cdot \sin(w \cdot \psi) =$$

$$= \frac{\vec{N}_a \cdot (\operatorname{ctg}\psi + \frac{z2}{z1}\left(\operatorname{ctg}(\eta \cdot \psi) - \operatorname{ctg}\psi\right)) + \vec{N}_k}{\sqrt{1 + \left[\operatorname{ctg}\psi + \frac{z2}{z1}\left(\operatorname{ctg}(\eta \cdot \psi) - \operatorname{ctg}\psi\right)\right]^2}}$$

Let us introduce the notation $b = \operatorname{ctg}\psi + \frac{z2}{z1}\left(\operatorname{ctg}(\eta \cdot \psi) - \operatorname{ctg}\psi\right)$, then

$$\vec{N}(w) = \frac{\vec{N}_a \cdot b + \vec{N}_k}{\sqrt{1 + (b)^2}}.$$

When applying spherical-angular interpolation, the time-consuming procedure of normalizing the normal vectors is excluded from the computational process. An important issue in this aspect is the need to normalize the vectors of normals obtained during the perspective-correct formation of images. Let us find the vector modulus according to the expression

$$\left|\vec{N}(w)\right|^2 = \left(\frac{\vec{N}_a(\operatorname{ctg}\psi + \frac{z2}{z1}\left(\operatorname{ctg}(\eta \cdot \psi) - \operatorname{ctg}\psi\right)) + \vec{N}_k}{\sqrt{1 + \left[\operatorname{ctg}\psi + \frac{z2}{z1}\left(\operatorname{ctg}(\eta \cdot \psi) - \operatorname{ctg}\psi\right)\right]^2}}\right)^2 =$$

$$=\left|\vec{N}_a\right|^2 \frac{\left(\operatorname{ctg}\psi+\frac{z2}{z1}\left(\operatorname{ctg}(\eta\cdot\psi)-\operatorname{ctg}\psi\right)\right)^2}{\left(\sqrt{1+\left[\operatorname{ctg}\psi+\frac{z2}{z1}\left(\operatorname{ctg}(\eta\cdot\psi)-\operatorname{ctg}\psi\right)\right]^2}\right)^2}+$$

$$+2\cdot\vec{N}_a\cdot\vec{N}_k \frac{\left(\operatorname{ctg}\psi+\frac{z2}{z1}\left(\operatorname{ctg}(\eta\cdot\psi)-\operatorname{ctg}\psi\right)\right)}{\left(\sqrt{1+\left[\operatorname{ctg}\psi+\frac{z2}{z1}\left(\operatorname{ctg}(\eta\cdot\psi)-\operatorname{ctg}\psi\right)\right]^2}\right)}+$$

$$+\left|\vec{N}_k\right|^2 \frac{1}{\left(\sqrt{1+\left[\operatorname{ctg}\psi+\frac{z2}{z1}\left(\operatorname{ctg}(\eta\cdot\psi)-\operatorname{ctg}\psi\right)\right]^2}\right)^2}.$$

The second term of the last expression is zero, since the vectors $\vec{N}_a$, $\vec{N}_k$ are orthogonal, and, as a result, their scalar product is zero. The vectors $\vec{N}_a$, $\vec{N}_k$ are unit, so $\left|\vec{N}_k\right|^2=\left|\vec{N}_a\right|^2=1$. Taking into account the above

$$\left|\vec{N}(w)\right|^2=\frac{1+\left(\operatorname{ctg}\psi+\frac{z2}{z1}\left(\operatorname{ctg}(\eta\cdot\psi)-\operatorname{ctg}\psi\right)\right)}{\left(1+\left[\operatorname{ctg}\psi+\frac{z2}{z1}\left(\operatorname{ctg}(\eta\cdot\psi)-\operatorname{ctg}\psi\right)\right]\right)}=1\,.$$

Thus, it can be stated that normalization is not required for perspective-correct object rendering.

The results obtained allow us to increase the realism of the generated images due to the fact that the color intensities of the corresponding surface points in the screen and object coordinate systems will coincide, that is, the adequacy of the generated image to the real object increases.

## 4.4. ANALYSIS OF METHODS FOR APPROXIMATING PERSPECTIVE-CORRECT TEXTURING

One of the approaches to forming highly realistic images is to use textures [1, 5–9], which are mapped on graphic objects to give the image relief. The use of textures in many cases allows to successfully solve problems that are extremely time-consuming to solve by direct methods and significantly reduce computational costs. In texturing tasks, a correlation is established between screen and texture coordinates.

Perspective-correct texturing (PCT) uses nonlinear functions, the calculation of which involves time-consuming pixel-by-pixel operations. In the vast majority of cases, perspective-correct texturing is implemented using the following formulas [7, 9]

$$u = \frac{ax + by + c}{gx + hy + i}, \quad v = \frac{dx + ey + f}{gx + hy + i}, \tag{4.8}$$

where $u$ and $v$ — texture coordinates (TC), $x$ and $y$ — screen coordinates of an object, $a...i$ — coefficients of the polygon to be textured.

As can be seen from formula (4.8), finding the TC is a time-consuming procedure, since for each pixel of the image, six multiplication and two division operations are required, which significantly affects the speed of generating graphic scenes. Therefore, the task of simplifying the computational process of perspective-correct texturing is quite relevant.

In order to simplify the computational process, scientists have proposed various approaches to approximating perspective-correct texturing.

The simplest approach is linear interpolation, but it provides low accuracy of texture coordinates calculation and introduces significant artifacts and distortions in perspective rendering, which leads to the fact that the perspective of the object is not accurately reproduced.

The quadratic approximation uses the equation

$$u(x) = A_1 \cdot x^2 + A_2 \cdot x + A_3,$$

$$v(x) = B_1 \cdot x^2 + B_2 \cdot x + B_3,$$

where $A_1$-$A_3$, $B_1$-$B_3$ are approximation coefficients that are constant for each rasterization line, and $x$-coordinate values are normalized.

The known formulas for calculating the coefficients are as follows [8]

$$A_1 = 2u_0 - 4u_1 + 2u_2, \; A_2 = -3u_0 + 4u_1 - 2u_2, \; A_3 = u_0, \tag{4.9}$$

where $u_0$, $u_1$ i $u_2$, and are the values of the texture coordinate $u$ at the start, middle, and end points of the rasterization line, respectively. The formulas for calculating the coefficients $B_1$ - $B_3$, however, the corresponding coordinate values are used instead of the $v$ -coordinate values.

The use of quadratic approximation gives a fairly realistic reproduction of the perspective with relatively simple calculations. The disadvantage of formulas (4.9) is that they can be used only in the case of normalized values of screen coordinates. The normalization procedure requires a division operation, which significantly affects the computational complexity.

The cubic approximation [7] uses the relationship

$$u(x) = C_1 x^3 + C_2 \cdot x^2 + C_3 \cdot x + C_4,$$
$$v(x) = D_1 x^3 + D_2 \cdot x^2 + D_3 \cdot x + D_4,$$

where $C_1$-$C_4$, $D_1$-$D_4$ are approximation coefficients that are calculated for each rasterization line, the $x$-coordinate values are also normalized. To calculate the approximation coefficients, you need to find the exact values of the texture coordinates at four anchor points: the start, end, and two interior points that divide the rasterization line into three equal segments. The cubic approximation provides a more realistic perspective than the quadratic approximation, but the computational complexity increases significantly, which limits the scope of this type of approximation in real-time computer graphics systems.

The bi-quadratic [8] approximation uses the equation of the form

$$u(x) = A_1 x^2 + A_2 y^2 + A_3 xy + A_4 x + A_5 y + A_6,$$
$$v(x) = B_1 x^2 + B_2 y^2 + B_3 xy + B_4 x + B_5 y + B_6,$$

where $A_1$-$A_6$, $B_1$-$B_6$ are the approximation coefficients.

To calculate the 12 bi-quadratic approximation coefficients, you need to know the exact values of a pair of texture coordinates at six points: at the vertices of a triangular polygon and at the midpoints of its edges.

The bi-cubic approximation [8] uses the equation

$$u(x) = A_1 x^3 + A_2 y^3 + A_3 x^2 y + A_4 xy^2 + A_5 x^2 + A_6 y^2 + A_7 xy + A_8 x + A_9 y + A_{10},$$
$$v(x) = B_1 x^2 + B_2 y^2 + B_3 x^2 y + B_4 xy^2 + B_5 x^2 + B_6 y^2 + B_7 xy + B_8 x + B_9 y + B_{10},$$

where $A_1$-$A_{10}$, $B_1$-$B_{10}$ are the approximation coefficients.

The bi-cubic approximation requires accurate values of a pair of texture coordinates at 10 control points and the calculation of 20 approximation coefficients. The result looks quite realistic, but the computational cost is very high, so this type of approximation is of limited use.

Another approach to approximation is to split the rasterization line into several segments and use linear, quadratic, or cubic interpolation to calculate texture coordinates within the resulting segments. In this case, you need to calculate the exact values of the texture coordinates at the points between which the interpolation is performed.

## 4.5. RECURRENT DETERMINATION OF TEXTURE COORDINATES

In paper [9], it is proposed to use the midpoint method to approximate the hyperbolic curve, which allows you to find not approximate, but exact values of texture coordinates. In this case, the value of each texture coordinate is checked for compliance with the condition [9] for the values of the object's screen coordinates. If the value of the texture coordinate meets the condition, it is considered that its valid value has been found, otherwise, the next value of the texture coordinate is checked for compliance. At the same time, instead of division operations, addition and comparison operations are used, which greatly simplifies calculations. The disadvantage of the method is the assumption that the texture coordinates are set to integers, which is only a special case of texturing.

According to the OpenGL and DirextX specifications, texture coordinates are specified in the range [0;1], i.e., they are fractional numbers, which necessitates the development of a new PCT method that would take into account the noninteger representation of texture coordinates.

Suppose it is needed to find the texture coordinates $u$ and $v$. Since the coordinates $u$ and $v$ are calculated similarly, then in the following we will consider only the coordinate $u$ with generalization of the results and on the coordinate $v$.

When perspective-correct texturing using the midpoint method, texture coordinates are determined that meet the condition [9]

$$\frac{ax+by+c}{gx+hy+i}-0{,}5\leq u<\frac{ax+by+c}{gx+hy+i}+0{,}5\ . \qquad (4.10)$$

Condition (4.10) is acceptable for the case of integer texture coordinates. For texture coordinates from the range $\left[0;1\right]$, this condition will take the form

$$\frac{ax+by+c}{gx+hy+i}-\frac{\Delta u}{2}\le u<\frac{ax+by+c}{gx+hy+i}+\frac{\Delta u}{2}, \tag{4.11}$$

where $\Delta u$ — the difference between neighboring values of a texture coordinate $u$, which is calculated by the formula $\Delta u = u_{max}/p$, $u_{max}$ — the maximum value of the texture coordinate $u$, $p$ — number of texture points along the axis $u$.

Let $E(x,y)=2(gx+hy+i)$. Multiplying inequality (4.11) by $E(x,y)$, we obtain

$$\begin{gathered}2(ax+by+c)-\Delta u(gx+hy+i)\le uE(x,y)< \\ <2(ax+by+c)+\Delta u(gx+hy+i)\end{gathered}.$$

Write the resulting inequality in the form of a system of inequalities

$$\begin{cases}uE(x,y)-2(ax+by+c)+\Delta u(gx+hy+i)\ge 0; \\ uE(x,y)-2(ax+by+c)-\Delta u(gx+hy+i)<0.\end{cases}$$

After simplification, we get

$$\begin{cases}uE(x,y)-x(2a-\Delta ug)-y(2b-\Delta uh)-(2c-\Delta ui)\ge 0; \\ uE(x,y)-x(2a+\Delta ug)-y(2b+\Delta uh)-(2c+\Delta ui)<0.\end{cases} \tag{4.12}$$

Let's introduce the following notation

$$W(x,y,u)=uE(x,y)-x(2a-\Delta ug)-y(2b-\Delta uh)-(2c-\Delta ui),$$

$$Q(x,y,u)=uE(x,y)-x(2a+\Delta ug)-y(2b+\Delta uh)-(2c+\Delta ui).$$

Then the condition (3.4) will take the form

$$\begin{cases}W(x,y,u)\ge 0; \\ Q(x,y,u)<0.\end{cases} \tag{4.13}$$

The system of inequalities (4.12) defines the conditions that the texture coordinate $u$ must meet for the given values of the screen coordinates $x$ and $y$. If the conditions (4.13) are not met for the current coordinate value $u$, then an increment $\Delta u$ is added (subtracted) to it. The increment $\Delta u$ is added (subtracted) until the conditions (4.13) are satisfied. When a texture coordinate value $u$ is found that satisfies the conditions (4.13), the next point in the rasterization line is moved to.

Consider how the values $E(x,y)$, $W(x,y,u)$ and $Q(x,y,u)$ will change when changing coordinates $x$ and $y$ by 1.

When $x$ is increased by 1, $E(x+1,y)=E(x,y)+2g$ . Hence

$$W(x+1,y,u)=W(x,y,u)+2gu-(2a-\Delta ug),$$

$$Q(x+1,y,u)=Q(x,y,z)+2gu-(2a+\Delta ug).$$

If we reduce $x$ by 1, then $E(x-1,y)=E(x,y)-2g$ . Hence

$$W(x-1,y,u)=W(x,y,u)-2gu+(2a-\Delta ug),$$

$$Q(x-1,y,u)=Q(x,y,z)-2gu+(2a+\Delta ug).$$

With an increase of $y$ by 1 $E(x,y+1)=E(x,y)+2h$ . Hence

$$W(x,y+1,u)=W(x,y,u)+2hu-(2b-\Delta uh),$$

$$Q(x,y+1,u)=Q(x,y,u)+2hu-(2b+\Delta uh).$$

Expressions $(2a-\Delta ug)$, $(2a+\Delta ug)$, $(2b-\Delta uh)$ and $(2b+\Delta uh)$ is enough to calculate once for the entire polygon. Then finding the conditions (4.13) for the next point in the rasterization line involves performing only two multiplication operations and four addition operations.

If the current value of the texture coordinate does not satisfy the conditions (4.13), then you need to add or subtract an increment to it $\Delta u$ . Let's define how the values $W(x,y,u)$ and $Q(x,y,u)$ change when changing a texture coordinate $u$ by $\Delta u$ .

Increasing $u$ by $\Delta u$

$$\begin{cases} W(x,y,u+\Delta u)=W(x,y,u)+\Delta u\cdot E(x,y); \\ Q(x,y,u+\Delta u)=Q(x,y,u)+\Delta u\cdot E(x,y). \end{cases} \tag{4.14}$$

Reducing $u$ by $\Delta u$

$$\begin{cases} W(x,y,u-\Delta u)=W(x,y,u)-\Delta u\cdot E(x,y); \\ Q(x,y,u-\Delta u)=Q(x,y,u)-\Delta u\cdot E(x,y). \end{cases} \tag{4.15}$$

The resulting formulas allow you to find the values of the conditions $W(x,y,u)$ and $Q(x,y,u)$ for the new value of the coordinate $u$ , by performing only two multiplication and two addition operations.

From formulas (4.14) and (4.15) it is clear that the expressions $W(x,y,u)$ and $Q(x,y,u)$ increase with increasing $u$ , and decrease with decreasing $u$ ,

because $\Delta u > 0$ and $E(x,y) > 0$. This allows to select the type of change of coordinate $u$. If $W(x,y,u) < 0$, then it is needed to add the increment $\Delta u$ to the current value of the coordinate $u$. If $Q(x,y,u) > 0$, then the increment $\Delta u$ is subtracted from the current value of the coordinate $u$.

The texture coordinate $v$ is calculated in a similar way.

The principle of the proposed method is shown in Fig.4.9.

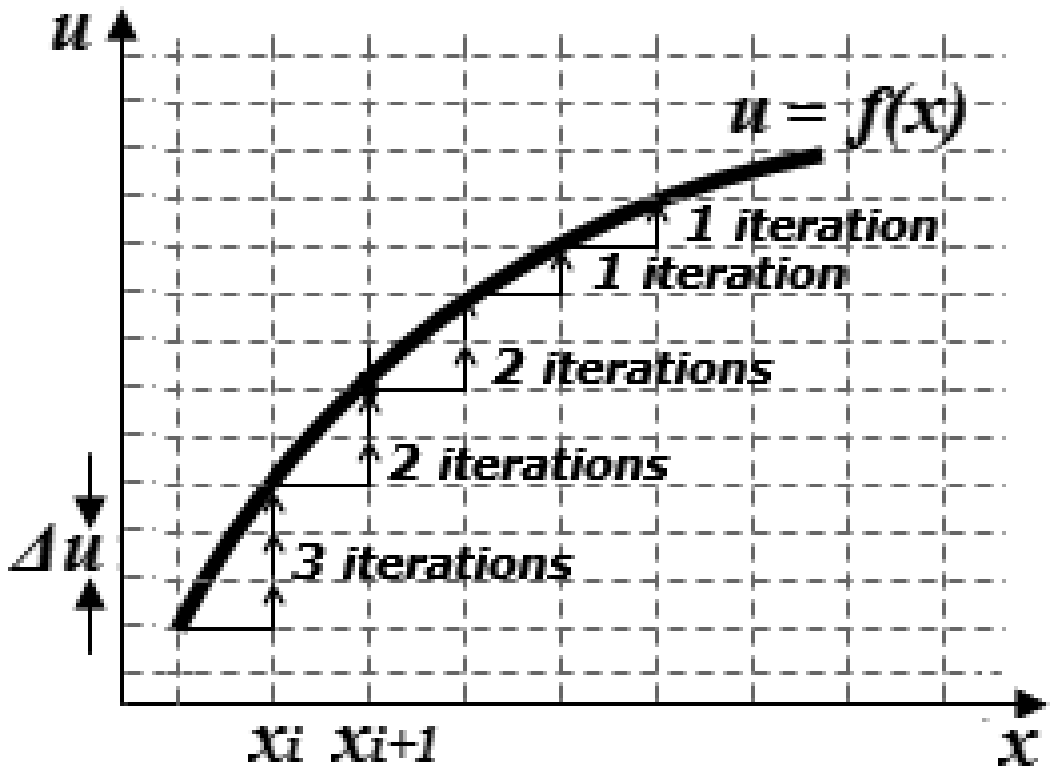

Fig. 4.9. Additive determination of a texture coordinate

From the example shown in Fig.4.9, it can be seen that to find the coordinate $u$ at a point $x_i$, it is needed to perform 3 iterations of $\Delta u$ incremental addition, and at a point $x_{i+1}$ — 2 iterations, etc.

In the proposed method of perspective-correct texturing, division operations are replaced by addition operations, which significantly reduces the amount of computation and speeds up the texturing process.

The method allows finding the exact values of texture coordinates. Unlike the proposed method, the classical PCT method requires rounding or discarding the fractional part of the texture coordinate, which leads to texturing errors.

Compared to the method in [9], the proposed method is universal, since it can be applied to both integer and fractional representations of texture coordinates.

## 4.6. USING QUADRATIC APPROXIMATION FOR PERSPECTIVE-CORRECT TEXTURING

The peculiarity of the known methods of approximating perspective-correct texturing is that the internal reference points of the segments are placed evenly between the start and end points of the rasterization line (RL). In particular, with the traditional quadratic approximation, the internal reference point is the midpoint of the RL, which divides the line into two equal parts. In this case, there is a symmetrical distribution of the absolute approximation error relative to the midpoint of the PL. At the same time, the relative error takes on maximum values when the texture coordinates are close to zero. The relative error can be reduced if the internal reference point is not the middle point, but a point offset from the middle of the RL, depending on the nature of the change in the texture coordinate values. If the values of the texture coordinate increase along the RL, then the internal anchor point should be shifted towards smaller coordinate values, and if they decrease, then towards larger values.

Let's derive general formulas for calculating the approximation coefficients. To do this, solve the system of equations:

$$\begin{cases} Ax_0^2 + Bx_0 + C = u_0; \\ Ax_{вн}^2 + Bx_{вн} + C = u_{вн}; \\ Ax_1^2 + Bx_1 + C = u_1, \end{cases} \tag{4.16}$$

where $A$, $B$ and $C$ — quadratic approximation coefficients, $x_0$, $x_{вн}$, $x_1$ — screen $x$-coordinates at the first, inner, and end points of the PL, respectively, $u_0$, $u_{вн}$, $u_1$ — texture $u$-coordinates at the first, inner, and end points of the PL, respectively.

Since the traditional quadratic approximation uses normalized values of the screen coordinate, $x_0 = 0$ and $x_1 = 1$. Then the system (4.16) will take the following form

$$\begin{cases} C = u_0; \\ Ax_{вн}^2 + Bx_{вн} + C = u_{вн}; \\ A + B + C = u_1. \end{cases} \tag{4.17}$$

Let's find the coefficient $A$ from the third equation of system (4.17) and substitute it into the second equation.

$$A = u_1 - C - B = u_1 - u_0 - B \ ,$$

$$\left(u_1 - u_0 - B\right)x_{вн}^2 + Bx_{вн} + u_0 = u_{вн} \ .$$

From the last equation, we find the coefficient $B$

$$B = \frac{x_{вн}^2\left(u_1 - u_0\right) + u_0 - u_{вн}}{x_{вн}^2 - x_{вн}} \ .$$

Hence

$$A = u_1 - u_0 - \frac{x_{вн}^2\left(u_1 - u_0\right) + u_0 - u_{вн}}{x_{вн}^2 - x_{вн}} = \frac{-\left(x_{вн}\left(u_1 - u_0\right) + u_0 - u_{вн}\right)}{x_{вн}^2 - x_{вн}} \ .$$

The formulas for calculating the coefficients $A, B, C$ are as follows

$$\begin{cases} A = -\left(x_{вн}\left(u_1 - u_0\right) + u_0 - u_{вн}\right) \big/ \left(x_{вн}^2 - x_{вн}\right); \\ B = \left(x_{вн}^2\left(u_1 - u_0\right) + u_0 - u_{вн}\right) \big/ \left(x_{вн}^2 - x_{вн}\right); \\ C = u_0. \end{cases} \quad (4.18)$$

Calculating coefficients $A$ and $B$ requires 2 division operations, 5 multiplication operations, and 8 addition operations. Let's introduce the following notation

$$r = x_{вн}\left(u_1 - u_0\right), \ s = u_0 - u_{вн}, \ q = \frac{1}{x_{вн}^2 - x_{вн}} \ . \quad (4.19)$$

Let us substitute the notation (4.19) into the system (4.18). Then we get

$$\begin{cases} A = -q\left(r + s\right); \\ B = q\left(r \cdot x_{вн} + s\right); \\ C = u_0. \end{cases} \quad (4.20)$$

The calculation of the approximation coefficients using formulas (4.20) requires 1 division operation, 5 multiplication operations, and 5 addition operations.

If the value of the internal reference point for the next rasterization line remains unchanged, it is obvious that the value of the coefficient $q$ will not change either, which allows to eliminate 1 division, 1 multiplication, and 1 addition operation.

Table 4.1 shows the formulas for calculating the coefficients $A$ and $B$ for different values of the internal point of the rasterization line

Table 4.1

**Formulas for calculating the coefficients of quadratic approximation for different values of the internal point**

| Value $x_{вн}$ | Formula for calculating the coefficient $A$ | Formula for calculating the coefficient $B$ |
|---|---|---|
| 0,75 | $4u_1 + 1{,}333u_0 - 5{,}333u_{0,75}$ | $-3u_1 - 2{,}333u_0 + 5{,}333u_{0,75}$ |
| 0,7 | $3{,}333u_1 + 1{,}429u_0 - 4{,}762u_{0,7}$ | $-2{,}333u_1 - 2{,}429u_0 + 4{,}762u_{0,7}$ |
| 0,6 | $2{,}5u_1 + 1{,}667u_0 - 4{,}167u_{0,6}$ | $-1{,}5u_1 - 2{,}667u_0 + 4{,}167u_{0,6}$ |
| 0,5 | $2u_1 + 2u_0 - 4u_{0,5}$ | $-u_1 - 3u_0 + 4u_{0,5}$ |
| 0,4 | $1{,}667u_1 + 2{,}5u_0 - 4{,}167u_{0,4}$ | $-0{,}667u_1 - 3{,}5u_0 + 4{,}167u_{0,4}$ |
| 0,3 | $1{,}4297u_1 + 3{,}333u_0 - 4{,}762u_{0,3}$ | $-0{,}429u_1 - 4{,}333u_0 + 4{,}762u_{0,3}$ |
| 0,25 | $1{,}333u_1 + 4u_0 - 5{,}333u_{0,25}$ | $-0{,}333u_1 - 5u_0 + 5{,}333u_{0,25}$ |

Similar formulas are used to calculate the coordinate $v$.

Fig. 4.10 shows graphs of the relative error of $u$ texture coordinate approximation for a specific rasterization line (polygon: (1,1), (32,96), (96,128), (128,32), $y = 32$) at different locations of the internal anchor points, provided that the texture coordinate values vary from 0 to 1. The maximum relative error occurs when the internal anchor point splits the rasterization line in half, i.e. $x_{вн} = 0{,}5$. When the internal reference point is shifted to the left, the value of the maximum relative error decreases, and when the value $x_{вн} = 0{,}25$ is reduced by almost half.

Shifting the internal anchor point to the left allows to find more accurate texture coordinate values in the left part of the RL due to a proportional increase in the error in the right part. This way, the relative error values are centered along the rasterization line.

Shift the internal anchor point to lower values of the screen coordinate $x$ is performed when the value of the texture coordinate along the rasterization line increases, i.e. $u_1 > u_0$, and towards higher values of the screen coordinate $x$ — when $u_1 < u_0$. Table 4.2 shows the recommended values of the internal reference point for different values of texture coordinates at the start and end points of the rasterization line.

The proposed approach reduces the relative error in determining texture coordinates by up to 2 times. The advantage of the approach is that the increase in accuracy is achieved not by using higher-order curves or by introducing additional reference points, but only by shifting the internal ref-erence point. The formulas for calculating the approximation coefficients for different values of the internal reference point are calculated once for the entire PL.

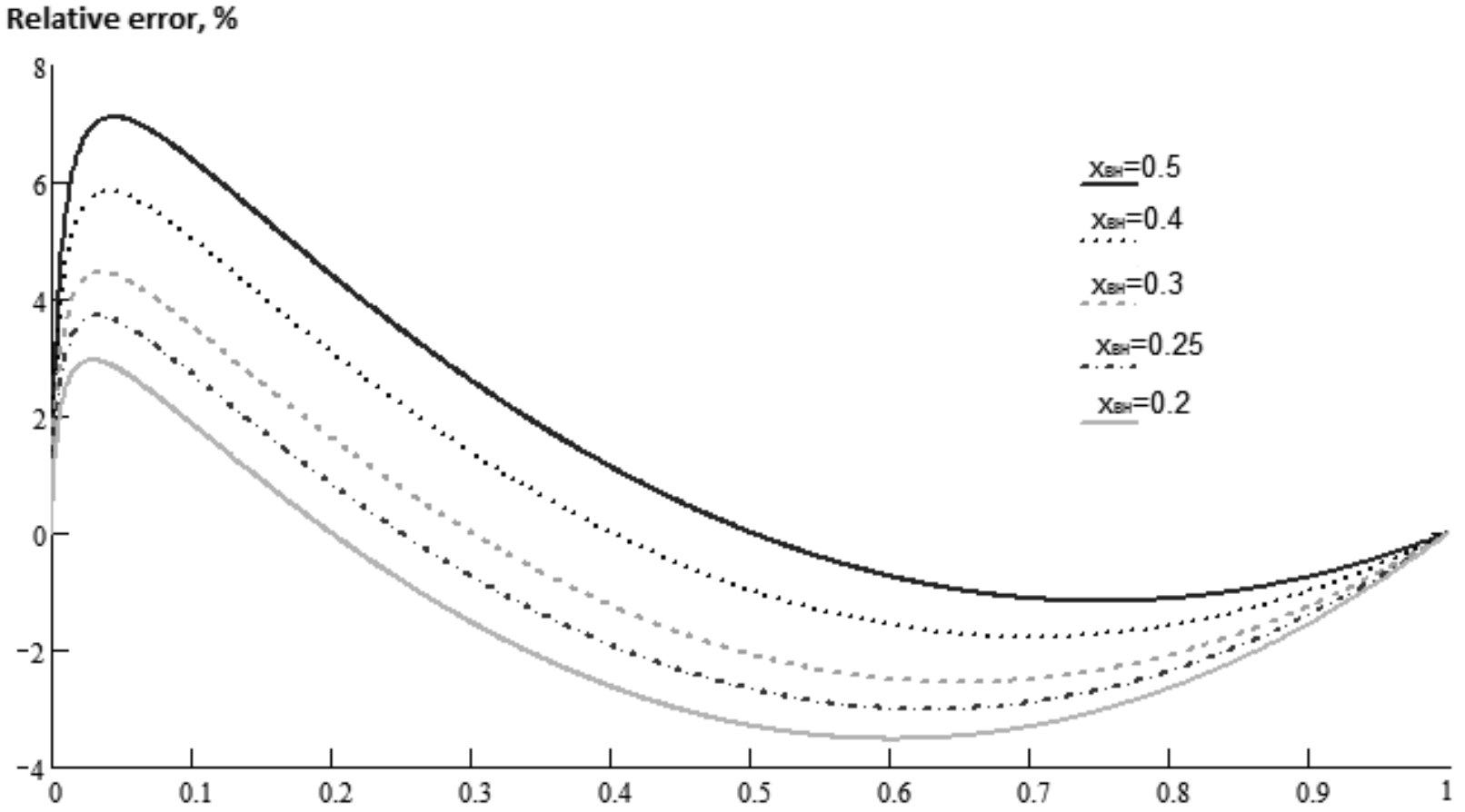

Fig. 4.10. Graphs of relative approximation errors at different locations of the internal reference point

Let's consider another possible approach to using approximation for texturing.

Traditional quadratic approximation [1, 7] uses normalized values of the object's screen coordinates for perspective-correct texturing. The normalization procedure involves performing a time-consuming division operation, which reduces the speed of texture overlay.

Let's find the formulas for calculating the quadratic approximation coefficients, provided that the object's screen coordinates are not normalized.

To determine the texture coordinate $u$ according to quadratic approximation, it is necessary to find the unknown coefficients $A_1$–$A_3$. To find them, we write the system of three equations in matrix form:

$$\begin{bmatrix} x_0^2 & x_0 & 1 \\ x_1^2 & x_1 & 1 \\ x_2^2 & x_2 & 1 \end{bmatrix} \times \begin{bmatrix} A_1 \\ A_2 \\ A_3 \end{bmatrix} = \begin{bmatrix} u_0 \\ u_1 \\ u_2 \end{bmatrix},$$

where ($x_0$, $x_1$, $x_2$) and ($u_0$, $u_1$, $u_2$) — values of coordinates $x$ and $u$ at the start, inner, and end points of the rectangular polygon rasterization line, respectively, $A_1$, $A_2$ and $A_3$ — quadratic approximation coefficients.

Table 4.2

**Recommended values for the internal reference point**

| The value of texture coordinates at the start and end points | Range of the difference between the texture coordinate values at the ends of the rasterization line | Value $x_{вн}$ |
|---|---|---|
| $u_0 \to 0$, $u_1 \in [0;1]$ | – | *0,25* |
| $u_1 \to 0$, $u_0 \in [0;1]$ | – | *0,75* |
| $u_0 \to 1$, $u_1 \in [0{,}05;1]$ | $(u_0 - u_1) \in [0;0{,}35]$ | *0,5* |
| | $(u_0 - u_1) \in [0{,}35;0{,}75]$ | *0,6* |
| | $(u_0 - u_1) \in [0{,}75;0{,}95]$ | *0,7* |
| $u_1 \to 1$, $u_0 \in [0{,}05;1]$ | $(u_1 - u_0) \in [0;0{,}35]$ | *0,5* |
| | $(u_1 - u_0) \in [0{,}35;0{,}75]$ | *0,4* |
| | $(u_1 - u_0) \in [0{,}75;0{,}95]$ | *0,3* |
| $u_0 < 0{,}5$, $u_1 < 0{,}5$ :<br>$u_0 < u_1$<br>$u_0 > u_1$ | <br>–<br>– | <br>*0,4*<br>*0,6* |
| All other cases | – | *0,5* |

Values $u_0$, $u_1$ and $u_2$ can be calculated using formula (4.9). The coefficients $A_1$, $A_2$ and $A_3$ will be found by Kramer's method [3]:

$$A_1 = \frac{\Delta_1}{\Delta}, \quad A_2 = \frac{\Delta_2}{\Delta}, \quad A_3 = \frac{\Delta_3}{\Delta},$$

where $\Delta$, $\Delta_1$, $\Delta_2$ and $\Delta_3$ — determinants of matrices, which are calculated as follows:

$$\Delta = \begin{bmatrix} x_0^2 & x_0 & 1 \\ x_1^2 & x_1 & 1 \\ x_2^2 & x_2 & 1 \end{bmatrix} = x_0^2(x_1 - x_2) + x_1^2(x_2 - x_0) + x_2^2(x_0 - x_1) \ ,$$

$$\Delta_1 = \begin{bmatrix} u_0 & x_0 & 1 \\ u_1 & x_1 & 1 \\ u_2 & x_2 & 1 \end{bmatrix} = u_0(x_1 - x_2) + u_1(x_2 - x_0) + u_2(x_0 - x_1) \ ,$$

$$\Delta_2 = \begin{bmatrix} x_0^2 & u_0 & 1 \\ x_1^2 & u_1 & 1 \\ x_2^2 & u_2 & 1 \end{bmatrix} = x_0^2\left(u_1 - u_2\right) + x_1^2\left(u_2 - u_0\right) + x_2^2\left(u_0 - u_1\right) \ ,$$

$$\Delta_3 = \begin{bmatrix} x_0^2 & x_0 & u_0 \\ x_1^2 & x_1 & u_1 \\ x_2^2 & x_2 & u_2 \end{bmatrix} = x_0^2\left(x_1u_2 - x_2u_1\right) + x_1^2\left(x_2u_0 - x_0u_2\right) + x_2^2\left(x_0u_1 - x_1u_0\right) \ .$$

Then the formulas for calculating the coefficients $A_1$, $A_2$, and $A_3$ will look like this

$$A_1 = d\left(u_0\left(x_1 - x_2\right) + u_1\left(x_2 - x_0\right) + u_2\left(x_0 - x_1\right)\right),$$

$$A_2 = d\left(x_0^2\left(u_1 - u_2\right) + x_1^2\left(u_2 - u_0\right) + x_2^2\left(u_0 - u_1\right)\right),$$

$$A_3 = d\left(x_0^2\left(x_1u_2 - x_2u_1\right) + x_1^2\left(x_2u_0 - x_0u_2\right) + x_2^2\left(x_0u_1 - x_1u_0\right)\right),$$

$$d = 1 \Big/ \left(x_0^2\left(x_1 - x_2\right) + x_1^2\left(x_2 - x_0\right) + x_2^2\left(x_0 - x_1\right)\right).$$

The coefficients $B_1 - B_3$ are calculated in a similar way, but the values $v_0$, $v_1$ and $v_2$ are used instead of the values $u_0$, $u_1$ and $u_2$. The approximation coefficients $A_1 - A_3$ and $B_1 - B_3$ are calculated once for each rasterization line.

Thus, instead of normalizing the screen coordinate value at each point of the rasterization line, only one division operation is performed per rasterization line to find the coefficient $d$ .

### 4.7. USAGE OF SECOND-ORDER BЙZIER CURVES TO SIMPLIFY THE CALCULATION OF TEXTURE COORDINATES

In computer graphics, the method of constructing Bézier curves [1] is widely used, which allows to form curves of any shape. For perspective-correct texturing, a function is used that can be approximated by a quadratic Bézier curve, the parametric representation of which is as follows

$$r(t) = (1-t)^2 p_0 + 2t(1-t)p_1 + t^2 p_2 , \tag{4.21}$$

where $p_0\left(x_0, u_0\right)$, $p_1\left(x_1, u_1\right)$, $p_2\left(x_2, u_2\right)$ — the anchor points, and $t \in [0,1]$.

In this case, the Bezier curve will be inscribed in a triangle $p_0 p_1 p_2$ (Fig. 4.11).

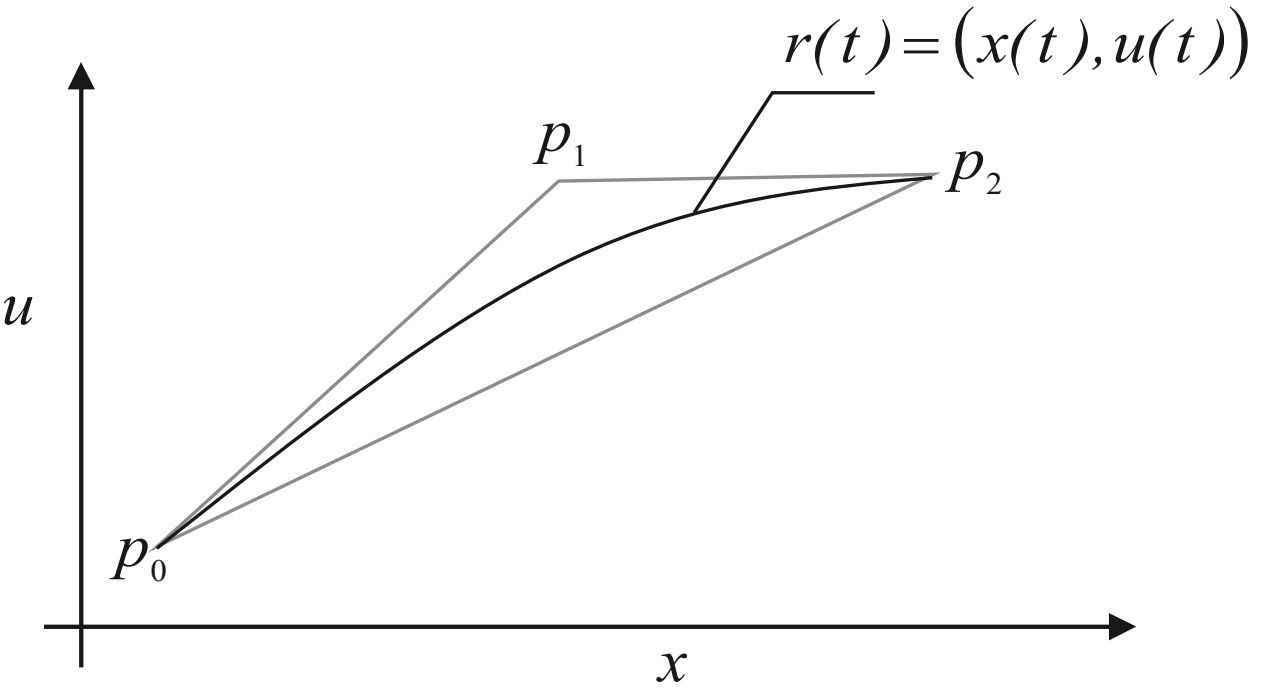

Fig. 4.11. Construction of a Bezier curve using three reference points

Since $r(t)=(x(t),u(t))$, then to build a Bézier curve, it is needed to solve two equations — relative to the screen $x$ and texture $u$ coordinates.

$$x(t)=(1-t)^2 x_0 + 2t(1-t)x_1 + t^2 x_2 , \tag{4.22}$$

$$u(t)=(1-t)^2 u_0 + 2t(1-t)u_1 + t^2 u_1 . \tag{4.23}$$

Equation (4.21) has a number of properties [1]:

1) $r(0)=p_0$, $r(1)=p_2$, that is, the first and last anchor points of the Bezier curve coincide with the corresponding points of the triangle in which it is inscribed. In order to find them, it is needed to find the exact values of the $u_0$ and $u_2$ texture coordinates for the first $x_0$ and last $x_2$ points of the rasterization line, respectively.

2) The tangent vectors at the ends of the Bezier curve coincide with the first and second sides of the triangle in which it is inscribed. That is, the point of intersection of these tangents will be the anchor point $p_1$.

Let's find the formulas for calculating the anchor point $p_1$.

The equation of the tangent for our case is

$$u = f(x_0) + f'(x_0)(x - x_0) , \tag{4.24}$$

where $f'(x_0)$ — the derivative of a function $f(x)$ at a point $x_0$.

To find the intersection point $p_1$, equate the equations of the tangents at points $x_0$ and $x_2$ and find the expression for the coordinate $x_1$

$$f(x_0)+f'(x_0)(x_1-x_0)=f(x_2)+f'(x_2)(x_1-x_2).$$

$$x_1=\frac{f(x_0)-f(x_2)-f'(x_0)x_0+f'(x_2)x_2}{f'(x_2)-f'(x_0)}. \qquad (4.25)$$

The value of the texture coordinate $u_1$ can be found by substituting the value $x_1$ into equation (4.24)

$$u_1=f(x_0)+f'(x_0)(x_1-x_0). \qquad (4.26)$$

In the case of perspective-correct texturing

$$f(x)=\frac{ax+by+c}{gx+hy+i}\text{, тоді } f(x_0)=u_0 \text{ і } f(x_2)=u_2. \qquad (4.27)$$

The expression for finding the derivative $f'(x)$ is

$$f'(x)=\frac{a(hy+i)-g(by+c)}{(gx+hy+i)^2}=\frac{A}{(gx+B)^2}, \qquad (4.28)$$

where coefficients $A=aB-g(by+c)$, $B=(hy+i)$ are constants for the rasterization line, since the screen coordinate $y$ is a constant value.

Given expressions (4.27), (4.28), formulas (4.25) and (4.26) can be written as follows

$$x_1=\frac{u_0-u_2-\dfrac{A}{(gx_0+B)^2}x_0+\dfrac{A}{(gx_2+B)^2}x_2}{\dfrac{A}{(gx_2+B)^2}-\dfrac{A}{(gx_0+B)^2}}.$$

After simplification, we get the following

$$x_1=\frac{(u_0-u_2)(gx_0+B)^2(gx_2+B)^2-A\left(x_0(gx_2+B)^2-x_2(gx_0+B)^2\right)}{A\left((gx_0+B)^2-(gx_2+B)^2\right)}.$$

Let's introduce the notation $C_0=(gx_0+B)^2$ і $C_2=(gx_2+B)^2$. Hence

$$x_1=\frac{(u_0-u_2)C_0C_2-A(x_0C_2-x_2C_0)}{A(C_0-C_2)}, \qquad (4.29)$$

$$u_1=u_0+\frac{A}{C_0}(x_1-x_0). \qquad (4.30)$$

When changing the coordinate $y$ by 1:

$$A(y+1)=a(h(y+1)+i)-g(b((y+1)+c)=$$
$$=a(hy+i)+ah-g(by+c)-gb=A(y)+D,$$

where $D=ah-gb$ — a constant for the entire polygon being textured.

$$B(y+1)=h(y+1)+i=hy+i+h=B(y)+h\,.$$

Thus, the calculation of the coefficients $A$ and $B$ for the next rasterization line requires only two addition operations.

Formula (4.22) contains 6 multiplication operations, 3 addition operations, and one shift operation. Let's derive recurrent formulas for calculating the screen coordinate $x$, which do not contain multiplication operations. To do this, we will use the method of finite differences [1, 3].

$$\Delta_x=x(t+dt)-x(t)=(1-t-dt)^2x_0+2(t+dt)(1-t-dt)x_1+(t+dt)^2x_2-$$
$$-(1-t)^2x_0+2t(1-t)x_1+t^2x_2,$$

where $dt$ — increment of parameter $t$ along the rasterization line.

Opening the brackets, we get

$$\Delta_x=2tdt(x_0-2x_1+x_2)+2dt(x_1-x_0)+dt^2(x_0-2x_1+x_2)\,. \quad (4.31)$$

Let's introduce the notation

$$S_x=x_0-2x_1+x_2 \text{ i } R_x=2dt(x_1-x_0)+dt^2S_x=dt(2(x_1-x_0)+dtS_x)\,.$$

The coefficients $S_x$ and $R_x$ are calculated once for the RL. The recurrent formula for calculating the screen coordinate $x$ along the RL will be

$$x(t+dt)=x(t)+2tdtS_x+R_x\,. \quad (4.32)$$

Formula (4.32) requires two addition operations, two multiplication operations, and one shift operation. To remove the multiplication operations, we apply the finite difference method for $\Delta_x$ one more time so that the equality $\Delta_x(t+dt)=\Delta_x(t)+d_x$.

Then we get

$$d_x=\Delta_x(t+dt)-\Delta_x(t)=2dt^2(x_0-2x_1+x_2)=2dt^2S_x\,. \quad (4.33)$$

The expression $2dt^2S_x$ does not depend on the value of the parameter $t$, so it is a constant for the entire rasterization string.

Taking into account formula (4.33), formula (4.30) can be written as follows

$$x(t+dt)=x(t)+\Delta_x(t+dt)\text{, де } \Delta_x(t+dt)=\Delta_x(t)+d_x. \qquad (4.34)$$

Similar formulas apply to the coordinate $u$.

The proposed recurrent formulas (4.34) allow to find the coordinate by performing only two addition operations.

In formulas (4.21) and (4.22), $x$ and $u$ depend on the parameter $t$, and the texturing task involves calculating the texture coordinate $u$ for a specific integer value of the screen coordinate $x$. Therefore, it is necessary to express the parameter $t$ in terms of the screen coordinate $x$.

Let's consider two approaches to solving this problem.

The first approach involves gradually increasing the value of the parameter $t$ until it provides an integer value of the screen coordinate $x$, which is larger than the previous value of the found screen coordinate by 1, that is $x_{i+1}=x_i+1$, where $i=\overline{0,\ m-1}$ is the ordinal number of the current point in the rasterization line, and $m$ is the number of points in the rasterization line.

Using a sufficiently small step of parameter $t$ does not always guarantee an integer value of the screen coordinate $x$. Therefore, it is advisable to make the assumption that

$$x_i(t_i)+1-\varepsilon \le x_{i+1}(t_{i+1}) \le x_i(t_i)+1+\varepsilon, \qquad (4.35)$$

where $\varepsilon$ — permissible deviation from the integer value.

If $\Delta t_{i+1}=t_{i+1}-t_i$, then

$$t_{i+1}=t_i+\Delta t_{i+1}. \qquad (4.36)$$

Since for the point $x_0$ the value of the parameter $t_0=0$, then $t_1=\Delta t_1$.

Let's write down the formula for the calculation $\Delta t_1$

$$\Delta t_1=\sum_1^k dt - t_{kor},$$

where $k$ — the number of iterations of adding a step, $t_{kor}$ – the corrective value.

The increment operation of step $dt$ is performed until the following condition is satisfied

$$x_1(\Delta t_1)\ge x_0+1-\varepsilon.$$

In the case, when $x_1(\Delta t_1) > x_0 + 1 + \varepsilon$, it is necessary to reduce the value $\Delta t_1$ by a certain amount $t_{kor}$, that is, adjust the value $\Delta t_1$.

The value $t_{kor}$ is calculated as follows. First $t_{kor} = \frac{dt}{2}$. If, after adjusting the value $\Delta t_1$ $x_1(\Delta t_1) > x_0 + 1 + \varepsilon$, then $t_{kor} = \frac{dt}{2} + \frac{dt}{4}$. The value $t_{kor}$ should be increased until $x_1(\Delta t_1) \le x_0 + 1 + \varepsilon$. In this case, the value $\frac{dt}{2^n}$ is added, when $n$ — sequence number of the adjustment iteration.

If during the adjustment the value $x_1(\Delta t_1)$ becomes less than the lower bound of condition (4.35), i.e. $x_1(\Delta t_1) < x_0 + 1 - \varepsilon$, then the sign for the current adjustment term changes to minus. The value $t_{kor}$ calculation procedure is completed when condition (4.35) is met for the current value $\Delta t_1$.

Thus, the formula for the calculation $t_{kor}$ is written as follows

$$t_{kor} = \pm\frac{dt}{2} \pm \frac{dt}{4} \pm \ldots \pm \frac{dt}{2^n}. \qquad (4.37)$$

To calculate the incremental value $\Delta t_2$, we use the incremental value $\Delta t_1$ and find only the corresponding value $t_{kor}$.

The general formula for calculating the increase $\Delta t_{i+1}$ can be written as follows

$$\Delta t_{i+1} = \Delta t_i - t_{kor}. \qquad (4.38)$$

The calculation according to formulas (4.36)-(4.38) is simple from a hardware point of view, since only addition operations are necessary, and division operations can be realized by addition and mounting shift.

Computer simulations have shown that when applying the proposed approach to calculating the parameter $t$ values, the maximum relative error does not exceed 0.7 %, which is 7 times less than when using the traditional quadratic approximation. Unlike traditional quadratic approximation, the proposed method does not require normalization of screen coordinates, which reduces computational complexity and, as a result, speeds up the texturing process.

The second approach to determining a parameter $t$ through a screen coordinate $x$ involves establishing a quadratic relationship

$$t = A_1x^2 + A_2x + A_3,$$

where $A_1, A_2, A_3$ — quadratic approximation coefficients.

The calculation of the approximation coefficients $A_1, A_2, A_3$ can be performed using formulas presented in the previous section. Given that $t_0 = 0$, and $t_2 = 1$, that formulas can be simplified as follows

$$A_1 = d\left(t_1\left(x_2 - x_0\right) + x_0 - x_1\right),\ A_2 = d\left(t_1\left(x_0^2 - x_2^2\right) - x_0^2 + x_1^2\right),$$

$$A_3 = d\left(x_0^2\left(x_1 - x_2 t_1\right) - x_0\left(x_1^2 + x_2^2 t_1\right)\right),$$

$$d = 1\Big/\left(x_0^2\left(x_1 - x_2\right) + x_1^2\left(x_2 - x_0\right) + x_2^2\left(x_0 - x_1\right)\right).$$

The results of computer modeling have shown that the maximum relative error when applying the quadratic dependence of the parameter $t$ on the screen coordinate $x$ does not exceed 1.7 %, which is almost 3 times less than the traditional quadratic approximation.

The proposed methods for calculating texture coordinates using the second-order Bezier curve, unlike the traditional quadratic approximation, do not require normalization of screen coordinates. This increases the accuracy of the approximation with a slight complication of the computational process.

## 4.8. NON-ORTHOGONAL RASTERIZATION METHODS

It is possible to reduce the number of division operations when texturing by rasterizing the object in the world coordinate system, provided that the rasterization lines are placed at a fixed distance from the observer. Let's find the slope coefficient of a line in the screen coordinate system, which corresponds to a line segment in the world coordinate system with a constant value of the coordinate z for the line (Fig.4.12). Let a triangle be given in the world coordinate system. It uniquely defines a plane whose equation is as follows

$$AX_w + BY_w + CZ_w = D, \tag{4.39}$$

where $A$, $B$, $C$ are the coefficients determined by the coordinates of the vertices of the triangle.

From the last equation we find that

$$Z_w = \frac{A \cdot X_w + B \cdot Y_w + D}{C}. \tag{4.40}$$

The following relations exist between the screen and world coordinates $X_v = \dfrac{X_w}{Z_w}$, $Y_v = \dfrac{Y_w}{Z_w}$. We write the last equations in the form

$$X_w = X_v \cdot Z_w, \quad Y_w = Y_v \cdot Z_w.$$

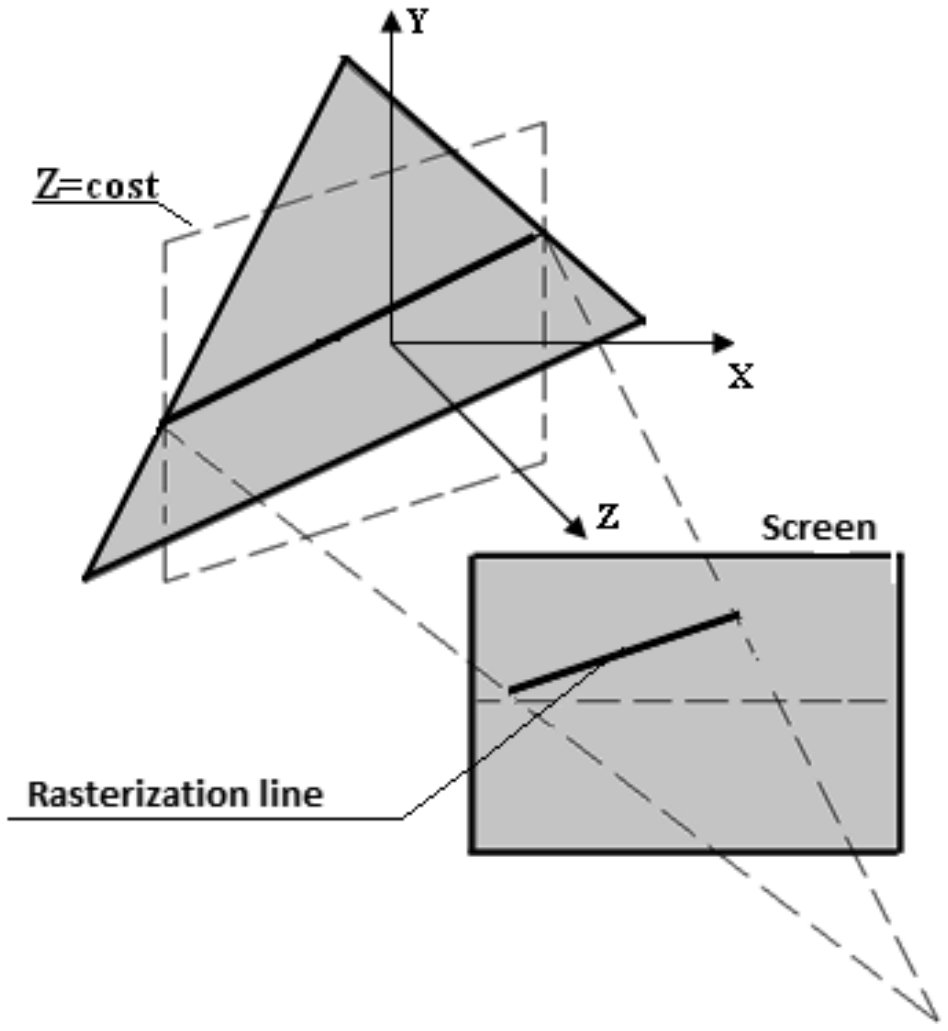

Fig. 4.12. Rasterization lines in the world and screen coordinate systems

Let's look for the angle of the line whose equation is $Y_v = k \cdot X_v + h$. With this in mind, we write that $Y_w = Y_v \cdot Z_w = (k \cdot X_v + h) \cdot Z_w$. Substituting into equation (4.40) and the value of $X_w, Y_w$, we obtain

$$A \cdot X_v \cdot Z_w + B \cdot (k \cdot X_v + b) \cdot Z_w + C \cdot Z_w = D.$$

From the last equation we find that

$$Z_w = \frac{D}{X_v \cdot (A + B \cdot k) + B \cdot h + C}. \quad (4.41)$$

Provided that for a triangle rasterization line in the on-screen coordinate system $Z_w = const$, then for any $j$ holds

$$\frac{D}{X_v \cdot (A + B \cdot k) + B \cdot h + C} = \frac{D}{(X_v + j) \cdot (A + B \cdot k) + B \cdot h + C}.$$

The last equation has a unique solution $k = -A / B$. Since $j$ and $X_v$ were chosen arbitrarily, it can be stated that the slope of the scanning rasterization line does not change for the entire triangle considered in the original coordinate system. The value of the coordinate $Z_w$ for a given rasterization line is easy to find by substituting the obtained value $k$ into equation (4.41).

$$Z_w = D / (B \cdot h + C).$$

The following relationship exists between the coordinates of the texture and screen spaces [1, 3]:

$$u = \frac{a \cdot X_v + b \cdot Y_v + c}{A \cdot X_v + B \cdot Y_v + C}, \quad v = \frac{d \cdot X_v + e \cdot Y_v + f}{A \cdot X_v + B \cdot Y_v + C}.$$

Denote the denominator of the above expressions by $T$ , and $1/T = \Re$ , then

$$u = (a \cdot X_v + b \cdot Y_v + c) \cdot \Re, \quad v = (d \cdot X_v + e \cdot Y_v + f) \cdot \Re .$$

The denominator for determining the texture coordinates $u,\ v$ is constant for the rasterization row, while in the conventional approach it is calculated for each point of the row. In the following, we will consider only one of the coordinates, for example, $u$ since the expressions for their calculation are similar. Let's express $Y_v$ through $k$ and substitute it into the previous expression. We get

$$u = (a \cdot X_v + b \cdot (k \cdot X_v + h) + c) \cdot \Re = [X_v \cdot (a + b \cdot k) + \ (b \cdot h + c)] \cdot \Re.$$

For the starting point of the rasterization line $X_v = 0$ . With this in mind $u_0 = \Re \cdot (b \cdot B + c)$. Consider how $u$ changes when coordinate $X_v$ changes by one

$$u_{i+1} = [(X_v + 1) \cdot (a + b \cdot k) + \ (b \cdot h + c)] \cdot \Re = u_i + \Re \cdot (a + b \cdot k).$$

The resulting ratio can be easily calculated in hardware, provided that $\Re$ is known.

The non-orthogonal direction of rasterization of the area bounded by the polygon allows to reduce the computational complexity of the process of applying textures to the surface of a three-dimensional graphic object. For a triangle containing *T* internal points*, (T-q)* division operations are performed, where *q* is the number of horizontal and vertical rasterization lines of the triangle.

The issue of triangle rasterization is important, since the direction of rasterization, which is not orthogonal to the coordinate axes, will inevitably lead to artifacts — the presence of "gaps" and duplicate points due to the offset of the starting points of the rasterization lines. This can be avoided if the leading edge of the triangle is parallel to the ordinate axis, but this requires a special surface triangulation and does not meet the requirements of graphic standards. Artifacts can be eliminated by adaptive phasing of the sequence of step increments of the scan line, which involves setting different initial values of the evaluation function during interpolation, which significantly complicates the linear interpolator.

The easiest way to solve the problem is to rasterize not the triangle, but the rectangle into which it is virtually inscribed (Fig. 4.13, a). You can determine the parameters of such a rectangle by comparing the co-ordinates of the vertices of the triangle (the left and rightmost vertices of the triangle represent the abscissa of the left and right sides of the rectangle, and the bottom one represents the ordinates of the bottom side of the rectangle). The parameter $r$ can be easily found by substituting the value of the abscissa of the top vertex of the triangle into the equation of the line used for rasterization. It is clear that the ordinate of the upper left vertex of the rectangle is equal to the sum of $r$ and the ordinate of the upper vertex of the triangle.

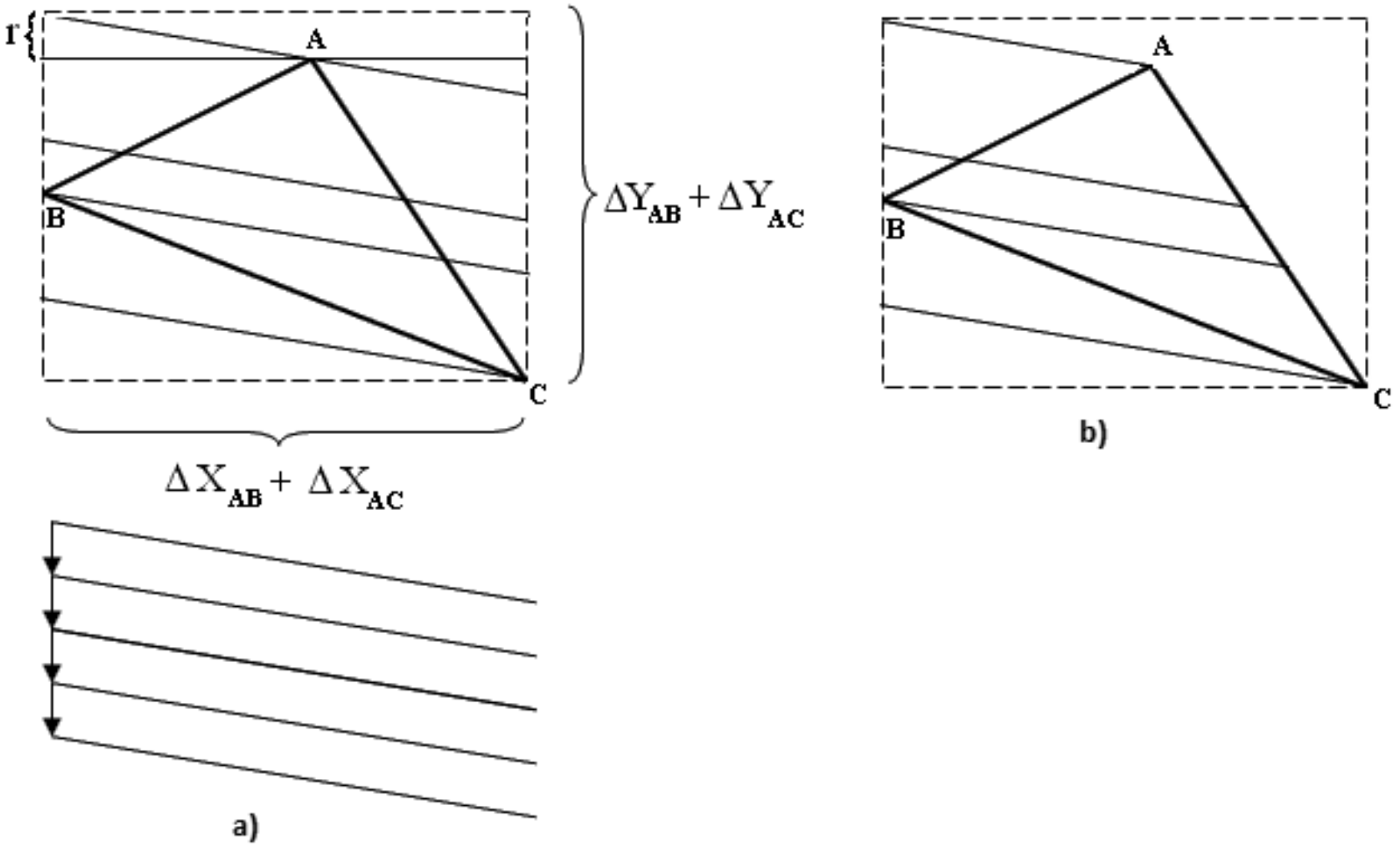

Fig. 4.13. Rasterizing an area bounded by a triangle

When rasterizing a rectangle, provided that the rasterization direction is not orthogonal, there will be no “cutting through” or “sticking” artifacts, since the step increments of neighboring lines will be identical along the ordinate directions.

Rasterization of a rectangle is performed to determine the coordinates of the left and right points of the triangle edges that intersect the rasterization line without calculations that require “long” operations. When the right edge of the triangle is reached, the transition to a new rasterization line of the rectangle is made, which is located one ordinal level lower (Fig.4.13, a), that is, the area of the triangle behind its right edge is not rasterized (Fig. 4.13, b). Determining the coordinates of the left and right edges of a triangle is achieved by comparing the foreground color of the triangle with a reference. In this case, one can, for example, use the principle of the parity criterion [4], according to which the number of intersections of a polygon is an even number. To improve performance, it is possible to rasterize the areas that have the background color with a pulse sequence of increased frequency. Direct texturing is performed at the clock frequency at which the video memory operates.

With anisotropic filtering [5–7], which significantly increases the realism of reproducing graphic scenes, the color of a pixel is determined by several textures. The shape of the light spot changes with the position of the polygon relative to the observer's point. The higher the level of anisotropic filtering, the more the performance of the graphic scene generation decreases. The high computational complexity of texturing using anisotropic filtering limits its widespread use in image formation, although modern graphics cards include such functions as basic ones.

With anisotropic filtering, the projection of a pixel onto the model surface is not seen as a circle, but as an elongated ellipse (Fig. 4.14). In order to correctly calculate the color of a pixel, it is necessary to take into account the colors of all texture samples that fall into the ellipse. This is a rather complicated procedure for generating images in real time, so a simplification is used — replacing the ellipse with a parallelogram (Fig. 4.15) or a rectangle that bounds it.

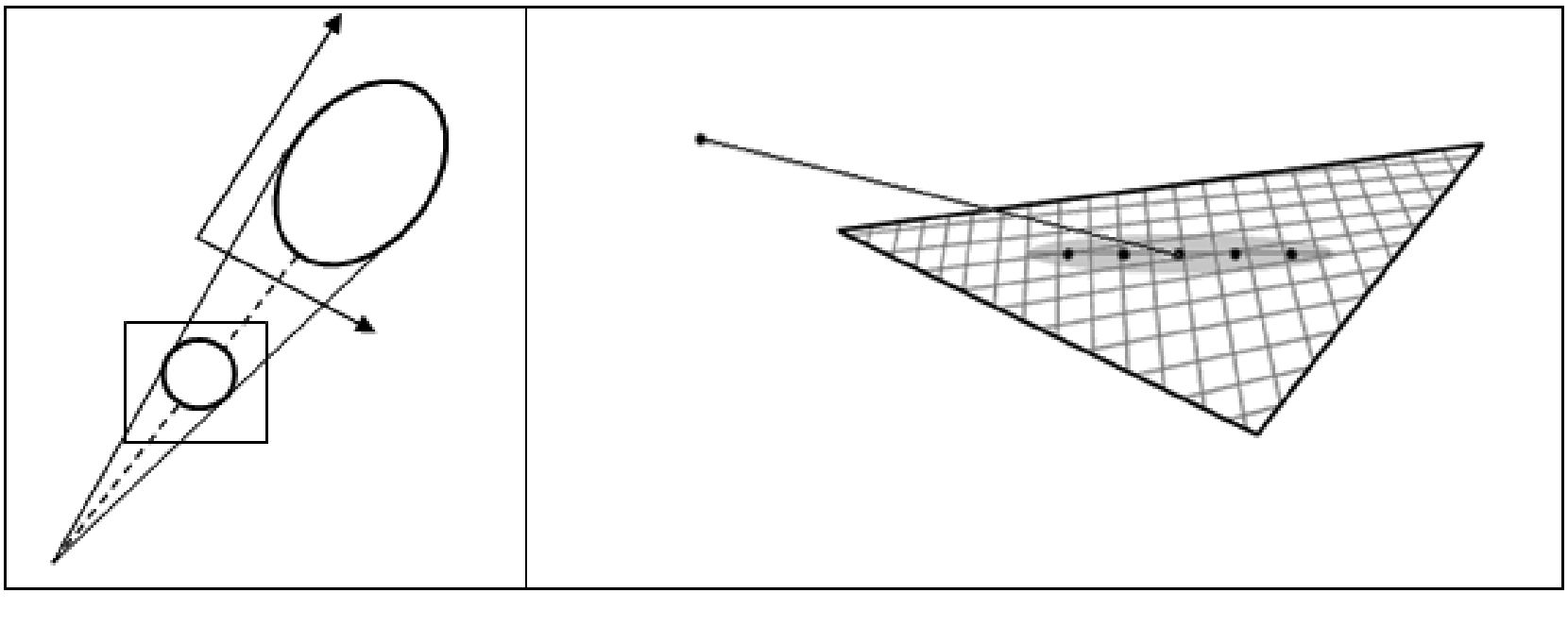

Fig. 4.14. Determination of the trace in anisotropic filtration

To calculate the principal axes of the ellipse, the functional determinant, the Jacobian, is calculated [3]:

$$j=\begin{vmatrix}\dfrac{\partial u}{\partial x} & \dfrac{\partial v}{\partial y}\\ \dfrac{\partial v}{\partial x} & \dfrac{\partial v}{\partial y}\end{vmatrix}=\Re^2\begin{vmatrix}aBY+aC-AbY-Ac & bAX+bC-BaX-Bc\\ dBY+dC-AeY-Af & eAX+eC-BdX-Bf\end{vmatrix}=$$

$$=\Re^2\begin{vmatrix}Y(aB-Ab)+aC-Ac & X(bA-Ba)+bC-Bc\\ Y(dB-Ae)+dC-Af & X(eA-Bd)+eC-Bf\end{vmatrix}.$$

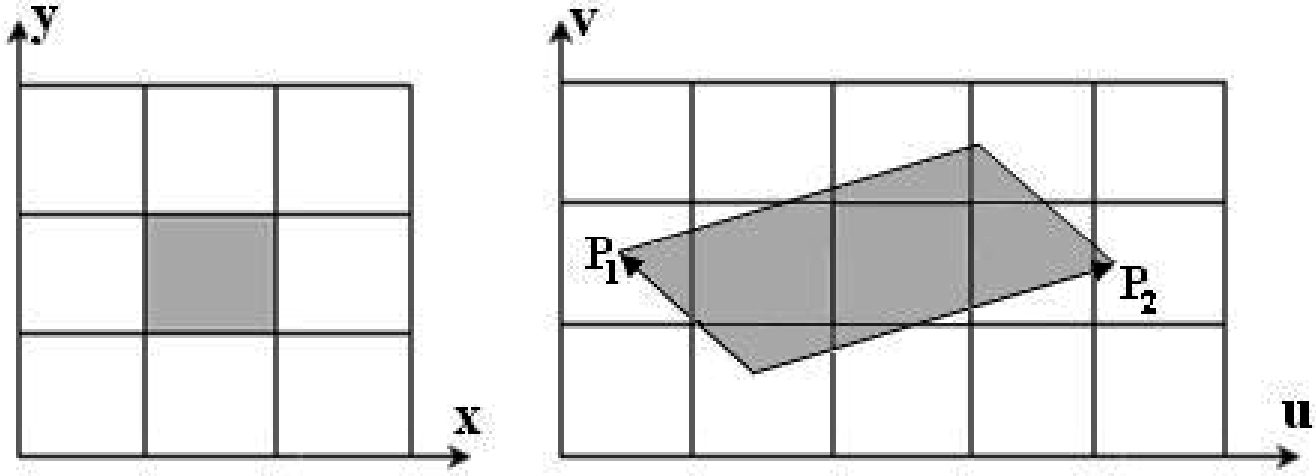

Fig. 4.15. Pixel trace in the form of a parallelogram with anisotropic filtering

The parameter $\Re$ is constant for the rasterization line. Unknown parameters $P_1, P_2$ can be determined by the formula

$$P_2=u((x+1),\ y)-u(x,y),\ v((x+1),y)-v(x,y),$$
$$P_1=u(x,\ (y+1))-u(x,y),\ u(x(y+1))-v(x,y).$$

$P_1, P_2$ can be calculated through the Jacobian, which defines the transformation from one coordinate system to another (in this case, from the coordinate system $OXY$ to the coordinate system $OVU$ ).

$P_1$, $P_2$ can be found through the derivatives in the directions *(1, 0)* and *(0, 1)*.

$$P_1=J\begin{bmatrix}1\\0\end{bmatrix}=\begin{bmatrix}Y(aB-Ab)+aC-Ac\\ Y(dB-Ae)+dC-Af\end{bmatrix},\ P_2=J\begin{bmatrix}0\\1\end{bmatrix}=\begin{bmatrix}X(bA-Ba)+bC-Bc\\ X(eA-Bd)+eC-Bf\end{bmatrix}.$$

In the case when $z=const$ , the sides $P_{1z}, P_{2z}$ of the parallelogram are found by derivatives in the directions *(0, 1), (1, k)*, respectively.

$$P_{1z}=J(0,\ 1)=P_1\quad P_{2z}=J(1,\ k)=\Re^2\begin{bmatrix}(aB-Ab)\\ (dB-Ae)\end{bmatrix}$$

The last expression shows that $P_{2z}$ is constant for the rasterization line, since

$$\Re^2 = \frac{1}{(AX_v + BY_v + C)^2} = \frac{1}{(AX_v + B(-\frac{A}{B}X_v + h) + C)^2} = \frac{1}{(Bh + C)^2}.$$

In most cases, anisotropic filtering uses a rectangular image window to rasterize a texture surface. In this case, the issue of fast rasterization of such a window, which is generally rotated relative to the coordinate axes, is important.

The authors have developed a method for forming a window at an arbitrary angle to the coordinate axes.

The vectors AB and AC (Fig. 4.16) have the same inclination angles with respect to the abscissa and ordinate axes. Thus, their component step increments relative to the given coordinate axes coincide and differ only in signs. Thus, when forming the output vector *AB* by memorizing its step increments, it is possible to implement the *AC* path. To do this, when forming the *i*-th line in the direction of *AB*, the type of the *i*-th step increment is memorized and formed after the completion of the development of the given vector in another orthogonal direction.

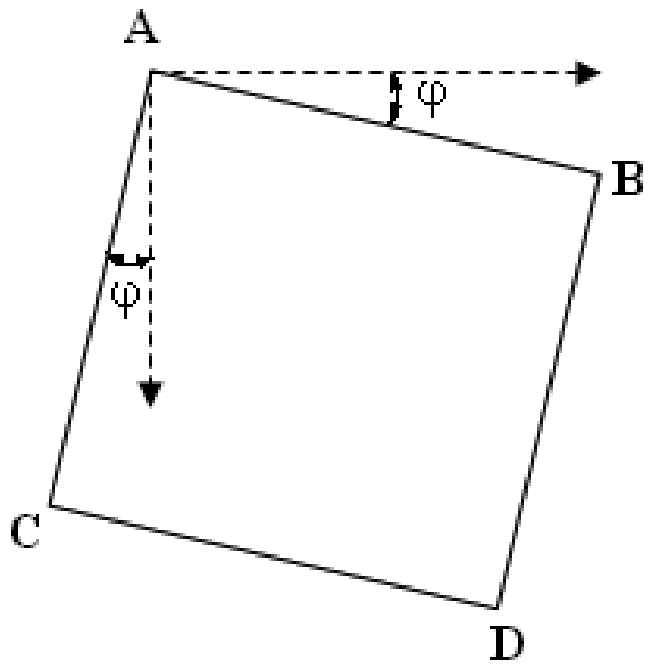

Fig. 4.16. Location of the window relative to the axes coordinates

To rasterize a rectangular image window, a loop traversal of the rectangular image window points is performed. During the loop traversal, the image points located in the first line *AB* of the image window are read (Fig. 4.17), after which a function of transition from point *B* to point *C* is formed for one discrete. After performing these actions, the direction of the traversal is reversed, that is, the traversal is performed along the *SD* vector, the signs of which are opposite to the *AB* vector. A similar traversal procedure takes place for all subsequent lines of the window.

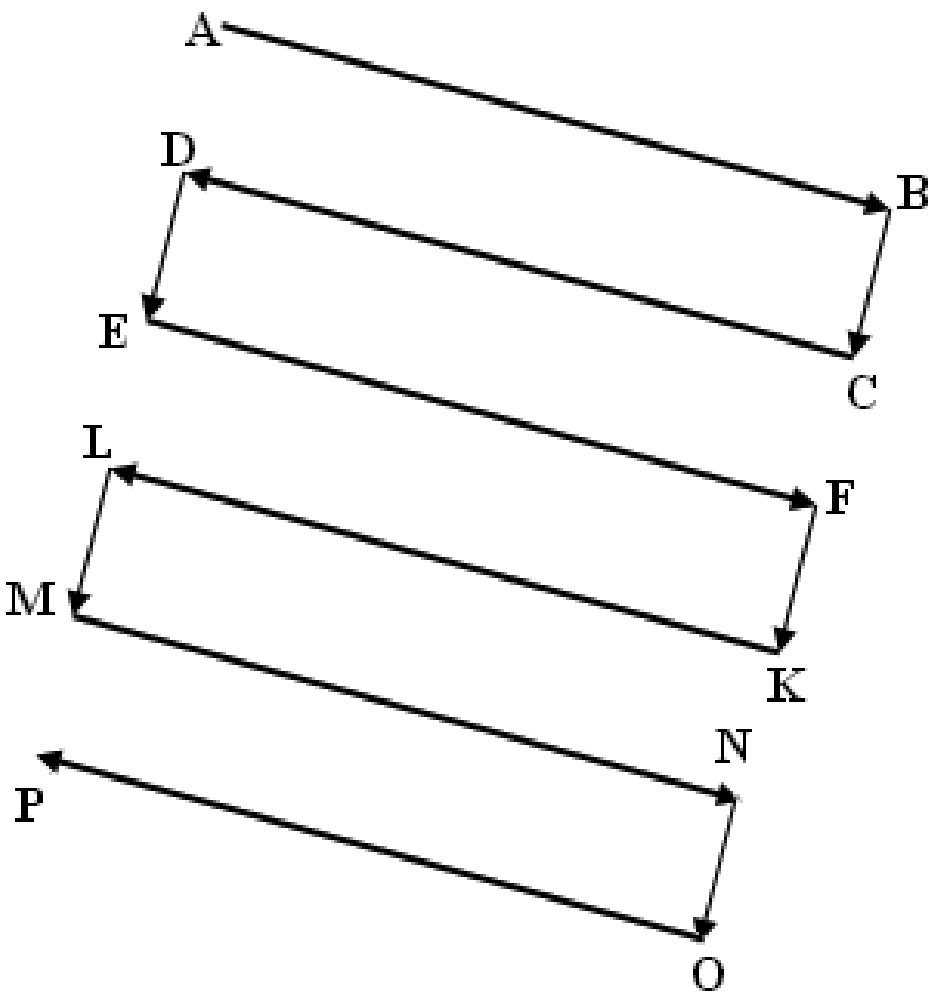

Fig. 4.17. Daisy-chain bypass of window rows

Suppose you need to rasterize a rectangular window *ABQP* (Fig. 4.18), the internal points of which define a certain texture image. To do this, perform the following steps.

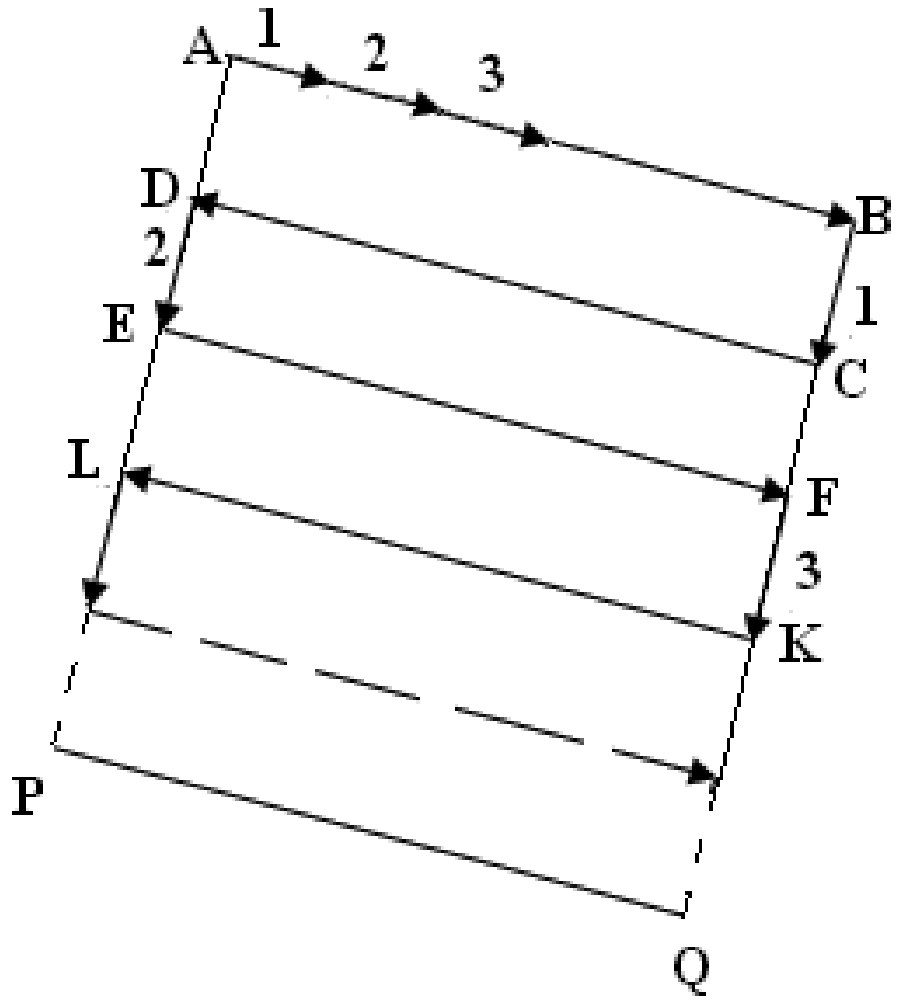

Fig. 4.18. Determining the coordinal displacements for the sides of the window

Determine the increments $\Delta X$, $\Delta Y$ of the base vector *AB*. Using the linear interpolation function, determine the coordinates of the image points in the direction of the base vector *AB*. When you reach point *B*, which corresponds to the end point of the vector *AB*, form a signal of transition from point *B* to point *C*. To do this, when forming a step trajectory of the base vector *AB*, the value of the step increment corresponding to point 1 of the *AB* vector is memorized, and when point B is reached, a step increment is formed with the opposite sign along the abscissa axis and the orthogonal components are swapped. This will ensure the transition to point *C*, which is located one discrete step away from point *B*. Then change the direction of the base vector to the opposite and repeat the above steps. Thus, when forming the *CD* vector, the step increment is memorized, which is formed in the second interpolation step. The recorded step increment is used at point *D* to move to point *E*. Similarly, when forming the third vector, the value of the step increment in the third cycle is memorized, which is used at point *F* to move to point *K*.

In general, when forming the *i*-th line, the type of the *i*-th step increment is memorized and formed relative to the ordinate axis after the specified vector is finished. The described actions are repeated until the image window is rotated.

During anisotropic filtering, the textures obtained during the rasterization of the image window are averaged according to the selected filtering method.

The proposed method of rasterizing a rectangular window formed at a certain angle to the coordinate axes has a simple hardware implementation and involves the use of not two, but only one linear interpolator, which allows reducing hardware costs by up to two times.

## 4.9. DETERMINING THE DIRECTION OF RASTERIZATION TO SPEED UP PERSPECTIVE-CORRECT TEXTURING

In texture mapping tasks, polygon points are traditionally processed along horizontal rasterization lines. Let's prove that for polygons whose two opposite sides are parallel to one of the screen coordinate axes, one of the texture coordinates remains constant along the rasterization line. Consider two cases of this dependence.

*Statement 4.1.* For polygons with two opposite sides parallel to the X-axis, the texture coordinate $v$ remains constant for the horizontal rasterization line when perspective-correct texturing is performed.

*Proof.* Let's assume a polygon with two opposite sides parallel to the X-axis and a sample texture to be applied to this polygon (Fig. 4.19). The coordinates of the vertices of the polygon are expressed in terms of $x$ and $y$; the coordinates of the vertices of the texture are expressed in terms $u$ of and $v$.

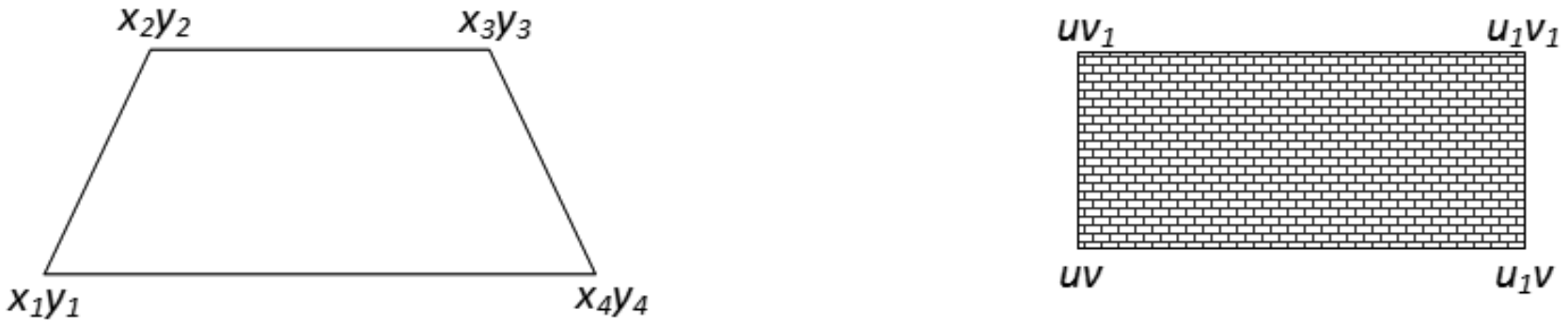

Fig. 4.19. A polygon with opposite sides parallel to the x-axis and a sample texture

According to formulas (3.1), the texture coordinate $v$ will remain constant for the horizontal rasterization line in the case when the coefficients $d$ and $g$ take zero values.

The coefficients $d$ and $g$ are calculated using the formulas [7]:

$$d = E \cdot I - F \cdot H\,,\; g = D \cdot H - E \cdot G\,, \tag{4.42}$$

$$H = \begin{vmatrix} \Delta x_1 & \sum x \\ \Delta y_1 & \sum y \end{vmatrix} \Big/ \begin{vmatrix} \Delta x_1 & \Delta x_2 \\ \Delta y_1 & \Delta y_2 \end{vmatrix},\; E = y_4 - y_1 + H y_3\,, \tag{4.43}$$

$$\sum y = y_1 - y_2 + y_3 - y_4\,,\; \Delta y_1 = y_2 - y_3\,,\; \Delta y_2 = y_4 - y_3\,. \tag{4.44}$$

From Fig. 4.19 it is obvious that $y_1 = y_4$ and $y_2 = y_3$. Then, according to the formulas (4.44)

$$\sum y = 0,\; \Delta y_1 = 0\,.$$

The obtained values allow to find the solution to the formulas (4.43):

$$\begin{cases} \sum y = 0, \\ \Delta y_1 = 0. \end{cases} \Rightarrow H = 0 \text{ and } \begin{cases} H = 0, \\ y_4 - y_1 = 0. \end{cases} \Rightarrow E = 0\,. \tag{4.45}$$

Substituting the values of (4.45) into formulas (4.42), we obtain

$$d = 0,\; g = 0. \tag{4.46}$$

Considering equation (4.46), the coordinates $v$ and $v_1$ are calculated as follows

$$v = \frac{ey + f}{hy + i}\,,\; v_1 = \frac{ey_1 + f}{hy_1 + i}\,. \tag{4.47}$$

The system of equations (4.47) shows that for a polygon with two opposite sides parallel to the X-axis, the texture coordinate $v$ does not depend on the value of the polygon's screen coordinate $x$, but only on the value of the screen coordinate $y$. That is, the texture coordinate $v$ for each horizontal rasterization line is calculated once and used for the entire RL.

Then the formula for determining the coordinate $u$ will be as follows

$$u = \frac{ax + by + c}{hy + i} = (ax + A_u)B_u,$$

where $A_u = by + c$, $B_u = \frac{1}{hy + i}$ — are the constants for the horizontal RL.

*Statement 4.2.* For polygons whose two opposite sides are parallel to the Y-axis, the texture coordinate $u$ remains constant for a vertical rasterization line when perspective-correct texturing is performed.

*Proof.* Let's take a polygon with two opposite sides parallel to the Y-axis and a sample texture that will be applied to this polygon (Fig. 4.20).

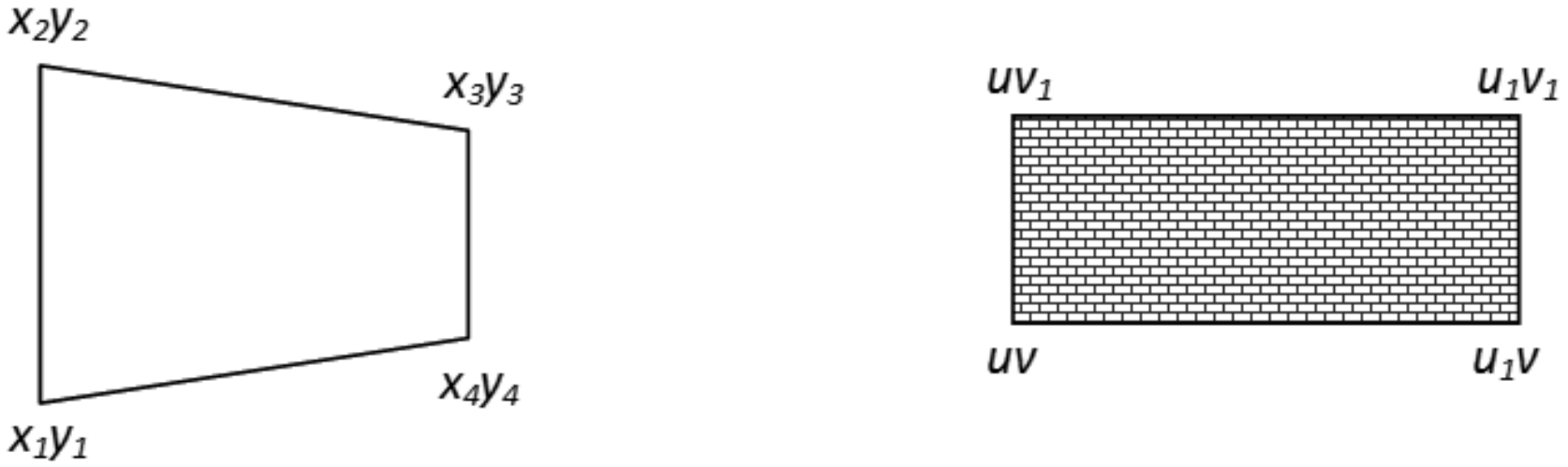

Fig. 4.20. A polygon with opposite sides parallel to the Y axis and a sample texture

Similarly to statement 4.1, we can prove that the coefficients $b$ and $h$ take on zero values, and the texture coordinate $u$ remains constant for each vertical rasterization line.

The formula for determining the coordinate $v$ is as follows

$$v = \frac{dx + ey + f}{gx + i} = (ey + A_v)B_v,$$

where $A_v = dx + f$, $B_v = \frac{1}{gx + i}$ — are constants for the vertical RL.

Polygons that have two opposite sides parallel to the X or Y axis are often found in software applications for visualizing room interiors or building exteriors, as well as in computer games where the main action takes place indoors. In particular, such elements of decor and furnishings as carpets, paintings, mirrors, doors, windows, and furniture are also described by such polygons.

A special case of perspective-correct texturing is affine texturing [1, 4], which occurs when the object to be textured is parallel to the projection plane, i.e., the screen. For affine texturing $g = h = 0$, $i = 1$. Then the texture coordinates $u$ and $v$ are defined as follows

$$u = ax + by + c \text{ and } v = dx + ey + f . \quad (4.48)$$

Formulas (4.48) do not contain division operations, which simplifies the texturing process. The constancy property of texture coordinates can also be applied to affine texturing, which will eliminate redundant operations when finding texture coordinates for cases where two opposite sides of a polygon are parallel to one of the coordinate axes.

If an object subject to affine texturing has two opposite sides parallel to the X-axis, the coefficient $d = 0$ and texture coordinate $v$ do not depend on the value of the screen coordinate $x$, but are defined as $v = ey + f$, that is, for each horizontal rasterization line, the value $v$ is calculated once and used along the entire line.

Similarly, the texture coordinate $u$ is calculated using the formula $u = ax + c$ if the object has two opposite sides parallel to the Y axis. In other words, the value $u$ is calculated once for each vertical RL.

The application of the proven properties to perspective-correct and affine texturing can significantly reduce the computational complexity of calculating texture coordinates without compromising the quality of texturing.

You can choose a particular direction of rasterization by checking the values of the coefficients in the Heckbert formula. If none of the proven properties can be applied, it is suggested to perform texturing along some non-orthogonal rasterization line (NRL). The denominator $gx + hy + i$ in formulas (4.8) is determined by the z-coordinate. Provided that in the object coordinate system we draw a plane parallel to the z-axis through the selected surface point, then, in general, we will get a RL, which in the screen system will correspond to a line with a non-orthogonal direction relative to the screen coordinate system. It is clear that in this case, the denominator in formulas (4.8) will have a constant value for the entire RL. Therefore, the task is to find a RL for which the expression $gx + hy + i$ will have a constant value.

Suppose that the coordinate $x$ is increased by 1, then the coordinate $y$ for the corresponding NRL will change by a certain value $\Delta y$. Let's find the value of the increment $\Delta y$, given that for each point of the NRL the value of the expression $gx+hy+i$ is a constant. Then, for two adjacent points of the NRL, we can write

$$gx+hy+i=g(x+1)+h(y+\Delta y)+i=gx+hy+i+g+h\Delta y .$$

From the last equation we find $\Delta y$

$$\Delta y=-g/h . \qquad (4.49)$$

As can be seen from formula (4.49), the value of the increment $\Delta y$ does not depend on the values of the coordinates $x$ and $y$, but only on the coefficients $g$ and $h$ of the polygon being textured. This indicates that the increment $\Delta y$ is constant for the entire polygon and it is enough to calculate it once.

Let's write down the general formula for calculating the y-coordinate for the NRL

$$y_i=y_p+\Delta y\cdot(x_i-x_p), \qquad (4.50)$$

where $(x_p,y_p)$ — coordinates of the starting point of the NRL that belongs to the polygon.

In the recurrent form, the coordinate $y$ for the NRL can be found by the formula

$$y_{i+1}=y_i+\Delta y . \qquad (4.51)$$

Formula (4.51) requires only one addition operation, which leads to its simple hardware implementation.

Taking into account the discrete nature of the RL formation (Fig. 4.21), the values of the coordinate $y$ must be integer, so we rewrite formula (4.50) as follows

$$y_i=y_p+\lceil \Delta y\cdot(x_i-x_p)\rceil , \qquad (4.52)$$

If you start counting NRL from the first point belonging to the polygon, you may encounter cutoff points between neighboring RLs, which will lead to artifacts. According to the property that cutoff points will not occur if the neighboring RLs are shifted relative to each other by one horizontal or one vertical step. This can be easily achieved by assuming that the initial points of all NRLs belong to the Y-axis. Let denote them by $(x_0,y_0)$, and for all NRLs $x_0=0$, and the values $y_0$ differ by 1. Then the formula (4.52) will take the following form

$$y_i=y_o+\lceil \Delta y\cdot x_i \rceil .$$

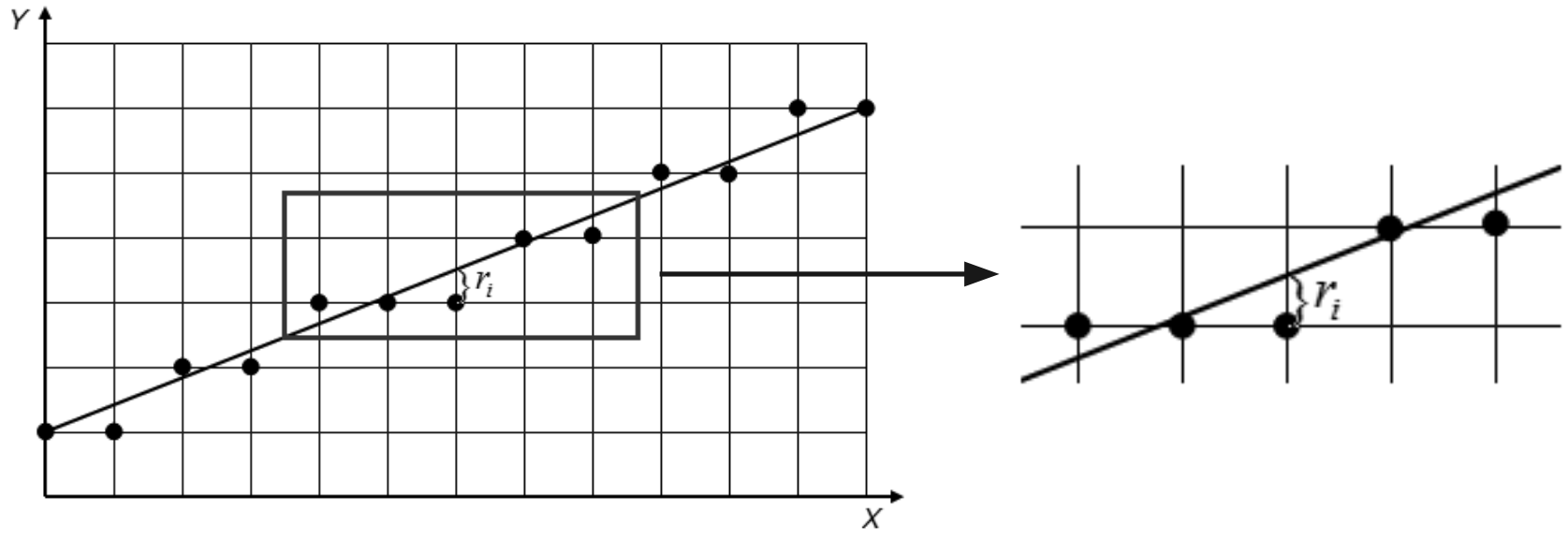

Fig. 4.21. Discretization of the NRL

The value $y_0$ can be found through the coordinate values of the point $(x_p, y_p)$

$$y_o = y_p - \lceil \Delta y \cdot x_p \rceil \ .$$

Consider the calculation of texture coordinates along the NRL.

Introduce the notation $k = \dfrac{1}{gx_p + hy_p + i}$.

Then the texture coordinates $u$ and $v$ along the NRL will be calculated as follows

$$u_i(x,y) = (ax_i + by_i + c)k, \ \ v_i(x,y) = (dx_i + ey_i + f)k.$$

Taking into account the discrete nature of the formation of rasterization lines, the coefficient $k$ must be adjusted.

To adjust the value $k$ , we find the first derivative $k'(y)$ of the coordinate $y$, then the formula for adjustment $k$ will be as follows

$$kor_i = k - r_i \cdot k'(y)\,, \tag{4.53}$$

where $r_i = \Delta y \cdot x_i - \lceil \Delta y \cdot x_i \rceil$ — is the vertical distance (Fig. 4.21) between the discretized and undiscretized NRL for the current point, and $k'(y)$ — is a derivative that characterizes the change in the value $k$ depending on the change in the value of $y$.

$$k'(y) = \frac{dk}{dy} = \frac{-h}{(gx + hy + i)^2} = -k^2 h\,. \tag{4.54}$$

Taking into account expression (4.54), formula (4.53) can be rewritten as follows

$$kor_i = k + r_i \cdot k^2 h \ .$$

The expression $k^2 h$ is a constant value for the NRL, so it is enough to calculate it once.

Thus, the proposed method of perspective-correct texturing along the NRL allows to remove division operations from the procedure for calculating texture coordinates. The number of division operations depends on the number of NRLs, since for each rasterization line it is necessary to find the coefficient $k$ . To find the increment $\Delta y$ , it is needed to perform only one division operation by polygon. The total number of polygon division operations is $(q+1)$, where $q$ is the number of NRLs. Calculating a pair of texture coordinates directly requires 8 multiplication operations and 7 addition operations.

## 4.10. CONCLUSIONS

The theoretical foundations of correct color reproduction in image formation have been developed, which has increased the realism of graphic scenes.

1. For the first time, methods of correct color reproduction when shading three-dimensional graphic objects based on linear interpolation of color component intensities, linear and spherical-angular interpolation of vectors and taking into account perspective projection are proposed, which allows to increase the realism of the formation of graphic scenes by establishing the color matching of surface points in the object and screen coordinate systems.

2. The method of Barenbrug perspective-correct texturing has been improved, in which, unlike the existing one, new formulas for calculating texture coordinates have been proposed. This made it possible to expand the scope of the method by both integer and fractional representation of texture coordinates and, as a result, to meet the requirements of the OpenGL and DirectX graphics standards.

3. It is proposed to use quadratic approximation for perspective-correct texturing without normalizing the values of screen coordinates and non-binary division of the rasterization line into segments. This made it possible to reduce the time for calculating the texture coordinates and reduce the relative error of their determination by up to two times.

4. A method for improving the performance of perspective-correct texturing is proposed, in which new recurrence relations are used for the first time to determine the texture coordinates of a streaming point in a rasterization line, which made it possible to exclude division operations from the texturing cycle and increase performance by simplifying the computational process. By using the second-order Bezier curve for approximation instead of quadratic interpolation, the accuracy of texture coordinates determination was improved.

5. An adaptive approach to perspective-correct texturing based on the analysis of coefficients in the Heckbert formula is proposed, which allows to choose the optimal orientation of the rasterization line, which will increase the performance of calculating texture coordinates by removing redundant division operations without compromising the accuracy of determining texture coordinates.

6. A method for improving the performance of perspective-correct texture application by using a non-orthogonal rasterization direction of a polygon-bounded area is proposed. Rasterization of an object in the world coordinate system is carried out under the condition that the rasterization lines are displaced at a fixed distance from the observer. For a triangle containing $T$ internal points, $(T\text{-}q)$ division operations are performed, where $q$ is the number of horizontal rasterization lines of the triangle.

7. A method for rasterizing a rectangular window placed at an arbitrary angle to the coordinate axes is developed, the peculiarity of which is that during the formation of the current rasterization line of the window, the step movement is remembered, which is used to move to the next rasterization line. This allows you to simplify hardware implementation by using one linear interpolator instead of two.

8. A method for increasing the performance of perspective-correct texturing for software components of graphic systems by means of non-orthogonal rasterization is proposed, which allows to reduce the number of division operations. The number of division operations per polygon is only $(q+1)$, where $q$ is the number of non-orthogonal rasterization lines.

# 5. ABOUT THE INFORMATIZATION OF PHD-STUDENTS' ACADEMIC AND RESEARCH WORK

***Svitlana Voinova***

*Informatization of educational space during postgraduate studies is a priority task of training applicants for the third educational and scientific level of higher education. A fundamentally important characteristic of the current stage of the development of science is its informatization. The process of informatization of society includes mediatization, computerization, and intellectualization. The Government of Ukraine has legislatively approved the task of informatization of society, thereby confirming the relevance of the issue. Among the scientific developments in open print on the issue of integration of academic and scientific activities of PhD-students, the question of the influence of informatization on this process is little researched. The purpose of the research is to study the impact of informatization on research and academic work of PhD-students and the possibility of using it as a tool for integrating these interrelated aspects of the activities of PhD-students. In the national doctrine of the development of the education system, the determining factor of the effectiveness of its informatization is the ability of teachers to carry out professional activities using information and telecommunication technologies. The development and informatization of science and education require universities to constantly adjust the methods of training PhD-students, taking into account new progressive methods of teaching and research. Educational researchers prone to innovation find it easier to introduce the results of modern research into the educational process and effectively use modern information technologies. It is useful to increase the attention of PhD-students to the fact that computer equipment and information technologies contribute to the automation and intensification of research. Informatization should become an essential factor in the integration of scientific and academic work of PhD-students.*

## 5.1. INTRODUCTION

In the conditions of globalization and acceleration of integration processes in higher education, one of the key elements of the innovative infrastructure of modern higher education is the research degree of professional training of highly qualified personnel. The key trends of modern higher education are changing the structure and methodical approaches to the organization of training of research specialists. Therefore, a pedagogical understanding of the processes taking place in this area is necessary.

In modern higher education, the need for a personally oriented, creative, competitive teacher, who is ready not only to reproduce acquired knowledge, formed abilities and skills, but also to independently design one's own activity, is growing with great progression [1].

Informatization of the educational space during postgraduate study is the priority task of training the students of the third educational and scientific level of higher education as a new generation of teachers capable of carrying out professional activities in the conditions of information and communication technologies.

In this way, the process of providing the education system with the theory and practice of developing and using new information technologies, focused on the realization of the goals of education and upbringing — informatization of education [2], is carried out.

On the other hand, at the turn of the 20th–21st centuries, the e-science paradigm replaced the empirical, theoretical, and computational paradigms, which could no longer provide the necessary rates of registration, accumulation, processing, and speed of exchange of the required volumes of scientific information using existing means, methods, and technologies. All over the world, appropriate infrastructures are being created that are able to ensure the rapid movement of both primary and processed data, as well as intensive scientific communication, based on the use of global networks and Web technologies [3].

Thus, a fundamentally important characteristic of the current stage of the development of science is the formation and rapid development of computer sciences and information technologies, that is, the informatization of science.

Informatization, according to the Ukrainian Wikipedia, is a policy and processes aimed at building and developing a telecommunications infrastructure that combines territorially distributed information resources. The informatization process is a consequence of the development of information technologies and the transformation of a technological, product-oriented way of production into a post-industrial one. Informatization is based on cybernetic methods and management tools, as well as information and communication technology tools.

It is common knowledge that scientific and academic work are the main types of activity of any PhD-student, and the necessity of their interconnec-tion is beyond doubt. At the same time, the necessary processes of integration of similar activities of PhD-students take place against the background of global processes of informatization of education and widespread dissemination of innovations.

It is necessary to search for significant factors of increasing the effectiveness of PhD-student training in higher education institutions, after which the training of the pedagogical elite is carried out in the graduate school — specialists who are able to develop and implement a large number of innovative approaches in the field of pedagogy. Graduate students are traditionally required to combine practical academic, research and innovation activities in conditions of widespread use of information technologies. With this very technology, there are many innovations.

The wide spread of modern information technologies has fundamentally changed the traditional ideas about the possibilities of development of human intelligence and led to the development of fundamentally new ways of organizing its educational and cognitive sphere, in particular at the third educational and scientific level.

The relevance of the use of new information technologies is dictated primarily by pedagogical needs to increase the effectiveness of learning, in particular, the need to develop skills for independent learning, a research, creative approach to learning, the formation of critical thinking, and a new culture. Nowadays, with the rapid increase in the volume of information, knowledge ceases to be an end in itself, they leave the conditions for the successful realization of an individual, his professional activity. In this regard, it is important to help graduate students become active participants in the learning process and to form their need for constant search. accordingly, the task remains to create such a model of the educational process that would allow revealing and developing their creative potential. Information and communication technologies contribute to the construction of a similar model created for an open information and educational space, based on which the principle of joint creative activity of those who study and are being studied is laid. The main goal of informatization is the creation of a unified information and educational space of the university, which is necessary to support the educational, research and organizational and commercial activities of the institution of higher education in the conditions of the introduction of modern information technologies.

In the new conditions of the formation of professionally significant qualities of the future of philosophy, education should be oriented not so much on the volume and completeness of specific knowledge, but on the ability to independently replenish knowledge, set and solve various tasks, put forward alternative solutions, develop criteria for selecting the most effective of them.

In general, there is a process of informatization of society, which includes at least three elements that complement one: mediatization as a process of improving the means of working with information, computerization as a process of improving the means of information processing, and intellectualization as a process of improving human knowledge and abilities to generation and perception of information [4].

A year ago, the government of Ukraine adopted Law 2807-IX (draft law No. 6241) "About the National Informatization Program", which demonstrates the relevance of the issue in modern Ukraine [5].

An important manifestation of the work of modern trends in the development of the educational space is the informatization of educational and scientific PhD-students [6, 7].

## 5.2. ANALYSIS OF LITERARY DATA AND STATEMENT OF THE PROBLEM

Many specialists investigate the issue of informatization of education and science in higher education, in particular, the activities of those who have obtained the third educational and scientific level of higher education. Thus, I. V. Oliinyk notes that informatization of the educational space of future doctors of philosophy can significantly affect the effectiveness of the pedagogical process, the improvement of the personality in professional and research aspects [8]. In the review of the study of world trends in the development of informatization of education, KNEU named after V. Hetman's attention is focused on the fact that the informatization of education requires the introduction into it of innovative methods, means and the formation of professional training of future specialists of the new education, the creation of a powerful information infrastructure in institutions of higher education with a developed information and computer educational environment, the introduction of the Internet — technologies., electronic learning, communication networks (global, national, local) [9]. O. O. Gagarin and S. V. Tytenko reveal the essence of the Web-system and the distance learning system regarding artificial intelligence in education, knowledge delivery model, site content management system, semantic content modeling, etc. [10]. A. Yatsyshyn emphasizes that in accordance with the new requirements for the training of PhD- and doctoral students in Ukraine, significant organizational and methodical changes must take place in the institutions that will further implement the educational program and the research aspect; the indicated innovations require justification and development of new educational programs, and for this need to coordinate and scientific and methodological support, which should be provided by the National Academy of Pedagogical Sciences of Ukraine [11]. Advanced countries recognize informatization as a factor of national development and create an appropriate legislative and regulatory framework on the basis of which policies (content, resources, finances) are implemented in this direction [12]. Aspects of informatization of science and education are considered in the works of V. Yu. Bykov [13, 14], V. P. Vember [15], R. S. Gurevich [16], and M. P. Shishkina [17]. The work of V. I. Lugovoi [18], I. Yu. Regheilo [19, 20] is devoted to issues related to the training of highly qualified scientific and scientific-pedagogical personnel.

However, the issue of integration of research and academic work of PhD-students in institutions of higher education for informatization of these areas of their activity is little researched.

### 5.3. THE PURPOSE AND TASKS OF THE RESEARCH.

The purpose of the study is to study the impact of informatization on research and academic work of graduate students and the possibility of using it as a tool for integrating these interrelated aspects of the activities of graduate students.

The objectives of the research are the following:

– to investigate the essence of the main types of activities of PhD-students in institutions of higher education;

– clarify the concept of informatization of academic and scientific activities in higher education;

– justify the need for the birth and development of the innovative component in the activities of PhD-students;

– to analyze approaches to the study of educational material by PhD-students, methodical principles of scientific thinking, logical means of cognition, stages of the cognitive process;

– clarify the components of the updated strategy for the development of the higher education system;

– to systematize approaches to the development of creativity in the scientific and academic activities of PhD-students.

Research methods and materials. The main method of the conducted research was the system method. Empirical methods such as observation and description are also used, and theoretical methods include analysis, generalization, induction, deduction, explanation, classification, etc.

## 5.4. RESEARCH RESULTS

Rapid and profound changes taking place in the modern world, which are often described as the formation of a global information society based on knowledge, have become the most important factors affecting the development of higher education in the late 20th and early 21st centuries. The role, organizational forms and methods of functioning of science are changing. The forms of life of the university as one of the main elements of the education system, which plays both a systemic and culture-forming role in it, are also changing. A modern institution of higher education increasingly finds itself at the forefront of innovative development, where it is required not only to perform educational functions, but also to create scientific divisions for the development of industry, conduct scientific expertise of business projects, develop science-intensive technologies and advanced theories that can be converted into marketable advantages for the research customer.

The vector of transformation of university education is accompanied by various organizational changes and a review of the mission of universities, in which the tasks of flexible management of intellectual and material resources, stimulation of innovations, positioning in the market of educational services, etc. become priorities. At the same time, universities remain a space where traditions of scientific knowledge are created and maintained.

Postgraduate studies were traditionally considered the pinnacle of the system of higher professional education. This degree, focused on the training of research specialists, focused on the values and content of the culture of university education itself, based on a combination of the traditions of scientific work and the values of knowledge and the pedagogical mission of preserving and transmitting cultural heritage. Remaining an integral part of the academic world and university education, the system of training specialists-researchers also faces the need to revise its orientations and forms of work.

One of the most important tasks of a higher school is the most complete disclosure of the intellectual potential of students, their abilities to generate and perceive new knowledge, the formation of skills to apply them in their everyday and professional activities, using modern information methods and tools. This task, namely the introduction of educational innovations, in particular information technologies, among other priority areas of state policy, is set in the national doctrine of education development. The determining factor in the effectiveness of informatization of the domestic education system is the ability of teachers to carry out professional activities using information and telecommunication technologies (Fig. 5.1).

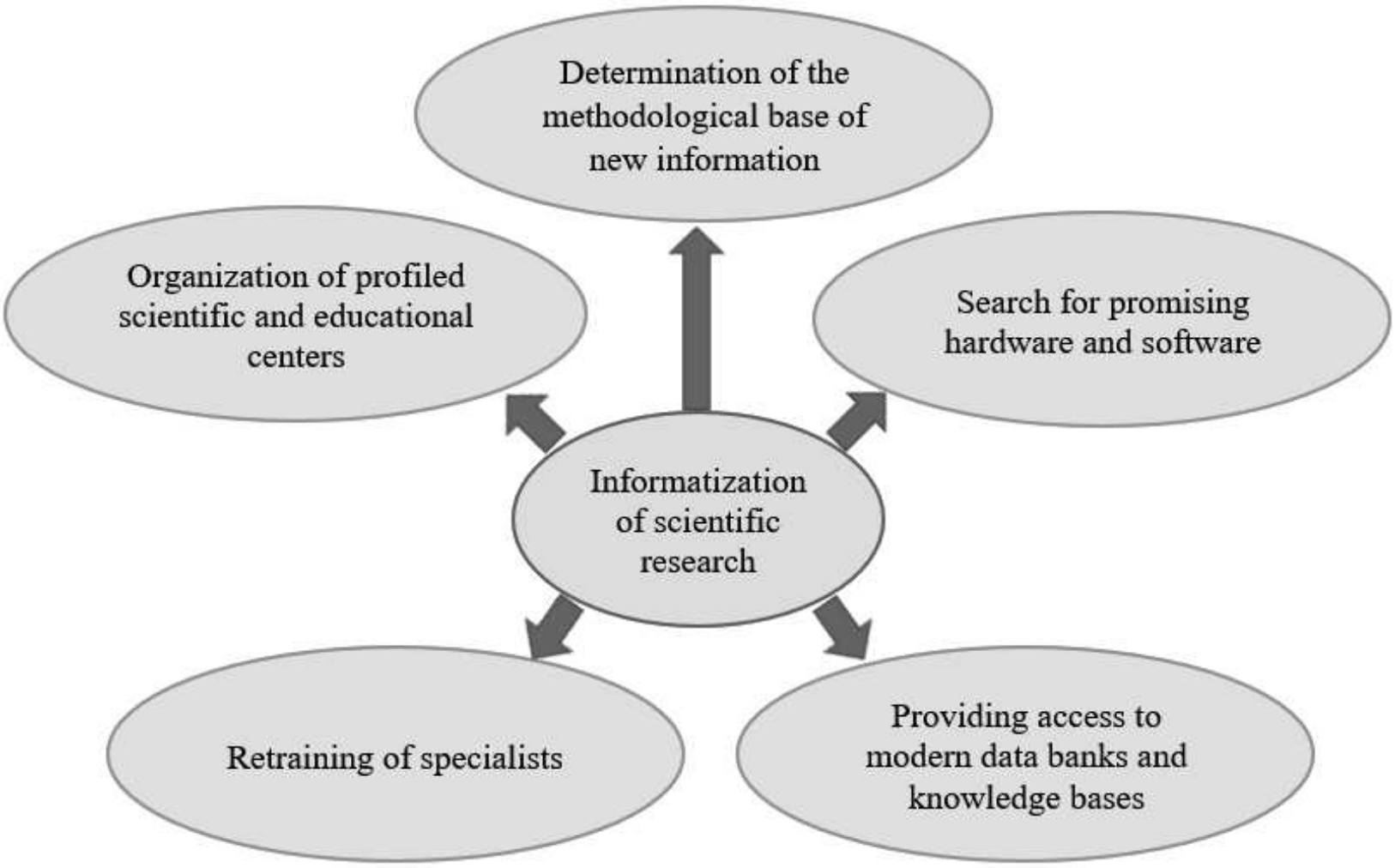

Fig. 5.1. Components of informatization of education

Against the background of global processes of informatization of education, which mean the processes of providing the education system with the theory and practice of developing and using new information technologies aimed at realizing the goals of education and training, the process of integration of research and academic work, which are the main activities of any — some graduate student.

PhD-students must master the methods of selection from the content, methods and means of science of basic ideas, patterns, and information technologies and build on this their innovative activity, and through it, the content and methodology of education. The constant development of science and the expansion of approaches to informatization of education require the institution of higher education to constantly adjust the methods of training PhD-students, taking into account new progressive methods and means of teaching and research.

Today's PhD-student tomorrow will be engaged in the training of higher education seekers in higher education, and only a teacher-researcher inclined to innovation can introduce the results of modern research into the educational process, share the most valuable scientific findings with students and effectively use modern information technologies. This will support the innovative atmosphere of classes, increase the attractiveness of scientific creativity, and introduce students of higher education to modern tasks facing science.

Informatization of education is a multifaceted process, which involves the PhD-student — the future teacher:

– management of the educational process based on the use of automated data banks of scientific and pedagogical information, informational and methodological materials, as well as communication networks;

– improvement of the methodology and strategy of content selection, methods and organizational forms of education, upbringing, relevant tasks of the student's personality development in modern conditions of informatization of society;

– creation of methodological training systems focused on the development of intellectual potential;

– formation of skills to independently acquire knowledge, to carry out experimental and research activities; various types of independent information processing activities;

– creation and use of computer testers, diagnostic methods of control and assessment of students' knowledge level.

The program of informatization of the educational process involves the introduction of new forms of work with the use of information technologies by a PhD-student — a future teacher. One of the effective ways of introducing new forms of work can be seen in the creation of a comprehensive system of ensuring the educational process.

Informatization of education is the process of changing the content, methods and organizational forms of training students at the stage of transition to life in the conditions of the information society.

Thus, the logical main directions of implementation of the educational process informatization program in modern conditions are:

– a systemic vision of the role of information computer technologies within the framework of informatization of education;

– designing and monitoring the development of the information and educational environment of the university at all levels of the educational process in it;

– formation of the readiness of the teaching staff to use new information technologies in education based on the system of supportive education, based on the continuous acquisition of new knowledge;

– development of the technical base; the use of telecommunications and their inherent technologies; development of educational information resources.

A great role in the formation of the innovative activity of a PhD-student belongs to his scientific supervisor, who should be a role model in conducting scientific works. He should lead the PhD-student to the conclusion that science is enriched by problems, through the solution of which new innovative ideas arise. At the same time, PhD-students develop creative thinking skills, the need for innovative activities and the use of information technologies develops.

In the process of training graduate students, special attention should be paid to the analysis of approaches to studying the material, their involvement in methodological principles of scientific thinking, arming with logical means of cognition, familiarization with the stages of the cognitive process. The cognitive result in this case can be new facts, laws, theories, innovations, methods of activity obtained by postgraduate students themselves, based on the use of new computer developments.

It is also necessary to develop in graduate students the idea that computer equipment and relevant technologies are capable of automating and intensifying work in many ways, increasing the effectiveness of the practical part of the research being conducted. Postgraduate students need to take into account that any research work begins with the study of domestic and foreign literature on the chosen topic. Such literature can be published in electronic resources of the Internet and be available for standard methods of searching for scientific sources.

Informatization of the educational activities of graduate students led to the creation of electronic textbooks, electronic planning, and electronic control. A characteristic feature of informatization at the university in general is the transition from fragmented to large-scale informatization, based on the creation of information resources (databases, knowledge bases, electronic libraries, etc.), the development of telecommunications, the creation of software for network information technologies, the development of conceptual and methodological foundations for the informatization of research. In addition to information technologies, which are the main elements of the actual educational process, the university began to pay more attention to the implementation and support of so-called service technologies (electronic textbook, multimedia, expert systems, publishing systems, video advertising).

The high efficiency of modern education can be ensured only if graduate students — future teachers create such computer packages (electronic textbooks, manuals, simulators, testers, etc.), the presence of which will ensure the same computer environment in a specialized classroom for practical classes, in computer classrooms of a higher education institution or dormitory, equipped for independent work of higher education students, PhD level students, as well as at home on a personal computer or on any gadgets.

Informatization of the educational process at the university is based on good basic computer training and implementation of the principle of continuous use of information technologies in the educational process. Therefore, it is impossible to talk about informatization without computer support.

Each study guide is a channel of the teacher's pedagogical influence on the student. Collectively, such influences merge into information noise. In this noise, it is almost impossible to single out a systematic and comprehensive source of information. A paradoxical situation arises: the large number of textbooks does not reduce, but rather increases the need of the teacher and student for a new textbook, which is as adequate as possible for the specific educational process in which they are involved.

An electronic textbook is a computer-based pedagogical software tool designed primarily for the presentation of new information that complements printed publications, serves for individual and individualized learning and allows to a limited extent to test the acquired knowledge and skills of the learner. Modification of the electronic textbook may be necessary, first of all, to adapt it to a specific curriculum that takes into account the specifics of the discipline studied at this university, the possibilities of the material and technical base, the personal experience of the teacher, the current state of science, the basic level of preparation of students, the amount of hours, allocated for the study of the discipline, etc.

It should be noted that the electronic textbook should not simply repeat printed editions, but should use all modern achievements of computer technologies.

An electronic textbook is necessary for independent work during face-to-face and, especially, distance learning, because it facilitates the understanding of the material studied due to methods of presenting the material other than in printed educational literature:

– inductive approach, impact on auditory and emotional memory, etc.;

– allows adaptation in accordance with the needs of the listener, the level of his training, intellectual capabilities and ambitions;

– frees you from cumbersome calculations and transformations, allowing you to focus on the essence of the subject, consider more examples and solve more problems;

– provides the widest opportunities for self-checking at all stages of work;

– provides an opportunity to carefully prepare the work and submit it in the form of a file or printout;

– performs the role of an endlessly patient mentor, providing an almost unlimited number of explanations, repetitions, tips, etc.

An electronic textbook is necessary for the student, because without it he cannot get solid and comprehensive knowledge and skills in this discipline.

The electronic textbook is useful in practical classes in specialized classrooms because it:

– allows you to use computer support to solve a larger number of tasks, frees up time for the analysis of the obtained solutions and their graphic interpretation;

– allows the teacher to conduct classes in the form of independent work on computers, retaining the role of manager and consultant;

– allows the teacher to quickly and effectively control the students' knowledge with the help of a computer, to set the content and level of difficulty of the test.

The electronic textbook is convenient for the teacher because it allows you to bring to lectures and practical classes the material at your own discretion, perhaps smaller in volume, but the most significant in content, leaving for independent work with it what turned out to be outside the framework of classroom classes, as well as :

– frees from tedious checking of individual tasks, typical calculations and control works, transferring this work to the computer;

– allows you to optimize the ratio of the number and content of examples and tasks that are considered in the classroom and assigned at home;

– allows you to individualize work with students, especially in the part related to homework and control measures.

Speaking about the control and systematization of the results of the innovative activities of graduate students, one cannot help but dwell on the regular reports, abstracts and reports that they prepare. During the implementation of such projects, there is an active process of consolidating scientific achievements, systematizing the knowledge obtained during the study of scientific literature and reference manuals, conclusions are drawn about the need to adjust the directions of experimental activity, new means of informatization of education are used more effectively.

One of the examples of Ukraine's involvement in the processes of globalization of education and science is its participation in the Bologna process, which from a separate political process aimed at improving the quality of training and mobility of qualified personnel in the European Union, became the basis for reforming the professional education system in Ukraine and other countries of the world. The Bologna Accords consider postgraduate studies (doctorates in the terminology of Western education) as the third degree of education, which must also build its work taking into account its basic principles. In addition to the European Union, the United States and countries of Anglo-Saxon culture, which accumulate more and more young researchers, exert a powerful influence on the development of university research.

Thus, Ukraine faces the influence of new global trends in the field of training of research specialists. Therefore, for the introduction of appropriate approaches to the use of information technologies in postgraduate studies, the experience of informatization of the training of future scientists, available in other countries, can be useful.

In general, the world has developed three main models of organizing the scientific and pedagogical life of a university, which differ in their priorities and attitudes regarding the role of science and education in their activities.

The German model envisages a merger of teaching and research in universities. The French model prefers to separate these two functions, leaving mainly pedagogical tasks to the university. The third, "Atlantic", British model is a combination of the first two, borrowing different experiences and traditions.

The French model is currently experiencing a deep crisis.

Two other types of organization of academic life, which were embodied in continental Europe, over the last 50 years have significantly evolved towards compromise solutions. The experience of countries — economic leaders shows that the leading role in the transition to an innovative economy belongs to universities, since the main components of success are concentrated here: training of highly qualified specialists; scientific and technical ideas and developments; opportunities to solve interdisciplinary problems.

At the top of the pyramid of the educational system of developed countries are universities of a special type, which have recently come to be called innovative universities, which do the most for the development of science, the invention of new technologies and the development of new markets and industries.

Doctoral students play a significant role in collaboration between universities and industry. Doctoral students perform three key functions in this collaboration:

– first, they act as producers of new knowledge within the framework of scientific creativity and the system of developing innovations and technologies;

– secondly, they contribute to the spread of knowledge in the wider social environment;

– thirdly, it is a connecting link in the configuration of partnership networks between universities and commercial enterprises.

Doctoral training allows a research specialist to acquire skills in critical thinking, scientific communication, and organization of research, which he can implement in those areas of activity where he continues his professional development. Doctoral students of universities create the infrastructure of innovation as a set of human and social capital within the framework of the organizations in which they conduct their activities. Thus, the training of researchers in higher education fulfills important functions of creating communities and social networks composed of people who are able to create new knowledge, perceive it and translate it into technologies that serve the public good and economic growth.

Distance learning in doctoral studies has become widespread in the USA. In Europe, universities in only a few countries, such as Great Britain, Spain, Cyprus, Switzerland, have online doctoral education programs.

According to the "Guide to Online Schools" collection, 274 accredited universities or their divisions offer online education in various subject areas aimed at further obtaining PhD degrees in the USA [21]. A total of 1,425 online doctoral programs are offered by American universities. The most popular distance learning programs are in the field of business (120 programs) and leadership (118 programs). They are offered by 14 and 52 American educational organizations, respectively. Physiology programs are also popular: 95 online PhD training programs are offered by 15 educational organizations. Doctoral programs in pedagogy (about 458 programs) are offered by 122 educational organizations.

These factors indicate a clear expansion of the scope of application of information technologies during the training of graduate students. In addition to the traditional use of computer technology in conducting research and processing the results of their experimental parts, informatization tools become an integral component of the system of educational tools in graduate school. In combination with new organizational approaches to the training of graduate students, informatization will have the corresponding effect and contribute to the growth of the number of highly effective scientific personnel.

It should not be forgotten that the scientific and closely related innovative work of graduate students is a connecting link between a higher education institution and socially useful professional activity, and its main goal is to consolidate theoretical knowledge and acquire skills in their practical application, to form the creative and innovative potential of future scientists, involving them in the use of advanced information technologies. Important for the development of such innovative work is the possibility of integration of all types of activities performed by graduate students: pedagogical, research and innovative against the background of widespread informatization of education.

Regarding the informatization of research conducted by graduate students, it should be noted that information technologies play a key role in the process of accumulation, dissemination and effective use of new knowledge (Fig. 5.2). Today, the traditional methods of information support for research, which consisted mostly in the computerization of mathematical calculations, the use of statistical modeling methods and the distribution of scientific and technical information through telecommunication networks, no longer satisfy scientists. They are being replaced by new methods based on the use of rapidly advancing capabilities of informatics and promising information technologies.

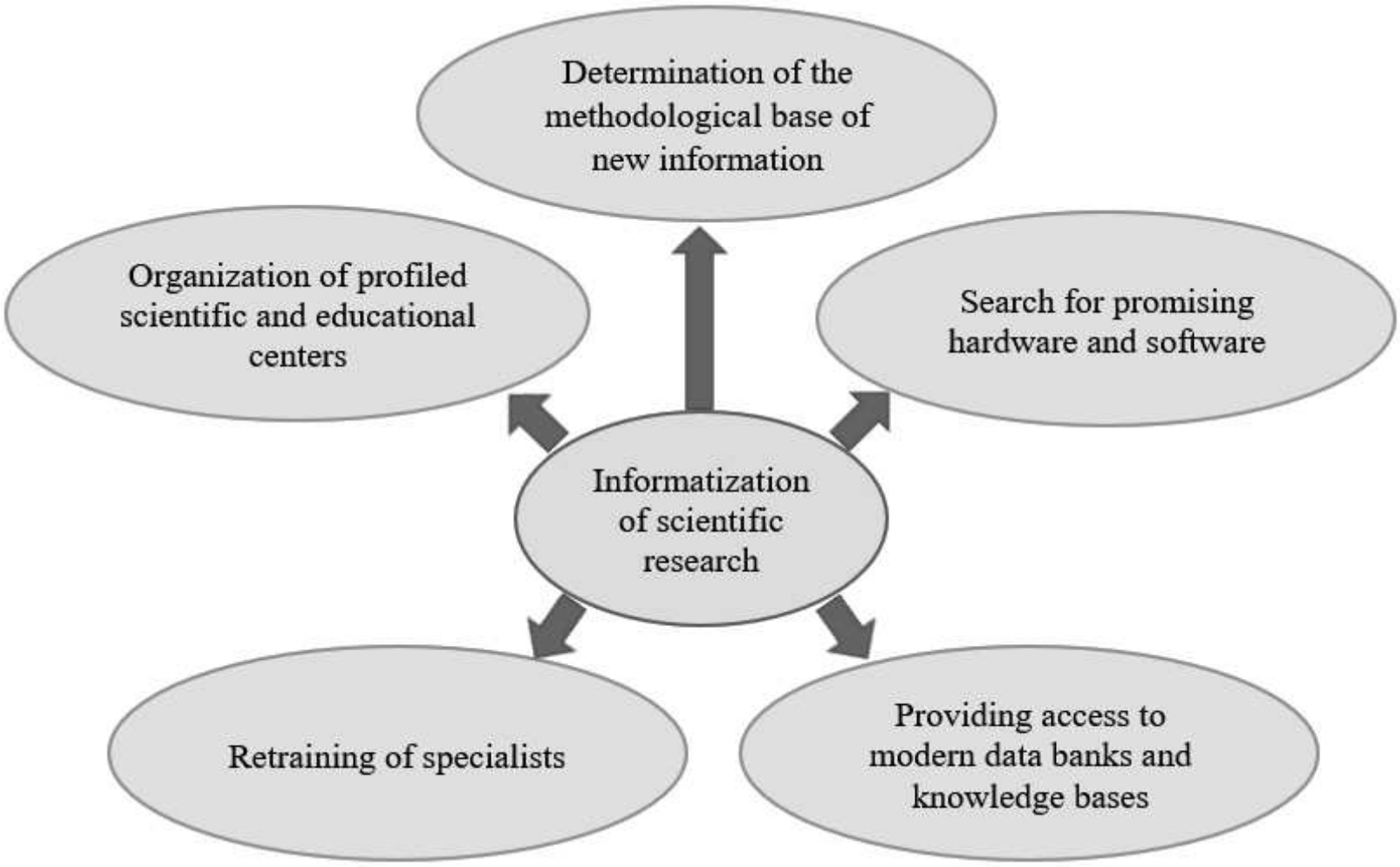

Fig. 5.2. Tasks of informatization of scientific and research activity

Clear examples can be teleconferences, distributed scientific teams united by a common information and telecommunications network, and even methods of complex information modeling of complex natural processes and phenomena; artificial intelligence methods that allow finding solutions to poorly formalized tasks, as well as tasks with incomplete information and vague initial data; methods of cognitive computer graphics, which allow to present various mathematical formulas and ratios etc. on the computer screen in a spatial form.

For the successful planned development and introduction of new information technologies in higher education, fundamental development of the scientific foundations of new information technologies is necessary for the following problems:

– systematic analysis of the development and implementation of new information technologies, timely clarification of selected priority directions, forecasting and prevention of possible negative trends;

– development of new principles of organization of computing processes, methods of presentation, processing and assimilation of data and knowledge;

– development of methods of description of subject areas and mathematical modeling;

– design and use of new information technology tools (interactive audio and video tools, computer and telecommunication environments).

The purpose of informatization of research activities is to accelerate the acquisition and deepening of scientific knowledge about phenomena and regularities in nature, technology and society due to the use of new information technologies at all stages of scientific work.

For this, it is necessary, in particular, to ensure the following tasks are solved:

– conducting research in fundamental fields that determine the methodological base of new information technologies in research;

– conducting research on the use of promising hardware and software tools;

– providing access to data banks and knowledge bases of leading scientific centers in the field of higher education of Ukraine and foreign countries using telecommunications;

– organization of profiled scientific and educational centers for the informatization of research, retraining of specialists taking into account these centers.

As mentioned, today the e-science paradigm dominates the world. Its characteristic features are as follows:

– automated registration and accumulation of observation and experiment data on electronic media;

– extensive use of computer resources and numerical methods for modeling phenomena, including the use of distributed computing environments;

– wide use of automated data processing and analysis methods to identify patterns and gain new knowledge;

– use of global networks (Internet) for exchange of research results, scientific communication, access to accumulated scientific results;

– cooperation of scientists and scientific resources in conducting research at different levels [22].

Carrying out research in such conditions requires the pooling of resources of the entire scientific community at various levels (institutions, states, international collaborations), the distribution of the entire volume of research between individual scientists and scientific teams, and the intensive exchange of research results. For this purpose, both in individual countries and in the whole world, appropriate infrastructures are being created that are able to ensure the rapid movement of both primary and processed data, as well as intensive scientific communication, based on the use of global networks and WEB technologies.

The transition to a new paradigm requires a corresponding restructuring of the material and organizational base of research, the mastering by each scientist of new methods of performing research, scientific communication and interaction in the scientific process.

The status of the National Academy of Sciences of Ukraine as the highest state scientific organization assumes its leading role in the organization and coordination of fundamental research, the implementation of scientific forecasting and expert assessment of the development of the economy, society, science and technology, active participation in the formation and implementation of state scientific and scientific and technical policy [23].

For the National Academy of Sciences of Ukraine, constant scientific and technological updating of the processes of informatization of the entire cycle of scientific activity — from scientific search and planning of scientific developments to the innovative implementation of results is vital. The strategic goal on this path is to support and develop the information infrastructure of the National Academy of Sciences of Ukraine with the connection of its institutions to national and international research and educational telecommunication networks.

For this purpose, a target program of the NAS of Ukraine was launched in 2004 — "The Informatization Program of the NAS of Ukraine" for the managed transition to a new paradigm, the creation of an appropriate technical base, information resources and software tools, and the implementation of the results in the everyday practice of the Academy.

To ensure a purposeful and controlled process of building, developing and supporting the information infrastructure of the NAS of Ukraine, the Institute of Software Systems of the NAS of Ukraine developed the Concept of the Informatization Program of the NAS of Ukraine, which was constantly refined and supplemented during the entire period of its operation. According to the concept, the goal of the Program is the wide implementation of new information technologies (NIT) in the scientific, scientific organizational and economic activities of scientific institutions and organizations, the Presidium of the National Academy of Sciences of Ukraine, increasing the productivity, quality and efficiency of research.

The problem of informatization of actual research is multifaceted, and its solution requires significant progress in the field of computer technology, programming, artificial intelligence, etc.

The specificity of the process of research requires a combination of computing and information resources of scientific communities and their joint use. Carrying out numerical theoretical calculations, processing the results of experiments requires enormous capacities, beyond the reach of a single scientist and even a separate scientific institution. At the same time, the load on computing power in the process of research is rather uneven. During the period of experiments, it is large, but the processes of analysis and preparation of results do not require such huge capacities. The combination of resources and their joint use allows to smooth out the unevenness of the resource load.

Back in 2009, implementation of the State target scientific and technical program "Implementation and application of grid technologies for 2009–2013" began and the Ukrainian National Grid was built [24]. The main achievement of the implementation of the programs is the creation of the Ukrainian national grid-infrastructure of the production type, integrated into the European grid-infrastructure, which unites more than 22 thousand researchers from all over the world. This provided the necessary services for Ukrainian scientists to conduct world-class digital research both independently and in collaboration with scientists from other countries.

Currently, higher education institutions are implementing an updated development strategy, which involves focusing on the formation of the student's creative personality, development of non-standard thinking, freedom of choice, needs and readiness for innovative activities in the conditions of informatization. In the set of means that ensure the solution of these tasks, a special role belongs to the accounting and analysis not only of the experience of modern higher education, but also of the positive historical experience of post-graduate studies and the activities of outstanding scientists and teachers. The need to study the experience accumulated by mankind is fully related to the process of training graduate students and to their integration of pedagogical, innovative and activities.

Graduate students in their activities need not only to rely on positive domestic and foreign experience, but also to look for their own ways of solving the problems facing the theory and practice of education, while realizing the relationship of educational, research and innovative components in the conditions of informatization of education. Analysis and characterization of the structure of pedagogical, research, and innovative activities of graduate students show that the optimality of the integration of these three main areas of training depends on various factors, such as specialty, age and year of study, general and pedagogical work experience, ability to use computer equipment in professional activity.

An important factor influencing the success of innovative activities of graduate students is the necessary level of professionalism in both teaching and research activities. It is formed in future scientists during training at a higher education institution in the conditions of independent practical activity, which is directed by a scientific supervisor when using information and telecommunication technologies. Moreover, the foundations of scientific, innovative and pedagogical creativity should be laid within the walls of the institution of higher education and serve as the main indicator of the quality of training of graduate students. An obvious conclusion is the desirability of increasing the share of post-graduate students who are professionally proficient in both pedagogical, research, and innovative aspects of activity in the context of informatization. Informatization of education and the correct application of appropriate technologies and tools can be considered as a significant factor in the integration of such activities.

The role of independent work in postgraduate studies in modern conditions tends to grow, based on the requirements for highly qualified specialists, where great importance is attached to the ability to navigate independently in the rapid flow of information and the need for constant improvement of professional growth and self-improvement.

One of the most revolutionary achievements of the last decades, which significantly influenced the educational process all over the world, was the creation of the worldwide computer network Internet. This factor led to new requirements for the technical equipment of educational institutions, their access to global information resources, and on the other hand, gave a powerful impetus to the development of the content of the teacher's activity, the use of new types, methods and forms of education, oriented to the active cognitive activity of graduate students. The Internet develops skills related to mental operations: analysis, synthesis, abstraction, comparison, juxtaposition, verbal and semantic prediction and prediction, etc.

The use of new information technologies in graduate studies is due to both the desire for novelty and the opportunity to implement a personally oriented approach to graduate students, which is the main conceptual direction of education in the 21st century. and provides individualization and differentiation of training taking into account the abilities of the students, their level of education, inclinations, etc.

Today's priority in the education of future doctors of philosophy is to focus on the formation of communicative competence, the rest of all goals (educational, upbringional, developmental) are realized in the process of implementing this main goal. Communicative dominant in education presents serious requirements for the content and forms of organization of the educational process. The communicative approach in its modern understanding implies learning to communicate and forming the ability for intercultural interaction, which is the basis of the functioning of the Internet.

The ability to learn is a by-product of any educational activity. The formation of self-learning skills (that is, the acquisition of the ability to learn) is an integral part of all educational goals. Therefore, the problem of formation of educational competence in graduate students in the process of their independent work is very relevant today.

In addition to communication needs, graduate students need to master the methodology of working on the Internet to be more responsible for their own learning.

The main task of information and communication technologies in the educational process is to develop in graduate students new cognitive capabilities necessary for people living in the age of digital technologies, to provide an opportunity for self-education to students with different strategies in learning.

The training of graduate students should be carried out taking into account many factors of their pedagogical, scientific and innovative activities with a focus on increasing the efficiency of teachers and scientists working in a higher education institution, as well as based on numerous aspects, trends and advantages of informatization of education. At the same time, innovations, their creation and promotion in combination with informatization can be considered as a basis for the desired integration, academical, educational and research work of graduate students (Fig. 5.3).

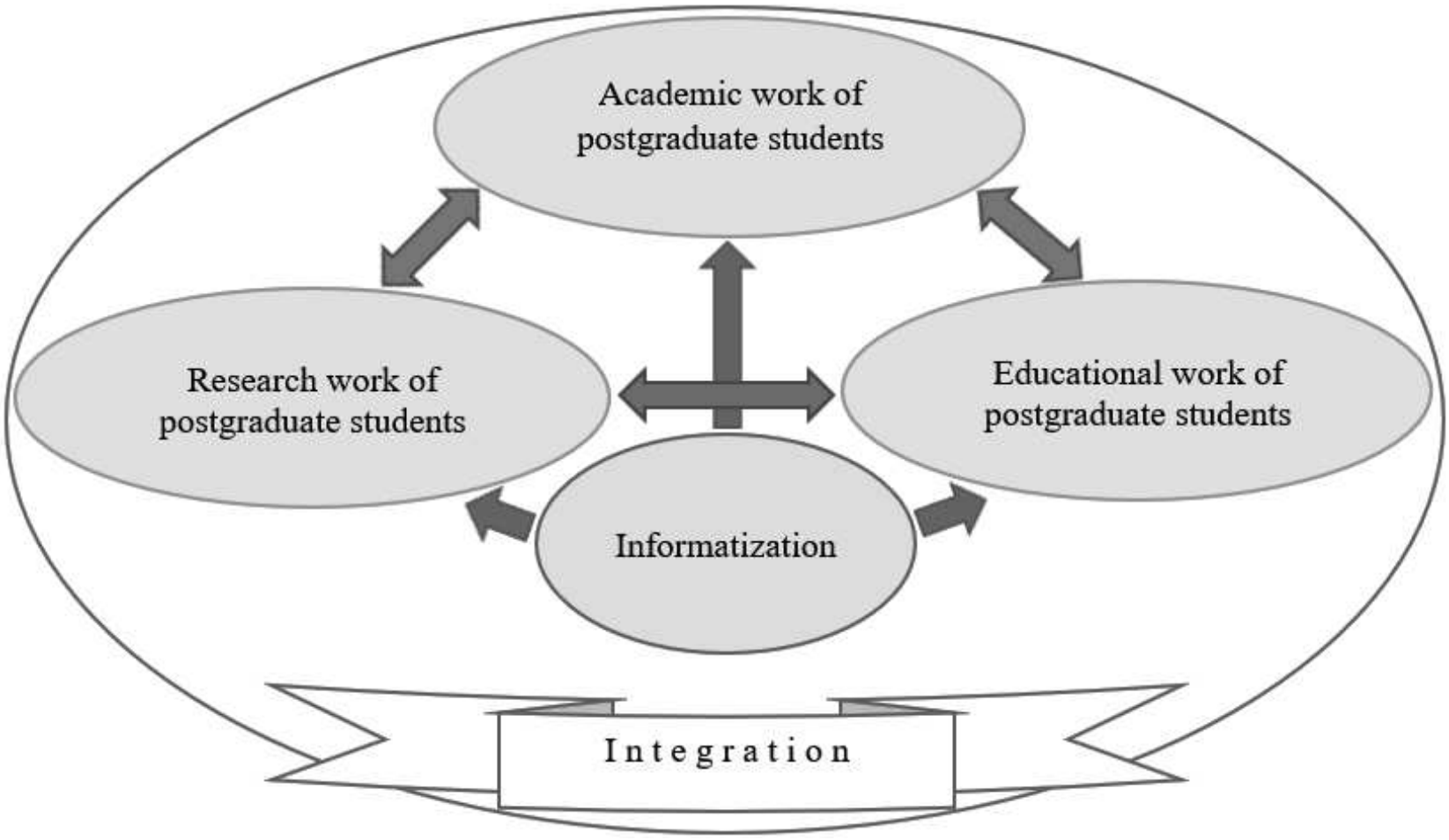

Fig. 5.3. Integration of the components of PhD-students' work

Undoubtedly, the development of the Internet became an objective condition for the actualization of the concept of electronic management of an educational institution.

Electronic management of an educational institution is based on the possibilities of information and telecommunication technologies and the values of an open society, characterized by a focus on the needs of the learner, economic efficiency, openness to control and initiative. As one of the consequences, it will ensure a new organization of the activities of the administrative divisions of the educational institution and a change in the entire complex of relations between the administration and students and teachers.

The principles of this type of management imply that the student, teacher and employee of the institution receive a real opportunity to access information or submit information in the shortest possible time, in an optimally convenient mode and in comfortable conditions. At the same time, financial management, personnel management, optimization of its distribution and use, management of training programs, schedule, etc. are simplified and made transparent.

In order to transition to full electronic control of the tower, a number of tasks should be solved. Including:

– creation of higher education management websites, their regular updating with the publication of basic information on key topics of higher education institution development;

– organization of interactive interaction of the higher education administration with higher education applicants, graduate students and employers. This includes, for example, providing electronic access to various document forms required by higher education applicants and employers, or searching for vacancies based on user-defined criteria;

– creation of an Internet portal with a wide range of higher education services (submission of documents for admission to a higher education institution, obtaining diplomas of higher education, exchange of lost documents, financial transactions by students, etc. in electronic form). At this stage, one of the main problems is ensuring the safety of working with official documentation and information.

These tasks have either already been partially solved, or are being partially solved today by the leadership of higher education institutions, or are considered as close prospects.

The development of interactive services and servers for students and employers should be a new concept for the websites of higher education institutions.

Among the program activities carried out within the framework of the concept of electronic management of a higher education institution, projects aimed at organizing administrative relations within it and improving interaction between the higher education institution and the student of higher education can be implemented.

Constant access to information about his success, which the student of education can obtain through the "Electronic Dean's Office", forms in him a sense of responsibility for the results of his studies. Thus, the introduction of IT forms a new concept of the relationship between the university and the student of higher education in a competitive educational environment and ensures a shift in emphasis in the organization of professional training for teaching an academic discipline for the active educational activity of the student, which involves the use of effective educational technologies, the ability of the student to independently assess the success or failure of his studies, to make corrections in time.

Today, the following educational models using IT are being implemented in the domestic system of higher education:

– independent activity of students: case technology and Internet technology (independent study of printed/electronic textbooks and manuals, completion of tasks, passing of self-tests);

– student interaction: Internet/Intranet technology (discussions and joint projects through computer conferences, audio and video conferences), chat, forum, audio and video communication programs;

– a model based on the teacher's pedagogical activity: telecommunication technology (video lectures and video conferences);

– a model of the context of the student's professional activity: any of the listed technologies (trainings, practical tasks, exercises, specific situations).

Today, the electronic space of the higher education institution opens up new opportunities for subjects of education: on the one hand, in remote access to get acquainted with the material and technical base and equipment of the educational process, on the other hand, to literally look into each classroom and laboratory, to understand where the student will "bite" granite of science.

It is intended to create the full effect of being present at the class both in the off-line and online mode, so that the user can not only fully feel the effect of being present at the class, but also repeatedly return to the material from one or another discipline, receive for personal use lecture content with both video signal and multimedia presentations, etc., and download the audio signal of the lecture for listening on iPode.

A single informational and educational space of a higher education institution can and should be built by ensuring high interactivity of the educational process, a uniform learning pace, and timely feedback.

The principle of interactivity is implemented through the joint activity of all participants of the educational process, as well as the interactive form of presentation of educational and methodical material. The principle of a single pace of learning allows to organize effective interaction of those who study with each other during the educational process. Consistency between the individual learning trajectories of individual students and the trajectories of a study group or discussion group gives students the opportunity to simultaneously participate in virtual discussions, carry out joint projects and carry out other types of educational activities. Implementation of the principle of timely feedback helps to strengthen the listener's motivation. In the system of traditional education, help to the student from the teacher's side is always late. The educational process within the framework of the analyzed model can be organized in such a way as to ensure timely feedback between the teacher (methodologist) and the student, and this allows him to be provided with timely psychological-pedagogical and technical support.

For this, there is no need to create special educational computer networks. Already existing ones can be used, in particular the Internet and services provided by the servers included in it. Such programs can be used for lecturing and conducting practical classes, similar in form to traditional ones. They allow you to organize a virtual educational session that exactly repeats the traditional one, but expands the audience of students quantitatively and spatially.

In such an educational model, it is possible to organize:

– the work of a teacher and a student with common applications and a common desktop;

– joint study of educational material by all students at the same time;

– teacher consultations in any of the permissible forms of interaction — text, audio or video;

– communication and interaction between students;

– transfer of files containing software, educational material or assignments.

A methodical solution to the problem of organizing independent search for information will be the creation of electronic workshops (electronic workplaces), which have a self-accumulating database of tasks and are used in the training of trainees in all specialties. For this, a software environment is organized, where a database works like a social network with a convenient interface that looks like an employee's workplace (desk, computer, webcam, calculator, pen, task and report sheets, etc.).

Performing tasks within the electronic workshop allows:

– to create conditions for the implementation of interdisciplinary connections in education;

– use the knowledge and skills acquired in the process of studying the basics of science;

– integrate knowledge from different blocks;

– to understand the role of knowledge in practical activities.

The interdisciplinary nature of the final result consists of step-by-step results achieved by individual interested groups of listeners — technologists, mechanics, automatists, specialists in the field of information technology, etc. The duration of the project is determined by the educational schedule.

The cyclical nature of actions related to the performance of tasks can be traced at the micro- (separate group of students within the specialty) and macro- (group of students within the same course, faculty/faculties, institute/institutes) levels. The cycle includes:

– definition of the topic;

– independent performance by the trainee and/or trainees of their part of the task;

– preparation of an intermediate product in accordance with the assigned task (programs, economic justification, business plan, etc.);

– creation of the final product;

– presentation of the received product;

– analysis of the work done.

Having completed a full cycle of work at the micro-level, the group passes the finished product to the next group. And the cycle repeats again. The procedure for working on the task and at the macro level is generally the same.

The main time of work on the task is allocated to the students' independent work with the involvement of various resources.

The effectiveness of the implementation of this stage depends on the availability of "excess information resources", such as modern technical means. As methodical support of the process, electronic educational and methodical complexes are used, which include electronic textbooks and video courses located in the electronic library of the higher education institution. A hardware and software complex is used for their development.

At the final stage of the task, the final product is presented, which is evaluated by teachers of various departments.

Thus, forming a self-filling database of tasks from one discipline of any cycle, using interdisciplinary connections and practical work of students, the departments form a common data bank of self-filling tasks for training in the specialty in general.

Created virtual workplaces with real practical tasks with proper and profitable interaction with partner companies (employers) can give students of higher education not only practical work experience, but also real earnings.

This approach to the organization of the training of future specialists in the higher education system has reason to be in demand for the reason that such an approach involves a change in teaching technology, focuses on the "modularity" of the educational process, on solving the problem of interdisciplinary "disconnection" and moving from knowledge with a separate discipline to the professional competencies of the future specialist.

## 5.5. CONCLUSIONS

1. The priority task of training applicants of the third educational and scientific level of higher education is the informatization of the educational space during graduate studies.

2. A fundamentally important characteristic of the current stage of the development of science is its informatization.

The main goal of informatization of the institution of higher education is the creation of a single informational and educational space in it.

3. The process of informatization of society includes mediatization, computerization, and intellectualization.

4. The Government of Ukraine has legislatively approved the task of informatization of society, thereby confirming the relevance of the issue.

5. Among the scientific developments in open print on the issue of integration of educational and scientific activities of graduate students, the question of the impact of informatization on this process is little researched.

6. In the national doctrine of the development of the education system, the determining factor of the effectiveness of its informatization is the ability of teachers to carry out professional activities using information and telecommunication technologies.

7. The development and informatization of science and education require universities to constantly adjust the methods of training graduate students, taking into account new progressive methods of teaching and research.

8. Educators-researchers prone to innovation find it easier to introduce the results of modern research into the educational process and effectively use modern information technologies.

9. It is useful to increase the attention of graduate students to the fact that computer equipment and information technologies contribute to the automation and intensification of research.

10. In addition to information technologies, it is important to introduce service technologies (electronic textbook, multimedia, expert systems, publishing systems, etc.) in universities.

11. The electronic textbook serves for individual and individualized training.

12. The implementation of the e-science paradigm requires the pooling of resources of the entire scientific community at different levels, the distribution of the entire volume of research between individual scientists and scientific teams, and the intensive exchange of research results.

13. The specificity of the process of research requires a combination of computing and information resources of scientific communities and their joint use.

14. The introduction of electronic management of a higher education institution will ensure a new organization of the activities of its administrative units and a change in the entire complex of relations between the administration and students and teachers.

15. A single informational and educational space of a higher education institution can and should be built by ensuring high interactivity of the educational process, a uniform learning pace, and timely feedback.

16. A methodical solution to the problem of organizing an independent search for information can be the creation of electronic electronic workplaces that have a self-accumulating database of tasks and are used in the training of students of all higher education specialties.

17. Informatization should become an essential factor in the integration of scientific and educational work of graduate students.

# 6. DEVELOPMENT OF ACCESSIBLE TECHNOLOGIES FOR CREATING THREE-DIMENSIONAL MODELS AND PRINTING THEM ON A 3D PRINTER

***Sergii Kotlyk, Oksana Sokolova***

*This article is devoted to the development of affordable, inexpensive technologies for scanning physical objects to obtain three-dimensional virtual 3D models for their subsequent printing on an FDM printer. The work shows that currently 3D printing has become an integral part of industrial production, provides an overview of existing 3D printing technologies, shows the need for rapid creation of 3D computer models, analyzes existing scanning technologies and analyzes industrial scanners.*

*The technology of affordable photogrammetry and scanning using an inexpensive Kinect device is discussed separately.*

*The first part of this section examines the problems and difficulties that can limit and complicate the scanning process with traditional technologies, formulates a step-by-step diagram of available photogrammetry technology for obtaining 3D models, shows areas of its application, and gives specific recommendations for implementation (lighting parameters, camera used, computer programs for photo processing, constructing a scanning area, features of setting up a slicer and a 3D printer). Experimental studies of the developed photogrammetry technology in domestic conditions are shown, and an example of its application for a test Fig.is given.*

*The second part of the section describes the technology for obtaining virtual 3D models using the Kinect device, describes its properties and its design. An accessible scanning technology using the Kinect device is proposed, recommendations are given for improving the scanning scheme, in particular, the features of installing light, the location of the sensor and the scanned object, the number of pictures taken, the exclusion of shiny objects, etc. In order to test the results of the proposed scanning technology using the Kinect device, a successful attempt was made to create a 3D model of the human torso and print it on a printer using plastic.*

## INTRODUCTION

In the modern world, three-dimensional (3D) models are playing an in-creasingly important role in various fields, ranging from design and indus-try to medicine and education. The development of accessible tools and resources for creating 3D models and then implementing them on 3D printers is opening up exciting opportunities for both professionals and enthusiasts in a variety of fields. However, despite their widespread use, the accessibility of creating and printing 3D models remains limited for many people.

This section is intended for those who are interested in the process of developing accessible technologies for creating three-dimensional 3D models and their subsequent printing in plastic at home. We will plunge into the world of innovations and developed methods that make the process of creating and implementing 3D ideas more fun and accessible. Developing affordable scanning technologies will not only increase opportunities for creativity and innovation, but also make them more inclusive, ensuring equal opportunities for everyone.

In this section, we will try to outline the basics of design for 3D printing, consider the printing process using the common FDM technology, provide a method for ensuring the availability of scanning objects when creating models for all users, and also talk about the possibilities and prospects for the development of this area.

There are a number of factors that contribute to the limited accessibility of creating and printing 3D models for many people. Certain barriers and limitations may hinder the full use of 3D modeling and 3D printing technologies, including the following:

***Complexity of the process:*** Traditional methods of creating 3D models require special skills in 3D modeling and the use of appropriate software. This can be difficult for people who do not have experience in this field or do not have the necessary technical skills.

***Software complexity:*** Creating complex 3D models requires the use of specialized software, which can be difficult to learn for beginners. Lack of experience or knowledge in working with such programs can be a serious barrier for many.

***Expensive Equipment:*** 3D printers, 3D scanners and related equipment can be quite expensive, especially for regular users. This places restrictions on the accessibility of the technology to a wider audience.

***Limited access to education:*** Successful creation and printing of 3D models requires specific technical skills and knowledge. An understanding of the basics of 3D modeling, design for 3D printing, and the ability to work with relevant programs and devices is required. The lack of available tutorials, online courses and comprehensive information can hinder people's ability to master the process of creating and printing 3D models.

***Materials and Environmental Issues:*** The use of different materials in 3D printing can also be a factor limiting the affordability of the process. The choice of quality materials, their availability and cost can influence the ability to create 3D models. Some materials used in 3D printing can be expensive or have a negative impact on the environment. This may also limit the availability of the technology and require the search for more environmentally sustainable alternatives.

***Lack of standards and regulations:*** There are still no universal standards and regulations in the field of 3D printing. This can create problems with the quality and safety of printed products, as well as impede accessibility to a wider audience.

***Difficulty in Setup and Calibration:*** Setting up a 3D printer and calibrating the printing process can be challenging for beginners. The need to understand the technical aspects of the 3D printing process can also create barriers for many interested people.

Removing these barriers and making 3D model creation and printing more accessible requires the development of simpler and more intuitive tools, educational programs, lower hardware costs, and the establishment of standards and regulations in this area. This section of the monograph is dedicated to the creation of such accessible technologies in the field of scanning and creation of 3D models.

Three-dimensional or 3D printing is the layer-by-layer creation of a physical object based on a virtual three-dimensional model. Unlike conventional printers, 3D printers do not print photographs and texts, but "things" — industrial and household goods [1, 2, 7].

3D modeling is one of the most relevant innovations of the 21st century. We, in most cases, associate this technology with the animation and film industry, but this technology covers much more spectrum of our lives. One such new industry in this latest technology is 3D printing. A 3D printer is a device that creates a three-dimensional object based on a virtual 3D model. Unlike a conventional printer, which outputs information onto a sheet of paper, a 3D printer allows you to output three-dimensional information, that is, to create certain physical objects. 3D printing technology is based on the principle of layer-by-layer creation (growing) of a rigid model.

Cheap desktop 3D printing is just in its infancy, but is rapidly progressing due to its limitless potential. This is far from an exaggeration, since thanks to its ability to reproduce 3D objects — from archaeological artifacts, complex mathematical surfaces, to medical prosthetics — the technology holds promising prospects for science, education and sustainable development [2, 6, 7, 12, 14, 17, 18].

With the constant growth of the computer industry, the requirements for specialists, including 3D engineers, are also growing. Creating modern models requires a huge investment of time and effort. Large companies employ dozens of people to create 3D objects.

Currently, sophisticated 3D graphics programs can be used to obtain naturalistic 3D objects, but the creation of such models using scanning or photography is becoming increasingly common [4, 5, 8, 10]. Scanning with professional scanners gives good results, but their cost (expressed in 6- and 7-digit figures in hryvnia) does not allow the use of such technologies for affordable household purposes.

The purpose of this section is to improve the technology for preparing virtual 3D models in everyday conditions by using several photographs of a real object (photogrammetry) and using available equipment, in particular, the Kinect device, as well as improving the use of technology for printing models on 3D printers printed with plastic using FDM technology. Based on the above, the development of such technology is a fairly relevant research topic in the field of information technology.

## 6.1. BENEFITS OF USING 3D PRINTING AND 3D MODELING

Currently, there are three main methods for creating industrial products, that is, moving from an idea or sketch to the physical production of a product [1, 2, 7, 17]. Subtractive creation is a product development process in which material is continuously cut from a rigid block (such as an alloy). This can be done manually (for example, using a lathe) or using computer numerical control (CNC) machines. Injection molding is a manufacturing process used to produce products in large volumes. Parts are produced by injecting heated material into a mold. Additive manufacturing, also known as 3D printing, which is the process of creating three-dimensional objects from a computer file in which a part is built by adding some material layer by layer.

An increasing number of companies across a variety of manufacturing industries are now using the 3D printing process because it offers significant advantages over more traditional manufacturing methods such as subtractive and injection molding. The following distinctive features and advantages can be identified [1, 2, 18]:

*Speed.* One of the trump cards of using 3D printing is rapid prototyping, that is, the ability to design, manufacture and test the desired part in the shortest possible time. This is achieved by introducing computer programs at all stages of model creation. In addition, if necessary, the design of the part can be changed without affecting the speed of the production process. Before the 3D printing industry took off, it took several weeks to make a prototype. Every time a change was made, more time was added to the process. Taking into account delivery times, complete product development could initially take up to a year. Today, with 3D printing techniques, a firm can design a part, make modifications, produce it on a professional 3D printer, and test it in a matter of days (or sometimes less). For small businesses or even individuals, this difference is quite significant. The freedom and creativity provided by 3D printing means that almost anything can be created without warehouses filled with expensive equipment. There are no long lead times associated with the need to outsource complex manufacturing projects. This means freedom from minimum order restrictions; the required products can be produced even in small batches. For small production runs, 3D printing is the best option in terms of speed.

*Price.* Using 3D printing for short production runs is the most cost-effective manufacturing process. Traditional prototyping methods, such as CNC machining and injection molding, require a large number of expensive machines, as well as a much higher cost of skilled labor, since they require experienced machine operators and machinery to run them. This is in contrast to the 3D printing process, which requires only 1 or 2 machines and far fewer operators to produce a part. At the same time, there is much less waste, since the part is created from scratch, rather than cut from a solid block, as in subtractive manufacturing, and does not require additional tools. Thanks to the ability to produce products with fewer dimensional parts and assembly, 3D printing allows you to save on the cost of tooling and the cost of purchasing all the necessary equipment. The entire process requires less time than conventional methods.

*Flexibility.* Any given 3D printer can create virtually anything that fits within its build volume, meaning there is no specialization in production. In traditional manufacturing processes, each new part or change in part design requires the manufacture of a new tool, mold, die, or fixture to create the new part. In 3D printing, the created virtual model is loaded into the slicer software, the necessary support supports are included where necessary, and then printed with little or no changes to the physical hardware. 3D printing allows you to create and produce geometric shapes that cannot be produced using traditional methods, either as a single part or in general. Such geometries include cavities within solid parts and parts within parts. This printing method, unlike traditional methods, can incorporate multiple materials into a single object, allowing a variety of colors, textures and mechanical properties to be mixed and matched. Because designs can be produced digitally, products need not be limited by geometric complexity. This means manufacturers can produce stronger, lighter components, something the aerospace industry in particular is taking advantage of.

***Availability and versatility.*** 3D printing allows anyone, even those with limited CAD experience, to edit designs as they see fit, creating unique, customized new parts. This also means that any design can be made from a wide range of different materials.

***Competitive advantages.*** Due to the high production speed and lower cost of 3D printing, the product life cycle is significantly reduced. Businesses can improve and improve the product, allowing them to produce better products in a shorter period of time. 3D printing allows you to physically demonstrate a new product to customers and investors rather than leaving it to their imagination, reducing the risk of misunderstanding or losing information during communication. It also allows for cost-effective market testing, obtaining feedback from potential customers and investors on the actual product without the risk of large up-front prototyping costs.

***Tangible design and improved product testing***. With 3D printing, it is not at all necessary to physically create a model; it is enough to demonstrate its prototype on a computer screen. Subsequently, the created physical prototype can be thoroughly tested, and if flaws are found, the CAD file can be changed in the desired direction, and a new version of the model printed the next day.

***High quality.*** Traditional manufacturing processes may result in some parts in a batch being defective or of inconsistent quality when compared to other parts. In 3D printing, parts are printed sequentially. Each subsequent individual part can be tracked, allowing errors to be detected in real time, reducing the overall number of failed parts and wasted materials while increasing the consistent quality of parts produced.

***Reducing the risk of investment loss.*** Benefits such as high quality and consistency of 3D printing allow businesses to reduce their production risks. The technology allows product developers to test product prototypes before committing to significant production investments that could be disastrous.

***Availability.*** 3D printing systems are much more affordable and can be used by a much wider range of people than traditional manufacturing set-ups. Compared to the huge costs associated with setting up a traditional manufacturing facility, a 3D printing setup costs much less. Additionally, 3D printing is almost completely automated, requiring virtually no additional personnel to run, monitor, or maintain the machine, making it much more affordable than other manufacturing systems.

***Ecology and energy saving.*** With 3D printing, fewer parts require outsourcing for production. This means less impact on the environment as fewer goods are shipped around the world and there is no need to operate and maintain an energy-intensive plant.

***Lowest material consumption.*** 3D printing creates much less waste per part, and the materials used in 3D printing are typically recyclable.

Currently, the use of 3D technologies is becoming comprehensive; it is difficult to find an area of technology or the household sphere where such technologies are not used. This proliferation is largely facilitated by the fact that three-dimensional hard objects are created from a digital file, and the virtual model can be changed, edited, improved by computer many times, improving the result in the form of a hard copy. As 3D printing advances, it is destined to transform almost every major industry and change the way we will live and work in the future. Whereas in the early stages of their development such approaches were only for prototyping and one-off production, they are quickly becoming a production technology.

Today, the most common areas of application of 3D technologies include [1, 2, 6, 7,]:

***Rapid prototyping and production.*** From idea to 3D model and creation of a hard copy prototype now takes days, not weeks and months as before. Iterations are easier and cheaper to do, and there is no need for expensive molds or tools.

***Automotive industry.*** Car manufacturers have been using 3D printing for a long time. Automotive companies print spare parts, tools, devices and end-use parts. 3D printing has enabled on-demand production, resulting in lower inventories and shorter design and production cycles. Car enthusiasts around the world are using 3D printed parts to restore used cars, often printing parts that have been out of production for decades.

***Aviation industry.*** The aviation industry uses 3D printing in a variety of ways. The following example marks a major milestone in 3D printing manufacturing: GE Aviation 3D printed 30,000 cobalt chrome fuel injectors for its LEAP aircraft engines. They reached this milestone in 2020, and since then have been producing 600 pieces per week using forty 3D printers.

About twenty separate parts that previously had to be welded together have been combined into one 3D printed component that weighs 25 % less and is five times stronger. The LEAP engine is the best-selling engine in the aerospace industry due to its high level of efficiency, and GE is saving $3 million per aircraft by 3D printing fuel injectors, so this single 3D printed part has a financial benefit of hundreds of millions of dollars.

***Construction.*** 3D printed houses are already available commercially. Some companies print prefabricated parts, others do it in-house. No one is surprised anymore by the huge residential complexes that are being built in record time. Large-scale concrete printing systems with large nozzles for high flow rates are more often used. It is excellent for laying layers of concrete quickly and repeatably.

***Consumer goods.*** There are now many examples of 3D printed consumer products. For example, the Adidas 4D range features a fully 3D printed midsole and is printed in high volumes. Adidas initially produced just 5,000 pairs of shoes for the general public, but in recent years this Fig. has reached 100,000 pairs of AM shoes. Moreover, the shoes are available all over the world in local Adidas stores, as well as in various third-party online stores.

***Healthcare.*** It's not uncommon to see headlines these days about 3D printed implants. Often these cases are experimental in nature, so it may seem that 3D printing is still a fringe technology in the medical and healthcare field, but this is no longer the case. In the last decade, GE Additive has 3D printed more than 100,000 hip replacements.

***Bioprinting.*** In the early 2000s, 3D printing technology was being explored by biotech firms and academia for possible use in tissue engineering applications, where organs and body parts are built using inkjet technology. Layers of living cells settle on the gel medium and slowly grow to form three-dimensional structures. Many experts argue that bioprinting will subsequently revolutionize medicine.

Currently, by 2024, many 3D technologies are known [1, 2, 14, 17, 18], which can be divided into several groups according to the principle of operation:

***Extrusion of material.*** A process in which a filament of solid thermoplastic material melts, settles, and cools to form a rigid object. There is only one type: fused deposition modeling (FDM) or sometimes called fused filament fabrication (FFF).

***Polymerization.*** The process is based on the use of a container containing a photopolymer resin that hardens when exposed to ultraviolet light. There are two types: stereolithography (SLA), which uses a point laser, and direct light processing (DLP), which uses a projector.

***Powder melting (polymers).*** A process in which a source of thermal energy selectively causes dust particles within a site to fuse together to create a solid object. There is only one type: selective laser sintering (SLS).

***Structuring the material.*** A process in which droplets of material are selectively applied to a building plate and hardened by exposure to light. The nature of the material inkjet printing process allows different materials to be printed on the same object. There are two types: material inkjet (MJ) and Drop on Demand (DOD), which uses a pair of inkjet printers: one with the printed material and the other with the support material (usually dissolvable).

***Linking structure.*** A process similar to SLS, but using a laser to sinter the powder. It uses two materials: a powder-based material (a building material) and a binder (usually in liquid form). Sand or metal dust is usually used as a building material.

***Powder melting (metals).*** A process that uses a heat source to induce fusion of metal powder particles (layer by layer). There are different variants of this technology that change the energy sources: direct metal laser sintering (DMLS) or selective laser melting (SLM) using lasers, and electron beam melting (EBM) using electron beams.

Obviously, many technologies are prohibitively expensive and unaffordable for most people. Therefore, at the consumer level, FDM (fused deposition modeling) and SLA (stereolithography) are usually considered, more or less available in the consumer market.

The most common technology is FDM, which mainly uses PLA and ABS plastics. Biodegradable PLA plastic is cheap, made from cornstarch, and doesn't smell much when heated. Meanwhile, ABS is tougher and can produce thinner models, but it is more expensive, requires higher temperatures and sufficient ventilation to get rid of the peculiar smell. The resin used in SLA printers is even more expensive, but can produce more accurate results. Some FDM printers allow you to print with multiple colors or even different types of material in the same print. To do this, they use more than one extruder and nozzle. You can also extrude metal paste, biogels, concrete, chocolate and a wide range of other materials, but plastics are the most common. Characteristics of FDM technology:

- materials used: plastic thread (PLA, ABS, PET, PETG, TPU, nylon, ASA, PC, HIPS, carbon fiber and many others);
- dimensional accuracy: ±0.5 % (lower limit ±0.5 mm);

• Common applications: electrical enclosures, form and fit testing, figures and fixtures, investment casting models.

• strengths: the cheapest 3D printing method, wide selection of materials used.

## 6.2. FEATURES OF CREATING 3D MODELS USING SCANNING

The constant growth of the 3D printing industry leads to increasing demands for both the quality of virtual models and an increase in the level of education of specialists. The creation of modern production (and not only) models requires a truly gigantic investment of time and effort, as well as highly qualified (and sometimes artistic talent) employees of companies. In large companies, dozens and hundreds of people work on creating 3D computer models. In this case, complex three-dimensional graphics programs are used, which require large amounts of computer memory and speed to function (not every computer supports modern applications such as 3D Max). Currently, the technology of creating such models using scanning or photography is becoming increasingly widespread, allowing one to quickly obtain a fairly clear digital model of the input object [3, 4, 5, 8, 10, 12, 13, 14, 22].

3D scanning is the process of analyzing an object from the real world to collect all the data to reproduce its shape and appearance digitally. Through this process, an object can become a 3D model, which will be the basis for a 3D project, and can later be useful for reconstruction, analysis or modeling ideas (Fig. 6.1).

In short, 3D object scanning is the process of converting the physical form of a real object into a digital form. 3D scanning of objects helps prepare the necessary model for 3D printing and in some cases can play a decisive role in building the resulting digital model.

There are different devices and methods for 3D scanning of objects, that is, different means of creating a digital version of a real object. Of all the types of 3D scanning technologies, there are four main ones, listed below. There is no “best” 3D scanning technology, each method solves unique problems, providing the best solution in different contexts.

**Laser 3D scanning.** 3D laser scanning is by far the most common 3D scanning method used. In this method, the digital shape of an object is captured using laser light to obtain a digital representation of the real model (Fig. 6.2).

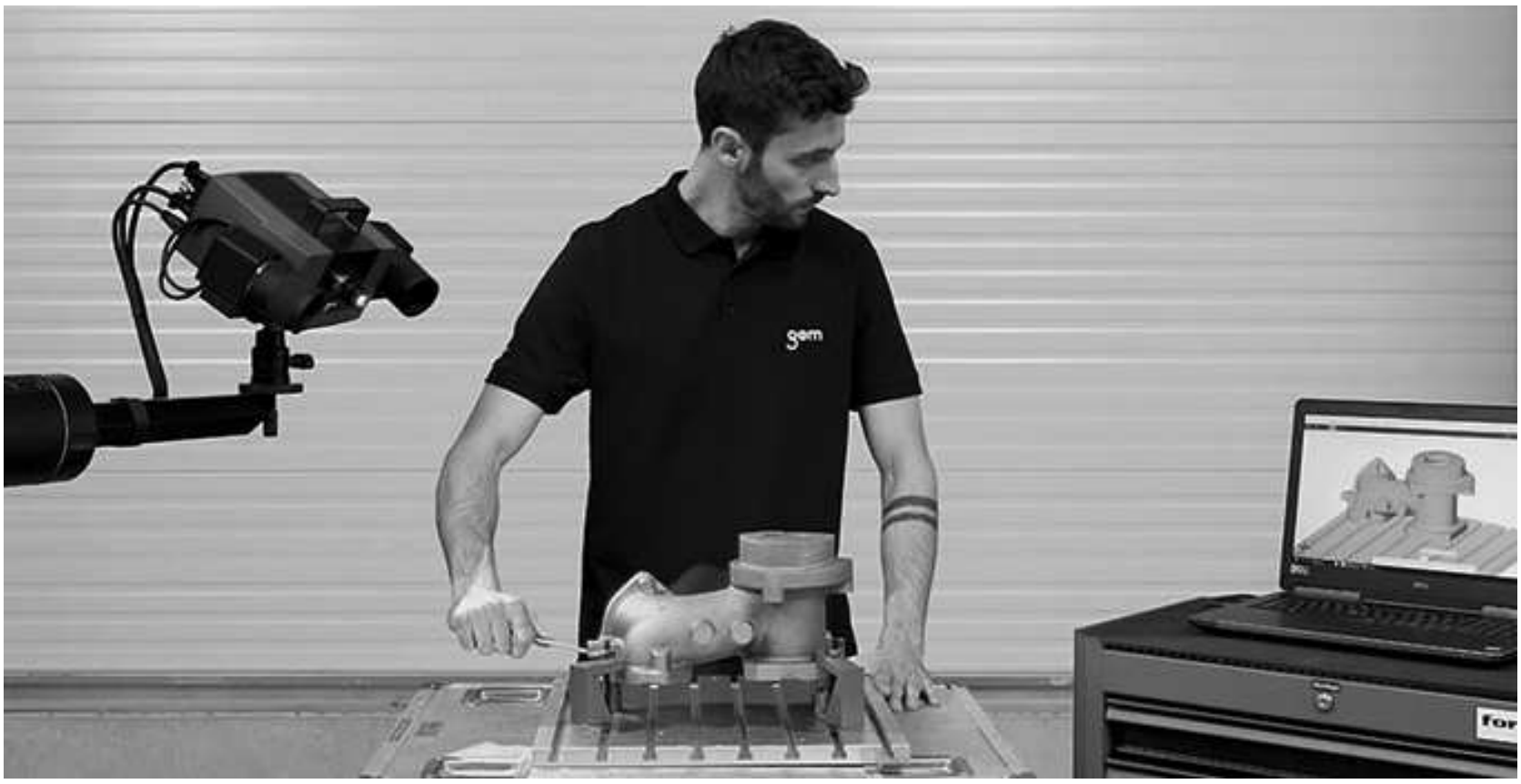

Fig. 6.1. Scanning a model with a compact scanner

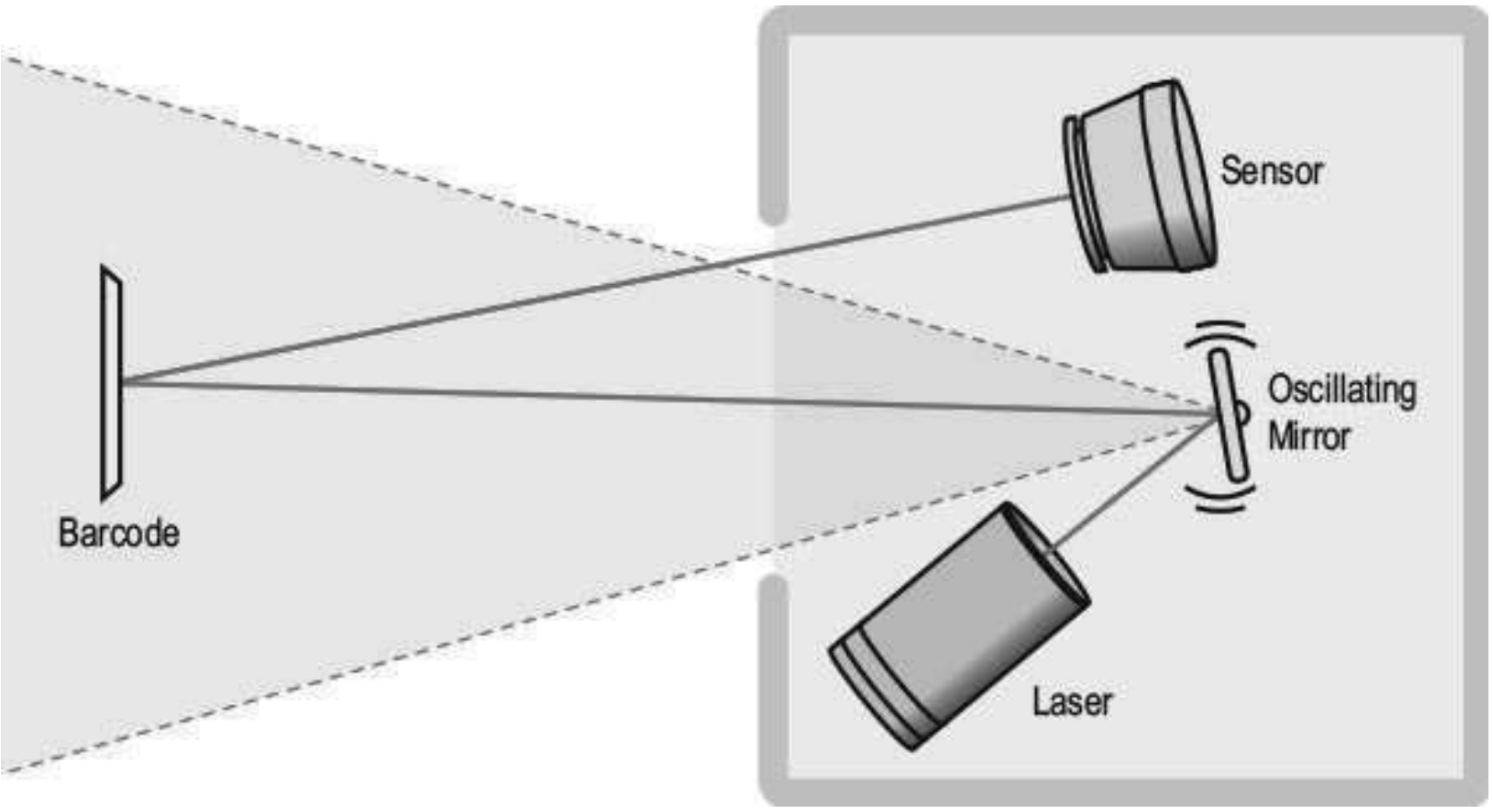

Fig. 6.2. Principle of scanning with a laser scanner

Laser scanning combines two sets of information to create a point cloud of an object's surface: data from a laser illuminating the object, and data from another sensor (usually a moving camera or two stationary ones). 3D scanning software stitches these data sets together using the known distance between the camera position and the laser source to create model points. To construct 3D geometry from laser scanning, it is necessary to determine where the laser line falls in the images captured by the camera during scan-ning. The laser line is usually the brightest pixel in the image, but sometimes other light sources can be captured. Captured points record everything from surface detail and texture to color, creating a direct representation of the scanned object.

These 3D scanners are capable of measuring fine details and capturing free-form objects to create highly accurate virtual shapes. This scanning method is ideal for measuring and monitoring objects of complex geometry. It allows you to obtain measurements and data from objects where this cannot be done using traditional methods. A scanner that uses laser light is a bit like a camera: it can only capture what is in its field of view. In this case, a laser point or line is projected from the device onto an object, and the sensor measures the distance to the surface of this object. By processing this data, it can be turned into a triangulated mesh and then into a CAD model.

**Photogrammetry**. Photogrammetry is the science of taking measurements from photographs. This method uses parallax obtained between multiple images taken from different perspectives. Photogrammetry can be used to record complex 2D and 3D motion fields. The method simulates the stereoscopy of binocular human vision and is used to obtain information about existing physical objects. This process collects data about the shape, volume and depth of the scanned object. This method is more error-prone and produces less consistent results, but it is more accessible and adaptable to different programs, as evidenced by the recent increase in its use in smartphone applications (Fig. 6.3).

This approach will not produce an accurate result, but with the help of good photogrammetry software it gives quite satisfactory results. Stereo viewing and photogrammetry techniques are used in robotics, 3D building mapping, and 3D film.

**Contact 3D scanning technology.** Contact 3D scanning is more invasive than the above processes. It uses a probe — stationary or mobile — to establish contact with the surface of an object (Fig. 6.4). Software used with the sensor determines how and where the sensor touches a surface and can record the three-dimensional location of the surface.

**Structured white light scanning.** In this structured light scanning method, one of the camera positions used in previous scanning methods is actually replaced by a projector, projecting different light patterns onto the surface of the object. The way objects distort these patterns is recorded, allowing a 3D image to be created. Model points are created using the known camera position and information about the incident light on the object. To be effective, there must be a clear relationship between the projector and the camera pixels. Structured light scanning is used in facial recognition technology, reverse engineering, design, manufacturing, and remote environmental sensing (Fig. 6.5).

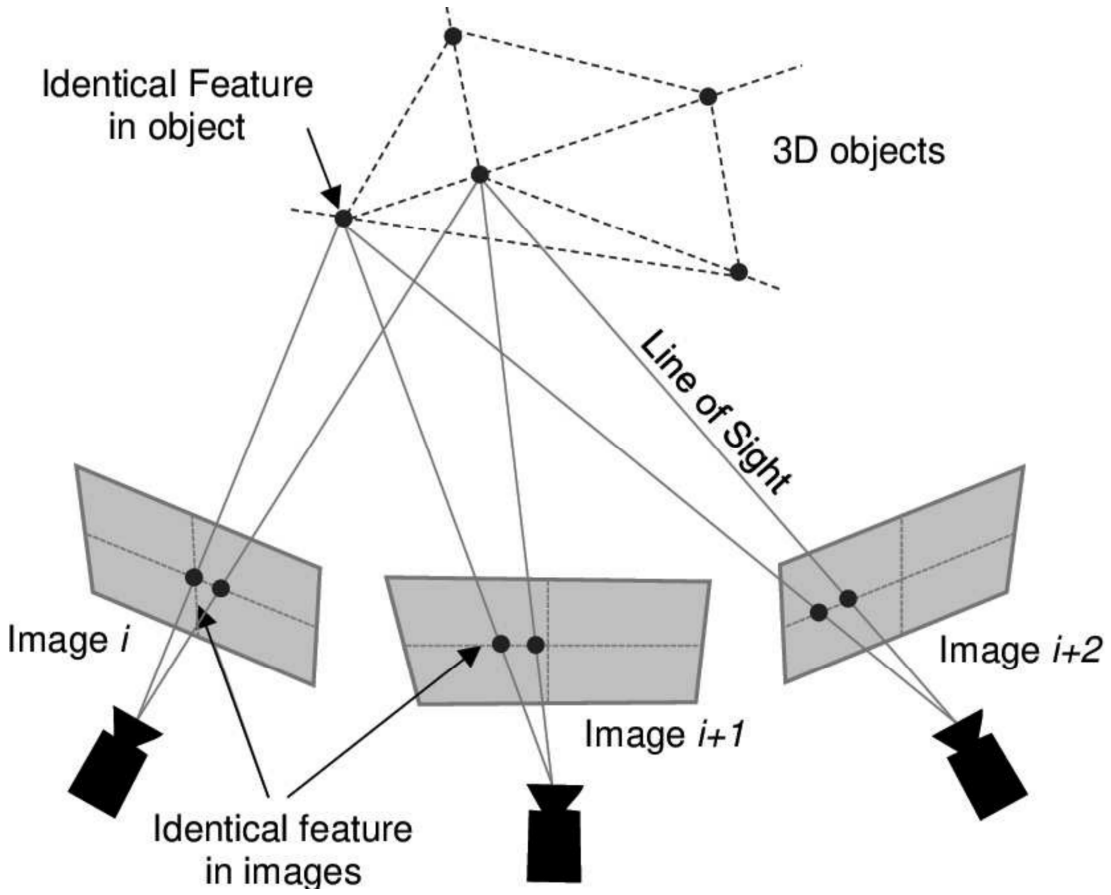

Fig. 6.3. Principle of scanning using photogrammetry

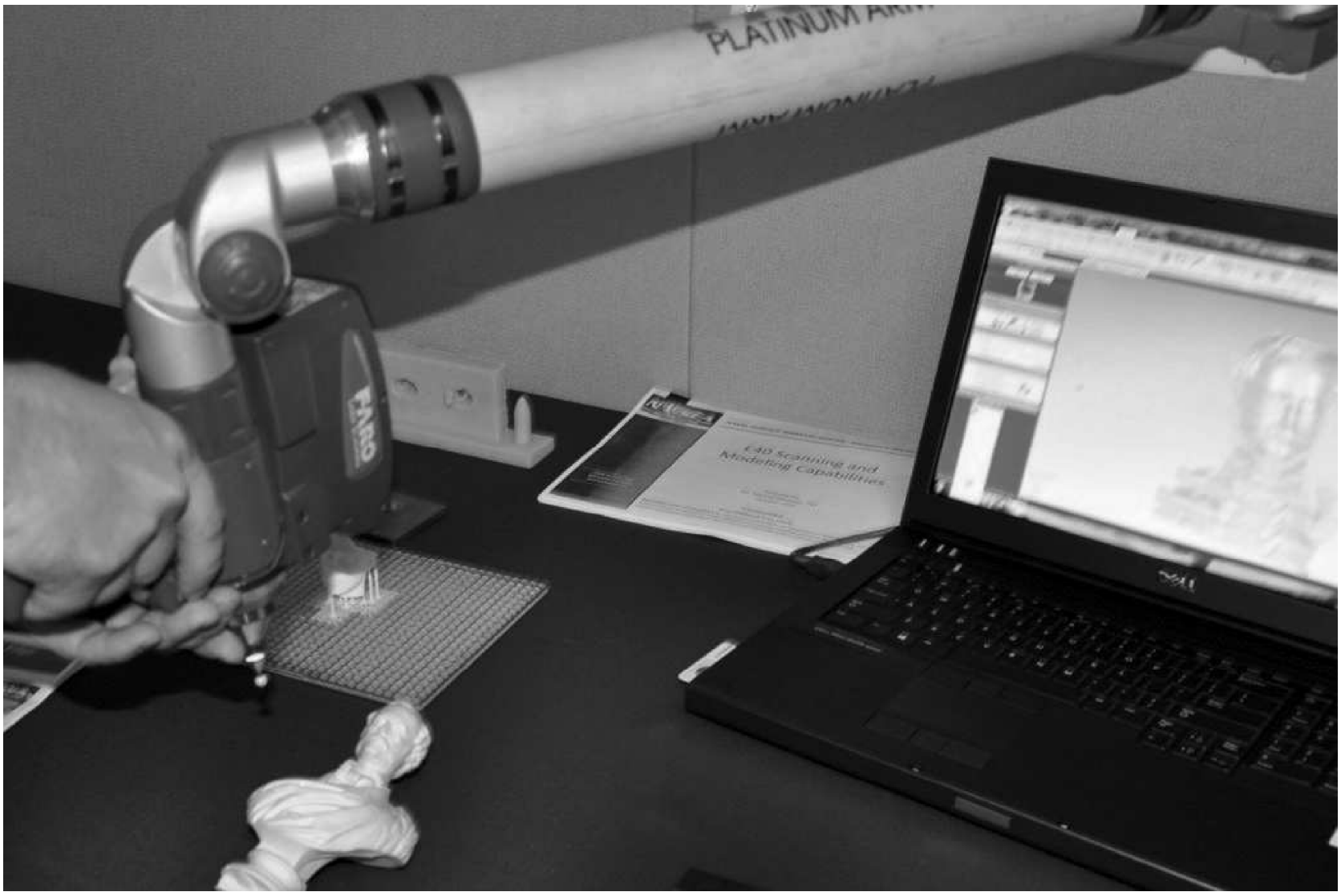

Fig. 6.4. Principle of contact 3D scanning

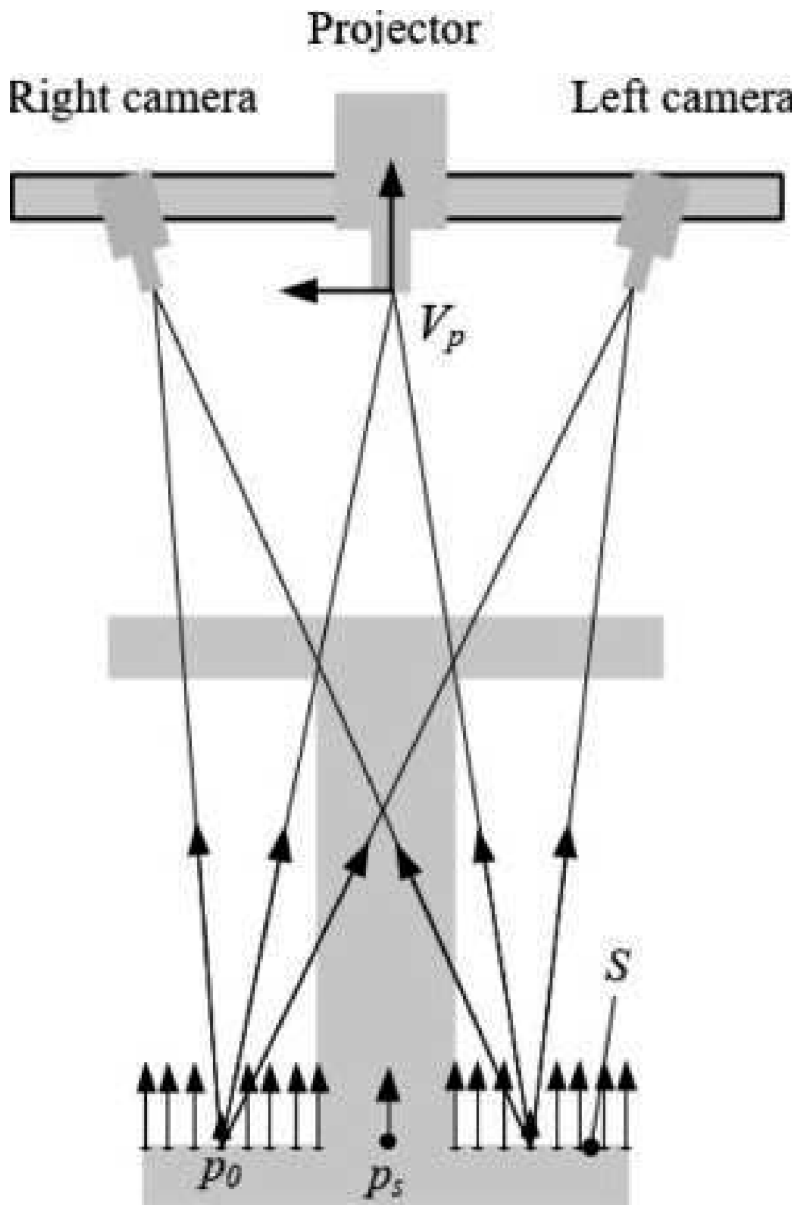

Fig. 6.5. Scanning principle using structured white color

## 6.3. CREATING 3D MODELS USING PHOTOGRAMMETRY TECHNOLOGY

### *6.3.1. Basics of photogrammetry*

Photogrammetry works by extracting geometric information from a two-dimensional image. By combining many pictures you can create a three-dimensional image. Indeed, it is possible to make a 3D model of an object with multiple images of that object taken from different perspectives to get a complete picture of the object. To do this, it is not necessary to use a 3D scanner; you can use a simple smartphone, tablet or a modern camera with good image stabilization [8, 9, 10, 11, 13, 15, 19].

Photogrammetry is almost as old as photography itself. Since its development approximately 150 years ago, photogrammetry has moved from a purely analogue, optical-mechanical technique to analytical methods based on the automated solution of mathematical algorithms, and finally to digital or electronic photogrammetry, based on digital images and a computer star, de-

void of any optics and mechanical equipment. Photogrammetry is primarily concerned with making precise measurements of three-dimensional objects and geographic areas from two-dimensional photographs. Programs include coordinate measurement; quantification of distances, heights, areas and volumes; preparation of topographic maps; creation of digital models of objects

Thanks to photogrammetry and a series of complex algorithms, it is possible to obtain a digital 3D model using multiple photographs of the original real-world object through a powerful reconstruction process. The more source photos, the more accurate the 3D design of the virtual model will be. It is also possible to turn one image into a 3D model, but, obviously, it is quite difficult to get the optimal result with only one image (a specialist needs to Fig.out the invisible planes in the photo).

Using photogrammetry allows you to scan large objects or landscapes that otherwise could not be digitized. Thanks to this method, maps can be obtained from aerial photographs. It is an essential tool for architects, engineers and anyone who needs to create topographic maps, architecture, geology. Photogrammetry is used in the military for topographic surveys or obtaining terrain. Archaeologists use it to create plans for complex and remote sites.

There are two main types of photogrammetry: aerial (with the camera in the air) and ground-based (with the camera in hand or on a tripod). Terrestrial photogrammetry, which works with object distances up to 200 m, is also called short-range photogrammetry. Small format aerial photography falls somewhere between these two types, combining an aerial viewpoint with close subject distance and high image detail. Aerial photography is technically a type of remote sensing that primarily involves visible light waves in the electromagnetic spectrum. There are several excellent applications for near-infrared. Aerial photography is perhaps the most widely used method for creating geographic databases in forestry and environmental management. Interpretive techniques, including geometry, trigonometry, optics, and familiarity with natural resources, allow us to identify and estimate the size, length, or height of objects on earth (Fig. 6.6).

Photogrammetry, as its name suggests, is a method of measuring three-dimensional coordinates that uses photographs as the primary means of metrology or measurement. The basic principle used in photogrammetry is triangulation. By taking photographs from at least two different locations, it is possible to develop what are called lines of sight from each camera to points on the subject. These lines of sight, sometimes called rays because of their optical nature, are mathematically intersected to produce the 3D coordinates of points of interest.

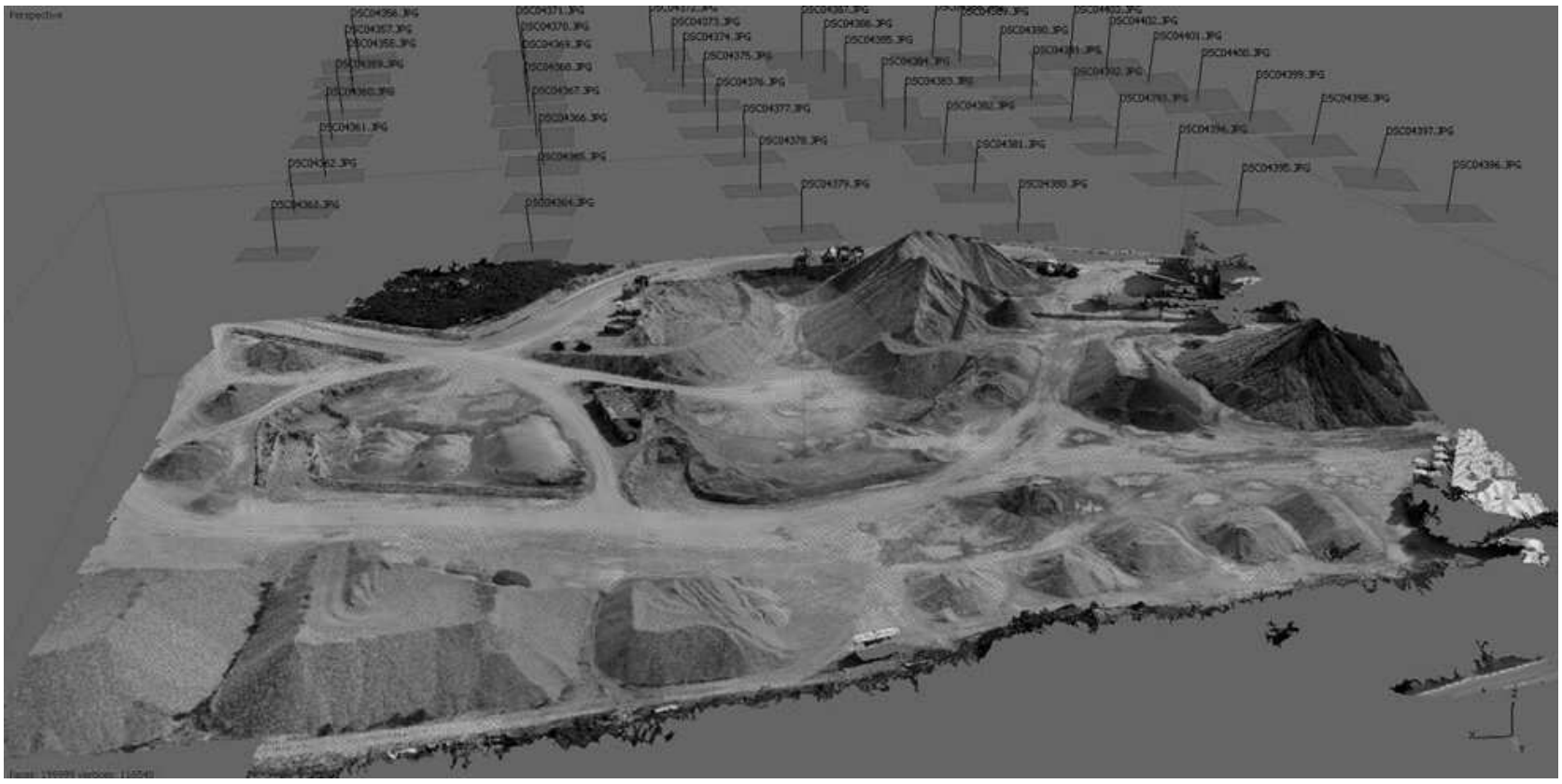

Fig. 6.6. Illustration of Photogrammetry as a type of remote sensing

In Photogrammetry, several coordinate systems are used to determine the spatial coordinates of the points of a scanned object: the main rectangular coordinate system, the Photogrammetric coordinate system, and the auxiliary coordinate system [9, 10, 20, 21, 22].

Photogrammetric coordinate system (in Fig. 6.7) the right coordinate system (OXYZ) in this case can be located arbitrarily, that is, the origin of the coordinate system and the directions of the axes of the Photogrammetric coordinate system can be any. In a particular case, the origin is combined with some point on the terrain or with the center of the image projection, and the XY plane is set horizontally (Fig. 6.7).

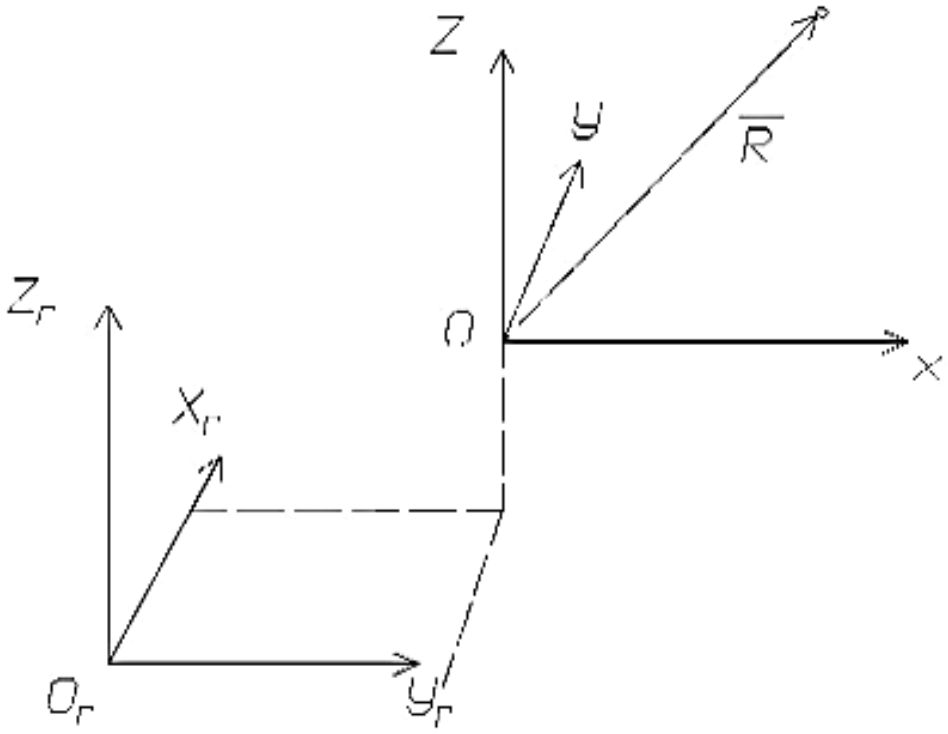

Fig. 6.7. Photogrammetric coordinate system

Auxiliary coordinate system Sẍỹẑ is a coordinate system in which the image of a point has three coordinates (Fig. 6.8). The ẍ and ỹ axes of this coordinate system are parallel to the corresponding axes of the oxy plane coordinate system, and the ẑ axis coincides with the main ray So (optical axis). The components of the vector that determines the position of the image point — the image of the object point, in the Sẍỹẑ system will be: ẍ, ỹ and ẑ = const = – f, where f is the focal length.

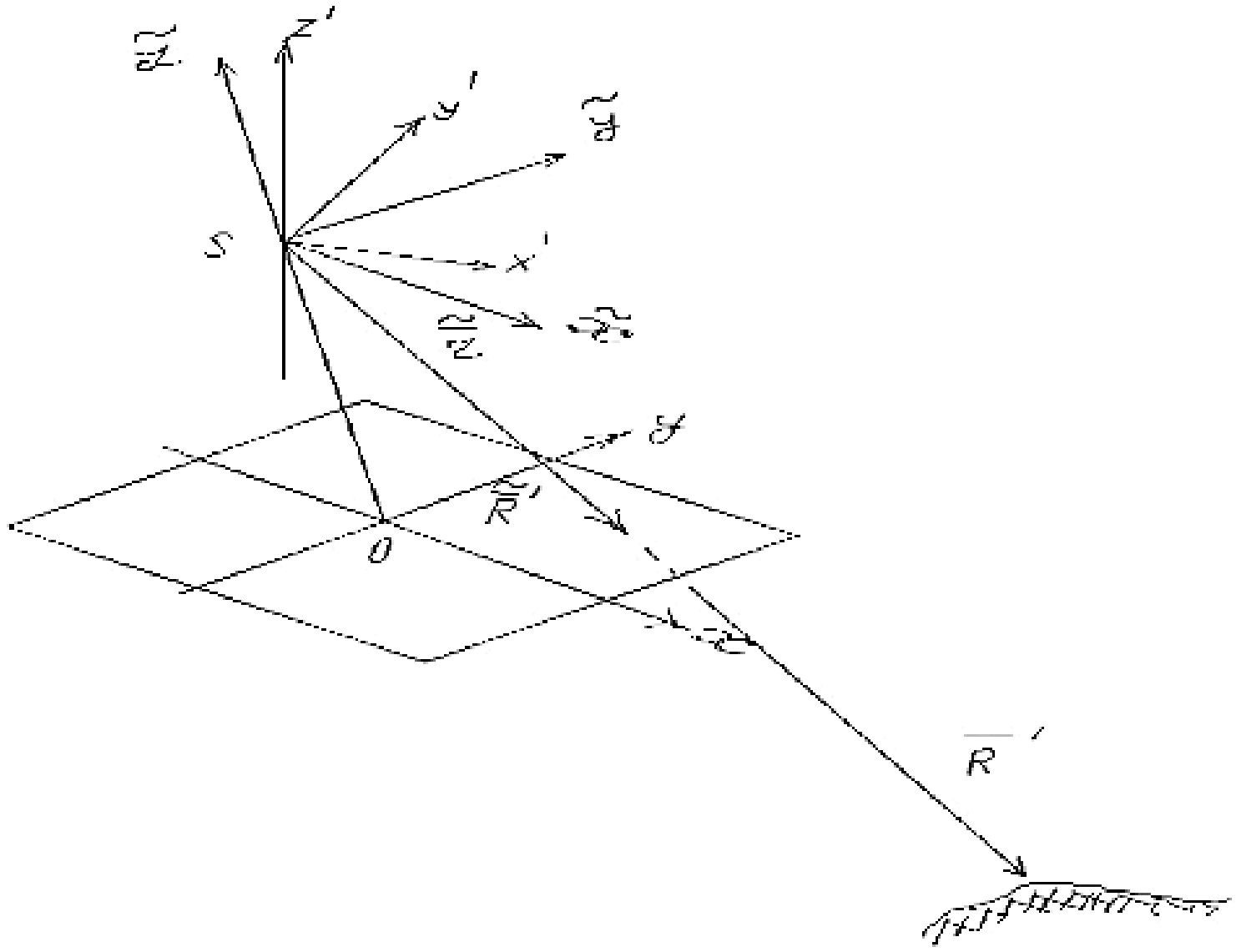

Fig. 6.8. Auxiliary coordinate system

The coordinate system plays an important role in establishing dependencies between the flat image coordinates of object points and the coordinates of object points in object space.

Photogrammetry also allows you to obtain three-dimensional coordinates from points in two two-dimensional images, which can be taken from different positions (for example, using charge-coupled device cameras (CCDs, Fig. 6.9). In this case, the points in each of the 2D images (i.e. e. points that are known to be the same in two images) were determined by the intersection of horizontal and vertical laser lines scanned across the patient's face (Fig. 6.9), and the mirror, in turn, is directed in such a way that it will see the patient's face.

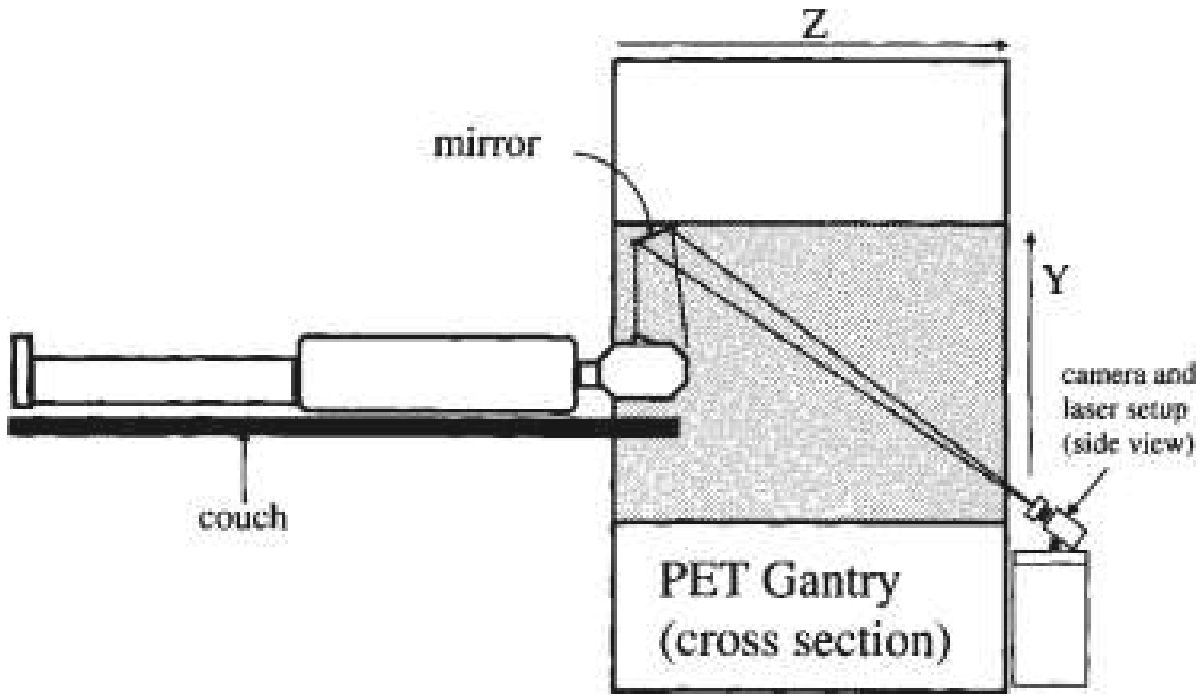

Fig. 6.9. Configuration for measuring photogrammetric information in the scanner

To digitize physical objects, photogrammetry must measure and record two different attributes. First, you need to measure the position of each surface point on the target object; second, the nature of each point (eg color, transparency, reflectivity) should be determined. Position estimation requires a series of images (at least two, but usually tens to several hundred) from different angles to collect spatial information. Algorithms compare different image segments and their relative movement from each other from image to image. To summarize, this process is comparable to human depth perception via motion parallax, which allows people to judge depth by the amount of movement in their field of view, with distant objects moving at a slower speed than near objects. Based on this, the relative position of each image is estimated according to the camera parameters (Fig. 6.10). The Fig. shows one of the possible uses of Photogrammetry — the digitization of historical artifacts for museums and educational purposes. Blue cones represent camera position in photographs of a physical object (left), previous coordinates are calculated (right), screenshots are taken in 3DF, and arrows are added to indicate camera orientation.

The resulting digital replica can be considered an exact digital copy of the physical object, the accuracy of which depends on several variables such as the quality of the original image, the number of vertices included, or algorithms. Subsequently, the entire appearance of the object is digitized, textures are generated from the original images and overlaid on the surface of the 3D mesh. Modern graphics engines allow the use of advanced textures with dynamic features such as transparency or mapping. However, these effects typically need to be applied manually as they are difficult to find automatically using Photogrammetry techniques.

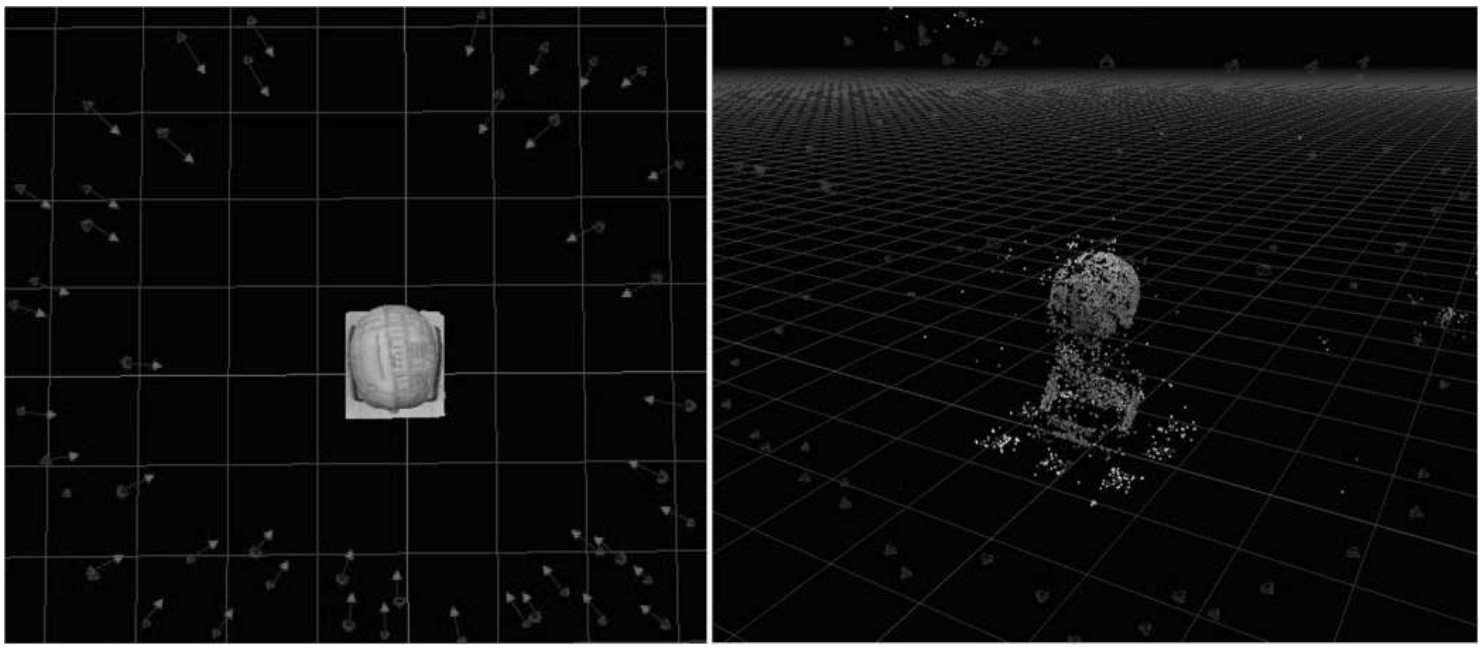

Fig. 6.10. Screenshot showing how a virtual model is created using a series of photographs

Not a single modern scanner (Fig. 6.1) can scan a complex model in such a way that it does not then require the intervention of specialized software. A good specialist will need several hours to post-process the information received by the scanner, which, of course, is much faster than modeling an object from scratch. But on the other hand, manual modeling is a completely controlled process, unlike the work of 3D scanners, which are unable to avoid mesh breakage. It is also difficult to work with complex surface topography of models, etc. As a result, many models are quite accurate in shape, but their further use requires the intervention of a specialist. In any case, even if the modeler has to create the model again, then having a "blank" at hand will speed up the work process.

It should also be taken into account that the use of professional scanners is a rather labor-intensive and costly process. The cost of modern scanners amounts to five or even six digits, which, naturally, makes them difficult to use in everyday situations. As a result, when scanning is necessary, many resort to Photogrammetry technology.

### *6.3.2. Use of photogrammetry in everyday conditions*

This work makes an attempt to substantiate the possibility of using photogrammetry technology in everyday conditions, without the use of any expensive devices and complex computer programs.

This technology will allow you to create a three-dimensional 3D model of a real physical object using a large number of its photographs. During the research process, only programs that were distributed free of charge (or shareware) were used; photography was taken using a household camera or telephone. Photogrammetry allows you to determine the size of an object, its position in space and reproduce the original model in virtual form in the smallest detail. The submission of the material is accompanied by photos and screenshots of the creation and printing of a 3D model of a real object (for which a souvenir bottle was chosen).

Photogrammetry involves the use of a large number of photographs of an object from different angles, which are then processed using special computer programs. In this work, the capabilities of the following programs using Photogrammetry technology were investigated: Agisoft Photoscan, RealityCapture, Meshroom [4, 8, 9, 10, 20].

Agisoft Photoscan: This is a very popular program that is suitable for both beginners and professionals. The software is paid. A stripped-down version, for those who are just learning to use this software product, costs $180; to create a higher-quality model, you will need a professional version of the program, its price reaches $3,500. The program does not place restrictions on the number and quality of photos, but the more photos, the more computer resources will be used.

Pros: the application creates a three-dimensional object without surrounding parts, which speeds up the program's running time and reduces the load on the computer's processing power.

Cons: to create an object, you need to create a mask for each photo in an editor like Photoshop, which greatly complicates the work, since creating complex objects requires 100 or more photos.

RealityCapture: An equally widely used software product. This software product processes images much faster than its competitors. It should be noted that the main disadvantage of this program is that it only works with GPUs manufactured by NVidia. The software is paid, the minimum price for this product is 99 € for 3 months. This version can process 2,500 images per project, which is more than enough to scan any object except large ones. There is also a large version that costs 7500 € per year, it is used to create huge models such as buildings, terrain and even mountain ranges.

Pros: the software has a simple interface that is understandable even to those who are starting to work in this direction. Allows you to use a large number of photos and process them at record speed.

Cons: The app is only suitable for those running NVidia GPUs.

Meshroom: unlike the previously presented programs, it is free for a small number of photos, which is suitable for creating a model of small objects. The program is the youngest of those reviewed and is updated almost every month. The interface is simple and clear, although it allows you to almost manually conFig.each process. The main problem of this program is its low performance; the program processes photos very slowly. The process of creating the final model can take from several hours to a whole day of continuous work

Pros: cost of the software product, ability to control the creation of the model at each stage.

Cons: the program is very slow.

To create a finished 3D model, a large number of photographs are required. There should be at least 20 pictures even for the simplest and smallest objects. There is no maximum number of photos. To create a model of a human head, it is recommended to use at least 600 photographs, while to create a model of a building may require at least 2000, or even 3000 photographs. To process such a number of photographs, you need powerful and modern computer equipment. Many authors recommend using a processor with I7-level performance with at least 4 physical cores. It is also recommended to use at least 12GB of RAM; the ideal option is 32GB of RAM. The GPU must be at least NVidia Geforce 1050.

In this work, we used a computer with the following characteristics:

Processor: Intel Core i5–8300H

GPU: NVidia GeForce GTX 1050 4Gb GDDR5

RAM: 8Gb DDR4

Before starting work using Photogrammetry technology, in order to build a high-quality model, it is necessary to conFig.the device with which photographs of a real object are taken. We recommend a household camera, all settings of which must be set manually, since when creating photos in automatic mode, the settings may differ depending on the angle and in the future the application will not be able to determine the location of the photos to each other. Recommended camera settings:

ISO: As low as possible, preferably no higher than 400.

Shutter speed: As high as possible, preferably at least 1/125.

Focal Length: Use a 50mm lens to minimize distortion.

Aperture: f8 recommended.

Also, for work you will need a space in which the object is located with the ability to move around it from all possible angles. The lighting should be the best, the object should be illuminated from all sides. Ideally, you need to use a tripod or stabilizer, this will help get rid of obstacles due to hand shake (since shooting is carried out at a low shutter speed — this is important). Even minor fluctuations affect the quality of the image, which in turn will affect the operation of the program and the object will be impossible to create (Fig. 6.11).

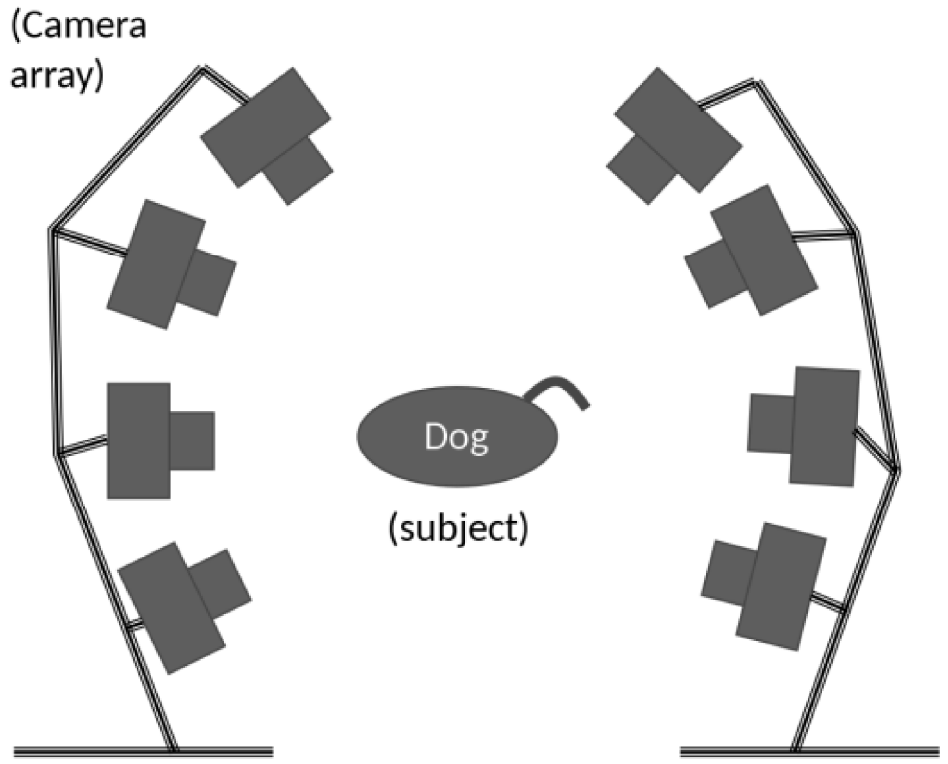

Fig. 6.11. Scheme of using cameras for photogrammetry

Photogrammetry uses the principles of optics and geometry in its base. Determining the coordinates of an object in space is carried out by calculating and comparing points on two or more photographs (first of all, 2 identical points are determined on two images). The next step is to draw a ray from the location of the camera to a point, roughly such rays can be called rays of vision, because their operating principle is similar to the way human vision works. The cross section of such rays will be the coordinates of the desired point in space. Modern applications can independently determine the position of the camera by comparing a large number of points on a larger number of photos. Because of such a huge number of calculations, such huge computing power of computer technology is required. The more photos taken, the more accurate the 3D model will be (but the more time and computing power is used). An example of using a camera in everyday photogrammetry conditions in this work is shown in Fig. 6.12.

It is almost impossible to take perfect photographs, so almost all programs that work with photogrammetry use so-called control points. Anchor points are placed on two photographs that are similar to each other, the position of which the program cannot determine on its own. For example, the Real Capture application, after aligning photos, creates groups of photos that were able to be compared with each other. Anchor points help connect photos from one group to another. If there is more than one group of photos, it will not be possible to create a model because the application will consider them as different objects. It is almost impossible to determine the position of all cameras, so the more photos you take, the better.

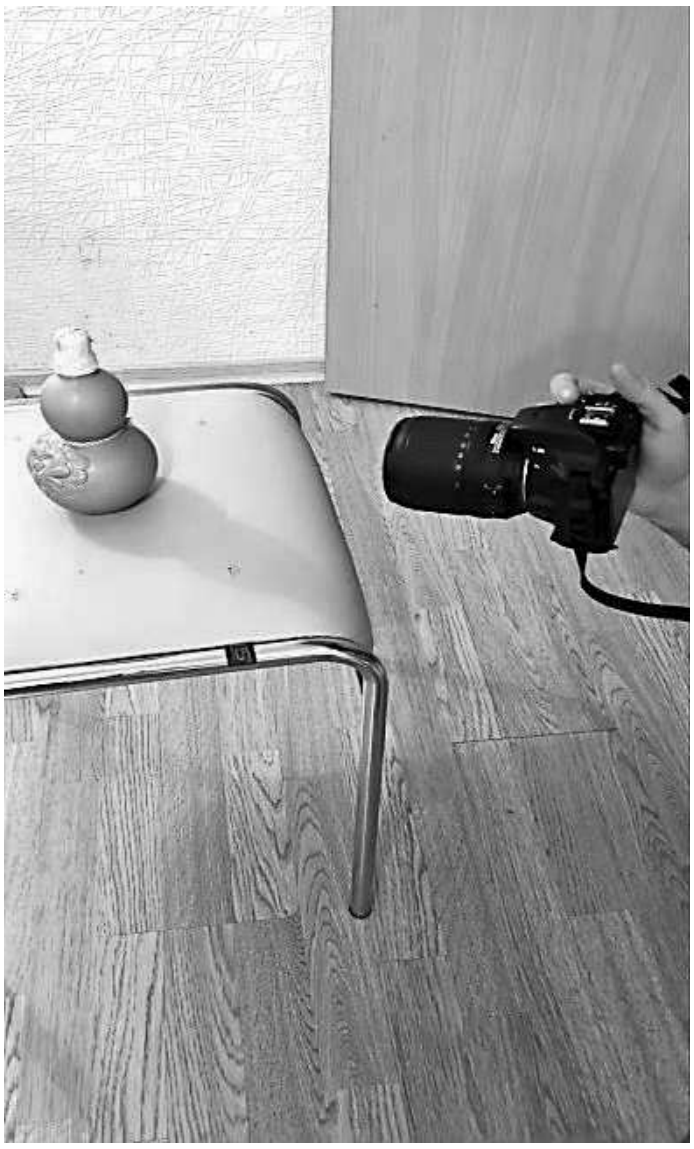

Fig. 6.12. Shooting with a household camera to obtain a model

Along with determining the camera positions, a cloud of points is created, through which the cameras are placed in the virtual space of the computer. Visually it looks like a large set of unconnected pixels, but in shape this cloud of pixels should already resemble the desired model. The number of such points can be set manually. Unfortunately, a large number of such points will not make the model much better, but the running time of the program will increase significantly.

After almost all cameras have been identified correctly in the application, with an error of 3 to 20 (depending on the number of source photos), you can begin to create a full-fledged model. The program begins to select the remaining points and create polygons between them. The number of such polygons can be set manually. A large number of polygons slows down programs, but improves the quality of the model.

Often, along with the model, part of her environment is created from the photographs used. This greatly slows down the program, increases the load on the computing power of computer equipment, and also increases the weight of the model. Unfortunately, this cannot be avoided, since surrounding objects help determine the object's coordinates in space. After creating the model, you can apply textures based on photographs. Texture overlay is not necessary if the models will be printed in solid colors. The sequence of actions of the application after receiving photos is shown in Fig. 6.13–6.16.

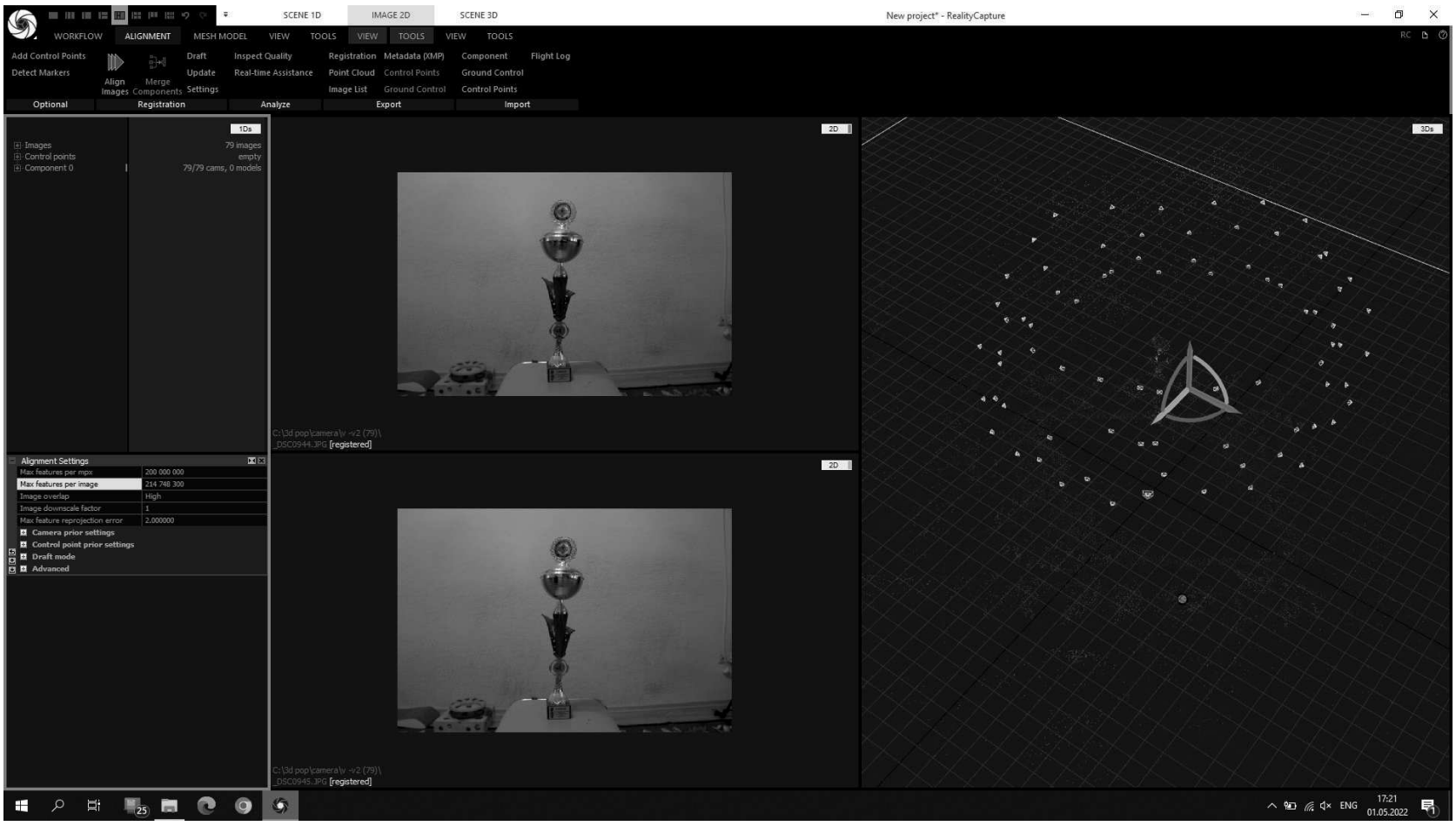

Fig. 6.13, Aligning a photo when setting the parameters of the RealityCapture program: Max features per mpx 200,000,000, Max features per image 214,748,300, operating time 25 sec

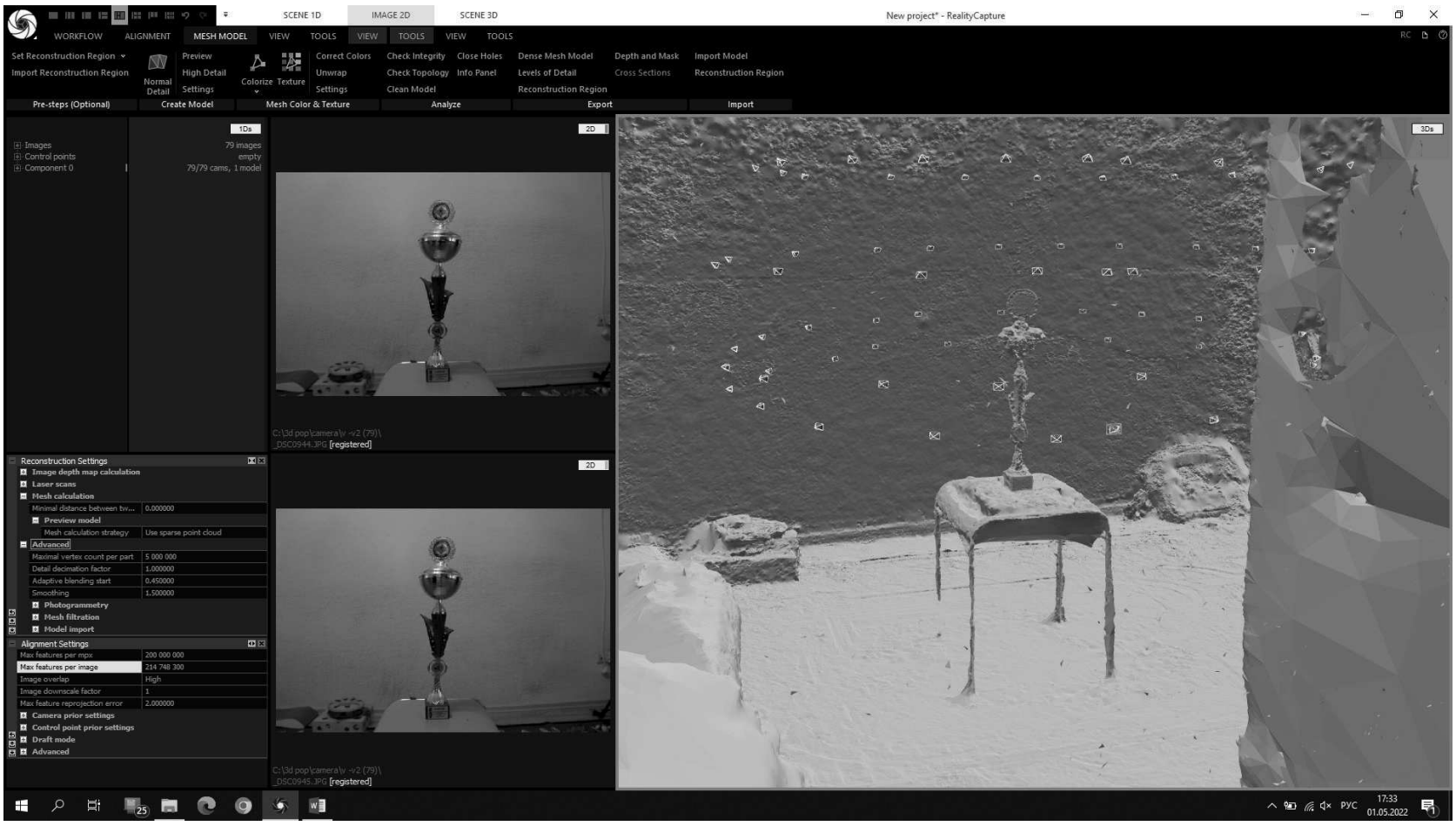

Fig.6.14. Creating a mesh with the Maximum vertex count per part settings of 5,000,000, operating time 7:54 minutes

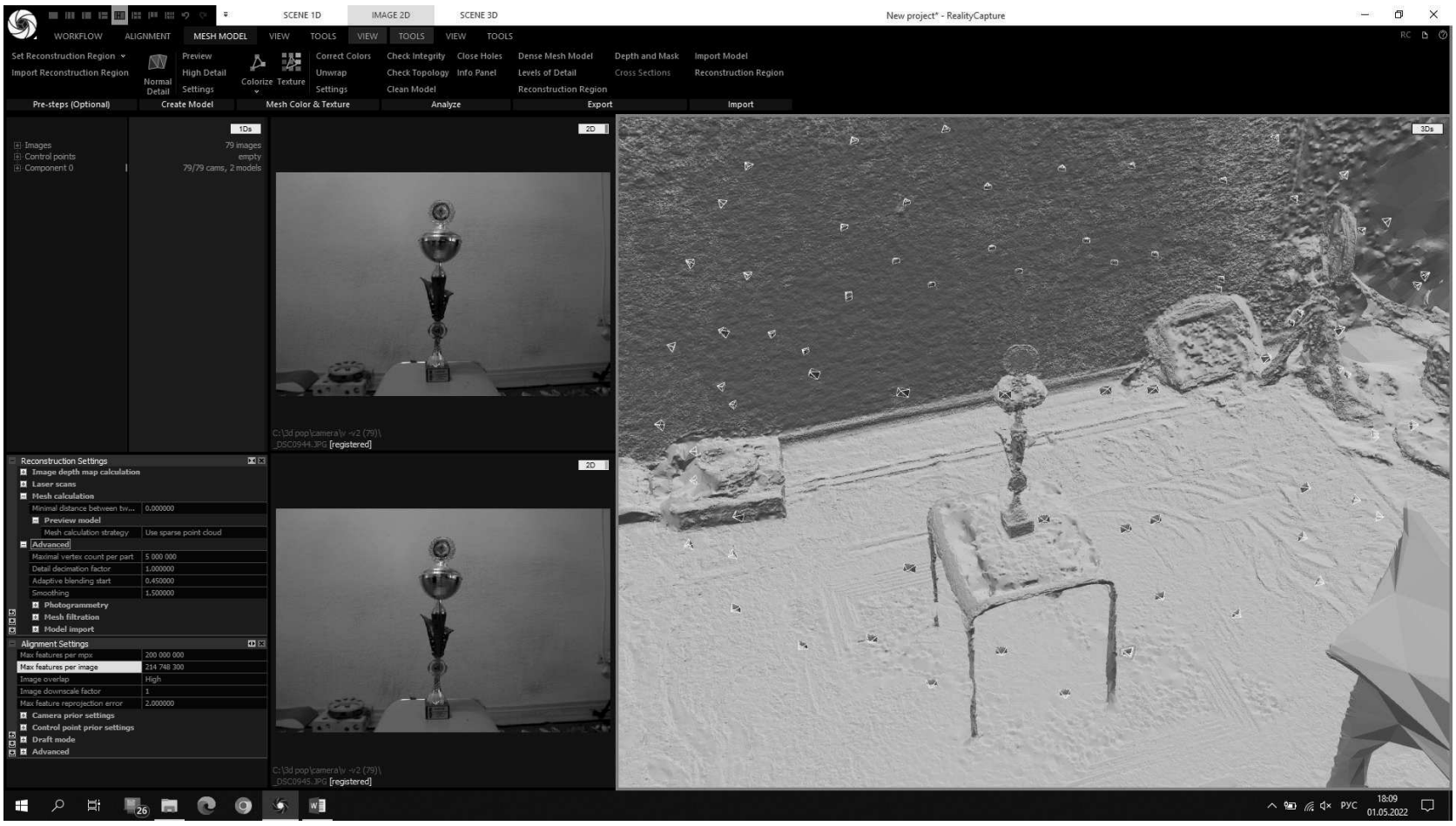

Fig. 6.15. Reconstruction and construction of small parts, operating time 32:15 minutes

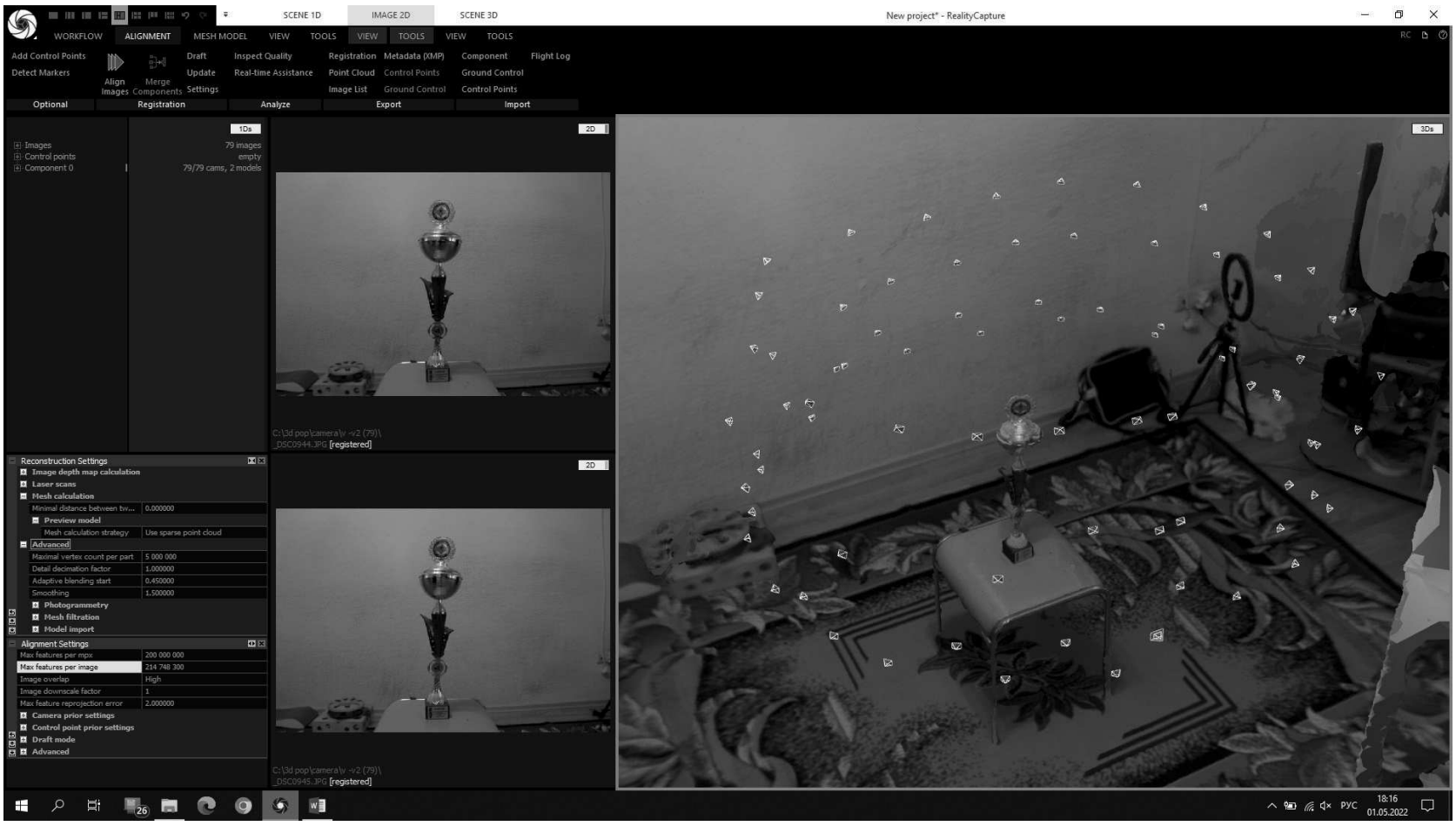

Fig. 6.16. Texture overlay, operating time approximately 5 minutes

Creating a perfect model is impossible even when using more than 1000 photos. Therefore, most often you have to modify the models manually, for which 3D modeling programs (such as 3D Max, Blender) are suitable.

Thus, the following low-cost photogrammetry technology in domestic conditions is proposed (Fig. 6.13–6.16):

- using a regular camera or phone in good lighting, several dozen photographs are taken (the more complex the original model, the more photographs you need to take);
- photographs are processed using one of the photogrammetry programs (for example, Reality Capture);
- if the application also creates a 3D model of part of its surroundings with a photograph, it is removed using a 3D modeling program (for example, 3D Max);
- if it is necessary to smooth the resulting model (inequalities can result from shadows when photographing), you can use their alignment using Blender or Zbrush;
- the resulting model in STL or OBJ format is processed in a slicer (for example, in the Cura program) to obtain code for a specific 3D printer;
- a real object is printed based on a previously built virtual 3D model on a household 3D printer.

#### *6.3.3. Experimental studies of the effectiveness of the proposed technology*

In this work, research was conducted to determine the best operating modes of equipment when creating virtual models. At first, there were attempts to create photographs with automatic camera settings, since the quality of such photographs is much higher than that of photographs taken using manual settings. The main problem with such photographs is that the programs cannot determine the correct position of the cameras in the work area. With manual settings, the pictures are of lower quality, but photogrammetry apps are better at determining the position of the cameras. In any case, the camera positions of the images are not perfectly determined, but this is not necessary. Usually from 3 to 10 photos remain without a specific place, this is one of the reasons why you need to take more photos to get a high-quality model.

Tests of creating virtual models were made in various programs for working with photogrammetry, but the best results came out in Reality Capture. This program is better than its competitors in determining the position of cameras in the work area. It turned out that to facilitate the work of this program, it is necessary to place reference points on photographs that are not defined in relation to each other. All further comments on the construction of the computer model will refer to the Reality Capture program.

In theory, there should be quite a lot of reference points, but in some cases only two or three will be enough. Experiments have shown that a large number of such reference points is also not very good. The reference points are correlated manually and it is almost impossible to get exactly pixel to pixel, as in the next photo, so a large number of such misses only confuse the program even more. Reality Capture groups photos together if they can be linked together. To create a model, there must be at least one group. Anchor points help you combine photos from different groups. It also turned out that if the photographs were taken with different camera settings, it is impossible to link them together.

Another problem that became clear while photographing the subject was adjusting the focus. Focus may vary depending on the distance to the subject, but when there are many photos, adjusting the focus for each photo takes a lot of time and effort. The only way to avoid this is to take photos at the same distance from the model. If the focus is not clear, the photo itself will be blurry. It is impossible to create a 3D model from blurry photographs.

The first attempts to create a finished model using photogrammetry technology were unsuccessful; it was very difficult to choose the right lighting at home. It turned out that in fact, even standard lighting available in any apartment will do, the main thing is that it is constant and does not change or changes slightly from photograph to photograph.

Photographs must be taken from all sides of the property. The more such photographs there are and the smaller the camera rotation step, the higher quality the virtual model will be created as a result.

After completing the shooting, all received images must be transferred to a program for working with photogrammetry (in this case, Reality Capture). Photos can be placed one at a time or in a whole folder at once (this is more productive). The next step is to level the camera. The Align settings menu has a large number of settings, the main ones being Max features per mpx, Max features per image, Image overlap. The Image overlap parameter is always used in the High value, since it is inappropriate to use lower quality, and this value is recommended on weak computers. The Max features per mpx and Max features per image parameters are set at the user's discretion. The higher they are, the higher quality the model will be. The values used in this work were 100000 and 400000, respectively, since the models are small and it makes no sense to increase these parameters. After this, you can start the alignment procedure. If after this procedure the application has created more than one Component, it is necessary to place reference points on them for syn-chronization. Data processing by the program results in a cloud of points appearing in the work area that vaguely resembles the required model. If most of the photos have been identified, you can proceed to the next step of the photogrammetry program.

The next step of creating a model is called Mesh Model. This stage also has many settings, but since the program is designed to create models of large objects and even entire reliefs, the settings can be left at default. This stage is the longest and resource-intensive. On a large number of photographs, computer operation time can take several hours and even days (if the object is an entire building or area). As a result of this stage, we obtain an almost finished three-dimensional model.

However, for the most part, it is not possible to immediately obtain an ideal model. The last step is to apply textures (if necessary). If all steps are completed correctly, we will have a finished model. You should keep in mind that the surrounding elements in it will be of low quality, since the photos were aimed at a specific object.

The resulting 3D model can be stored in different formats, but it must be assumed that it may be subject to further editing. The folded model can weigh a lot (in GB of memory), but after the unnecessary elements are removed, the weight of the file will be several times less.

Removing polygons may be necessary in most cases to obtain an accurate model. This can be done in 3D max or Blender programs by selecting the appropriate menu item. The process is not the most difficult, but it can take a long time if the computer's power parameters are not too high.

### *6.3.4. Features of processing the test model and its FDM printing*

To confirm the effectiveness of the proposed household 3D scanning technology, a souvenir bottle was used to create a virtual three-dimensional model (Fig. 6.12) — it has an irregular relief, which is difficult to create using geometric shapes in the 3D Max program.

According to the proposed photogrammetry technology, the Fig.was photographed along the entire perimeter with a household camera (about 125 pictures) — (Fig. 6.12).

The resulting array of photographs from one folder was transferred to the Reality Capture program (Fig. 6.17) for further processing. At this step there were 125 photos, all of them are displayed in the left window as a list, you can also display them in one of the other windows in the form: 2D, 2D list or 3D.

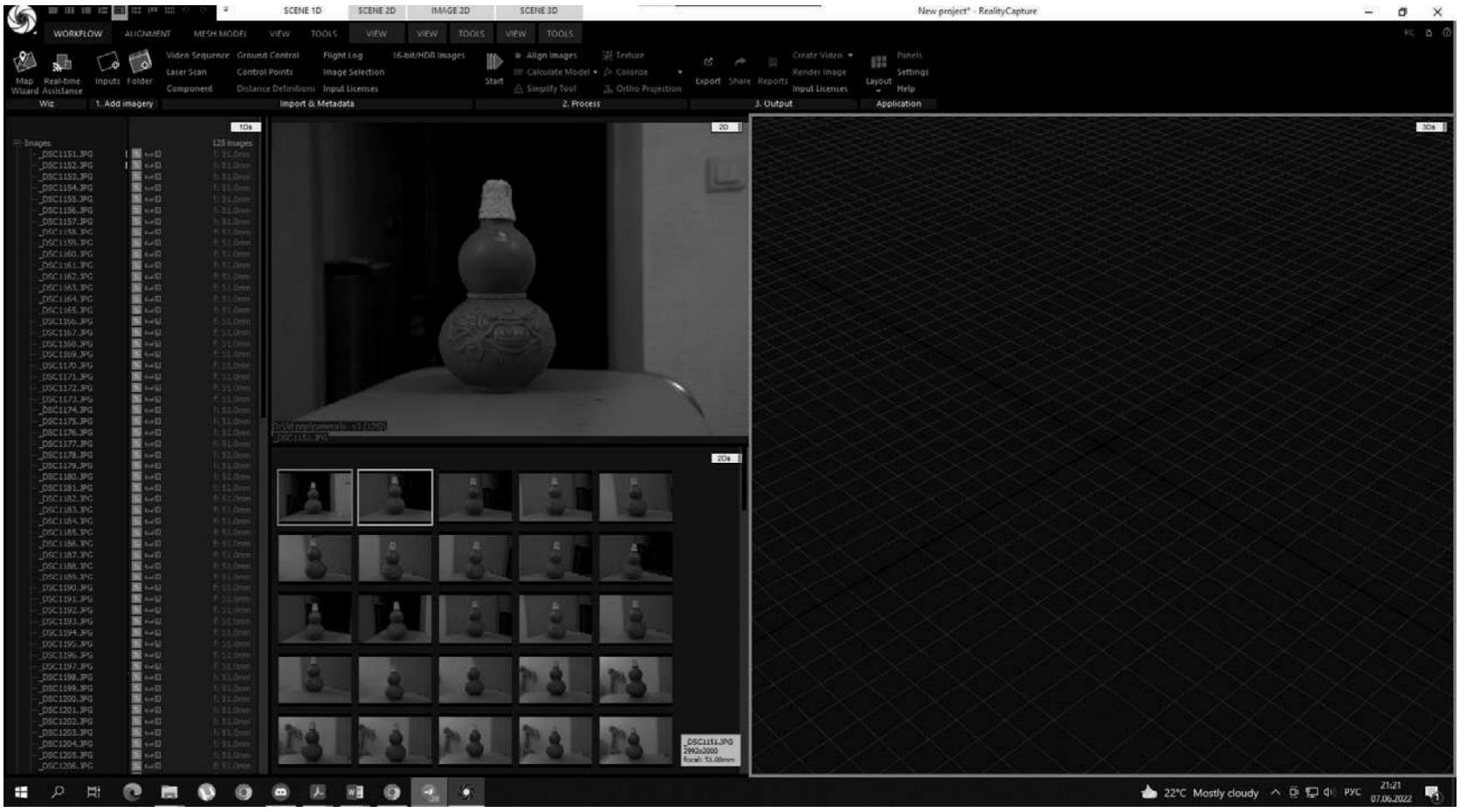

Fig. 6.17. Reality Capture program window after adding photos

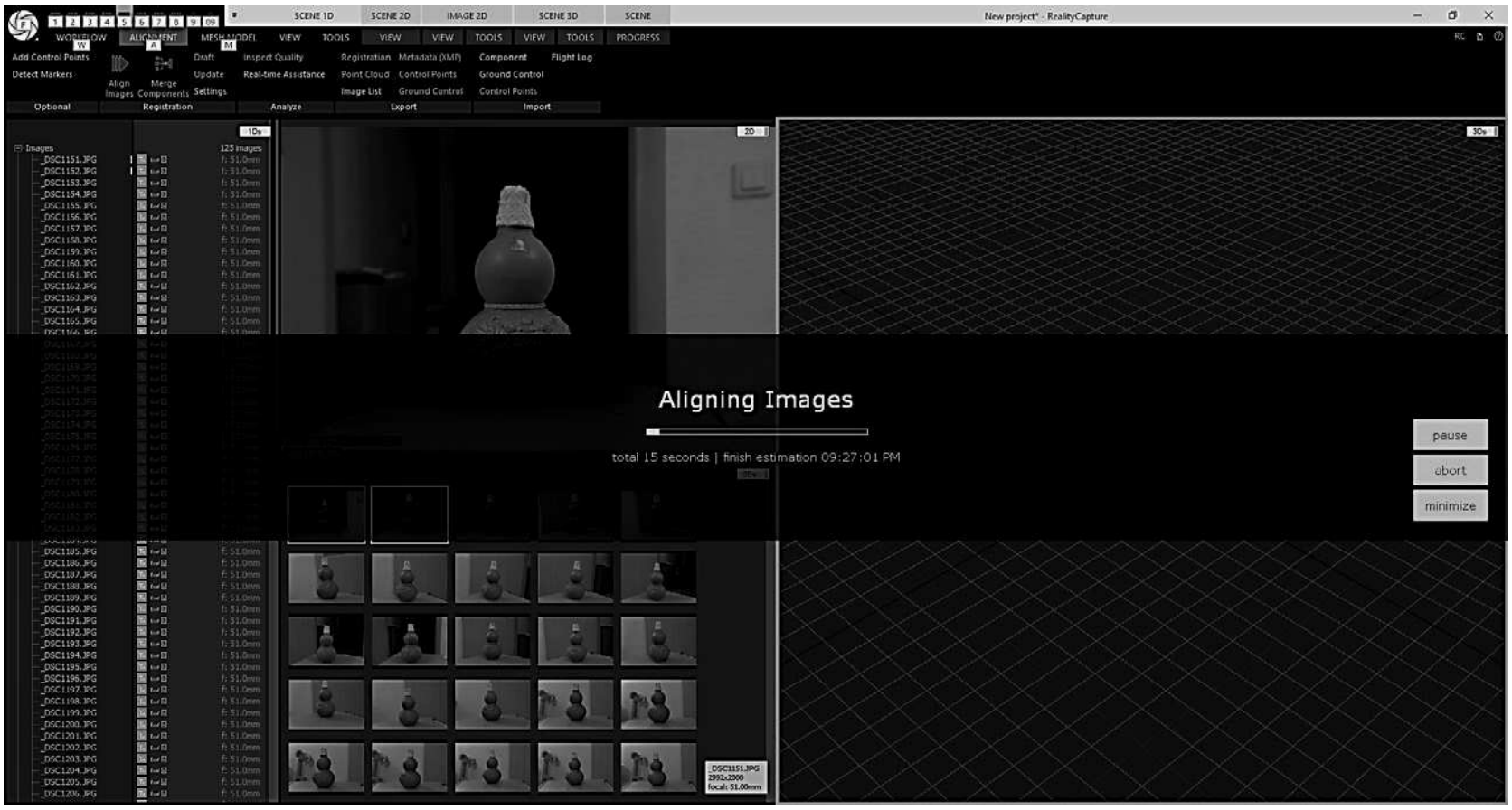

Fig. 6.18. Determining camera positions in Reality Capture

The next step was to determine the positions of the cameras in space using the Aligning Image function (Fig. 6.18). The following alignment settings were used:

- Max features per mpx: 200,000
- Max features per image: 400,000
- Image overlap: High

After alignment, several groups (Component) were defined — Fig. 6.19. Many of these groups include a small number of photographs—less than five. But one of the components includes 82 photographs, which is enough to create a small model. Therefore, other groups were simply removed from consideration as unnecessary.

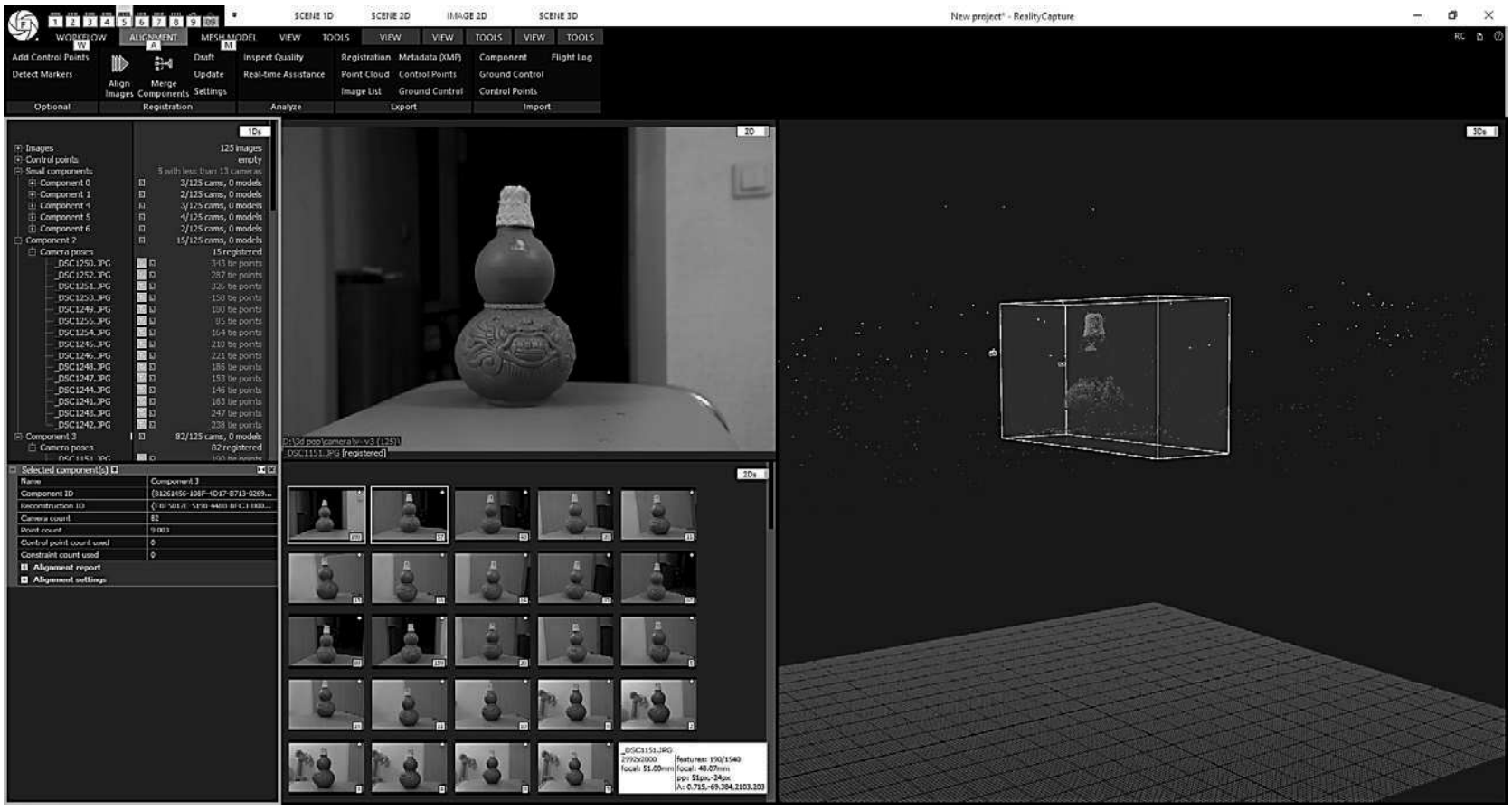

Fig. 6.19. Illustration of highlighting groups of components

After this, the program launched the process of reconstruction into a normal model (Reconstruction in Normal Detail) — Fig. 6.20. This process is the longest; for the model we tested, it took 36 minutes. Due to the fact that a large number of photographs were not determined in the settings for the Mesh Model, the amaximal vertex count per part value was chosen equal to 5,000,000. Therefore, the final model turned out to be of higher quality, but larger in size (i.e., more computer memory is required to accommodate it).

As a result of this processing, a three-dimensional model was obtained (together with the furnishings) — Fig. 6.21.

The program also makes it possible to view the position of the resulting model in space relative to the initial photographs — Fig. 6.22 (if some are not suitable, they can be excluded and the previous step of the program can be repeated to obtain higher quality).

If necessary, to improve the quality of the resulting model, textures can be applied, as noted above in the given Photogrammetry algorithm — Fig. 6.23.

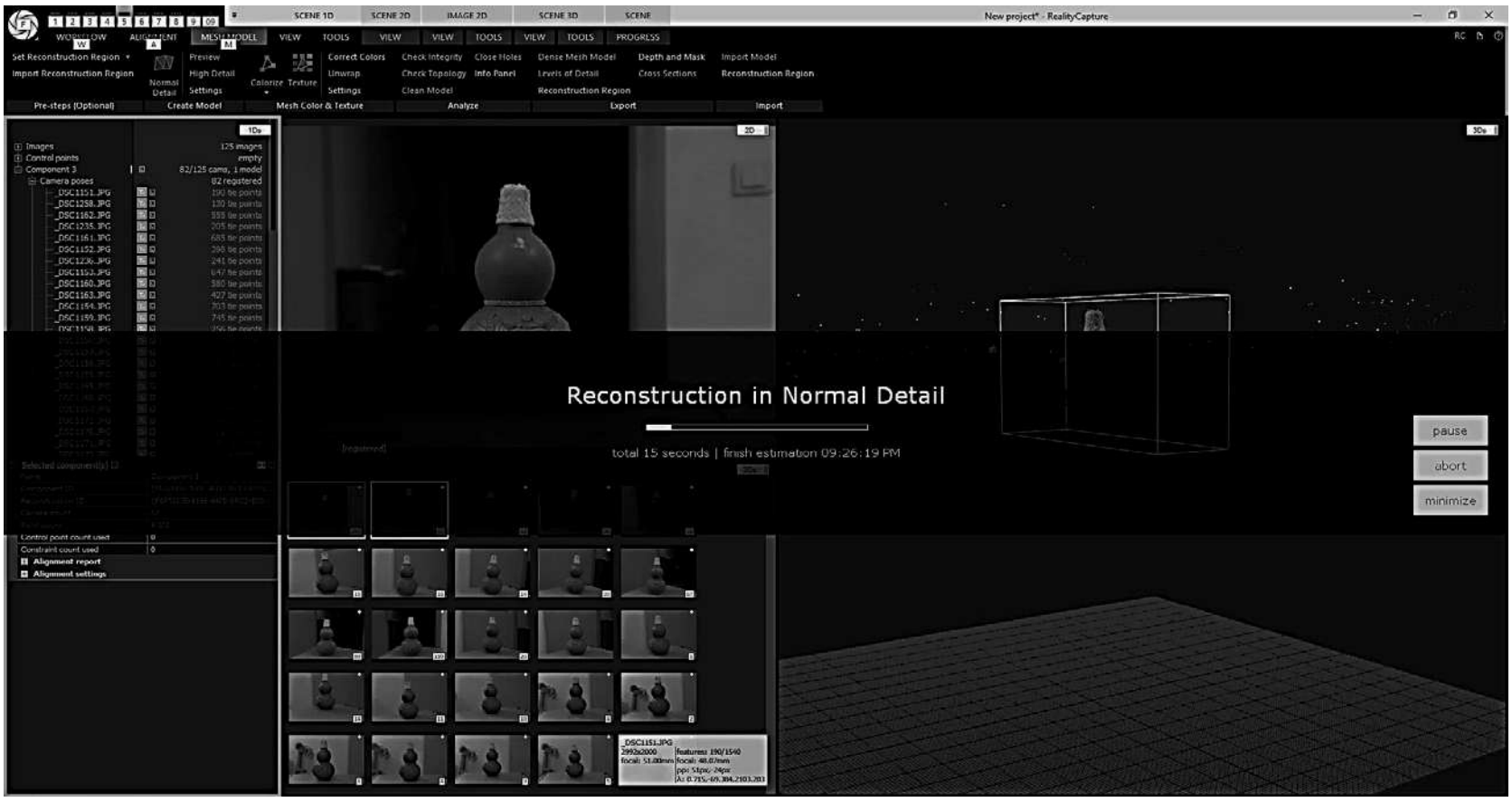

Fig. 6.20. Model reconstruction mode window

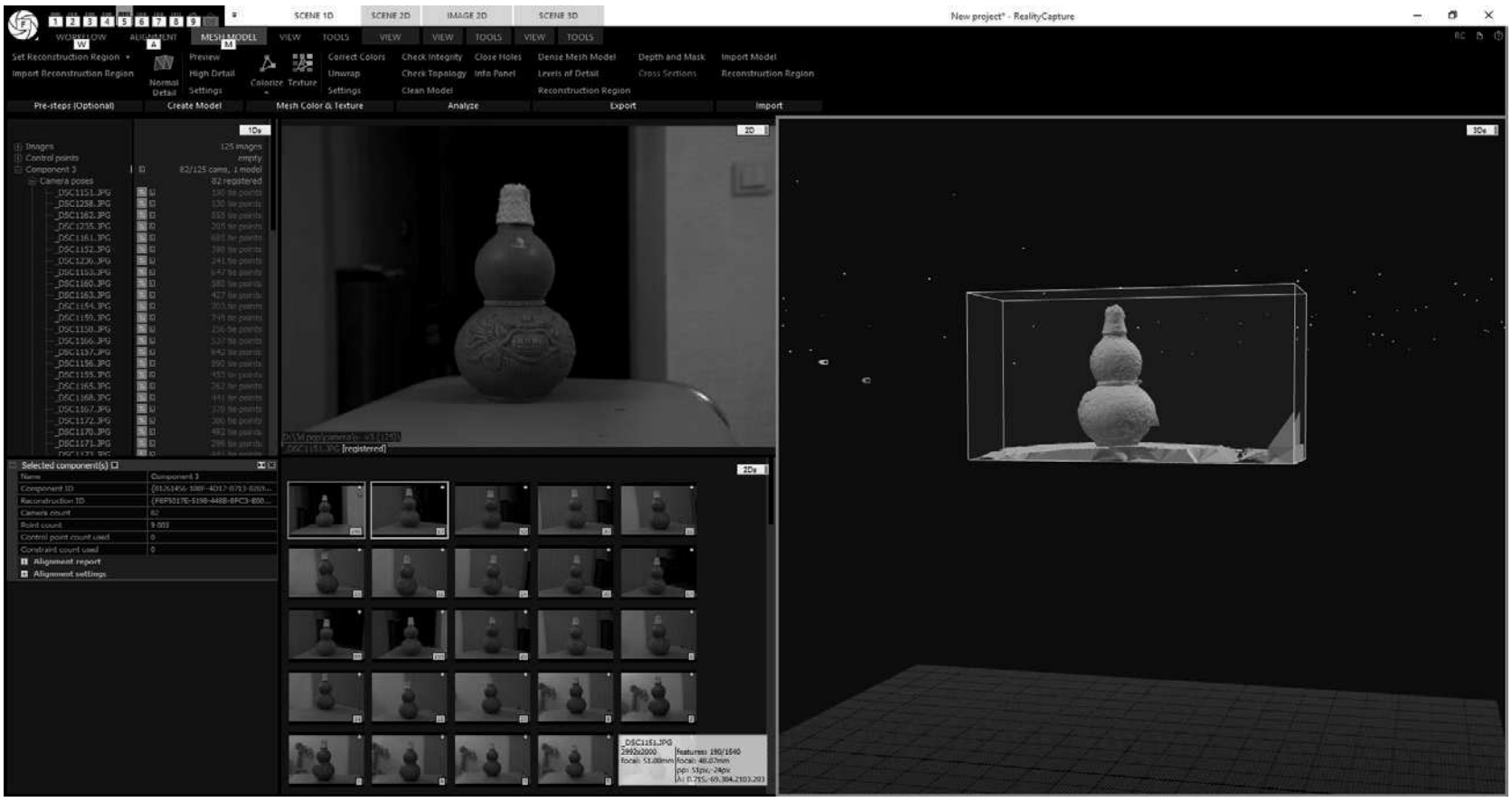

Fig. 6.21. Image of the resulting model without texture overlay

As a result of these manipulations, a three-dimensional image of the original model (with surrounding objects) was obtained — Fig. 6.24.

If necessary, if there are unnecessary details of the furnishings, surroundings, or hanging structures, they can be easily removed from the resulting three-dimensional model using modeling programs such as 3D max or Blender. So, this was done in 3D max, the result in the form of a model with the STL extension is presented in Fig. 6.25.

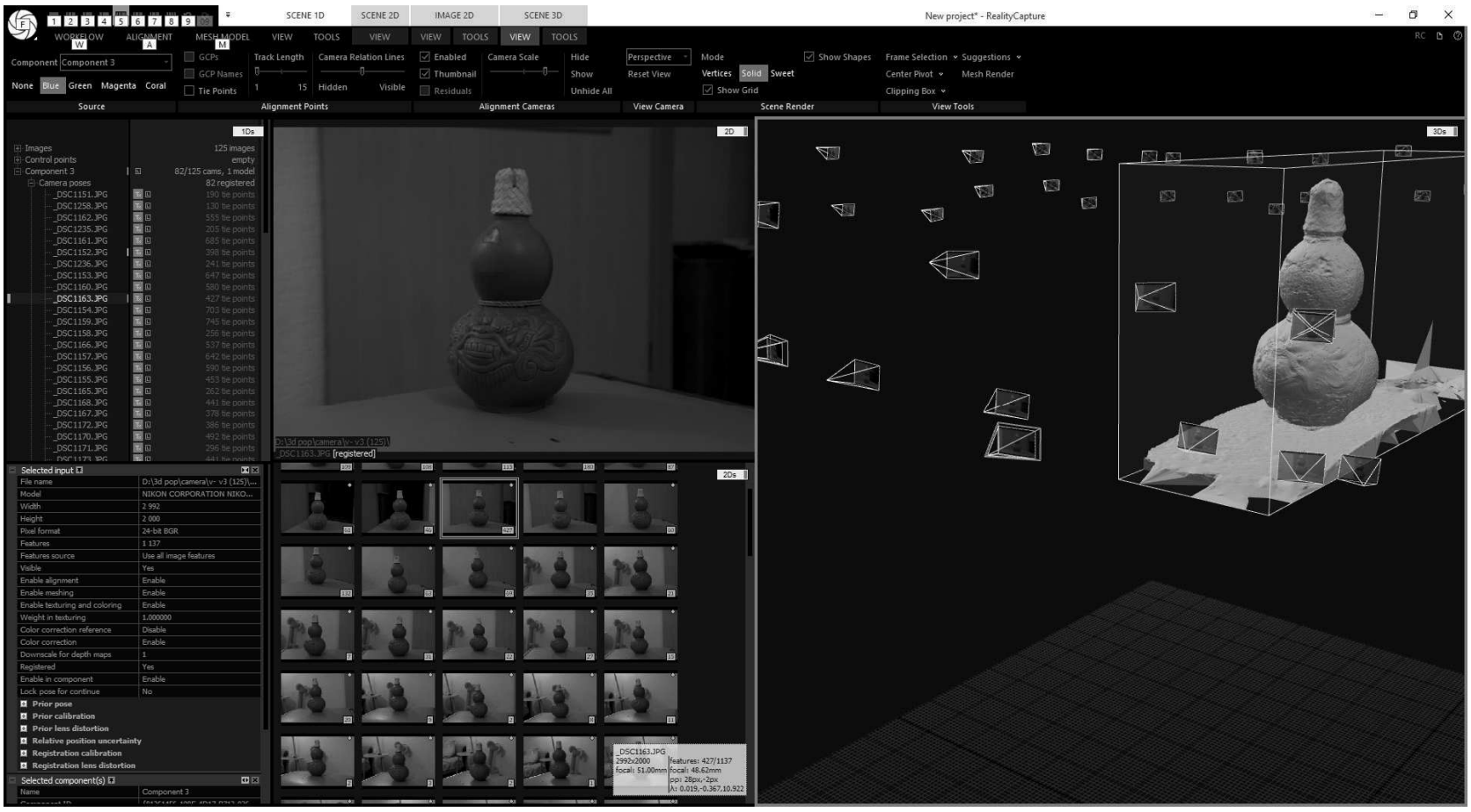

Fig. 6.22. Image of the resulting model from photographs in space

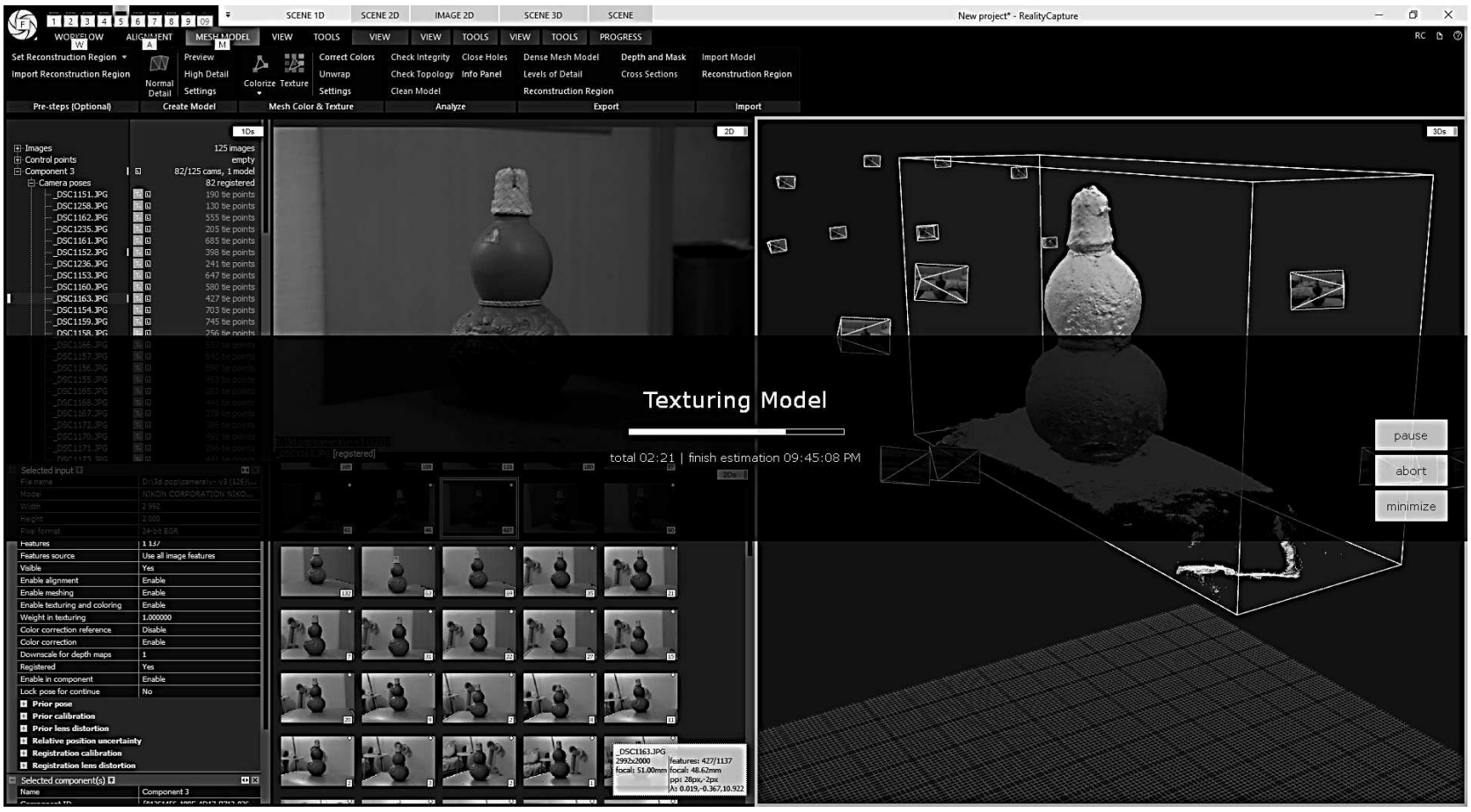

Fig. 6.23. The process of applying textures

Due to the above circumstances (disadvantages of lighting, shadows and hand tremors when photographing), the resulting model has a number of shortcomings that can be easily removed using the Sculptring mode in the Blender program (Fig. 6.26). The result of processing in the Blender program is presented in Fig. 6.27.

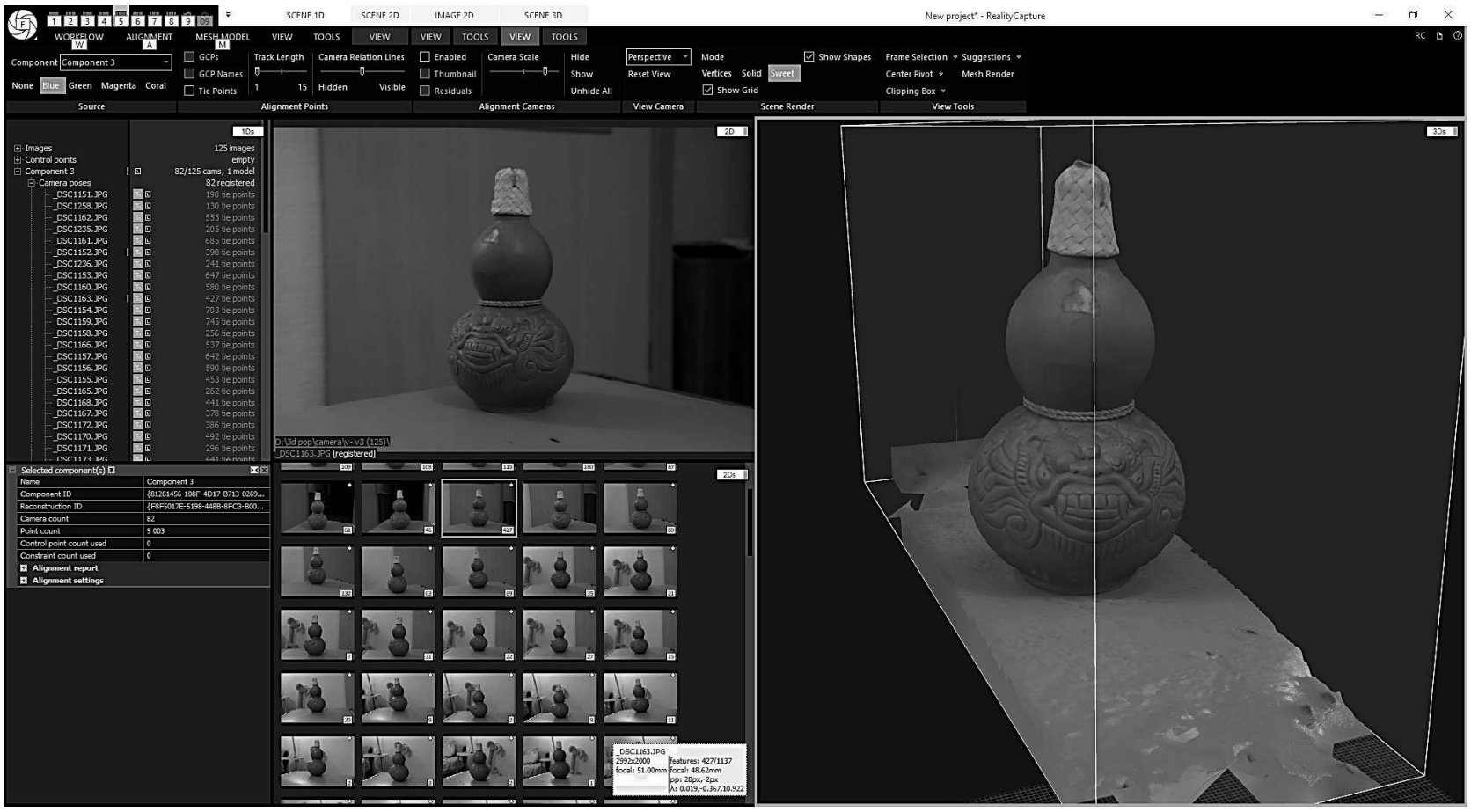

Fig. 6.24. Image of a three-dimensional model with textures

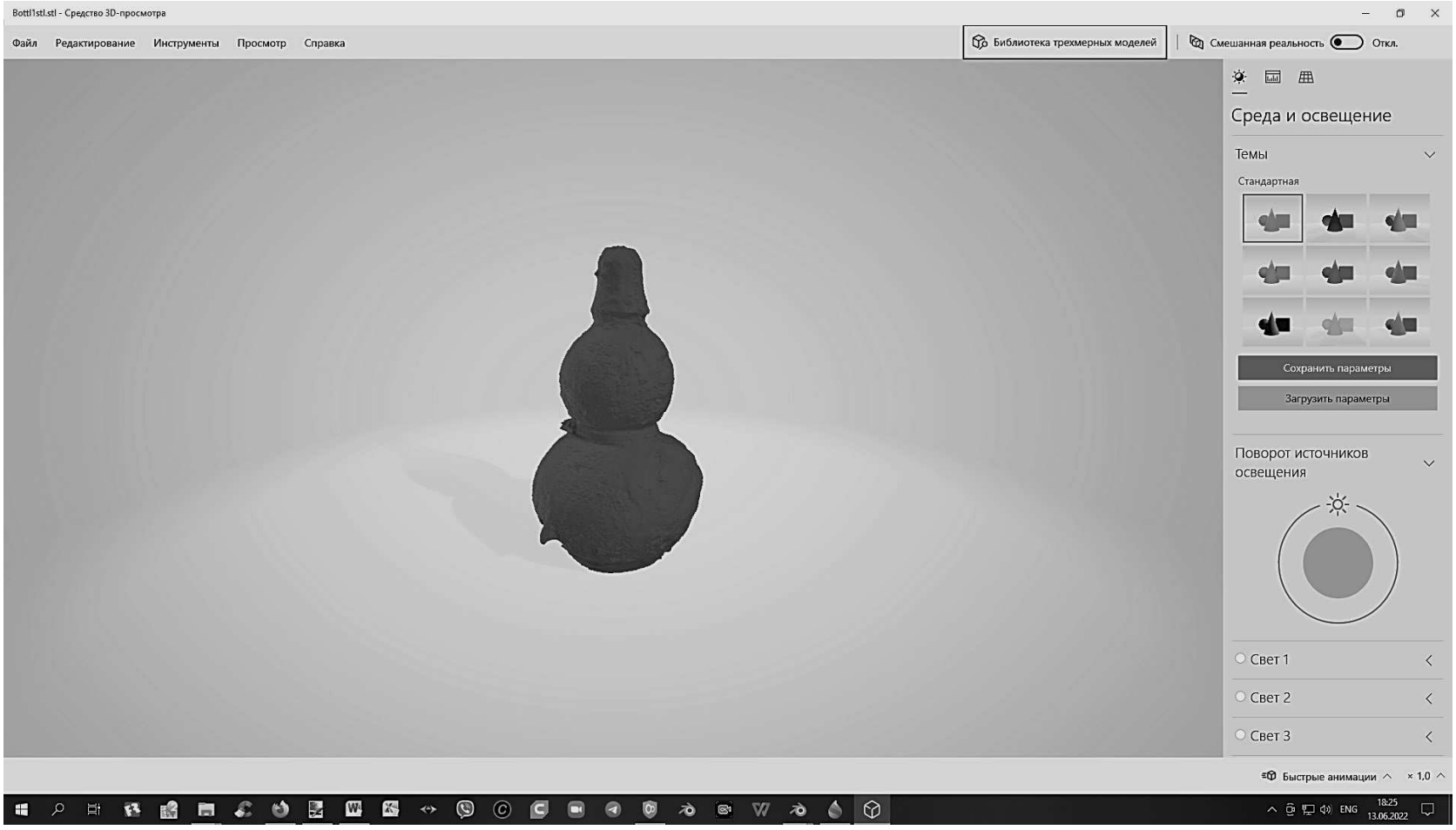

Fig. 6.25. Three-dimensional model after processing in the Reality Capture program

Now the resulting three-dimensional model should be printed on a 3D printer. An FDM printer is designed in such a way that in order to operate, it must receive a program in the form of a special G-Code, which is a sequence of instructions for moving the printer extruder and extruding plastic at certain intervals. G-Code is a standard printing language used by many 3D printers to control the printing process. G-Code files can be opened using various 3D printing applications such as Simplify3D, GCode Viewer, and also with a text editor since the content is plain text.

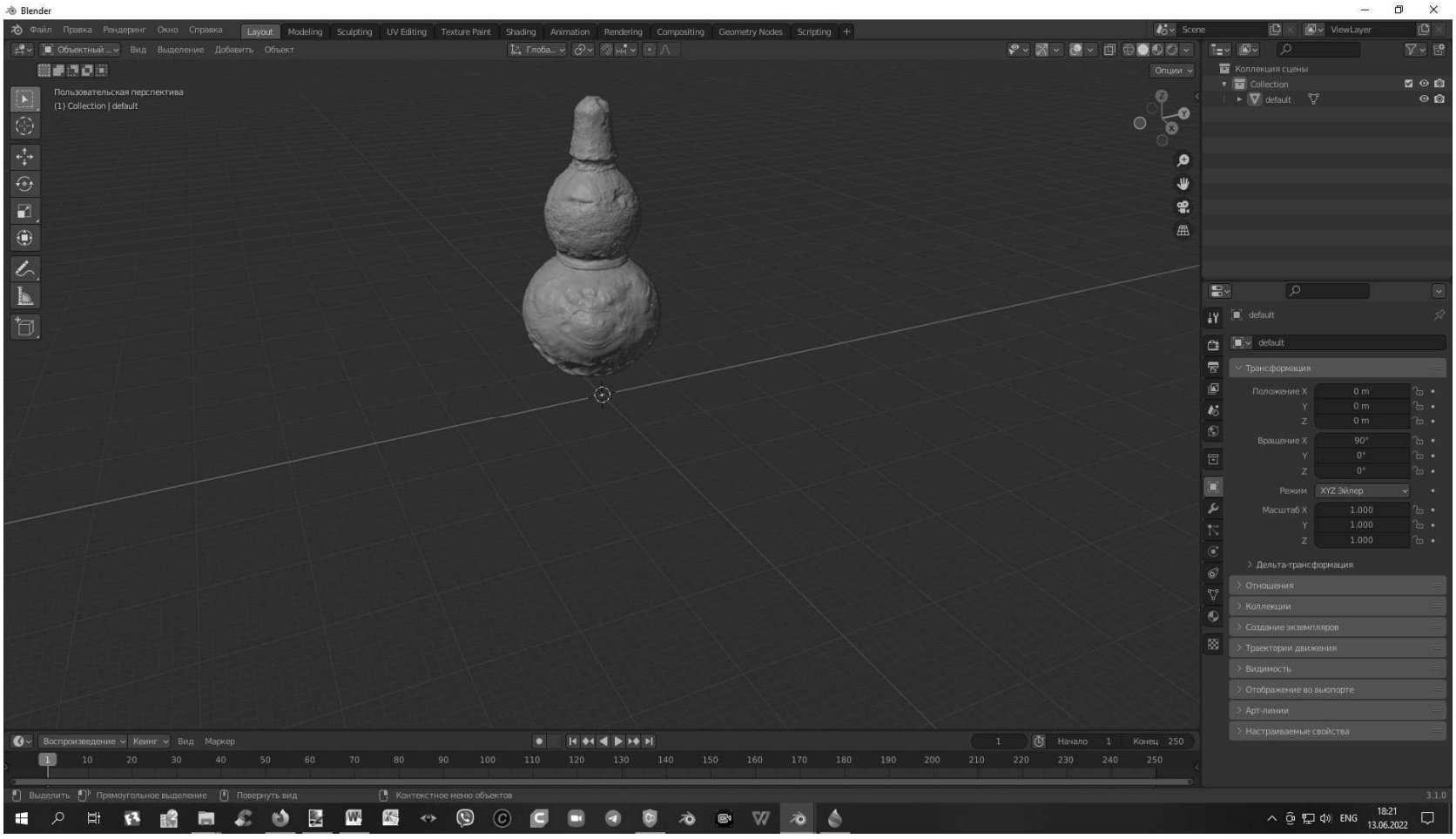

Fig. 6.26. Processing the resulting model in Blender

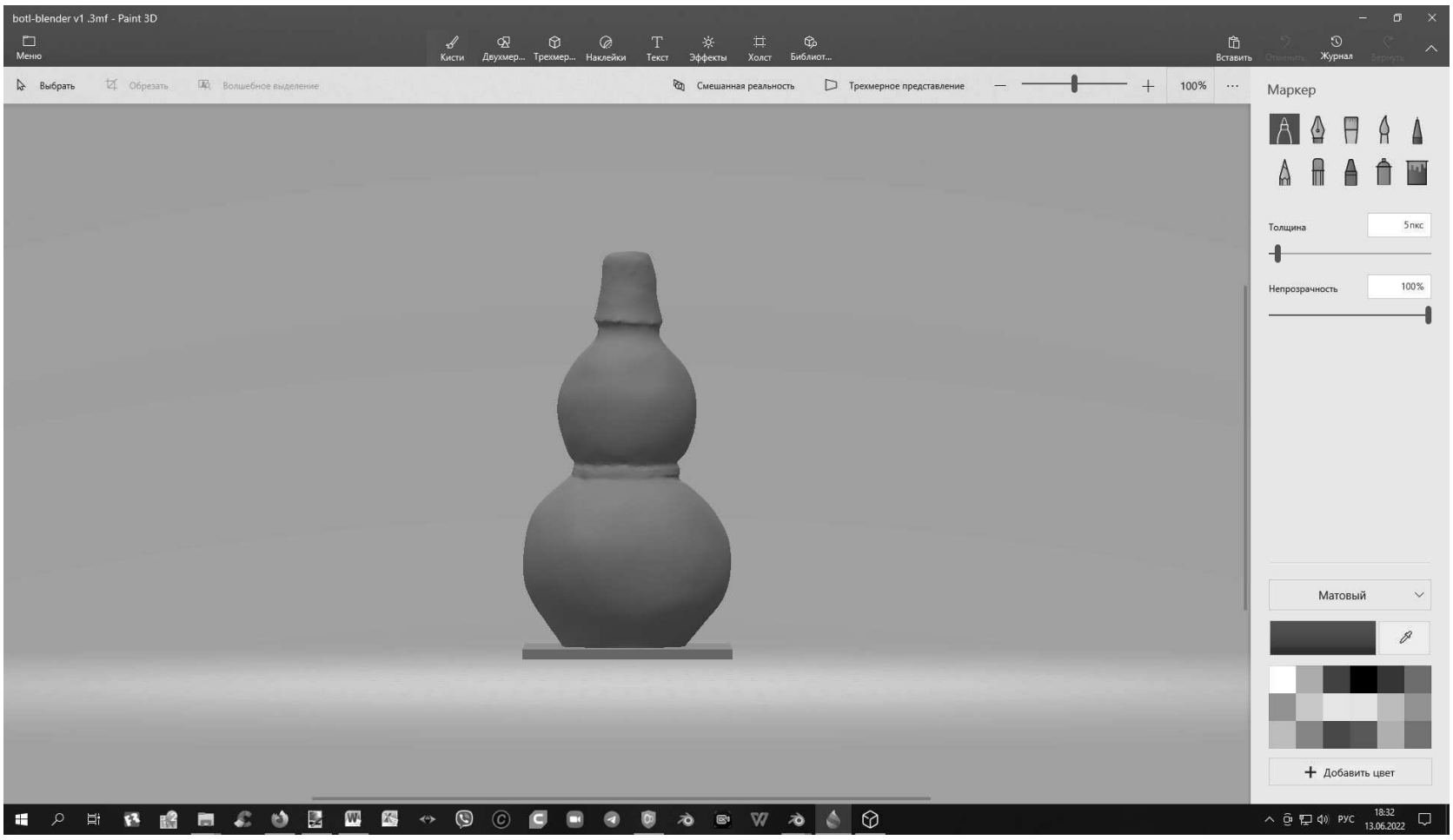

Fig. 6.27. The result of using photogrammetry technology in the form of a three-dimensional figure

A program written using G-Code has a rigid structure (Fig. 6.28):

```
;FLAVOR:Marlin
;TIME:19806
;Filament used: 12.7166m
;Layer height: 0.1
;MINX:-61.927
;MINY:-34.156
;MINZ:0.3
;MAXX:62.295
;MAXY:30.565
;MAXZ:35.5
;Generated with Cura_SteamEngine 4.9.0
M104 S230
M105
M109 S230
M82 ;absolute extrusion mode
M136
M73 P0
M103
G21
G90
M320
```

Fig. 6.28. Fragment of the beginning of G-Code for printing the resulting model

Comments to the program are placed in parentheses. The comment can be placed either on a separate line or after the program codes. It is not acceptable to use multiple lines as a comment, enclosed by a pair of parentheses. All control commands are combined into frames — groups consisting of one or more commands. The maximum number of elementary commands and setting coordinates in one frame depends on the specific interpreter of the device control language, but for most popular interpreters (control racks) does not exceed 6. The frame ends with a line feed character (PS/LF) and has a number, with the exception of the first frame of the program. The first frame contains only one “ %” character. The program ends with command M02 or M30.

Coordinates are specified by specifying the axis followed by a numeric coordinate value. The integer and fractional parts of the coordinate number are separated by a decimal point. Minor zeros may be omitted or added. Also, in the vast majority of interpreters it is acceptable not to add a decimal point to integers. For example: Y0.5 and Y. 5, Y77, Y77. and Y077.0

There are so-called modal and non-modal commands. Modal commands change a certain parameter/setting, and this setting affects all subsequent program frames until they are changed by the next modal command. Modal commands, for example, include extruder movement speeds, spindle speed control, coolant supply, etc. Modeless commands have effect only within the frame in which they are found.

The language allows for repeated execution of a once-recorded sequence of commands and tool movements called from different parts of the program, for example, cutting out many holes with the same complex contour in a sheet blank, located in different places of the future part. In this case, the body of the subroutine describes the trajectory of the tool for cutting one hole, and the program calls the subroutine multiple times. In the body of the subroutine, tool movements are specified in relative coordinates — coordinates associated with the shape of the hole, the transition to the relative coordinate system (sometimes such a coordinate system is called "incremental") is made by the G91 command at the beginning of the subroutine body, and the return to the absolute coordinate system by the G90 command is at the end subroutine bodies.

The translation of the constructed three-dimensional model into G-Code for printing it on a 3D printer is carried out by special programs — slicers, the most famous of which is Cura. A slicer is a computer program that turns a virtual 3D model layer by layer into machine code (G-Code) that allows an additive automated device to produce a part from a specific material. Ultimaker Cura is an easy-to-use slicer that generates G-Code for various 3D printer models. Open source 3D printing software — it works with almost all desktop 3D printers available.

Convenience, reliability and open source code are the reason for the popularity of the Cura slicer in many countries. Many 3D printer manufacturers such as LulzBot, Intamsys and SolidPrint use Cura to create their own software.

Advantages of this slicer:

- Software from a reliable developer;
- Constant update of the slicer;
- Convenient user interface;
- Open source code;
- Compatible with printers from many manufacturers;
- Built-in support for printed materials from famous brands.

Flaws:

- Setting up support is not always obvious; for better results you will have to understand the experimental functions;
- Low speed on weak computers.

The settings of the Cura program for building a test model are presented in Fig. 6.29.

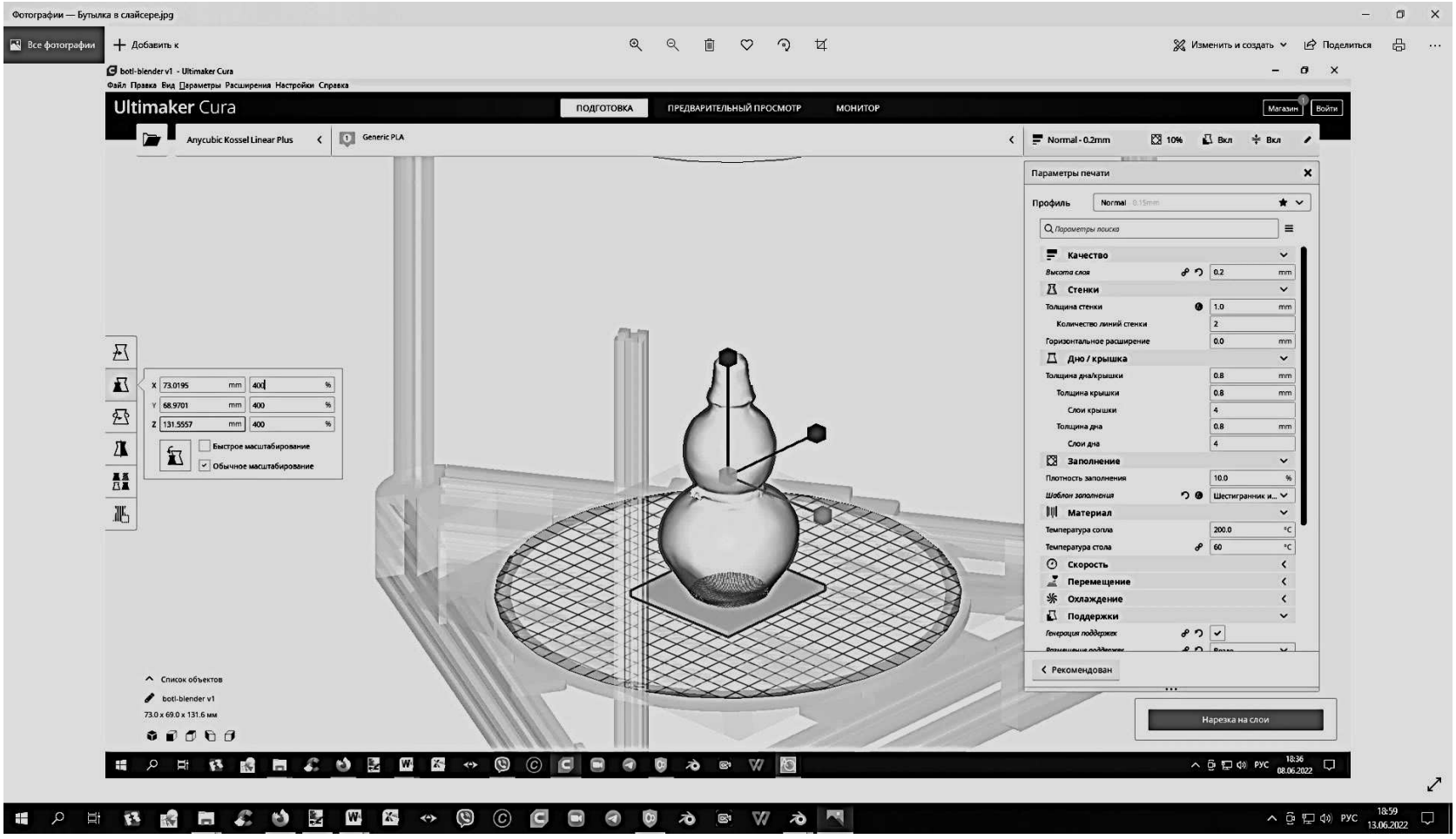

Fig. 6.29. Processing a test model in the Cura slicer

Since the resulting Fig.does not contain overhanging structures greater than 70°, it can be printed without supports and the "Supports" mode is not selected in the program. Red PLA type plastic was chosen as the material as it is cheaper (melting point 200° C). Now you need to print the G-Code obtained as a result of the Cura program in the form of a plastic figure.

For this purpose, the Anycubic Kossel 3D printer was used in the work. The delta kinematics printer design was created in 2012 by Johan Rocholl as part of the RepRap project. The idea initially led to the creation of the Kossel Pulley printer — the arms of the 3D printer moved in directions on rollers. The reel is hung on a bracket mounted on the right post. The filament is passed through a feeder and then inserted into a bowden tube and fed into the extruder.

You can transfer the finished G-Code to the printer either through a direct USB connection or using an SD card, which is more convenient. The use of rack and pinion guides and the replacement of rollers with carriages significantly increases print stability and repeatability compared to the previous model. It is believed that a similar design increases the noise level when operating a 3D printer. Indeed, Kossel makes sense to install in work areas, and not in sleeping or relaxation areas. The printing result is shown in Fig. 6.30.

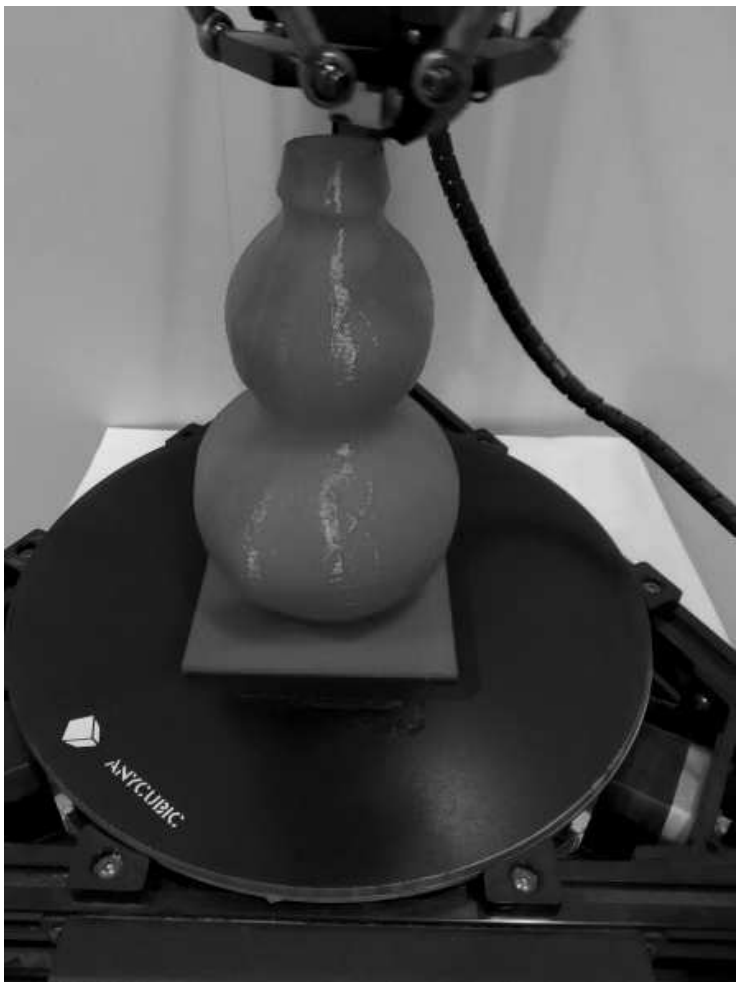

Fig. 6.30. The result of using photogrammetry technology is a three-dimensional plastic figurine corresponding to the initial model of a souvenir bottle (Fig. 6.12)

## 6.4. CREATING 3D MODELS USING THE KINECT DEVICE

As already described in Section 6.2., modern technologies for creating 3D models are of great importance in various industries, such as engineering design, architecture, medicine and game development. One of the important aspects of successfully creating 3D models is the accuracy and quality of the data obtained when scanning objects. In this context, scanning technology using Kinect is of significant interest.

The Kinect device is a depth sensor developed by Microsoft, which is used in Xbox game consoles [23, 25, 28, 29]. However, due to its motion recognition and 3D model creation capabilities, Kinect has found application in the field of 3D scanning. It allows you to quickly and accurately capture data about the geometry and textures of objects, opening up new possibilities for creating high-quality 3D models.

Kinect has an intuitive interface and does not require special skills or knowledge to operate; it uses motion recognition technology, allowing you to control the process of creating models without the need for physical contact with the device. This is convenient and allows you to maintain the cleanliness and integrity of the model. It should also be noted that this device has a fairly high accuracy in reproducing movements and shapes. It is capable of capturing fine details and textures, creating realistic 3D models [23].

In the literature, many authors testify to the wide range of applications of Kinect in various fields such as the gaming industry, medicine, architecture and design, it can be used to create animations, virtual reality, simulation and much more.

One of its main advantages is its availability and reasonable price. This makes Kinect suitable for a wide range of users who want to create 3D models at home.

Overall, the Kinect is an excellent choice for creating 3D models at home due to its ease of use, high accuracy, and variety of applications.

The following outlines the capabilities and limitations of using the Kinect to scan objects. Factors such as resolution, accuracy, scanning speed, ability to capture color and texture, and compatibility with data processing software are important when choosing a tool for creating quality 3D models.

***How the Kinect works:*** Research confirms that the Kinect uses infrared depth vision and a depth camera to create accurate 3D models of objects. This allows you to achieve high accuracy and detail of models.

***Data processing algorithms:*** Various data processing algorithms used to improve the quality and accuracy of 3D models are presented in the literature. Examples of such algorithms include noise filtering, object segmentation, point cloud alignment, and point cloud merging.

***Application in various fields:*** one of the main advantages of the developed technology is its versatility. Applications of Kinect have been explored in various fields such as computer vision, medicine, gaming, archeology and industrial applications. This confirms the wide range of capabilities and potential of this device.

***Limitations and Challenges:*** Some publications have noted several limitations and challenges associated with the development of Kinect scanning technology. This includes limited accuracy in certain lighting conditions, difficulty in processing large volumes of data, and limited compatibility with some software platforms.

***Comparison with other methods:*** Comparative analyzes have been made in the literature between Kinect scanning technology and other methods such as laser scanning and stereo vision. These studies showed the advantages and disadvantages of each method, as well as possible areas of their application.

A review of publications in recent years shows that the development of technology for scanning objects using the Kinect device to create high-quality 3D models is an actively researched area. Further research and development in this area could lead to new innovations and applications of Kinect in various fields.

### *6.4.1. Comparison of the Kinect with industrial scanners*

Creating 3D models by scanning in manufacturing plants is often associated with significant costs. This process may include the use of specialized equipment, payment for the services of professional scanners, and the cost of software for processing and analyzing the obtained data [5, 6, 28, 29].

The main factors that make scanning to produce 3D models an expensive process include:

1. Specialized Equipment: High-precision 3D scanners and image capture equipment require significant financial investment to purchase and maintain.

2. Skilled Personnel: Ensuring high-quality scans requires experienced technicians, which may incur additional costs for salaries and training.

3. Technical support and maintenance: regular maintenance and upgrades of equipment, as well as software updates, require additional costs.

4. Time and resources: The process of scanning, processing and analyzing data requires a significant amount of time and human resources, which also affects costs.

5. Additional costs: in addition to the main costs, there are also additional costs for renting premises, equipment and other related costs.

There are many industrial 3D scanners on the market today, their cost varies significantly depending on technical characteristics, performance and functionality [3, 5, 8]. Here are some examples of well-known brands of industrial 3D scanners that tend to be expensive:

***FARO:*** FARO offers a wide range of high-precision 3D scanners aimed at industrial and engineering applications. For example, the FARO FocusS 350 scanner, designed for scanning large objects, has high accuracy and performance, but its cost can reach several tens of thousands of dollars (Fig. 6.31).

***Hexagon:*** Hexagon, which owns brands such as Leica Geosystems, produces high-precision 3D scanners for a variety of industrial applications (Fig. 6.32). For example, the Leica RTC360 is a high precision scanner, but its cost can also be significant.

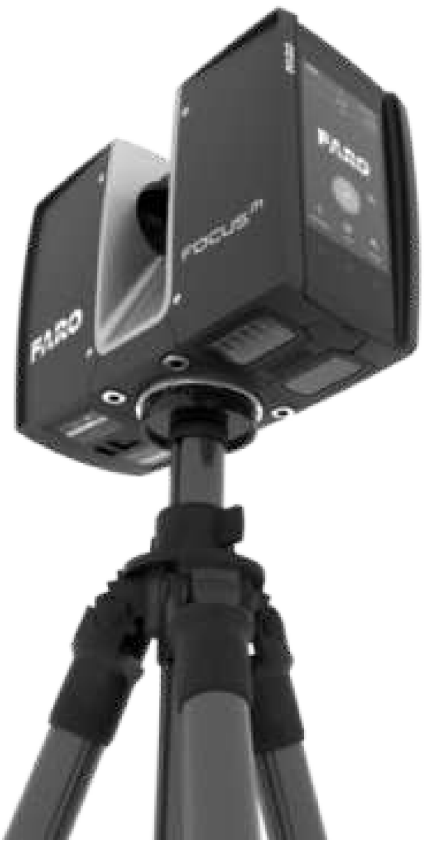

Fig. 6.31. FARO Focus — a powerful 3D laser scanner

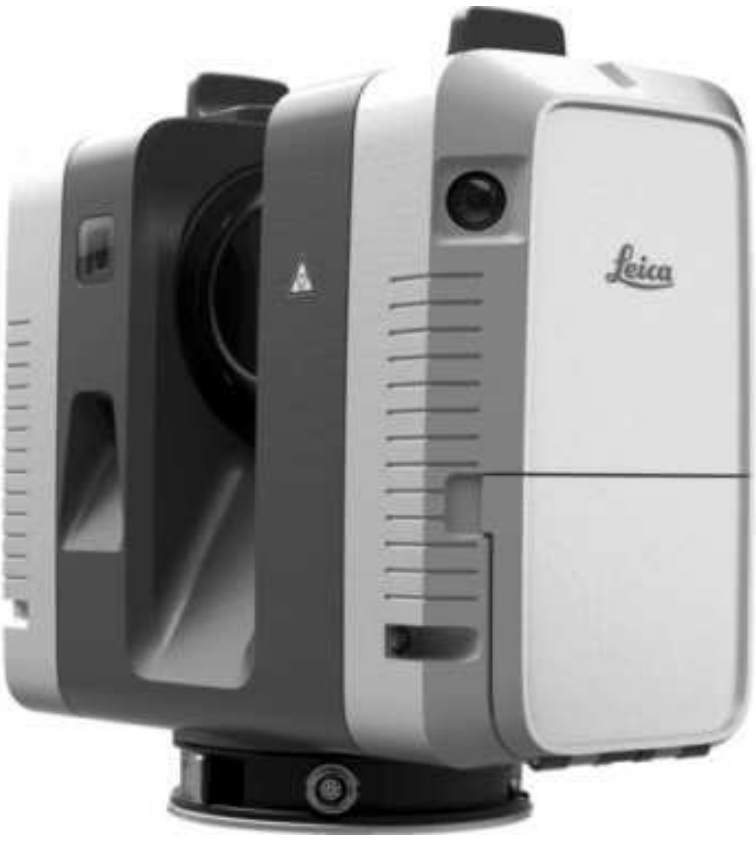

Fig. 6.32. Leica RTC360 laser scanner

***Artec 3D:*** Artec 3D specializes in the production of portable 3D scanners for creating high-quality 3D models. Their products, such as Artec Eva or Artec Space Spider, are highly accurate and designed for engineering and industrial applications (Fig. 6.33).

***Creaform:*** Creaform specializes in the development of high-resolution portable 3D scanners. For example, HandySCAN series scanners can be used to scan parts and process the surfaces of various objects, but their cost is also significant (Fig. 6.34).

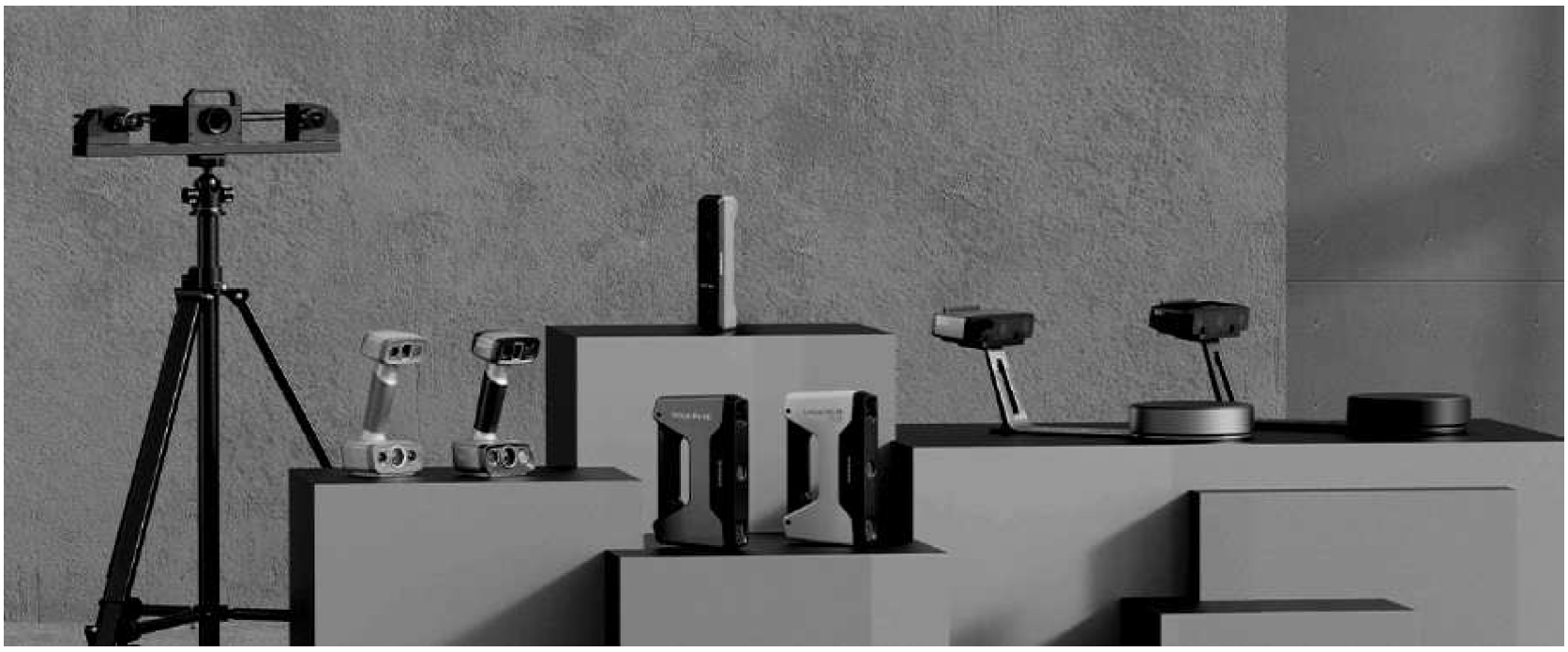

Fig. 6.33. Line of Artec Space Spider scanners

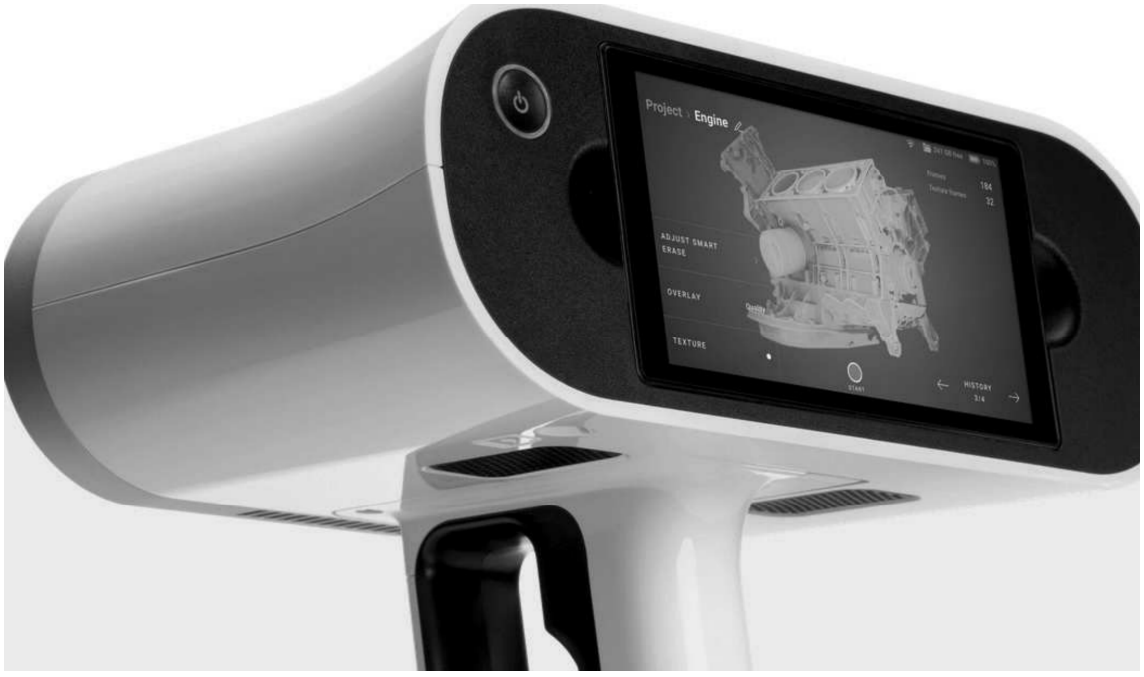

Fig. 6.34. HandySCAN series laser scanner

***EinScan:*** Shining 3D produces the EinScan series of scanners, including the EinScan Pro 2X and EinScan Pro 2X Plus, which range in price from \$5,000 to \$8,000 (Fig.6.35). These scanners are designed for a wide range of applications, from scanning objects to creating 3D models for industrial, design and educational use.

These companies offer high-quality industrial 3D scanners that can be used to scan objects in industrial, engineering, architectural and other fields. However, it is worth noting that in addition to the cost of scanners, additional costs may arise when purchasing the necessary software, training staff and maintaining equipment. Below are approximate prices for this equipment (Laser Scanners, Structured Light):

– Artec Eva Lite: from \$5,000 to \$7,000.

– Artec Eva: from \$15,000 to \$18,000.

– FARO Focus S150: from $30,000 to $35,000.
– Einscan Pro 2X: $4,000 to $5,000.
– FARO Focus S350: from $50,000 to $60,000.
– Leica BLK360: from $15,000 to $20,000.
– Leica RTC360: from $50,000 to $70,000.
– FARO Freestyle 3D: from $15,000 to $20,000.

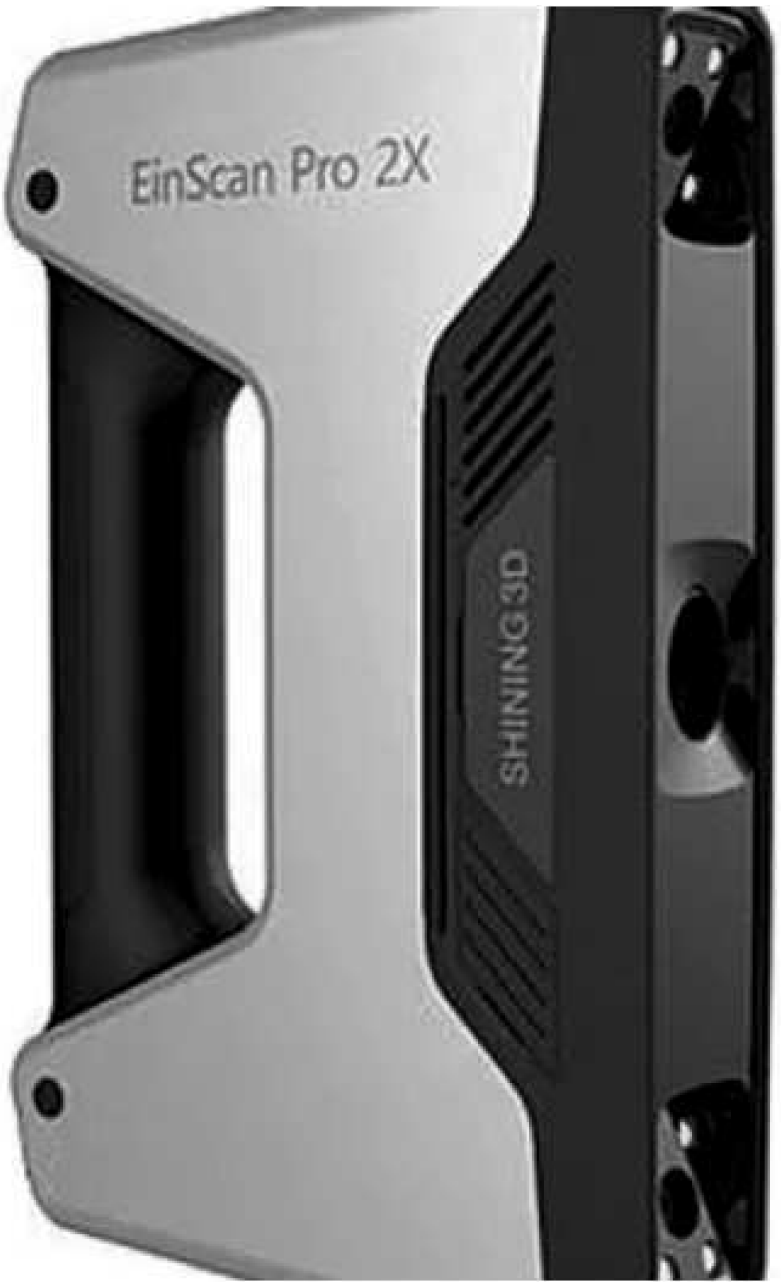

Fig. 6.35. EinScan Pro 2X portable scanner

The use of traditional 3D scanning devices can encounter a number of difficulties that can limit and complicate the scanning process [5, 8, 13, 19]. Here are some of them:

***Difficult to set up:*** Traditional 3D scanning methods often require complex setup and calibration of devices before use. This may involve installing special equipment, determining reference points, setting scanning parameters, etc. This requires specialized knowledge and can be time consuming.

***Limitations in Object Scanning:*** Some traditional 3D scanning methods may have limitations in what objects can be successfully scanned. For example, certain surfaces, textures, or materials may make it difficult to scan accurately or cause data corruption. This may limit your ability to scan certain types of objects.

***Limited Accuracy:*** Traditional 3D scanning techniques such as stereo photogrammetry or structured light may have limited accuracy in capturing 3D information. This may lead to inaccuracies and distortions in the resulting model.

***Time and complexity:*** Some traditional methods require a lot of time and effort to obtain 3D data. For example, scanning using stereo photogrammetry requires photographing an object from different angles, processing the images, and calculating reconstruction points. This can be a time-consuming and costly process.

***Limited mobility:*** Some traditional 3D scanning devices require the object or scanner to be fixed in a specific position. This may limit the ability to scan large objects or objects in hard-to-reach places.

***Dependency on environmental conditions:*** Traditional 3D scanning methods can be sensitive to lighting conditions and background, which can lead to problems in capturing accurate data. For example, using structured light may be difficult in bright sunlight or when there are reflective surfaces.

***Limited scanning area size:*** Some traditional 3D scanning methods may have limitations on the size of the object or scanning area. For example, scanning using stereo photogrammetry may be limited by the size of the camera or the distance between cameras.

However, 3D scanning technologies are constantly evolving, and modern devices such as laser scanning or photogrammetry-based methods can overcome some of these difficulties.

From the above it is clear that the listed equipment is not applicable at home, primarily for financial reasons. Because of this, due to the development of technology and the advent of more affordable equipment (for example, the Kinect device), scanning should gradually become more competitive, which will make this process more accessible to small-scale enterprises.

### *6.4.2. Using the Kinect device to scan models*

The Kinect device, developed by Microsoft, was originally intended for use in the Xbox gaming console. It is equipped with a depth camera, an infrared projector and microphones that allow it to recognize the user's movements and voice. Due to its spatial recognition capabilities, questions arise about its applicability for 3D scanning [23, 25, 28, 29, 31].

Although Kinect was not specifically designed for 3D scanning, its functionality can be used in this area. A depth camera, operating on the principle of structured light, allows you to obtain data about the geometry of objects in space. This data can be used to create 3D models of objects.

The idea of using the Kinect device to scan objects when creating 3D models is attracting wide interest due to a number of its features:

***Availability and Cost:*** Developed by Microsoft for the Xbox video game console, Kinect is relatively affordable and widespread, making it attractive as a potential 3D scanner for non-commercial use and small projects.

***Technical capabilities:*** Kinect has high resolution capabilities combined with a depth camera, making it suitable for scanning 3D shapes in high detail, including faces, objects and spaces.

***Software:*** There are various software solutions that allow the Kinect to be used as a 3D scanner. Programs like Skanect can turn Kinect into a powerful 3D model creation tool.

***Ease of Use:*** Kinect, thanks to its motion sensor concept, is available for home use, making it easier for hobbyists and enthusiasts to obtain 3D models.

Despite its advantages, it should be noted that the quality of scanning using Kinect may differ from more technologically advanced industrial scanners. However, for certain purposes and projects, Kinect can provide a cost-effective and easily accessible solution for creating 3D models.

However, it is important to note that the quality of 3D scanning using Kinect may be limited [30, 31]. The resolution and accuracy of the data obtained depend on the lighting conditions, the distance of the object, as well as the device itself. As a result, Kinect scanning may be more suitable for non-critical tasks, such as creating simple 3D models for gaming or educational purposes. Overall, we can say that the Kinect device has quite a lot of potential for use in 3D scanning, especially for simple tasks and general purposes.

It is advisable to use the Kinect device for 3D scanning in the following cases [25, 28, 31]:

***Basic 3D models:*** If you need to create basic 3D models for gaming, educational, or simple visualization purposes, the Kinect device can be a useful tool. For example, you can use it to create a 3D model of the human body or simple objects.

***Interactive development:*** If you are working on a project that requires fast and interactive 3D scanning, a Kinect device may be a convenient option. For example, to create virtual reality prototypes or gaming environments with 3D interaction.

***Educational Purposes:*** The Kinect device can be useful for teaching and practicing 3D modeling. It can be used in educational settings to introduce students to the basics of 3D scanning and 3D model creation.

***Home projects and hobbies:*** If you need an affordable and playful solution to fix a problem, a Kinect device may be the right choice. It can also be used to create 3D models for custom projects or for fun and experimentation.

In these cases, the Kinect device can provide a convenient and affordable 3D scanning solution. However, its limitations in data accuracy and resolution, as well as possible limitations in software and support, must be taken into account.

The scanning technology used to create 3D models using the Kinect device has some features that may prevent high-quality results from being achieved in certain scenarios. Here are some reasons why this technology may be limited in creating quality 3D models [23, 24, 26, 27, 29]:

***Limited Accuracy and Detail:*** Built-in Kinect sensors may not always provide sufficient scanning accuracy, especially for objects with fine detail or complex textures. This results in a loss of detail and quality in the resulting 3D models.

***Limited Resolution:*** The Kinect, as a gaming console device, was designed for different purposes. Its resolution may not be sufficient to scan small objects or reproduce fine details with high accuracy.

***Limited dynamic range:*** Kinect typically has limited dynamic range, which can affect the quality of scanning objects with varying shades and brightness.

***Lighting and Environment Complications:*** The Kinect may have difficulty scanning in low-light conditions, changing backlighting, or complex background structures, which can lead to distortions and errors in the generated 3D model.

So, depending on your needs and quality standards, Kinect scanning technology may not always provide the required level of accuracy and detail for certain applications and scenarios that require high-quality 3D models.

### *6.4.3. Kinect device design*

The Kinect device consists of several key components that work together to provide spatial scanning, motion tracking, and data collection functionality (Fig. 6.36). Here are the main parts of the Kinect device [23, 28]:

***Infrared Projector:*** The infrared projector in Kinect emits infrared rays, which are used to create a depth map of the environment. This allows the device to determine the distance to objects in the field of view.

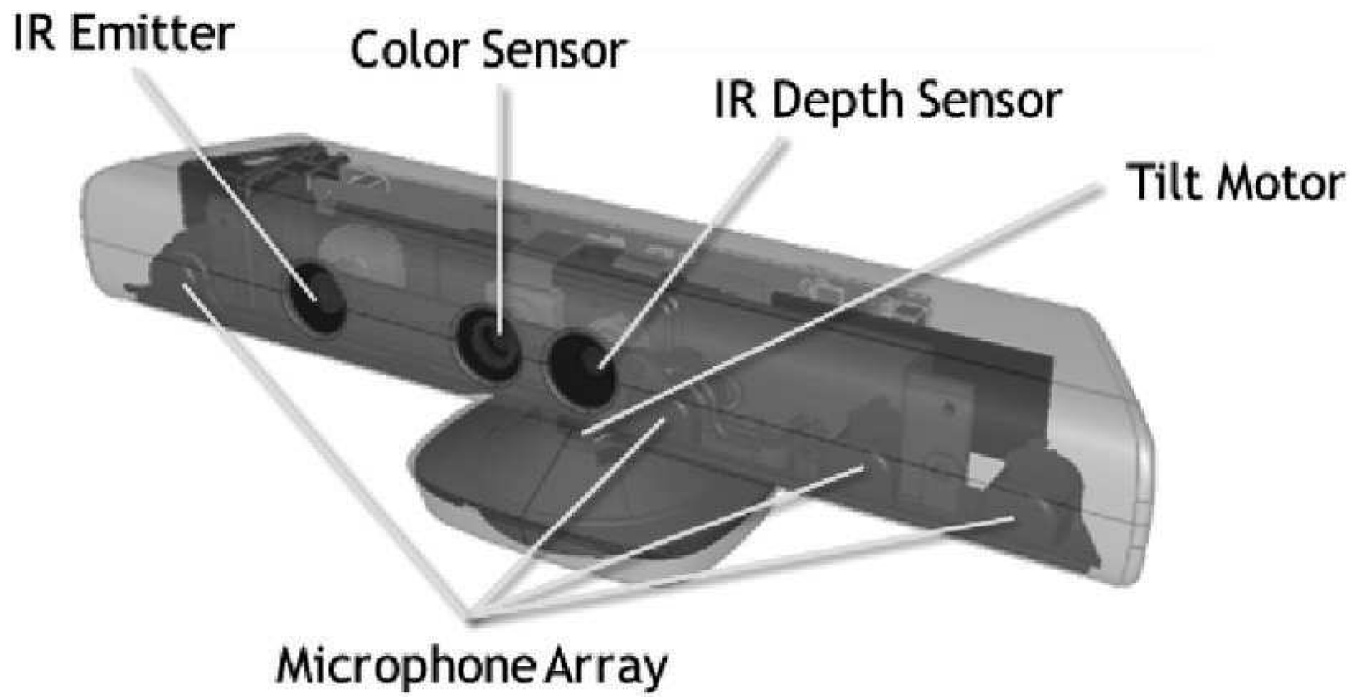

Fig. 6.36. Kinect device structure

***Infrared Camera:*** Kinect also has an infrared camera that captures reflected infrared rays. This data is used to create a depth map.

***RGB Camera:*** The Kinect also includes an RGB camera that captures a color image of objects in the device's field of view. This allows the device to obtain information about the color and texture of objects.

***Infrared motion sensor:*** To track the movements of the human body and other objects in space, Kinect also includes an infrared motion sensor. It provides the ability to determine position and movements in real time. Microphones: Kinect is also usually equipped with microphones, which allow you to read audio data and use voice commands to control the device.

How the Kinect device works. These internal devices work together to read information about the shape, depth, color and movements of objects in space. Once the data is received, it is processed by software that converts the data into a format suitable for specific purposes such as virtual reality, 3D modeling, gaming and many other applications.

When working, Kinect collects data from an infrared camera, an infrared projector and microphones, and then processes it using special computer vision and deep learning algorithms. The result is a three-dimensional model of the scene, detection and tracking of user movements, and recognition of voice commands.

Using these devices together allows the Kinect device to collect information about the environment and objects in real time, create depth maps, track movement, and provide data for a variety of applications and purposes, including scanning to create 3D models (Fig. 6.37).

Although marketed as a "sensor" for the Xbox, the Kinect is much more than that, actually containing an RGB video camera with a resolution of 640x480 pixels; in combination with an additional video camera with a lower resolution (equal to 320x240 pixels), that is, sensitive only to infrared radiation. This second video camera is used to detect images created by displaying infrared rays projected onto the scene viewed through a special infrared projector built into the front panel of the device (Fig. 6.37). In addition to the Kinect video sensors, there is a three-axis accelerometer to detect possible vibrations and movements to which it is exposed. But that's not all, as Kinect offers four microphones: these are arranged radially, working by using the mapping of sound waves in a room to calibrate the environment in which it is placed. In addition to so many sensors, Kinect integrates the mechanism used to move up and down to better track the movements of the people operating it. As for communication with the outside world, the Microsoft Kinect uses a cable similar in shape to a USB cable, from which it differs in the number of connections. In fact, a USB cable usually consists of 4 wires (two for power and the other two for serial communication), while a Kinect cable consists of as many as 9 poles; therefore, to connect to it, you must use an adapter to provide an auxiliary 12V power supply in addition to the regular USB connections.

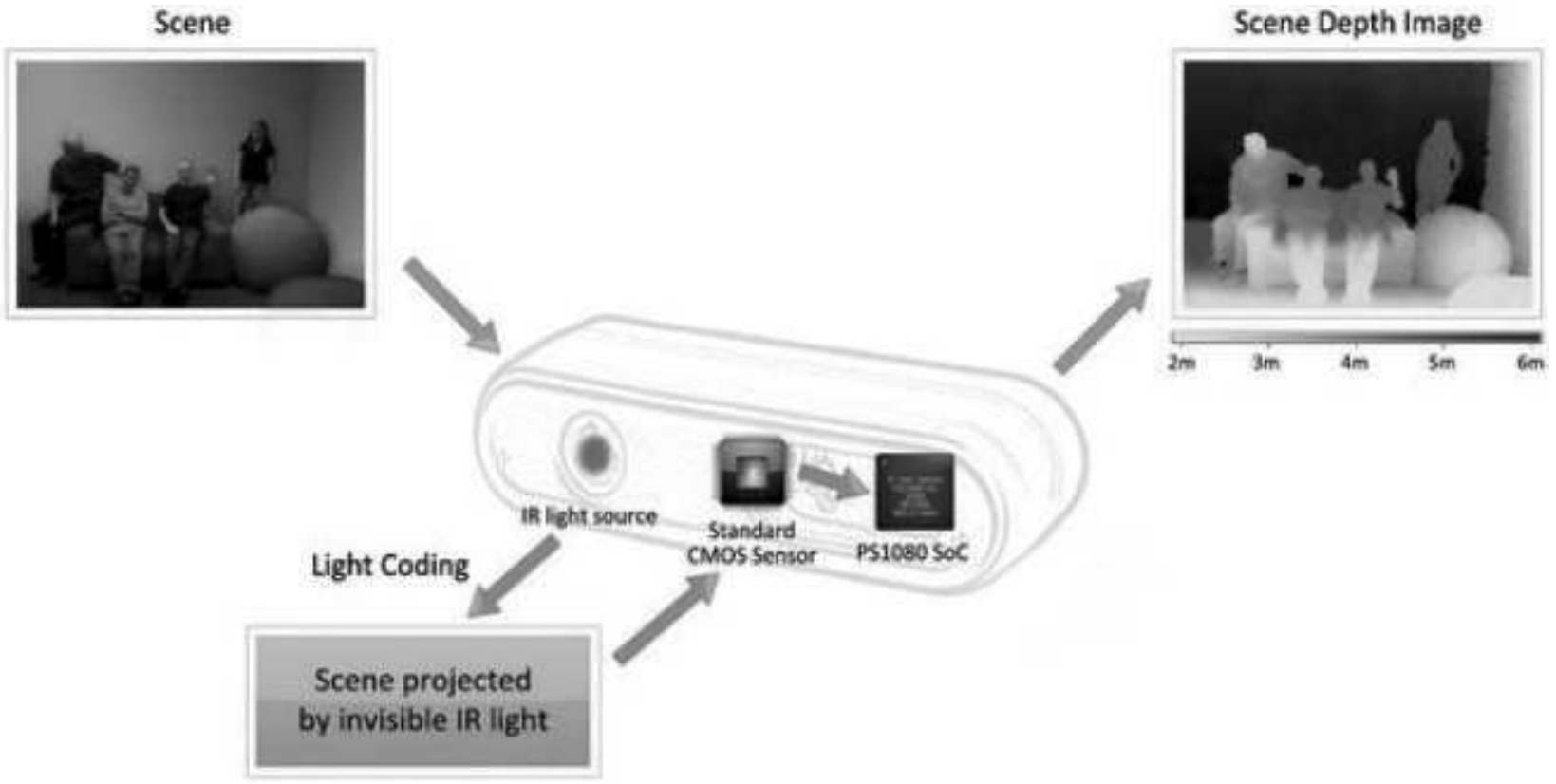

Fig. 6.37. Illustration of the principle of operation of the Kinect device

Kinect has the unique ability to "see" in 3D. Unlike most other computer vision systems, the Kinect system is able to build a depth map of the area in front of it. This map is created entirely on the touch panel and then transmitted over a USB cable to the host in the same way as a regular camera image, except that instead of the color information of each pixel in the image, the sensor transmits a distance value.

You might think that the depth sensor uses some kind of radar or ultrasonic transmitter to measure how far things are from the sensor bar, but that's not the case. This would be difficult to do over a short distance. The sensor uses a clever technique consisting of an infrared projector and a camera that can see the tiny dots that the projector creates (Fig. 6.37).

Kinect works like this: an infrared projector casts a very precise pattern made up of many equally spaced points into the space in front of it. By analyzing an image captured by a camera sensitive to such IR rays, Kinect determines the distances between different points, thereby obtaining the distance and tilt of the illuminated object. Since the IR beam comes out from a limited area and expands as the distance increases, where the points appear very close, it means that the object is close to the sensor, and vice versa, if the distance between the points is significant, it means the object is further away. By summing up all the information, one can obtain the spatial configuration of the cordon, that is, the shape of the object.

In its operation, Kinect uses a method called "structured light", described in section 6.2 (Fig. 6.5). The idea is simple — if you have a light source offset a small distance from the detector, then the projected spot of light is offset according to the distance from which it is imaged. So, by projecting a fixed grid of dots onto the scene and measuring how much each dot moved when viewed with a video camera, you can determine from what distance each dot was displayed. The actual details are more complex because the IR laser in Kinect uses a hologram to project a random pattern onto the scene. It then measures the displacement of each point to create an 11-bit depth map. A special chip performs the calculations involved in turning a point map into a distance map.

In general, Kinect v1 and Kinect v2 are a line of devices developed by Microsoft, designed for motion recognition and interactive gaming. Both devices offer unique features and differ in some specifications.

Kinect v1 was released in 2010 and was the first commercial product capable of detecting user movements without the use of a controller. It is equipped with a camera, an infrared projector and a microphone, which allows it to detect a person's position and gestures in three-dimensional space. Kinect v1 has been widely used in the gaming industry, providing a unique gaming experience on the Xbox 360 (Fig. 6.38).

Kinect v2, released in 2013, was a vastly improved version over its predecessor. It had higher camera resolution, improved motion recognition accuracy, and an expanded feature set. Kinect v2 was also supported on the Xbox One platform, providing players with an even more interactive gaming experience (Fig. 6.39).

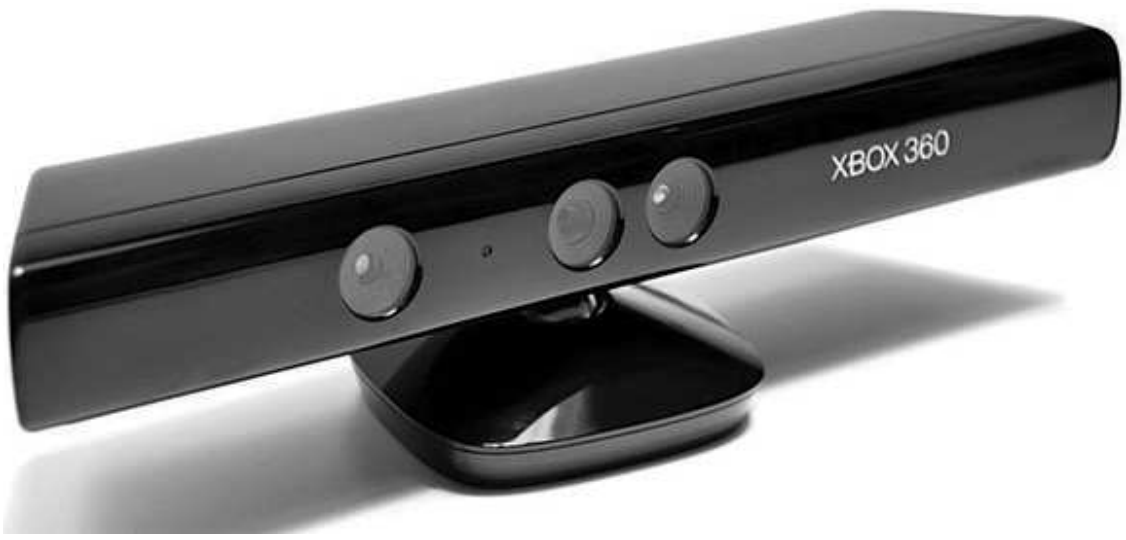

Fig. 6.38. View of the Kinect v1 device

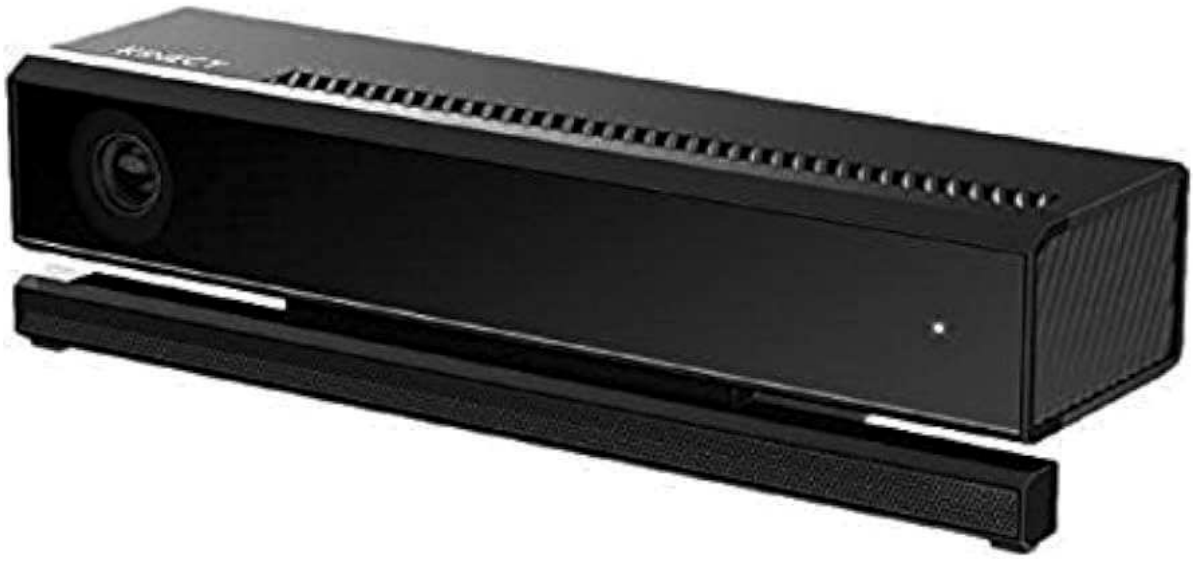

Fig. 6.38. View of the Kinect v2 device

Both Kinect devices offer unique gaming and entertainment experiences. They allow users to control game characters, interact with virtual objects, and perform various physical exercises without requiring the use of controllers. This has brought new opportunities to the gaming industry and has become one of the important innovations in the entertainment industry.

Although Kinect v1 and Kinect v2 are no longer manufactured or supported by Microsoft, their influence on the development of motion recognition technologies remains significant. The main differences between the Kinect v2 and Kinect v1 models:

***Resolution and Image Quality:*** Kinect v2 has improved resolution and image quality compared to Kinect v1. Kinect v2 has an RGB camera with a resolution of 1920x1080 pixels and can record video at 1080p resolution. While Kinect v1 has a resolution of 640x480 pixels and records video at 640x480 resolution.

***3D scanning accuracy:*** Kinect v2 provides more accurate 3D scanning compared to Kinect v1. It uses advanced Time-of-Flight depth technology with higher resolution and advanced point tracking system.

***Skeletal tracking support:*** Kinect v2 features more advanced skeletal tracking technology. It allows you to accurately determine a person's posture and recognize his movements in real time with higher accuracy than is possible with Kinect v1.

***Improved voice recognition:*** Kinect v2 has more advanced voice recognition, which improves the quality and accuracy of commands recognized by the device.

***Performance:*** Kinect v2 has more powerful hardware and improved processing for better performance than Kinect v1.

Overall, Kinect v2 is a significant improvement over Kinect v1 in specifications, resolution, and accuracy. However, both devices have their own advantages and can be used in various fields, including gaming, research, virtual and augmented reality application development, as well as in medical and industrial fields. In this work, the Kinect v2 device was used to build the required scanning technology.

### *6.4.4. Scheme of using Kinect for scanning purposes*

When using Kinect in 3D scanning, the device captures data about the shape and depth of objects, then this data is processed by special programs (for example, scanning programs mentioned earlier, such as Skannect, ReconstructMe and 3D Builder), which convert this data into point clouds or three-dimensional models of objects in digital form (Fig. 6.38).

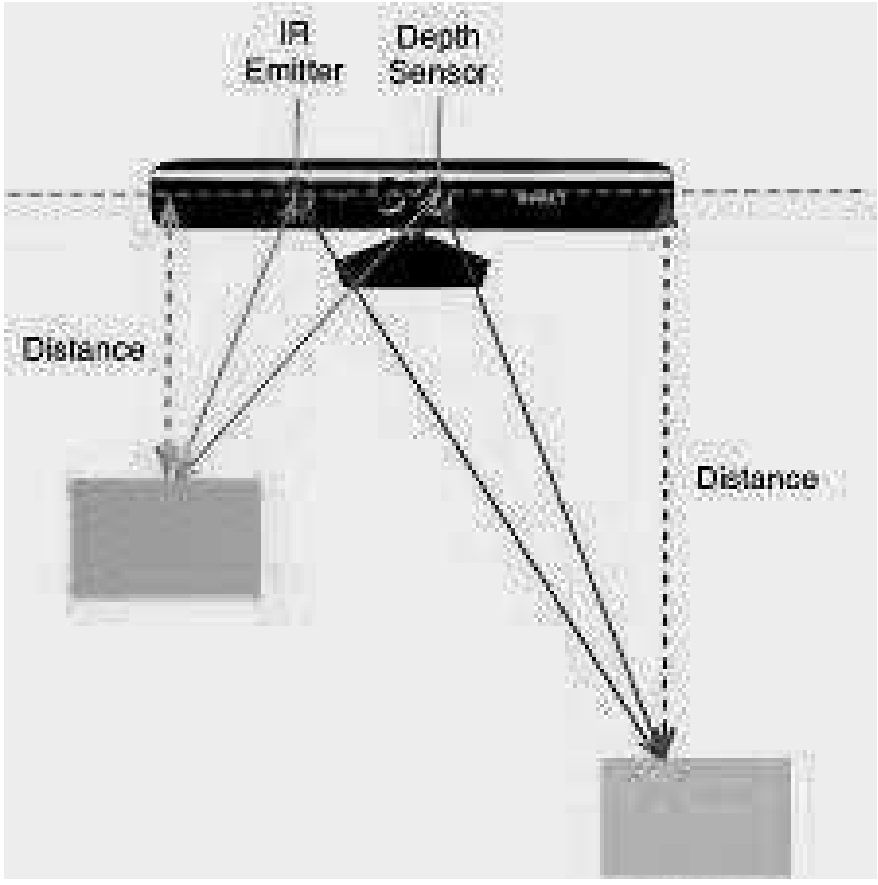

Fig. 6.38. Scheme of using Kinect for scanning

Due to its broad functionality and relative affordability, Kinect has become a popular tool for 3D modeling and scanning enthusiasts, and has found widespread use in academic and professional environments for a variety of purposes, including student projects, medical research, virtual reality, and entertainment applications.

Here is a general detailed diagram of using Kinect v2 [23, 30, 31] for scanning purposes (Fig. 6.38):

***Initialization:*** The Kinect device is connected to a computer or other device using a USB cable.

***Depth Sensing:*** The Kinect uses an infrared projector and an infrared camera to measure the depth of a scene. The projector projects a grid of infrared dots onto objects in the scene, and the camera records their distortion, allowing you to determine the distance to each dot.

***Color detection:*** The Kinect device also uses an RGB camera to capture color information about the scene. An RGB camera records the color of every point in the scene.

***Data Synchronization:*** Depth and color data are synchronized to create a 3D model of the scene. Each point in the model has XYZ coordinates (depth, width, height) and color information.

***Point cloud generation:*** The acquired depth and color data are combined into a single point cloud representing a 3D model of the scene.

***Data Processing:*** The Kinect device can apply data processing algorithms to remove noise, level, and smooth the model.

***Model Export:*** The processed model can be exported to various file formats such as.obj or.stl for further use in 3D modeling or printing software.

The framework outlined provides general steps for using a Kinect device for 3D scanning and developing interactive applications, but the specific steps and requirements may vary depending on the software used and the purpose of use [5, 28, 30, 31]. Here are some of the schemes used:

1) ***Scheme of using Kinect for 3D scanning in medicine:*** a doctor or medical specialist installs the Kinect device in the required place in a room or operating room, the patient poses in front of the Kinect, standing or sitting in a certain position. The Kinect scans the patient, collecting data about their bodily structure and shape, which is transmitted to a computer or medical equipment for analysis and interpretation. The medical professional uses the resulting 3D scans for diagnosis, surgical planning, or to create models for the design of specialized medical devices such as prosthetics or orthoses.

2) ***Scheme of using Kinect for 3D scanning in architecture and construction:*** an architect or engineer installs a Kinect device in the desired loca-tion, for example, on a construction site or indoors. The Kinect scans the environment, collecting data about the size, shape and texture of objects, which is transferred to a computer or CAD program to create an accurate 3D model. An architect or engineer uses a 3D model to design and visualize a building or structure; a 3D model can be used for calculations or planning construction work.

3) ***Scheme of using Kinect for 3D scanning in the entertainment industry:*** Kinect devices are installed in an entertainment center or game studio to create an interactive gaming experience. Users stand or move in front of a Kinect device, which scans their movements and body shape, which is used by software to create characters or images that mimic the users' movements and shape in real time. Players can interact with created characters or images by playing games, dancing or participating in other entertaining activities.

To work with scanning objects when using the Kinect device, you can use several computer programs that allow you to process the data obtained using the Kinect and create three-dimensional models. Here are a few programs that are often used to work with Kinect when scanning:

***Skannect:*** Skannect is software that allows you to use your Kinect to scan objects and create 3D models. The program has a simple and intuitive user interface that allows scanning and processing of data.

***ReconstructMe:*** This is a program that allows you to use your Kinect to scan objects and create 3D models. It has extensive functionality, including the ability to create high-quality 3D models from scanned data.

***3D Builder:*** 3D Builder is a free application from Microsoft that includes tools for working with 3D models. Using Kinect, you can scan objects and transfer the resulting data to 3D Builder to create and edit 3D models.

***MeshLab:*** MeshLab is free and open source software for processing and editing 3D meshes. It can be used to process data captured by Kinect and create complex 3D models.

***Blender:*** Blender is a powerful and free 3D modeling and animation tool. Using plugins or add-ons for Blender, you can integrate the data captured by Kinect and use it to create 3D models and animations.

These programs provide the ability to process Kinect data and create high-quality 3D models. Each of them has its own unique functions and capabilities, so the choice of program depends on the specific needs and requirements of the project.

Using a Kinect scanning device has several potential disadvantages that should be considered when choosing this device for creating 3D models. Here are a few main disadvantages:

***Limited Accuracy and Detail:*** Built-in Kinect sensors may not always provide sufficient scanning accuracy, especially for objects with fine detail or complex textures. This can result in a loss of detail and quality in the resulting 3D models.

***Limited Resolution:*** The Kinect, as a gaming console device, was designed for different purposes. Its resolution may not be sufficient to scan small objects or reproduce fine details with high precision.

***Limited dynamic range:*** Kinect typically has limited dynamic range, which can affect the quality of scanning objects with varying shades and brightness.

***Lighting and Environment Complications:*** The Kinect may have difficulty scanning in low-light conditions, changing backlighting, or complex background structures, which can lead to distortions and errors in the generated 3D model.

***Limited scene size:*** Kinect has limitations on the size of the scene that can be scanned, which means that scanning large objects may require multiple scans and then merging them together, which can make the process more complex.

***Limitations with transparent objects:*** The Kinect has difficulty scanning transparent or mirror-like objects due to reflections and opacity.

All these factors can affect the quality and accuracy of the resulting 3D models, especially in the case of projects that require detail and accuracy. In such cases, it may be preferable to use specialized industrial 3D scanners with high accuracy and resolution to achieve the required level of quality.

### *6.4.5. Kinect software analysis*

As already noted in subsection 6.4.3, a number of programs can be used to process data coming from the Kinect device: Skannect, ReconstructMe, 3D Builder, MeshLab, KScan3D. A comparative analysis of them leads to the conclusion that the most acceptable and simplest program for scanning purposes is the KScan3D program [30, 31]. KScan3D is software designed to scan objects with a 3D scanner and create 3D models. Here are some ways to use KScan3D:

***Create digital models of objects:*** KScan3D allows users to scan physical objects and create high-quality 3D models of those objects. This can be useful in areas such as engineering design, medical modeling, prototyping, and animation character creation.

***Recreating objects for virtual reality and animation:*** Using KScan3D, you can scan real objects for subsequent recreation and use in virtual environments and animation projects. This can be useful for virtual museums, computer games, animated films and visualization projects.

***Sculpting and anatomy exploration:*** KScan3D allows you to create digital sculpture models and study the anatomy of various objects, including the human body. This can be useful for artists, students and educational institutions.

***Restoration and Preservation of Cultural Heritage:*** In the field of cultural heritage, KScan3D can be used to restore and document historical artifacts, monuments and architectural structures for the purpose of preserving and researching historical sites.

***Education and Training:*** The program can be used for educational purposes to teach students the basics of 3D modeling, scanning and visualization. Students can use KScan3D to create educational projects and research.

In addition, KScan3D can be used in combination with other software and hardware to further analyze, visualize and develop projects that require 3D models and object scanning. KScan3D is a program for scanning 3D objects using multiple Kinect sensors. Here are some of its features:

***Multiple Kinect Sensors:*** KScan3D supports the use of multiple Kinect sensors simultaneously, which allows you to get a more complete and accurate 3D model of an object.

***Real-time scanning:*** the program allows you to scan objects in real time, which allows you to see the scan results directly as you work.

***Automatic Reconstruction:*** KScan3D has an automatic reconstruction feature that allows you to quickly and easily create 3D models without having to manually adjust scanning parameters.

***Editing and optimizing models:*** the program offers tools for editing and optimizing created models, including removing noise, filling holes, smoothing the surface and other operations.

***Data Export:*** It is possible to export scan data in formats such as PLY, STL and OBJ, allowing scan results to be used in other applications and software environments.

***Point cloud import and fusion:*** KScan3D allows you to import point clouds from other programs or scans, and then combine them into one model to get a more complete representation of the object being scanned.

***Calibration and Tuning:*** The program offers tools to calibrate and tune Kinect sensors to get the best scanning results.

These are just some of the features of KScan3D, which provides tools for accurate and high-quality scanning of 3D objects using Kinect sensors.

The main screen of the KScan3D program is shown in Fig. 6.39.

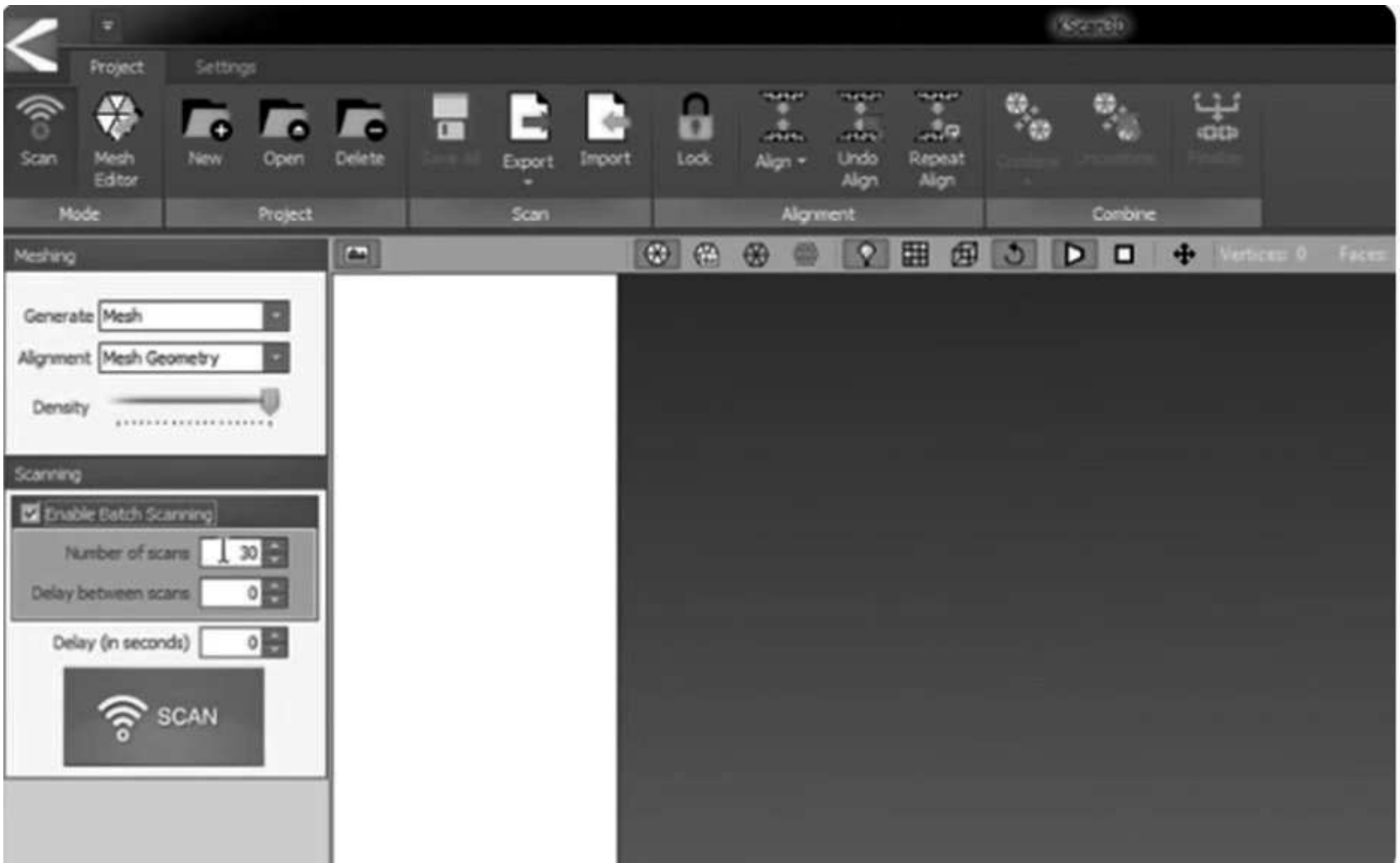

Fig. 6.39. View of the main screen of the KScan3D program

The following menu options provide access to various functions and settings in the KScan3D software, allowing users to control scanning devices, process data and manage projects efficiently.

The File menu contains options such as Open, Save, and Export. They allow you to manage processed projects and export models to various file formats for use in other applications or for sharing.

The Edit menu may contain tools related to editing processed models, such as optimization options or additional adjustments to processed data. It can provide options for improving the 3D model after the scanning and initial processing steps.

View menu options can allow you to visualize rendered 3D models from different angles, adjust rendering settings, or explore the model using specialized viewing options.

The 3D Model menu may contain options related to 3D mesh post-processing, such as mesh simplification, smoothing, or other modifications to improve the quality and detail of the rendered model.

The Tools menu can include a variety of post-processing tools, such as alignment and registration tools, measurements, or other utilities to help improve your rendered 3D models.

The Help menu typically provides access to a program's user guide, technical support, or other resources that offer guidance on how to effectively use processing tools and functions.

Using these menu options, users can efficiently process scanned data, enhance 3D models, and prepare them for use in a variety of applications, including 3D printing, rendering, or integration into other design workflows.

To effectively use the KScan3D program and perform the process of scanning and processing 3D models, it is recommended that the computer meet certain minimum and recommended specifications [30,31]. Here are the general requirements for computer characteristics when using the KScan3D program:

***Minimum requirements:***

1. Processor: Multi-core processor such as Intel Core i5 or equivalent.
2. Memory: 8 GB of random access memory (RAM).
3. Hard disk: 25 GB of free hard disk space.
4. Graphics Processing Unit: Discrete Graphics Processing Unit (GPU) supporting DirectX 11 or higher.
5. Ports: USB ports for connecting a scanning device.
6. Operating system: Windows 7, Windows 8, Windows 10.

***Recommended requirements:***

1. Processor: Intel Core i7 or more powerful.
2. Memory: 16 GB of random access memory (RAM) or more.
3. Hard Drive: SSD for faster loading and processing of data.
4. Graphics Processing Unit: A discrete graphics processing unit (GPU) with higher performance and more memory, such as NVIDIA GeForce or AMD Radeon professional-grade series.
5. Monitor: It is recommended to use a high resolution monitor for easier viewing of 3D models and scan data.
6. Ports: USB 3.0 ports for faster data transfer when using more modern scanning devices.
7. Operating system: Windows 10 for optimal compatibility and performance.

Specifications may need to be adapted depending on the specific KScan3D software configuration, version and the requirements of your specific scanning project.

### *6.4.6. Available scanning technology using a Kinect device*

To check the top technology scanning of fixed models with powerful equipment The Kinect system was then used to create 3D models of human breasts and e and print to the printer. Before this, the experiments must be carried out before the careful parameters are determined Bots Kinect, powerful programming tools, display and program slicer (to set the optimal properties of 3D printing with plastic). Also, we recommend the creation of 3D human body models, This can be used in the production cells (napprimer, for and the birth of individual clothing).

As we have seen in in subsection 6.2, the advanced 3D scanning feature is available Or if you choose any object in the short term or with good quality Detailed details can be used to create a high-quality scanner with high-quality equipment With scanning modules with high sizing and speed analysis in the result Scanning takes place in the same way A unicorn form a cloud that can produce manipulation throughout the world The 3D editor is installed or in the previous program.

To obtain three-dimensional computer 3D models (mainly large objects, for example, the human body), available scanning technology is offered:

• One of the Kinect-compatible image acquisition programs is installed on the computer, as well as the corresponding drivers (KScan3D v1.2 x64 bit was used in this work);

• The Kinect device is installed at a height of 1-1.5 m, suitable for a person Scanning should be less than 1.5 m (Fig. 6.40);

• If you are a human being, you will work on your own, and the assistant will give you the tips for you all the angles of the body;

• As a result of scanning in the program, all measurement data in the form of a point cloud are analyzed and displayed on the screen as a three-dimensional image;

• The partial display can be viewed from the selected point in the program menu;

The image as a file is transferred to a graphics editor (for example, Blender) for further finishing processing;

• The result in the form of an stl file is transferred to a slicer (for example, Cura), where the image is translated into G-code to control printing on a 3D printer;

• Printing an object on an inexpensive printer using FDM technology (for example, as in this work, on an Anycubic Kossel printer).

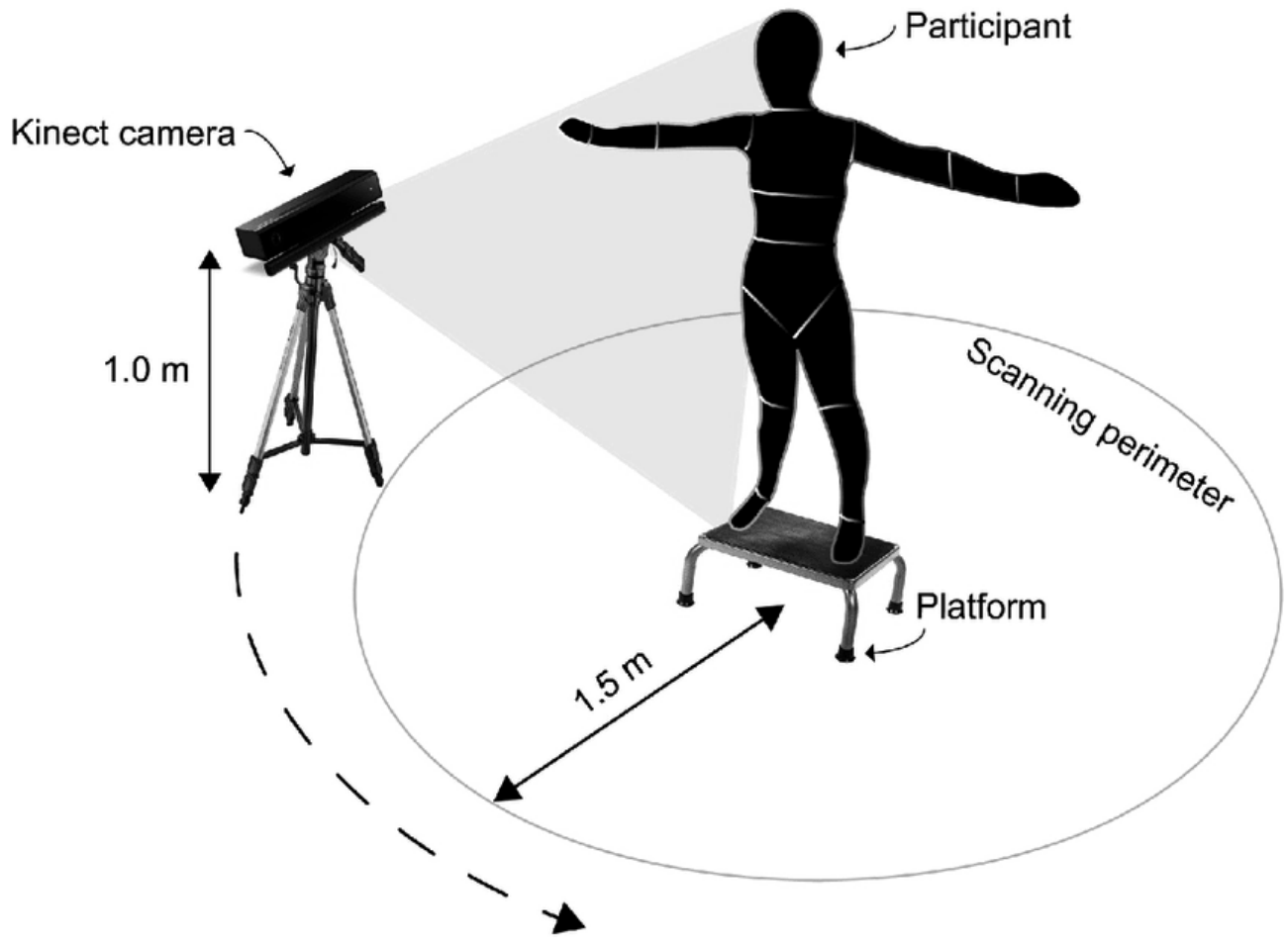

Fig. 6.40. Scheme for the distribution of the Kinect device and the human being before scanning

The launch of Kinect in the scanner has a detailed description [23, 25, 28, 31], so please Do not use different types of devices and technology. In this work, an attempt is made to modify a fairly cheap technology for creating 3D models using the Kinect device and shareware available programs. The definition is based on the primary scanning of the breasts of a person, in the future Illuminated 3D models and printed on a 3D printer. Please note that the Kinect scanner is used for this purpose All of these products, it will be difficult to obtain a three-dimensional high-quality image of a pen or key fob.

Because the Kinect was designed to capture the entire figure to detect its movements, the optics and firmware that produced the depth data were calibrated for volume. Kinect allows you to grow the face, but not only the characteristics of the face, but also the characteristics of the face This means that when the cards are expanded, they store 320x240 pixels. This is also a driving trail with a view of the screens, and in the same place The creation of the program, the efficient scanning, was written to Tom, the application This is how the graphic processors are presented to the video card board, and then cut This means that you can have the device with the eyes of the collar in the second. The software essentially has to deal with tracking objects, adding new points to the point cloud and linking them to previous ones. In practice, when the object is created, the program expresses the main point

Because of the power of the new glubin card, it has a “connection” to it This means that it is not a fixed folder.

If you pay attention to the scan, you should assemble the location with the instructions provided aracteristics. The Kinect needs to be placed on a stable tripod so that it stands at face height or at least bust height. The person should take care of each other's axis (or improve) as well Listen to the eyes of your body in all the stories (Fig. 6.40). The scanning of the Fig.can be realized by two people: it is located between Kinect and When moving around with the controller around the non-moving object, the tool is used this is the program.

As a result of experiments, the general scheme of scanning using Kinect was supplemented with some recommendations outlined below

When scanning an object using the proposed technology, it is important to take into account certain requirements for the location of the object and the Kinect device itself. This will help ensure the accuracy and quality of the data obtained.

The object must be within the Kinect device's field of view, which means that it must be positioned in such a way that the entire object is visible and scannable. It is important that the subject is well lit, preferably natural light or a uniform light source. This will make the scanning process easier and provide better results. When scanning, you should avoid strong reflections from the surface of the object or the presence of shadows, as this may affect the accuracy of the scan and the quality of the resulting data.

The Kinect device must be mounted in a stable position to ensure stable scanning and minimize potential data corruption. The distance between the Kinect device and the scanned object is also important, a distance of 1.5 - 2 m is recommended. You also need to ensure that the Kinect is located in a suitable angle and at such a height to cover the entire scanned object and from different angles.

Following these site and Kinect placement requirements will help ensure efficient and accurate scanning, ultimately resulting in high-quality 3D object models.

You don't need any special lighting per se to successfully scan objects with your Kinect device, but there are some guidelines that can help ensure a more efficient and accurate scanning process. It is advisable that the subject be illuminated with uniform light without strong shadows, since uneven lighting or strong shadows can lead to distortions in the resulting scans and degrade the quality of the data. In addition to shadows, strong glare or reflections on an object can cause problems for the Kinect device as it uses infrared light to create depth maps, so you should avoid using objects with shiny surfaces or where there is strong light reflection. Also, using natural light can make the scanning process easier, as natural light often provides uniform, pleasant illumination without direct glare. In general, a well-lit room with a minimum of strong shadows and glare can make the scanning process using your Kinect device much easier and better.

When scanning objects using the proposed technology, the number of images required depends on several factors, including the size of the object, the desired detail of the scan and the specific requirements of the process.

Typically, Kinect devices provide real-time scanning, which means that as the device and object move, data about the shape and depth of the environment is collected, creating a 3D model. Therefore, the specific number of shots may be difficult to determine accurately.

However, if we take a more traditional approach, it is usually recommended to collect data from different angles and angles to create an accurate and detailed 3D model of an object using a Kinect device. This may involve rotating an object or device around an object to ensure complete coverage of the object on all sides.

The total number of images may vary depending on the specifics of the object, detail requirements and features of the data processing software. In most cases, the number of images will be determined during the scanning process to ensure that enough data is obtained to create a high-quality 3D model. It is generally recommended to scan from multiple angles and angles to ensure that you have enough data to create a complete and accurate 3D model of the object.

After scanning an object using the Kinect device and KScan3D software, you may need to perform some additional steps to improve the quality of the resulting 3D model. Here are a few possible modifications that may be needed.

The resulting scan data may contain noise or unwanted artifacts, especially if the scan was performed in conditions with limited visibility or lighting. To improve the quality of the model, it may be necessary to use specialized software to remove noise, smooth the surface, and eliminate unwanted elements.

If the scan was carried out from different angles or in several parts, a procedure may be required to combine the data and align individual scan fragments. This will allow you to create a single and complete 3D model of the object.

Sometimes the resulting model may have an excessive number of polygons (triangles), which can slow down processing and creation of visualizations. In this case, a retopology process will be required to optimize the model's geometry, reducing the number of polygons, while maintaining its shape and details.

If texture or color data of an object was captured during the scanning process, additional processing of this data may be required to improve the quality of the texturing and complete the appearance of the model.

Sometimes scanning may result in defects, such as a missing part of an object or distortion. In such cases, it will be necessary to check the model for defects and their subsequent correction using specialized tools.

It is important to understand that each scan may have unique features, and not all of these additional steps will always be necessary. However, given the potential for artifacts, noise and imperfections in the scanning process, these additional improvements can significantly improve the quality and accuracy of the resulting 3D model of the object.

### *6.4.7. Creating a test 3D model of a human torso using Kinect*

As already indicated, in this work, in order to verify the scanning results using the Kinect device, an attempt was made to create a 3D model of the human torso.

The KScan3D program was used for scanning; pictures were taken when a person rotated around an axis in front of the Kinect device (Fig. 6. 41). The man sat on a chair, turning around his axis, and an assistant used a device to take pictures.

One thing to remember when shooting is that depth is controlled by designing a very complex and dense dot matrix using infrared light. If something is not displayed clearly, the matrix will therefore create problems with data collection (this explains why you don't want to have reflective or bright objects on your clothing).

In experiments to obtain test data, the Kinect scanning device was mounted rigidly using a tripod at the level of a person's head (Fig. 6.42), at a height of approximately 1.4 m, with an inclination of approximately 10 degrees (when the Kinect is slightly at an angle, the cloud points turns out to be more detailed).

After starting a scan with the KScan3D application you are using, you need to start rotating slowly until the rotation is complete: the program will detect this and give a combination of different notes to indicate that it has examined the entire surface. The most important thing at this stage is to find a position where the sensor works best and can reproduce reasonable detail of what was in the frame. In order to find the best position for the Kinect, I had to take several series of pictures at different positions of the sensor, choosing the best angle. During scanning, the person rotated around its axis — 18 degrees per rotation, so 20 images were taken per full rotation. These movements were recorded using images in the KScan3D program (Fig. 6.43).

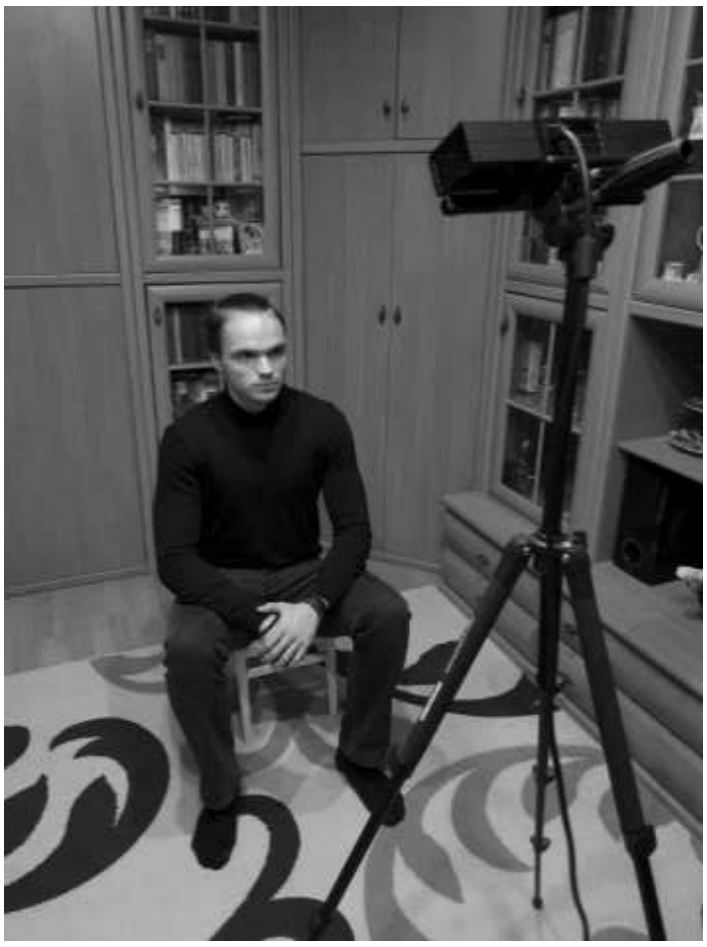

Fig. 6.41. Location of the object and Kinect device during scanning

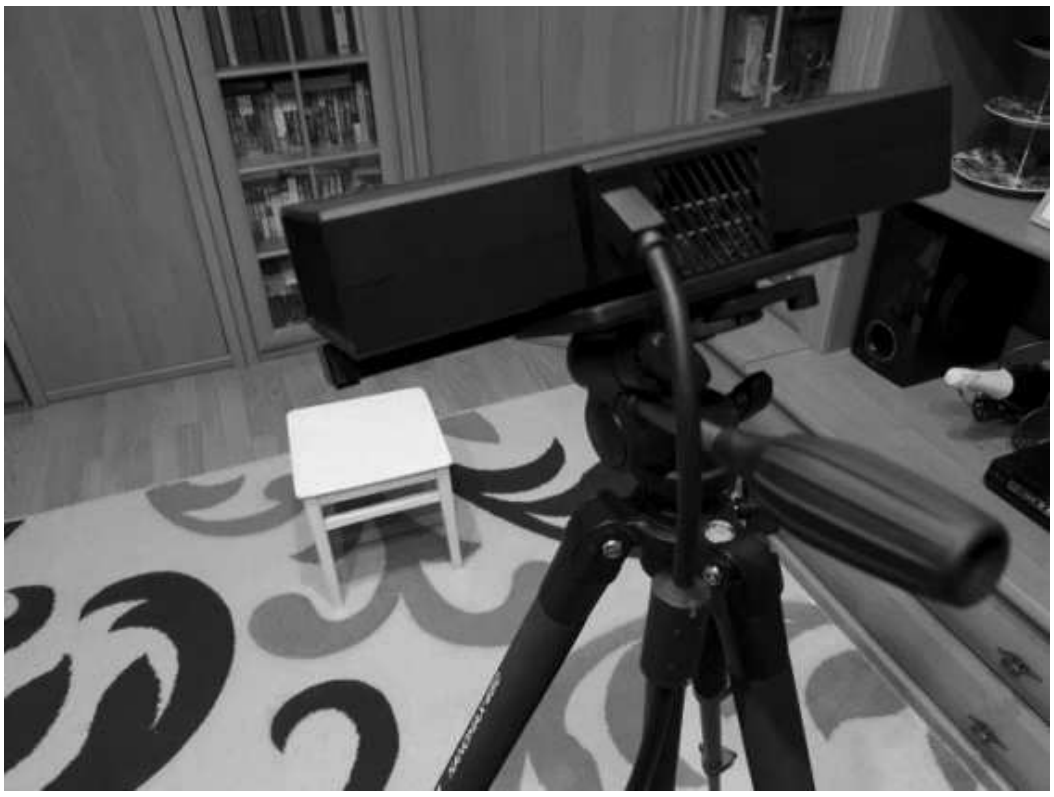

Fig. 6.42. Rigidly securing the Kinect device using a tripod

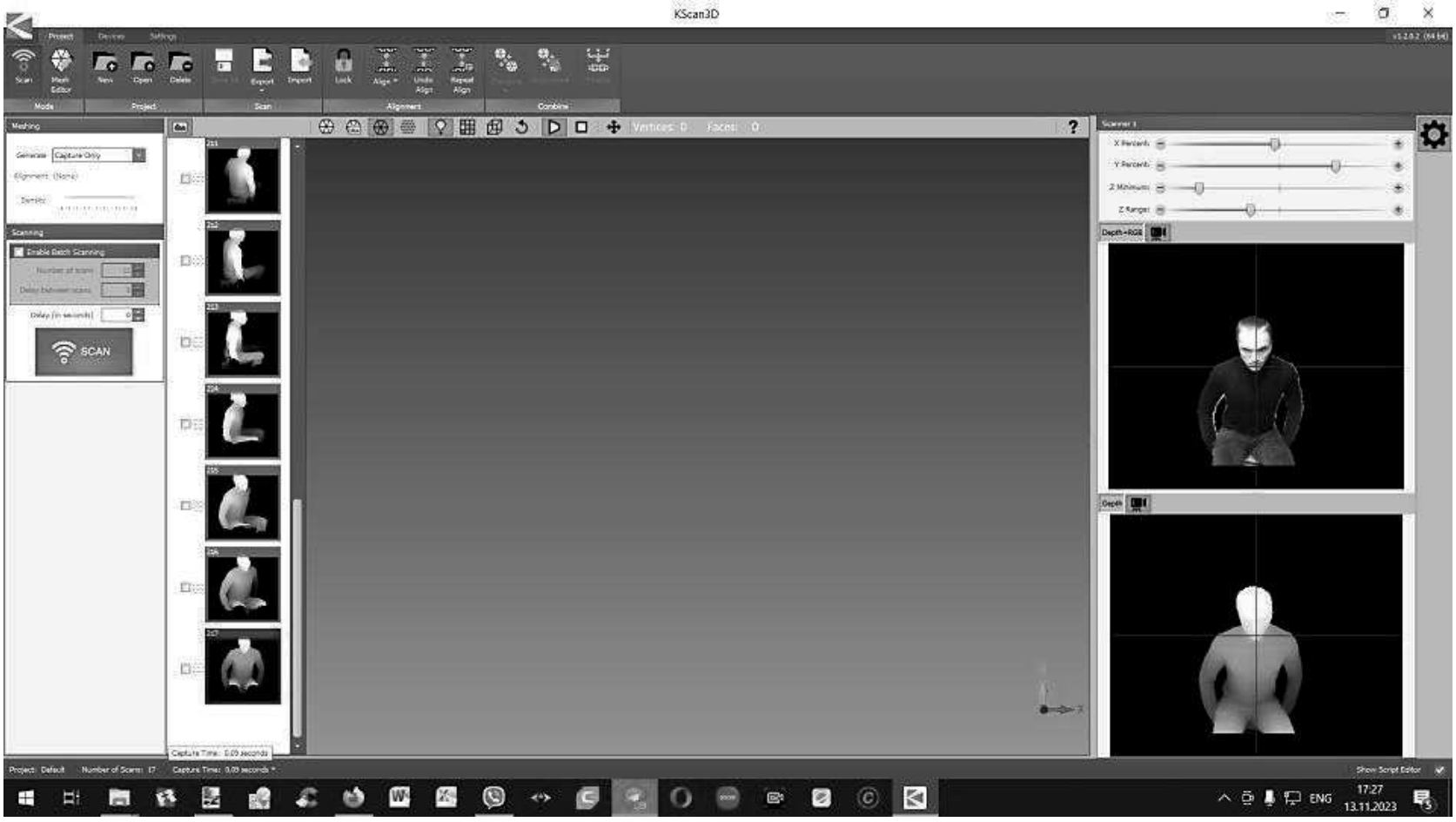

Fig. 6.43. Frame capture in the KScan3D program during scanning

After obtaining a satisfactory result (Fig. 6.44), the model can be edited based on the capabilities of the KScan3D program used. To do this, after scanning, you need to select all the images and go to the meshbuilder tab, tighten the holefilling slider a little, by 30 percent, so that when building the model, all open surfaces are closed (Fig. 6.44).

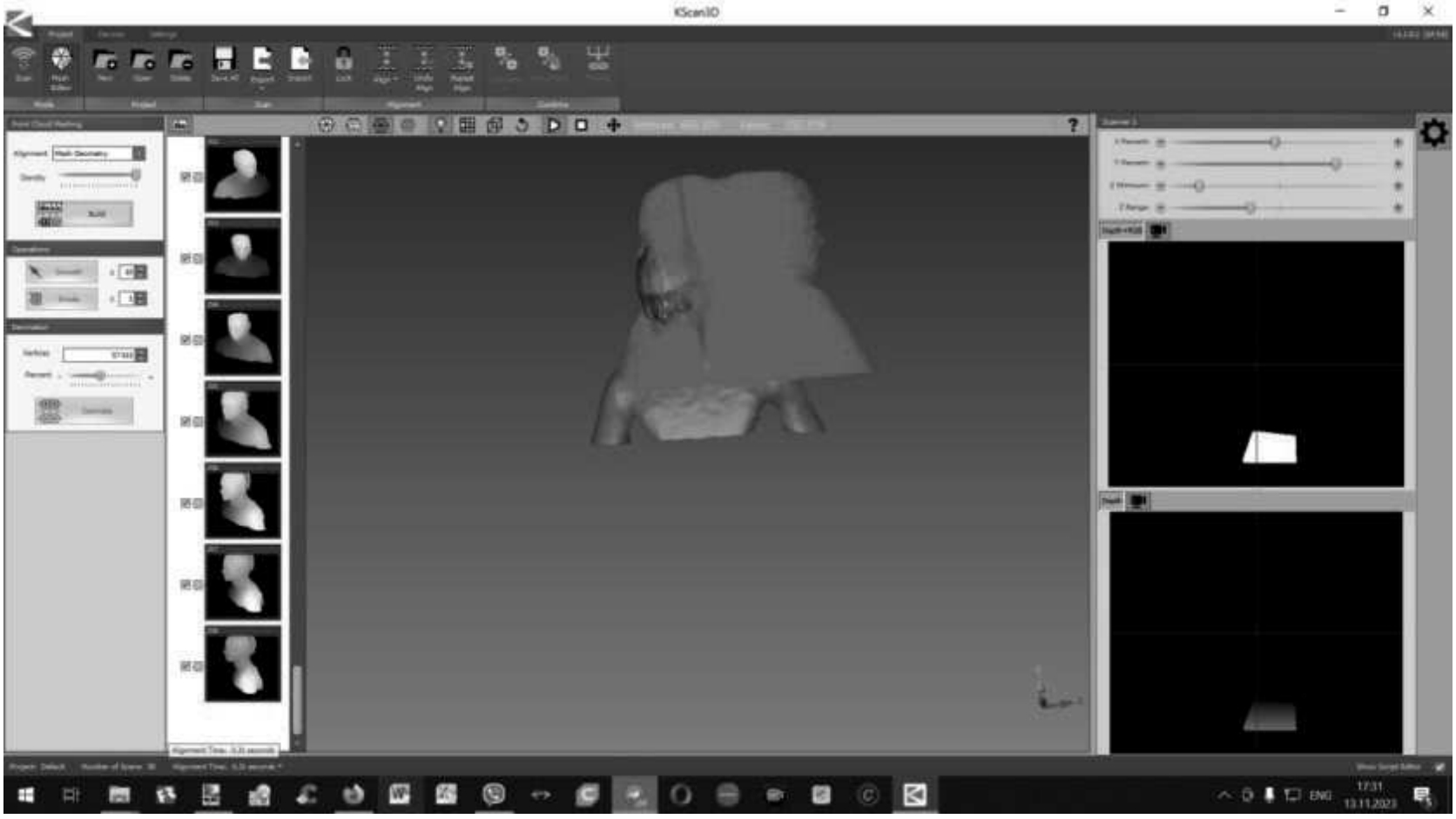

Fig. 6.44. Creating a 3D model from a point cloud in the KScan3D program

As a result of the KScan3D program, a three-dimensional model of the human torso was obtained, shown in Fig. 6.45.

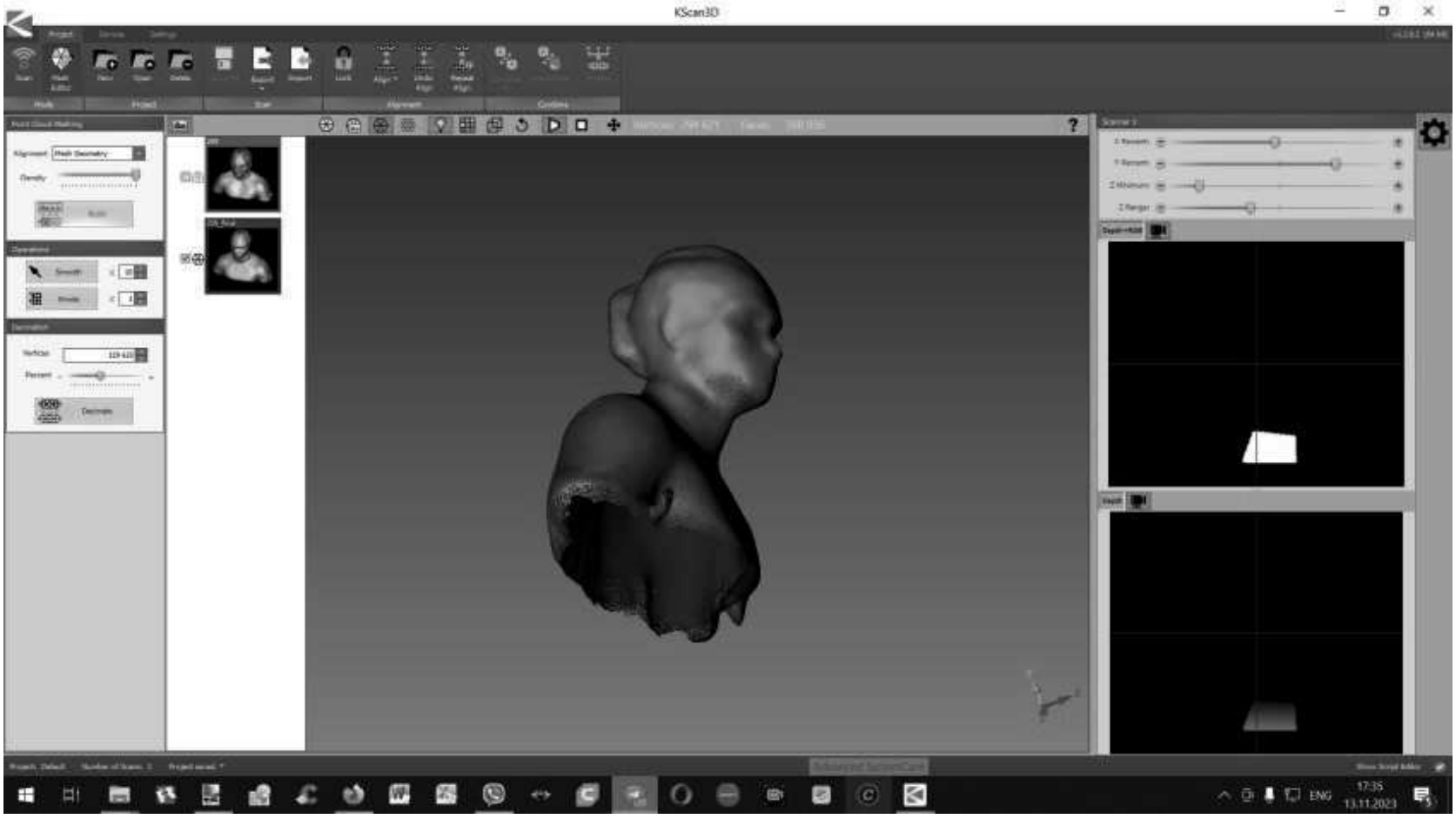

Fig. 6.45. Three-dimensional 3D model of a person obtained in the KScan3D program

The resulting 3D model of a person (Fig. 6.45) can be saved in a form accessible to other graphics programs (in this case, as a file with the stl extension). You can also edit it, remove unnecessary details, make the bottom of the figure, all this is done in the Blender program. In the 3D Viewer program for three-dimensional images in Windows 10, the resulting finish looks like in Fig. 6.46.

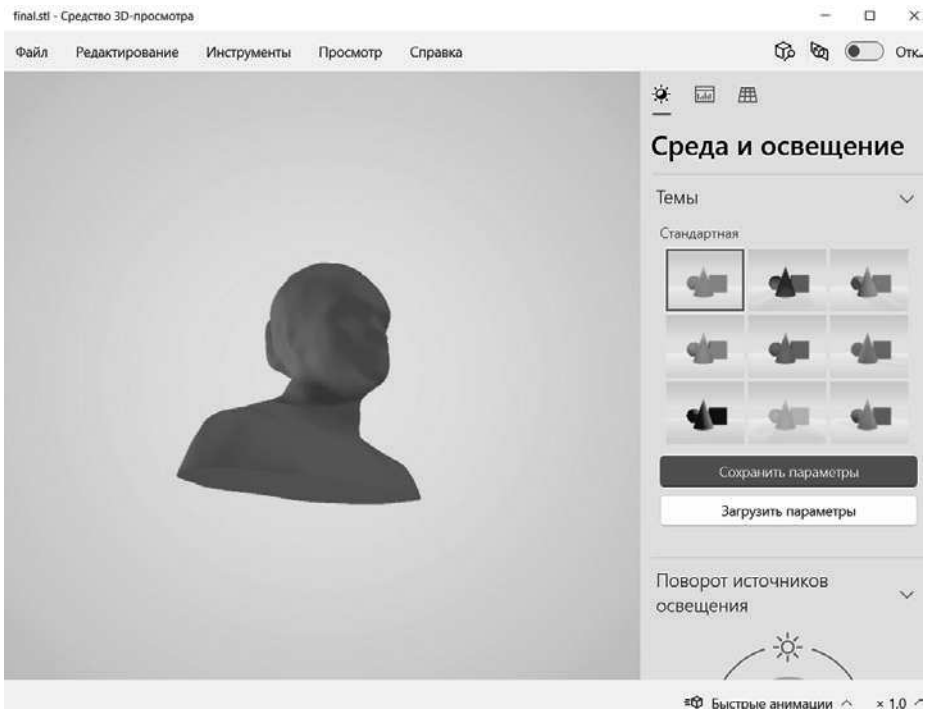

Fig. 6.46. View of the constructed 3D model of a person in the “3D Viewer” program

### *6.4.8. Printing a model on a 3D printer using FDM technology*

This subsection presents the results of printing the resulting model on a 3D printer using PLA plastic on an Anycubic Kossel printer. Subsection 6.3.4 described in detail typical operations for using the CURA slicer and the Anycubic Kossel 3D printer for 3D printing. In most cases at home, researchers turn to the CURA slicer for the following reasons:

- CURA allows you to import 3D models in various formats (for example, STL, OBJ) and prepare them for subsequent 3D printing. As part of this process, CURA provides a wide range of tools for adjusting model orientation, scaling, placing multiple models on the same print bed, and more.
- CURA supports many 3D printer models, making it a versatile tool for owners of various devices. This includes support not only for officially supported manufacturers, but also the ability to manually conFig.settings for unofficial printers.
- CURA provides extensive customization capabilities for print parameters such as print speed, extruder and bed temperature, advanced material settings, pattern fill, support, and many other parameters, allowing users to fine-tune the print process to their specific needs and requirements. For models that require support during printing, CURA can automatically generate optimal support structures, which helps avoid distortions and ensure successful printing of complex models.
- CURA allows you to visualize the process of printing a model, taking into account all specified parameters, which allows you to preliminary evaluate the result and identify possible problems before the physical printing process begins.
- After preparing the model, CURA allows you to export a print-ready file or even, with the appropriate drivers, directly control the printing process on a connected printer.

These capabilities make CURA a powerful tool for preparing 3D models and managing the 3D printing process, providing extensive customization options and excellent print quality. The slicer settings for working with the previously obtained 3D model are shown in Fig. 6.47.

To print the model, we used an Anycubic Kossel printer, a photo of which is shown in Fig. 6.48.

As a result of using scanning technology using the Kinect device, using the KScan3D data processing program, the CURA slicer and the Anycubic Kossel 3D printer, a model of the human torso was obtained, shown in Fig. 6.49.

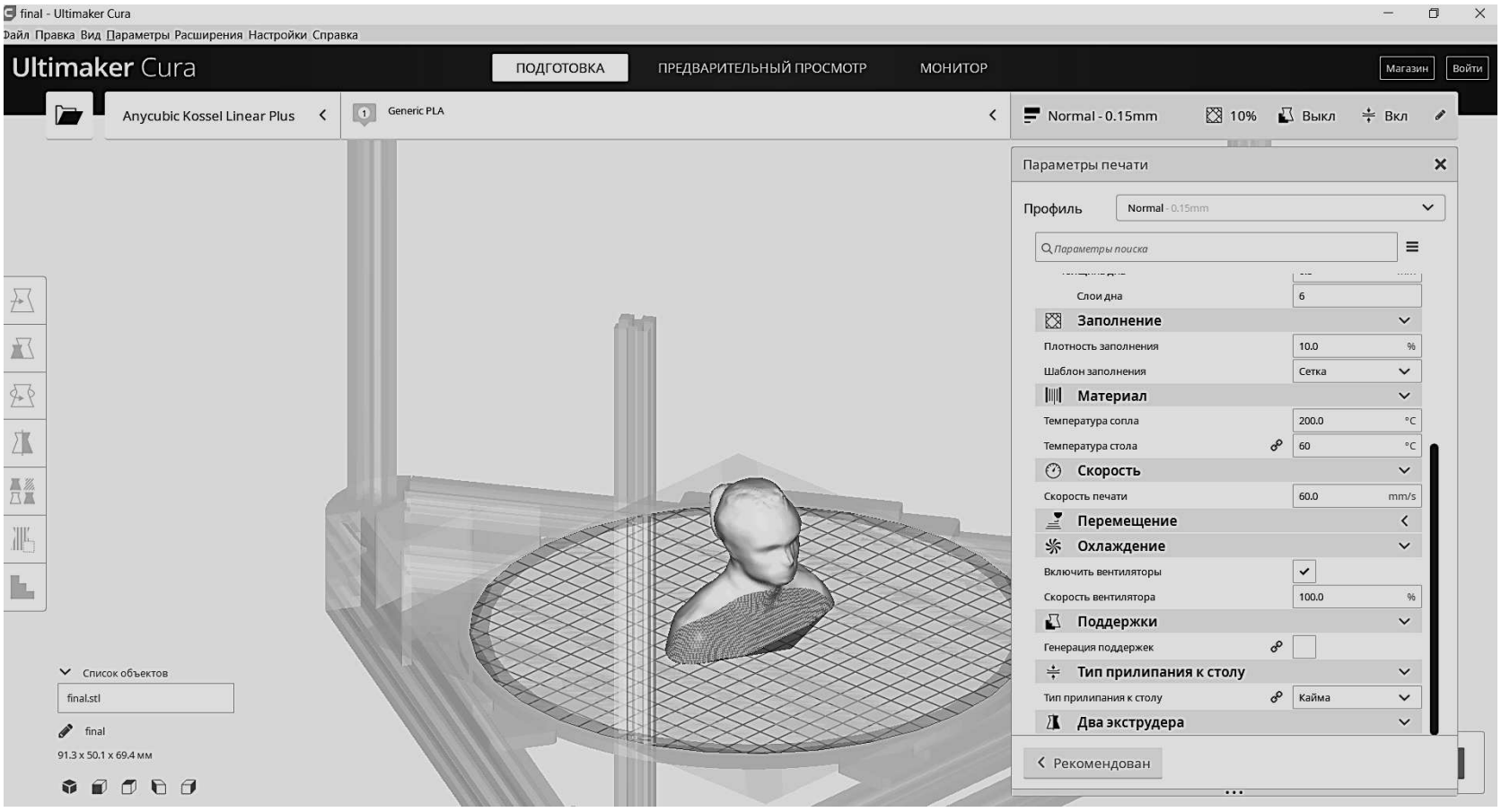

Fig. 6.47. Model processing in the CURA slicer

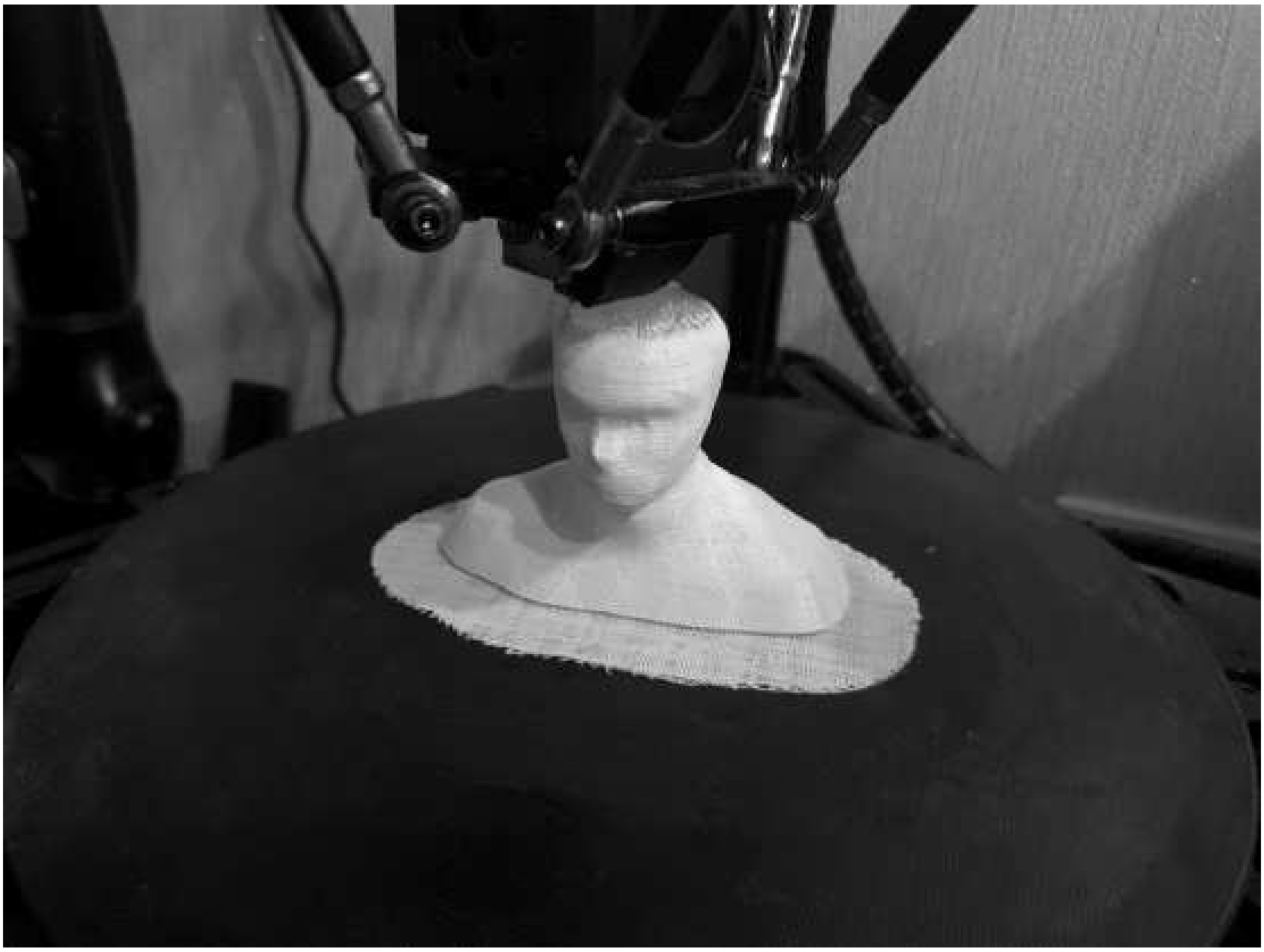

Fig. 6.48. Process of printing the developed model on the Anycubic Kossel 3D printer

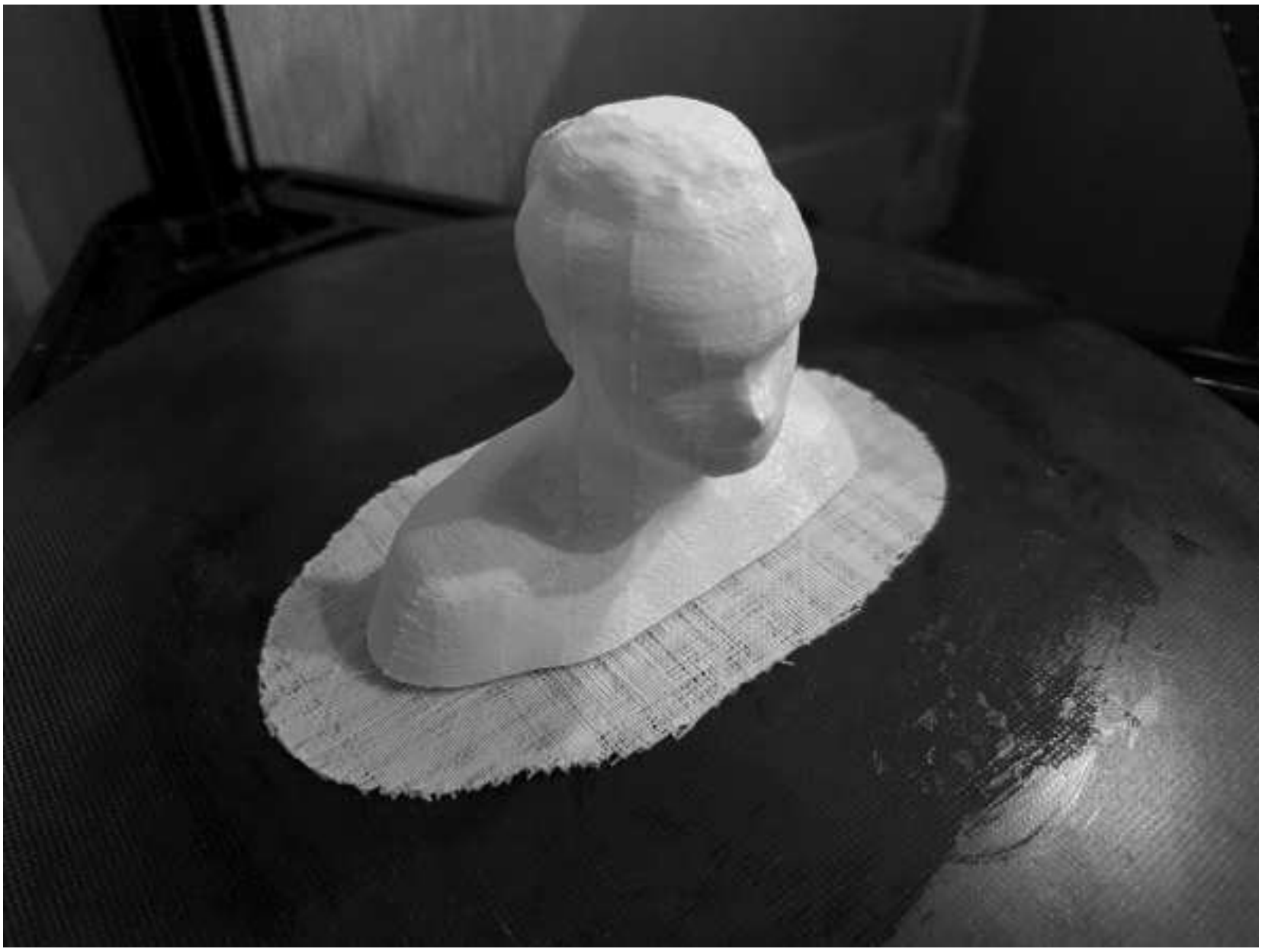

Fig. 6.49. Printed figurine of a person, the model of which was obtained using a Kinect device

## 6.5. CONCLUSION

The section shows that 3D printing has now become an integral part of industrial production, and provides an overview of existing technologies and methods of scanning objects to obtain a three-dimensional 3D model. The areas of use of scanning technology are given and its advantages are shown. The problems and difficulties that can limit and complicate the scanning process using traditional technologies are analyzed. It has been shown that existing scanning methods are poorly applicable at home due to high financial costs.

Affordable household photogrammetry technology has been developed to create three-dimensional virtual models of real physical objects for subsequent printing on a 3D printer; the results of processing a test Fig.using this method are shown. A 3D prototype created using this technology (saved in the form of mathematical dependencies as a file with the STL or OBJ extension) can also be used to improve the parameters of the initial object, to increase the efficiency of 3D printing by reducing the number of supports in slicer calculations, for volume editing of the model to improve consumer performance. A detailed analysis of the possibility of using photogrammetry in everyday conditions to obtain virtual three-dimensional models is carried out, and the results of a number of experiments are presented to obtain the optimal characteristics of the equipment and computer programs used. Based on the analysis, a detailed step-by-step technology for household photogrammetry for obtaining three-dimensional models is formulated, an example of creating a three-dimensional model of a test object using this technology is given, and the process of creating a plastic figurine using the resulting model using a 3D printer is presented.

The technology for obtaining virtual 3D models using the Kinect device is also presented, its properties and its structure are described. The characteristics of the Kinect device are given, its capabilities and areas of applicability are described. A comparative analysis of existing scanning technologies is carried out, their positive aspects and disadvantages are noted. It is shown that each of them has its own unique advantages and limitations, however, in most cases, the cost of such alternatives and existing analogue scanners is much higher than the use of a Kinect device.

The areas in which the use of scanning using Kinect is justified and is a convenient option are listed, and the limitations of this technology are listed. The software used to work with Kinect was analyzed and a choice was made about the preference of the KScan3D application. The requirements for computer characteristics when using the KScan3D program are given.

An accessible scanning technology using the Kinect device is proposed, recommendations are given for improving the scanning scheme, in particular, the features of installing light, the location of the sensor and the scanned object, the number of pictures taken, the exclusion of shiny objects, etc. The possibility of finalizing the model obtained after scanning is given — removing noise, irregularities, and an excessive number of polygons.

In order to check the scanning results using the Kinect device, an attempt was made to create a 3D model of the human torso and print it on an Anycubic Kossel 3D printer, taking into account all the previously made recommendations and suggestions.

# List of authors

**Kyrylo Malakhov**, Master's degree in Computer Science, Scientific associate, V. M. Glushkov Institute of Cybernetics, The National Academy of Sciences of Ukraine, Kyiv, Ukraine;
ORCID 0000–0003–3223–9844

**Vladislav Kaverinskiy**, PhD, Department of Wear-Resistant and Corrosion-Resistant Powder Construction Materials, Frantsevic Institute for Problems in Material Science of the National Academy of Sciences of Ukraine, Kyiv, Ukraine
ORCID 0000–0002–6940–579X

**Liliia Ivanova**, PhD (Candidate of Technical Sciences), director,
Separated structural subdivision «Odesa Technical Professional College of Odesa National University of Technology», Odessa, Ukraine
ORCID 0000–0003–1738–7697

**Oleksandr Romanyuk**, Dr. Sci. (Engin.), Professor, Vinnytsia National Technical university, Vinnytsia, Ukraine
ORCID 0000–0002–2245–3364

**Oksana Romaniuk,** Ph.D. (Engin.), Associate Professor, Vinnytsia National Technical university, Vinnytsia, Ukraine
ORCID 0000–0003–0235–8615

**Svitlana Voinova**, PhD, (Candidate of Technical Sciences), Associate Professor, Odesa National University of Technology, Odessa, Ukraine
ORCID 0000–0003–0203–0599

**Sergii Kotlyk**, PhD, (Candidate of Technical Sciences), Director of the Institute of computer engineering, automation, robotics and programming, Odesa National University of Technology, Odessa, Ukraine
ORCID 0000–0001–5365–1200

**Oksana Sokolova**, Master's degree in Computer Engineering, Senior Lecturer, Odesa National University of Technology, Odessa, Ukraine
ORCID 0000–0001–9224–6734